\documentclass[a4paper,11pt]{article}
\pdfoutput=1 

\usepackage{jheppub} 
\usepackage{amsmath,amssymb,amsthm,amscd,graphicx}
\input epsf.sty
\theoremstyle{definition}

\newcommand{\CJ}{{\cal J}}

\newcommand{\CO}{{\cal O}}
\newcommand{\CP}{{\cal P}}

\newcommand{\CW}{{\cal W}}

\def\IZ{{\mathbb Z}}

\def\IC{{\mathbb C}}

\newcommand{\tr}{{\rm Tr}}
\newcommand{\re}{{\rm e}}
\newcommand{\ri}{\mathsf{i}}
\newcommand{\rd}{{\rm d}}

\newcommand{\mA}{\mathsf{A}}

\newcommand{\mO}{{\mathsf{O}}}
\newcommand{\mP}{\mathsf{P}}

\newcommand{\mx}{\mathsf{x}}

\newcommand{\im}{\mathsf{i}}

\newcommand{\be}{\begin{equation}}
\newcommand{\ee}{\end{equation}}
\newcommand{\ba}{\begin{aligned}}
\newcommand{\ea}{\end{aligned}}
\newcommand{\ben}{\begin{eqnarray}\displaystyle}
\newcommand{\een}{\end{eqnarray}}

\newdimen\tableauside\tableauside=1.0ex
\newdimen\tableaurule\tableaurule=0.4pt
\newdimen\tableaustep
\def\phantomhrule#1{\hbox{\vbox to0pt{\hrule height\tableaurule width#1\vss}}}
\def\phantomvrule#1{\vbox{\hbox to0pt{\vrule width\tableaurule height#1\hss}}}
\def\sqr{\vbox{%
  \phantomhrule\tableaustep
  \hbox{\phantomvrule\tableaustep\kern\tableaustep\phantomvrule\tableaustep}%
  \hbox{\vbox{\phantomhrule\tableauside}\kern-\tableaurule}}}
\def\squares#1{\hbox{\count0=#1\noindent\loop\sqr
  \advance\count0 by-1 \ifnum\count0>0\repeat}}
\def\tableau#1{\vcenter{\offinterlineskip
  \tableaustep=\tableauside\advance\tableaustep by-\tableaurule
  \kern\normallineskip\hbox
    {\kern\normallineskip\vbox
      {\gettableau#1 0 }%
     \kern\normallineskip\kern\tableaurule}%
  \kern\normallineskip\kern\tableaurule}}
\def\gettableau#1{\ifnum#1=0\let\next=\null\else
\squares{#1}\let\next=\gettableau\fi\next}

\def\({\left(}
\def\){\right)}

\usepackage{amsmath,amssymb,graphicx,epstopdf,subfigure}
\numberwithin{equation}{section}

\usepackage{mathtools}

\usepackage{caption}
\usepackage{tabulary}
\usepackage{longtable}
\usepackage{breqn}
\usepackage{xcolor}
\usepackage{indentfirst}
\usepackage{setspace}

\newcommand{\bs}{\begin{split}}
\newcommand{\es}{\end{split}}
\newcommand{\cycle}[1]{\mathbf{r}^{#1}}
\newcommand{\cz}{\mathbb{C}^3/\mathbb{Z}_5}

\newcommand{\bdm}{\begin{dmath*}}
\newcommand{\edm}{\end{dmath*}}

\preprint{USTC-ICTS-16-12}

\title{Exact Quantization Conditions, Toric Calabi-Yau and Non-perturbative Topological String}
\author[\dagger]{Kaiwen Sun,}
\author[\S]{Xin Wang,}
\author[\S]{Min-xin Huang}
\affiliation[\dagger]{Department of Mathematics, University of Science and Technology of China, 96 Jinzhai Road,  Hefei, Anhui 230026, China}
\affiliation[\S]{Interdisciplinary Center for Theoretical Study, Department of Modern Physics, University of Science and Technology of China, 96 Jinzhai Road, Hefei, Anhui 230026, China}

\emailAdd{skw@mail.ustc.edu.cn}
\emailAdd{wxin@mail.ustc.edu.cn}
\emailAdd{minxin@ustc.edu.cn}
\abstract{We establish the precise relation between the Nekrasov-Shatashvili (NS) quantization scheme and Grassi-Hatsuda-Mari\~no conjecture for the mirror curve of arbitrary toric Calabi-Yau threefold. For a mirror curve of genus $g$, the NS quantization scheme leads to $g$ quantization conditions for the corresponding integrable system. The exact NS quantization conditions enjoy a self S-duality with respect to Planck constant $\hbar$ and can be derived from the Lockhart-Vafa partition function of non-perturbative topological string. Based on a recent observation on the correspondence between spectral theory and topological string, another quantization scheme was proposed by Grassi-Hatsuda-Mari\~no, in which there is a single quantization condition and the spectra are encoded in the vanishing of a quantum Riemann theta function. We demonstrate that there actually exist at least $g$ nonequivalent quantum Riemann theta functions and the intersections of their theta divisors coincide with the spectra determined by the exact NS quantization conditions. This highly nontrivial coincidence between the two quantization schemes requires infinite constraints among the refined Gopakumar-Vafa invariants. The equivalence for mirror curves of genus one has been verified for some local del Pezzo surfaces. In this paper, we generalize the correspondence to higher genus, and analyze in detail the resolved $\mathbb{C}^3/\mathbb{Z}_5$ orbifold and several $SU(N)$ geometries. We also give a proof for some models at $\hbar=2\pi/k$.}
\begin{document}
\maketitle
\bibliographystyle{unsrt}
\tableofcontents
\section{Introduction}\label{sec:intro}
The non-perturbative completion of topological string theory has drawn much attention in recent years. Due to its various dual facets, topological strings may have many different non-perturbative definitions. For example, based on the localization calculation in superconformal field theories, Lockhart and Vafa proposed a non-perturbative partition function of refined topological string which makes the $SL(3,\mathbb{Z})$ modular invariance manifest \cite{Lockhart:2012vp}. The closed B-model amplitudes can be described as matrix model partition functions \cite{Dijkgraaf:2002fc}, which makes the non-perturbative techniques in matrix model available to study the non-perturbative features of topological string \cite{Marino:2008ya}. The ABJM theory on three-sphere is known dual to the topological string theory on local $\mathbb{P}^1\times \mathbb{P}^1$ Calabi-Yau threefold \cite{Marino:2009jd}. This duality with the knowledge of exact ABJM partition function from localization \cite{Kapustin:2009kz} provides significant insights on the non-perturbative topological string \cite{Hatsuda:2013oxa}. Other relations with Chern-Simons theory \cite{Gopakumar:1998ki}\cite{Aganagic:2002qg}\cite{Krefl:2015vna}, supersymmetric gauge theories \cite{Katz:1996fh}\cite{Iqbal:2003zz}, string/M theory \cite{Gopakumar:1998jq}\cite{Katz:1999xq}\cite{Dijkgraaf:2006um}, quantum integrable systems \cite{Aganagic:2003qj}\cite{Aganagic:2011mi}, conformal blocks \cite{Cheng:2010yw}, OSV formula for black holes \cite{Ooguri:2004zv}\cite{Aganagic:2004js}, and techniques like resurgence \cite{Santamaria:2013rua}\cite{Couso-Santamaria:2014iia}\cite{Hatsuda:2015owa} and Mellin-Branes representation \cite{Krefl:2015qva} all shed some light on this issue. Each of these descriptions captures an aspect of non-perturbative topological string.\\
\indent The quantization of the mirror curve of local Calabi-Yau threefold has turned out to be an excellent testing ground to study the non-perturbative effects of topological string. Due to the various correspondences, the mirror curve is equivalent to the Seiberg-Witten curve in $N=2$ supersymmetric gauge theories \cite{Katz:1996fh}, the spectral curve in matrix model \cite{Dijkgraaf:2002fc} and in integrable system \cite{Aganagic:2003qj}. With appropriate parameterization, the mirror curve itself can also be promoted to some quantum mechanics satisfying the traditional Heisenberg relation \cite{Aganagic:2003qj}\cite{Aganagic:2011mi}. Because the dual theories are more familiar and often equipped with known techniques, and different perspectives usually open different avenues to quantize a curve, they may well result in non-trivial observations for the Calabi-Yau and topological string. Besides, the quantization of mirror curves is closely related to the quantization of A-polynomials of knots \cite{Gukov:2011qp}\cite{Aganagic:2012jb} and the mathematical theory of $\mathcal{D}$-modules \cite{Dijkgraaf:2008fh}. See also a recent paper \cite{Hatsuda:2016mdw} for an elegant relation between the Hofstadter's Butterfly and the quantum mirror curve of local $\mathbb{P}^1\times \mathbb{P}^1$ Calabi-Yau.\\
\indent In general, there exist two quantization schemes for the mirror curve of local Calabi-Yau. One originates from the Bethe/gauge correspondence proposed by Nekrasov-Shatashvili, which says the NS limit of Nekrasov partition function of $N=2$ gauge theories is the Yang-Yang function of some quantum integrable systems while the supersymmetric vacua become the eigenstates and the supersymmetric vacua equations become the thermodynamic Bethe ansatz \cite{Nekrasov:2009rc}. Via geometric engineering, this correspondence can be rephrased as a direct relation between the phase volumes of mirror curve and the Nekrasov-Shatashvili free energy of topological string. Now the Bethe ansatz is just the traditional Bohr-Sommerfeld or Einstein-Brillouin-Keller quantization conditions for the phase volumes of mirror-curve quantum mechanics \cite{Aganagic:2011mi}. We call this approach as Nekrasov-Shatashvili quantization scheme.\\
\indent However, the original Nekrasov-Shatashvili proposal could not be the whole story. In fact, it is easy to show the phase volumes derived from NS free energy contain poles at rational Planck constant $\hbar=2\pi p/q$. This indicates the original NS quantization conditions must receive non-perturbative contributions to be well-defined. The first non-perturbative completion of NS quantization was proposed in \cite{Wang:2015wdy} for toric Calabi-Yau with genus-one mirror curves, and soon be generalized to $SU(N)$ geometries with Chern-Simons number $m=0$ which correspond to Toda systems \cite{Hatsuda:2015qzx}, and to arbitrary toric cases in \cite{Franco:2015rnr} using the elegant Goncharov-Kenyon construction which associates an integrable
system to an arbitrary toric CY \cite{Goncharov:2011hp}. It turns out the non-perturbative terms in phase volumes are quite simple, and resemble the BPS part of NS free energy merely with a transformation on K\"{a}hler parameters and Planck constant. In fact, the total phase volumes including both the perturbative and non-perturbative contributions enjoys a self-S duality with respect to Planck constant \cite{Hatsuda:2015fxa}. We call the quantization of total phase volumes as exact Nekrasov-Shatashvili quantization conditions. These exact NS quantization conditions are actually directly ascribed to the Lockhart-Vafa partition function of non-perturbative strings, as suggested in \cite{Hatsuda:2015fxa} and will be fully proved in this paper. In \cite{Kashani-Poor:2016edc}, it is also argued the exact NS quantization condition arises upon imposing single-valued conditions on open topological string partition in the NS limit, combined with the requirement of smoothness in Planck constant. See also \cite{Krefl:2016svj} for the discussion on the Stokes phenomena of the quantum periods from NS quantization conditions.

On the other hand, Grassi-Hatsuda-Mari\~no proposed an entirely different approach from Nekrasov-Shatashvili quantization in \cite{Grassi:2014zfa}. Their approach takes deep root in the study on the non-perturbative effects in ABJM theories. ABJM theory is a three dimensional $N=6$ superconformal Chern-Simons matter theory describing the M2 branes \cite{Aharony:2008ug}. In \cite{Kapustin:2009kz}, the exact ABJM partition function on three-sphere was obtained by localization method. As we have mentioned, ABJM theory on three sphere is dual to topological string on local Hirzebruch surface $\mathbb{F}_0=\mathbb{P}^1\times \mathbb{P}^1$ \cite{Marino:2009jd}. Therefore the known form of exact partition function of ABJM is extremely helpful to understand the non-perturbative topological string on local $\mathbb{P}^1\times \mathbb{P}^1$. After a series of quest on the non-perturbative effects in ABJM theory \cite{Drukker:2010nc}\cite{Drukker:2011zy}\cite{Marino:2011eh}\cite{Calvo:2012du}\cite{Hatsuda:2012hm}\cite{Hatsuda:2012dt}\cite{Hatsuda:2013gj}, a general form of non-perturbative topological string free energy for an arbitrary local CY was proposed in \cite{Hatsuda:2013oxa} as an analogy from local $\mathbb{P}^1\times \mathbb{P}^1$ model, and also shown to be roughly consistent with the Lockhart-Vafa partition function. In the spirit of this non-perturbative completion of topological string, a non-perturbative quantization condition for spectral curve was proposed in \cite{Kallen:2013qla}. However, it was later found in \cite{Huang:2014eha} and \cite{Wang:2014ega} that the non-perturbative quantization conditions in \cite{Kallen:2013qla} are not accurate and submit to some higher order corrections. Finally, in \cite{Grassi:2014zfa} Grassi-Hatsuda-Mari\~no proposed the complete form of what we now call the GHM conjecture. In that paper, they established an elegant relation between the spectral determinant (Fredholm determinant) $\Xi$ of the quantum mirror curve and the modified grand potential $J$ defined from the topological string partition function and NS free energy,
\be
\Xi(\mu,\xi,\hbar)=\sum_{n=-\infty}^{+\infty}\exp(J(\mu+2\pi\ri n,\xi,\hbar)),
\ee
where $\mu$ is the true moduli and $\xi$ are the mass parameters of a local del Pezzo. In particular, the quantization condition is given by the vanishing of the quantum (generalized) Riemann theta function $\Theta$ defined by
\be
\Xi(\mu,\xi,\hbar)=\exp(J(\mu,\xi,\hbar))\Theta(\mu,\xi,\hbar).
\ee
With this relation, they successfully reproduced all the corrections obtained in \cite{Huang:2014eha}. The different points in the moduli space lead to different expansions for the spectral determinant, which all contain rich information of the topological string and pass nontrivial tests, for a review see \cite{Marino:2015nla}. The theory proposed in \cite{Grassi:2014zfa} was only for the genus-one mirror curve and only checked for local $\mathbb{P}^2$, $\mathbb{F}_0$ and $\mathbb{F}_1$, it was soon be generalized to higher genus in \cite{Codesido:2015dia} and checked for more local del Pezzo surfaces in \cite{Gu:2015pda}. The spectral determinant (spectral zeta function) was also used to study the non-perturbative effects in ABJM Fermi-gas \cite{Hatsuda:2015oaa}. For the discussion on the trace class property of some relevant operators, see \cite{Kashaev:2015kha}. For the relation of GHM quantization scheme with matrix models, see \cite{Marino:2015ixa}\cite{Kashaev:2015wia}. Due to complicated expression of the modified grand potential, so far it still out of reach to grasp a rigorous proof on the GHM conjecture of any specific Calabi-Yau. Nevertheless, using the $\tau$ function of Painlev\'{e} $\mathrm{III}_3$ equation, it was proved in \cite{Bonelli:2016idi} that under some limit, the GHM conjecture holds for local $\mathbb{F}_0$. See also a recent paper \cite{Grassi:2016vkw} on the symplectic properties of the spectral determinants around the maximal supersymmetric point $\hbar=2\pi$.\\
\indent Albeit both the two quantization schemes are well-defined and have passed numerous nontrivial tests, their relation is still waiting to be unraveled. The main purpose of this paper is to establish the precise relation between them for arbitrary toric Calabi-Yau threefold. Our starting point is the observation that the K\"{a}hler parameters in the Grassi-Hatsuda-Mari\~no conjecture are not the same with the K\"{a}hler parameters in NS quantization conditions, but with a shift of a constant integral vector, which we call the $\mathbf{r}$ field. In some sense, $\mathbf{r}$ field reflects the sign-changing of complex moduli in mirror curve like the $\mathbf{B}$ field, but with subtler meaning. The $\mathbf{r}$ field of a toric Calabi-Yau is not arbitrarily chosen but indeed with quite some freedom. The primary condition for $\mathbf{r}$ field is that it must be a $\mathbf{B}$ field, which means they share the same parity. Besides this, we define the $\mathbf{r}$ fields as equivalent if they result in the same quantum Riemann theta function or generalized spectral determinant, and non-equivalent if they do not. Surprisingly, we find there always are only finite non-equivalent $\mathbf{r}$ fields for a toric Calabi-Yau, and the number is no less than the genus of the mirror curve. In the single quantization condition of the original GHM conjecture, one choose one modulus of the mirror curve to be quantized and leave others arbitrarily chosen. The quantization condition draw a divisor in the moduli space, which is just the theta divisor of the quantum Riemann theta function. This viewpoint actually regards the mirror curve as a normal quantum mechanics but not a quantum integrable system, therefore doubtlessly losses much information of the integrability. We will show in this paper that in the GHM quantization scheme, we can as well exhaust the full quantum integrability of mirror curve and determine the discrete spectra merely from the quantum Riemann theta function they have defined. The non-equivalent $\mathbf{r}$ fields will be the crucial ingredient to achieve this goal.\\
\indent The main result of this paper is as follows: \emph{For the mirror curve $\Sigma$ of an arbitrary toric Calabi-Yau threefold $X$ with K\"{a}hler moduli $\mathbf{t}$, suppose the non-perturbative phase volumes of quantum mirror curve in Nekrasov-Shatashvili quantization are denoted as $\mathrm{Vol}_i(\mathbf{t},\hbar)$, $i=1,\dots,g_{\Sigma}$, and the quantum Riemann theta function defined by Grassi-Hatsuda-Mari\~no is denoted as $\Theta(\mathbf{t},\hbar)$, then there exist a set of constant integral vectors $\mathbf{r}^a$, $a=1,\cdots,w_{\Sigma}$, where $w_{\Sigma}\geq g_{\Sigma}$, such that the intersections of the theta divisors of all $w_\Sigma$ quantum Riemann theta functions $\Theta(\mathbf{t}+\ri\pi\mathbf{r}^a,\hbar)$ coincide with the spectra determined by the exact Nekrasov-Shatashvili quantization conditions.}\footnote{Strictly speaking, to make the coincidence precise, one need to take account of the imaginary periodicity of K\"{a}hler moduli $\mathbf{t}\rightarrow\mathbf{t}+2\pi\ri\mathbf{n}\cdot\mathbf{C}$, which is manifest in the GHM approach but obscure in the NS approach. We will address this issue in Section \ref{sec:rem}.}
\be\label{equiv1}
\bigg\{\Theta(\mathbf{t}+\ri\pi\mathbf{r}^a,\hbar)=0,\ a=1,\cdots,w_{\Sigma}.\bigg\} \Leftrightarrow\left\{\mathrm{Vol}_i(\mathbf{t},\hbar)=2\pi\hbar \left(n_i+\frac{1}{2}\right),\ i=1,\cdots,g_{\Sigma}.\right\}
\ee
\emph{In addition, all the vector $\mathbf{r}^a$ are the representatives of the $\mathbf{B}$ field of $X$, which means for all triples of degree ${\bf d}$, spin $j_L$ and $j_R$ such that the refined BPS invariants $N^{{\bf d}}_{j_L, j_R}(X) $ is non-vanishing, they must satisfy}
\be
\label{ra-prop}
(-1)^{2j_L + 2 j_R-1}= (-1)^{{\bf r}^a \cdot {\bf d}},\quad a=1,\cdots,w_{\Sigma}.
\ee
In fact, we find a set of novel identities which guarantee the above equivalence,
\be\label{check}
 \sum_{\mathbf{n}\in\mathbb{Z}^g}\exp\left(\sum_{i=1}^{g}n_i\pi\ri+F_{\text{unref}}\left(\mathbf{t}+ \ri\hbar \mathbf{n}\cdot C+\frac{1}{2}\ri\hbar\mathbf{r}^a,\hbar\right)
 -\ri n_iC_{ij}\frac{\partial }{\partial t_j}F_{\text{NS}}\left(\mathbf{t},\hbar\right)\right)\equiv0,
 \ee
in which $F_{\text{unref}}$ is the traditional topological string partition function, $F_{\text{NS}}$ is the Nekrasov-Shatashvili free energy, $C$ is the charge matrix of toric Calabi-Yau and $a=1,\cdots,w_{\Sigma}$. These identities impose infinite constraints among the refined BPS invariants.\\
\indent Besides the main result, we also close some gaps in the previous study on quantum curves. For example, the existence of $\mathbf{B}$ field was only established for the toric Calabi-Yau with genus-one mirror curve \cite{Hatsuda:2013oxa}. In this paper, we will prove its existence for arbitrary toric Calabi-Yau and propose an effective method to determine its value merely based on the toric data. By modifying the existing techniques, we also provide two derivations of the exact Nekrasov-Shatashvili quantization conditions from Lockhart-Vafa partition function of non-perturbative topological string.\\
\indent This paper is organized as follows. In section \ref{sec:review}, we review some basics on local Calabi-Yau and refined topological string theory. In section \ref{sec:NS}, we analyze the exact NS quantization conditions and provide two derivations from Lockhart-Vafa partition function and from Mellin-Branes representation. We also give an effective method to determine the $\bf{B}$ field for an arbitrary toric Calabi-Yau. In section \ref{sec:ghm}, we introduce the concept of $\mathbf{r}$ field, extend the GHM conjecture and construct the nonequivalent quantum Riemann theta functions. In section \ref{sec:equivalence}, we propose the precise relation between the two quantization schemes and provide a general proof for some geometries at $\hbar=2\pi/k$. In section \ref{sec:examples}, we test our conjecture for the resolved $\mathbb{C}^3/\mathbb{Z}_5$ orbifold and several geometries engineering $SU(N)$ gauge theories and obtain remarkable coincidences. In section \ref{sec:outlook}, we conclude with some interesting questions for the future study. The Appendix provides some properties of Riemann theta function, the refined topological vertex of $SU(N)$ geometries, some relevant mirror maps and the table of the refined BPS invariants of the Calabi-Yau threefolds we consider in the paper.
\section{Basics of Toric Calabi-Yau and Topological String}
\label{sec:review}
Toric (non-compact) Calabi-Yau threefold is the local model of compact Calabi-Yau. Its construction and mirror symmetry was first studied in \cite{Chiang:1999tz}. In this section, we review some well-known facts about toric Calabi-Yau and local mirror symmetry and set up the conventions. Similar reviews can be found in \cite{Codesido:2015dia}\cite{Marino:2015ixa}\cite{Klemm:2015iya}.
\subsection{Toric Calabi-Yau and Local Mirror Symmetry}
A toric Calabi-Yau threefold is a toric variety given by the quotient,
\be
M=(\mathbb{C}^{k+3}\text{\textbackslash} \mathcal{SR})/G,
\ee
where $G=(\mathbb{C}^*)^k$ and $\mathcal{SR}$ is the Stanley-Reisner ideal of $G$. The quotient is specified by a matrix of charges $Q_i^\alpha$, $i=0, \cdots, k+2$, $\alpha=1, \cdots, k$. The group $G$ acts on the homogeneous coordinates $x_i$ as
\be
x_i \rightarrow \lambda_\alpha^{Q^{\alpha}_i} x_i,\quad\quad i=0, \cdots, k+2,
\ee
where $\alpha=1,\ldots,k$,  $\lambda_\alpha \in \mathbb{C}^*$ and $Q^{\alpha}_i \in \mathbb{Z}$.

The toric variety $M$ is the vacuum configuration of a two-dimensional abelian $(2,2)$ gauged linear sigma model. Then $G$ is gauge group $U(1)^k$. The vacuum configuration is constraint by
\be
\sum_{i=1}^{k+3}Q_{i}^\alpha |x_i|^2=r^\alpha,\ \ \ \alpha=1,\ldots,k,
\ee
where $r^{\alpha}$ is the K\"{a}hler class. In general, the K\"{a}hler class is complexified by adding a theta angel $t^{\alpha}=r^{\alpha}+i\theta^\alpha$. In order to avoid R symmetry anomaly, one has to put condition
\be
\sum_{i=1}^{k+3}{Q^{\alpha}_i}=0,\ \ \ \alpha=1,\ldots,k.
\ee
This is the Calabi-Yau condition in the geometry side.

The mirror to toric Calabi-Yau was constructed in \cite{Chiang:1999tz}. We define the Batyrev coordinates
\begin{equation}\label{w_1}
z_\alpha=\prod_{i=1}^{k+3} x_i^{Q_i^{\alpha}},\ \ \ \alpha=1,\ldots,k,
\end{equation}
and
\begin{equation}\label{w_2}
H=\sum_{i=1}^{k+3}x_i.
\end{equation}
The homogeneity allows us to set one of $x_i$ to be one. Eliminate all the $x_i$ in (\ref{w_2}) by using (\ref{w_1}), and choose other two as $e^x$ and $e^{p}$, then the mirror geometry is described by
\begin{equation}\label{w_3}
uv=H(e^x,e^p;z_{\alpha}),\ \ \ \alpha=1,\ldots,k,
\end{equation}
where $x,p,u,v\in\mathbb{C}$. Now we see that all the information of mirror geometry is encoded in the function $H$. The equation
\be
H(e^x,e^p;z_{\alpha})=0
\ee
defines a Riemann surface $\Sigma$, which is called the mirror curve to a toric Calabi-Yau.

The form of mirror curve can be written down specifically with the vectors in the toric diagram. Given the matrix of charges $Q^\alpha_i$, we introduce the vectors,
\be
 \nu^{(i)}=\left(1, \nu^{(i)}_1, \nu^{(i)}_2\right),\qquad i=0, \cdots, k+2,
\ee
satisfying the relations
\be
\sum_{i=0}^{k+2} Q^\alpha_i \nu^{(i)}=0.
\ee
In terms of these vectors, the mirror curve can be written as
\be
\label{coxp}
H(\re^x, \re^p)=\sum_{i=0}^{k+2}x_i  \exp\left( \nu^{(i)}_1 x+  \nu^{(i)}_2 p\right).
 \ee
The holomorphic 3-form for mirror Calabi-Yau is
\be
\Omega=\frac{du}{u}\wedge dx \wedge dp
\ee
At classical level, the periods of the holomorphic 3-form are
\be
t_i=\oint_{A_i}\Omega,\ \ \ \ \ \mathcal{F}_i=\oint_{B_i}\Omega.
\ee
If we integrate out the non-compact directions, the holomorphic 3-forms become meromorphic 1-form on the mirror curve \cite{Katz:1996fh}\cite{Chiang:1999tz}:
\be
\label{diff-la}
\lambda=p\, \rd x.
\ee
The mirror maps and the genus zero free energy $F_0 ({\bf t})$ are determined by making an
appropriate choice of cycles on the curve, $\alpha_i$, $\beta_i$, $i=1, \cdots,s$, then we have
\be
t_i = \oint_{\alpha_i} \lambda, \qquad \qquad  {\partial F_0 \over \partial t_i} = \oint_{\beta_i} \lambda, \qquad i=1, \cdots, s.
\ee
In general, $s\ge g_\Sigma$, where $g_\Sigma$ is the genus of the mirror curve. The $s$ complex moduli can be divided into two classes, which are $g_\Sigma$ true moduli, $\kappa_i$, $i=1, \cdots, g_\Sigma$, and
$r_\Sigma$ mass parameters, $\xi_j$, $j=1, \cdots, r_\Sigma$.  The true moduli can be also be expressed with the chemical potentials $\mu_i$,
\be
\kappa_i =\re^{\mu_i}, \qquad i=1, \cdots, g_{\Sigma}.
\ee
Among the K\"{a}hler parameters, there are $g_\Sigma$ of them which correspond to the true moduli, and their mirror map at large $\mu_i$ is of the form
\be
\label{tmu}
t_i \approx \sum_{j=1}^{g_\Sigma} C_{ji} \mu_j + \sum_{j=1}^{r_\Sigma} \alpha_{ij} t_{\xi_j}, \qquad i=1, \cdots, g_\Sigma,
\ee
where $t_{\xi_j}$ is the flat coordinate associated to the mass parameter $\xi_j$ by an algebraic mirror map.

The $g_\Sigma\times s$ matrix $C_{ij}$ can be read off from the toric data of $X$. It was shown in \cite{Aganagic:2011mi} that the classical mirror maps can be promoted to quantum mirror maps. Note only the mirror maps for the true moduli have quantum deformation, while the mirror maps for mass parameters remain the same.
\subsection{Topological String}
Here we briefly review some well-known definitions in (refined) topological string theory. We follow the notion in \cite{Codesido:2015dia}. The Gromov-Witten invariants of Calabi-Yau are encoded in the genus $g$ free energies $F_g({\bf t})$, $g \ge 0$. At genus zero,
\be
\label{gzp}
F_0({\bf t})={1\over 6} a_{ijk} t_i t_j t_k  +P_2(t)+ \sum_{{\bf d}} N_0^{ {\bf d}} \re^{-{\bf d} \cdot {\bf t}},
\ee
where $a_{ijk}$ denotes the classical intersection and $P_2(t)$ is ambiguous.\footnote{For compact Calabi-Yau, this term has the structure of $\frac{-\chi}{2}\zeta(3)-4\pi^2 b_i t_t+A_{ij}t_it_j$. But in most circumstances, this term can be omitted. We will show in Section \ref{sec:LV} that if we include this term with a special choice for local Calabi-Yau, then the formulas for GHM conjecture and the identities (\ref{check}) can be written in a compact and elegant way.} At genus one, one has
\be
\label{gop}
F_1({\bf t})=b_i t_i + \sum_{{\bf d}} N_1^{ {\bf d}} \re^{-{\bf d} \cdot {\bf t}},
\ee
At higher genus, one has
\be
\label{genus-g}
F_g({\bf t})= C_g+\sum_{{\bf d}} N_g^{ {\bf d}} \re^{-{\bf d} \cdot {\bf t}}, \qquad g\ge 2,
\ee
where $C_g$ is the constant map contribution to the free energy.
The total free energy of the topological string is formally defined as the sum,
\be
\label{tfe}
F^{\rm WS}\left({\bf t}, g_s\right)= \sum_{g\ge 0} g_s^{2g-2} F_g({\bf t})=F^{({\rm p})}({\bf t}, g_s)+ \sum_{g\ge 0} \sum_{\bf d} N_g^{ {\bf d}} \re^{-{\bf d} \cdot {\bf t}} g_s^{2g-2},
\ee
where
\be
F^{({\rm p})}({\bf t}, g_s)= {1\over 6 g_s^2} a_{ijk} t_i t_j t_k +b_i t_i + \sum_{g \ge 2}  C_g g_s^{2g-2}.
\ee
The BPS part of partition function (\ref{tfe})
can be resummed with a new
set of enumerative invariants, called Gopakumar-Vafa (GV)
invariants $n^{\bf d}_g$, as \cite{Gopakumar:1998jq}
\be
\label{GVgf}
F^{\rm GV}\left({\bf t}, g_s\right)=\sum_{g\ge 0} \sum_{\bf d} \sum_{w=1}^\infty {1\over w} n_g^{ {\bf d}} \left(2 \sin { w g_s \over 2} \right)^{2g-2} \re^{-w {\bf d} \cdot {\bf t}}.
\ee
Then,
\be
\label{gv-form}
F^{\rm WS}\left({\bf t}, g_s\right)=F^{({\rm p})}({\bf t}, g_s)+F^{\rm GV}\left({\bf t}, g_s\right).
\ee
For local Calabi-Yau threefold, topological string have a refinement correspond to the supersymmetric gauge theory in the omega background. In refined topological string, the Gopakumar-Vafa invariants can be generalized to the refined BPS invariants  $N^{\bf d}_{j_L, j_R}$ which depend on the degrees ${\bf d}$ and spins, $j_L$, $j_R$ \cite{Iqbal:2007ii}\cite{Choi:2012jz}\cite{Nekrasov:2014nea}. Refined BPS invariants are also integers and are closely related with the Gopakumar-Vafa invariants,
\be
\label{ref-gv}
\sum_{j_L, j_R} \chi_{j_L}(q) (2j_R+1) N^{\bf d} _{j_L, j_R} = \sum_{g\ge 0} n_g^{\bf d} \left(q^{1/2}- q^{-1/2} \right)^{2g},
\ee
where $q$ is a formal variable and
\be
\chi_{j}(q)= {q^{2j+1}- q^{-2j-1} \over q-q^{-1}}
\ee
is the $SU(2)$ character for the spin $j$. Using these refined BPS invariants, one can define the NS free energy as
\be
\label{NS-j}
F^{\rm NS}({\bf t}, \hbar) ={1\over 6 \hbar} a_{ijk} t_i t_j t_k +b^{\rm NS}_i t_i \hbar +\sum_{j_L, j_R} \sum_{w, {\bf d} }
N^{{\bf d}}_{j_L, j_R}  \frac{\sin\frac{\hbar w}{2}(2j_L+1)\sin\frac{\hbar w}{2}(2j_R+1)}{2 w^2 \sin^3\frac{\hbar w}{2}} \re^{-w {\bf d}\cdot{\bf  t}}.
\ee
in which $b_i^{\rm NS}$ can be obtained by using mirror symmetry as in \cite{Huang:2010kf}. By expanding (\ref{NS-j}) in powers of $\hbar$, we find the NS free energies at order $n$,
\be
\label{ns-expansion}
F^{\rm NS}({\bf t}, \hbar)=\sum_{n=0}^\infty  F^{\rm NS}_n ({\bf t}) \hbar^{2n-1}.
\ee
The BPS part of free energy of refined topological string is defined by refined BPS invariants as
\be
\ba
F_{\text{ref}}^{\text{BPS}}(\mathbf{t},\epsilon_1,\epsilon_2)
=\sum_{j_L,j_R}\sum_{w, d_j \geq 1}
\frac{1}{w} N_{j_L,j_R}^{\bf{d}}
\frac{\chi_{j_L}(q_L^w)\chi_{j_R}(q_R^w)}{(q_1^{w/2}-q_1^{-w/2})(q_2^{w/2}-q_2^{-w/2})}
\re^{-w \bf{d} \cdot \bf{t}},
\ea
\label{eq:F-ref}
\ee
where
\be
\ba
\epsilon_j=2\pi \tau_j,\quad q_j=\re^{2\pi \ri \tau_j},\quad (j=1,2),\qquad q_L=\re^{\pi \ri (\tau_1-\tau_2)},\qquad
q_R=\re^{\pi \ri (\tau_1+\tau_2)}.
\ea
\ee
The refined topological string free energy can also be defined by refined Gopakumar-Vafa invariants as
\be
\label{ref-free-2}
F_{\text{ref}}^{\rm BPS}(\mathbf{t}, \epsilon_1, \epsilon_2)=
\sum_{g_L, g_R \ge 0 }\sum_{w\ge 1}\sum_{{\bf d}}
{1\over w} n_{g_L, g_R}^{\bf d}
{ \left(q_L^{w /2} - q_L^{-w /2} \right)^{2g_L} \over q^{w/2} - q^{-w/2} }
{ \left(q_R^{w /2} - q_R^{-w /2} \right)^{2g_R} \over t^{w/2} -t^{-w/2} }
e^{-w\mathbf{d}\cdot\mathbf{t}},
\ee
where
\be
\label{qt}
q=\re^{ \im \epsilon_1},   \qquad t=\re^{-\im \epsilon_2}.
\ee
The refined Gopakumar-Vafa invariants are related with refined BPS invariants,
\be
\label{change-basis}
\ba
\sum_{j_L, j_R\ge 0}  N_{j_L, j_R}^{\bf d}
\chi_{j_L}(q_L) \chi_{j_R}(q_R)=\sum_{g_L, g_R \ge 0 }  n_{g_L, g_R}^{\bf d}
\left(q_L^{1 /2} - q_L^{-1 /2} \right)^{2g_L}
\left(q_R^{1 /2} - q_R^{-1 /2} \right)^{2g_R}.
\ea
\ee
The refined topological string free energy can be expand as
\be
F(\mathbf{t},\epsilon_1,\epsilon_2)=\sum_{n,g=0}^{\infty}(\epsilon_1+\epsilon_2)^{2n}(\epsilon_1\epsilon_2)^{g-1}\mathcal{F}^{(n,g)}(\mathbf{t})
\ee
where $\mathcal{F}^{(n,g)}(\mathbf{t})$ can be determined recursively using the refined holomorphic anomaly equations.

With the refined free energy, the traditional topological string free energy can be obtained by taking the unrefined limit,
\be
\epsilon_1=-\epsilon_2=g_s.
\ee
Therefore,
\be
F_{\rm GV}\left({\bf t}, g_s\right)=F(\mathbf{t},g_s,-g_s).
\ee
The NS free energy can be obtained by taking the NS limit in refined topological string,
\be
F^{\rm NS}({\bf t}, \hbar)=\lim_{\epsilon_1\rightarrow 0}\epsilon_1F(\mathbf{t},\epsilon_1,\hbar).
\ee
The full refined topological string refined energy of toric Calabi-Yau can be obtained using the refined topological vertex in the A-model side \cite{Iqbal:2007ii}\cite{Taki:2007dh}\cite{Iqbal:2012mt}, or equivalently, using the topological recursion in the B-model side \cite{Eynard:2007kz}\cite{Bouchard:2007ys}. See the detailed calculations on various toric Calabi-Yau threefolds in \cite{Huang:2010kf}\cite{Huang:2013yta}\cite{Klemm:2015iya}.

\section{Nekrasov-Shatashvili Quantization Scheme}
\label{sec:NS}
In this section, we focus on the Nekrasov-Shatashvili quantization scheme. First, we review the original conjecture of Nekrasov and Shatashvili and its physical meaning. Second, we introduce the exact NS quantization conditions which involves the non-perturbative effects. This is first proposed in \cite{Wang:2015wdy} for the mirror curve of genus one, and soon be generalized to higher genus in \cite{Franco:2015rnr}. Then we show the pole cancellation in the exact quantization conditions. By modifying the existing techniques, we provide two derivations of the exact NS quantization conditions from Lockhart-Vafa partition function and Mellin-Barnes representation. Finally, we establish the existence of the $\bf{B}$ field for arbitrary toric Calabi-Yau and propose an effective method to determine its value.
\subsection{Bethe/Gauge Correspondence}
The correspondence between $N=2$ supersymmetric gauge theories and integrable systems dates back to 1990s. It was first observed on the classical level in \cite{Gorsky:1995zq} for $SU(N)$ Seiberg-Witten theories and soon be generalized to arbitrary Lie algebras in \cite{Martinec:1995by}. See \cite{D'Hoker:1999ft} for an excellent review. The simplest example is the correspondence between pure $SU(2)$ Seiberg-Witten theory and sine-Gordon model, where the Seiberg-Witten differential serves as the Bohr-Sommerfeld periods. More general, the pure $SU(N)$ Seiberg-Witten theories correspond to periodic Toda system with $N$ particles.\\
\indent In 2009, Nekrasov and Shatashvili promoted this correspondence between $N=2$ supersymmetric gauge theories and integrable systems to the quantum level \cite{Nekrasov:2009rc}, see also \cite{Nekrasov:2009uh}\cite{Nekrasov:2009ui}. The correspondence is usually called Bethe/Gauge correspondence. As it is not proved in full generality yet, we also refer the correspondence on quantum level as Nekrasov-Shatashvili conjecture. The $SU(2)$ and $SU(N)$ cases are soon checked for the first few orders in \cite{Mironov:2009uv}\cite{Mironov:2009dv}. For a proof of the conjecture for $SU(N)$ cases, see \cite{Kozlowski:2010tv}\cite{Meneghelli:2013tia}. This conjecture as a 4D/1D correspondence is also closely related to the AGT conjecture \cite{Alday:2009aq}, which is a 4D/2D correspondence. In fact, the duality web among $N=2$ gauge theories, matrix model, topological string and integrable systems (CFT) can be formulated in generic Nekrasov deformation \cite{Dijkgraaf:2009pc}, not just in NS limit. See \cite{Rim:2015aha} for a proof of Nekrasov-Shatashvili conjecture for $SU(N)$ quiver theories based on AGT correspondence.\\
\indent It is well-known the prepotential of Seiberg-Witten theory can be obtained from the Nekrasov partition function \cite{Nekrasov:2002qd},
\be
{\mathcal{F}}(\vec {a})=\lim_{\epsilon_1,\epsilon_2\rightarrow0} \epsilon_1\epsilon_2\log Z_{\text{Nek}}(\vec{a}; \epsilon_1,\epsilon_2),
\ee
where $\vec{a}$ denotes the collection of all Coulomb parameters. In \cite{Nekrasov:2009rc}, Nekrasov and Shatashvili made an interesting observation that in the limit where one of the deformation parameters is sent to zero while the other is kept fixed ($\epsilon_1\rightarrow 0,\,\epsilon_2 = \hbar$), the partition function is closely related to certain quantum integrable systems. The limit is usually called Nekrasov-Shatashvili limit in the context of refined topological string, or classical limit in the context of AGT correspondence. To be specific, the Nekrasov-Shatashvili conjecture says that the supersymmetric vacua equation
\be\label{qc1}
\exp {(\partial_{a_I}{\mathcal{W}}(\vec{a};\hbar))}=1\,,
\ee
of the Nekrasov-Shatashvili free energy
\be\label{NSfree1}
\mathcal{W}(\vec{a};\hbar)=\lim_{\epsilon_1\rightarrow0} \epsilon_1\log Z_{\text{Nek}}(\vec{a}; \epsilon_1,\epsilon_2=\hbar)
\ee
gives the Bethe ansatz equations for the corresponding integrable system. The Nekrasov-Shatashvili free energy (\ref{NSfree1}) is also called effective twisted superpotential in the context of super gauge theory. According to the NS conjecture, it serves as the Yang-Yang function of the integrable system. The quantized/deformed Seiberg-Witten curve becomes the quantized spectral curve and the twisted chiral operators become the quantum Hamiltonians. Since it is usually difficult to written down the Bethe ansatz for general integrable systems, this observation provide a brand new perspective to study quantum integrable systems.\\
\indent The physical explanations of Bethe/Gauge correspondence was given in \cite{Nekrasov:2010ka}\cite{Aganagic:2011mi}. Let us briefly review the approach in \cite{Aganagic:2011mi}, which is closely related to topological string. In the context of geometric engineering, the NS free energy of supersymmetric gauge theory is just the NS limit of the partition function of topological string \cite{Katz:1996fh},
\be\label{FNS5}
\mathcal{W}(\vec{a};\hbar)=F^{\rm NS}({\bf t}, \hbar).
\ee
See \cite{Huang:2012kn} for the detailed study on the relation between gauge theory and topological string in Nekrasov-Shatashvili limit.

Consider the branes in unrefined topological string theory, it is well known \cite{Aganagic:2003qj} that for B-model on a local Calabi-Yau given by
\be
uv+H(x,p)=0
\ee
the wave-function $\Psi(x)$ of a brane whose position is labeled by a point $x$ on the Riemann
surface $H(x,p)=0$ classically, satisfies an operator equation
\be
H(\mx, \mathsf{p})\Psi=0,
\ee
with the Heisenberg relation\footnote{In general, this relation only holds up to order $g_s$ correction.}
\be
[\mx, \mathsf{p}]= \im g_s.
\ee
In the refined topological string theory, the brane wave equation is generalized to a multi-time dependent Schr\"odinger equation,
\be
H(x,p)\Psi =\epsilon_1\epsilon_2\sum f_i(t){\partial \Psi \over \partial t_i},
\ee
where $f_i(t)$ are some functions of the K\"{a}hler moduli $t_i$ and the momentum operator is given by either $p =i \epsilon_1 \partial_x$ or $p= i\epsilon_2 \partial_x$, depending on the type of brane under consideration.\\
\indent In the NS limit $\epsilon_1\rightarrow0,\,\epsilon_2=\hbar$, the time dependence vanishes, and we simply obtain the time-independent Schr\"odinger equation
\be
H(\mx, \mathsf{p})\Psi=0,
\ee
with
\be
[\mx, \mathsf{p}]= \im \hbar.
\ee
To have a well-defined wave function we need the wave function to be single-valued under monodromy. In unrefined topological string, the monodromy is characterized by taking branes around the cycles of a Calabi-Yau shifts the dual periods in units of $g_s$. While in the NS limit, the shifts becomes derivatives. Therefore, the single-valued conditions now are just the supersymmetric vacua equation (\ref{qc1}). In the context of topological string, we have\footnote{Here the definition of $F^{\rm NS}({\bf t}, \hbar)$ is slightly different from the one in Equation (\ref{FNS5}) by some linear $\bf t$ terms. This modification is made to be consistent with the quantum zero point energy of vacuum. In fact, the zero point energy for quantum mirror curves emerges exactly in the same way as the quantum harmonic oscillator. See the analysis in \cite{Kashani-Poor:2016edc} for example.}
\be\label{qc2}
C_{ij}\frac{\partial{F^{\rm NS}({\bf t}, \hbar)}}{\partial{t_j}}=2\pi \left(n_i+\frac{1}{2}\right),\ i=1,\cdots,g.
\ee
In fact these conditions are just the concrete form of Einstein-Brillouin-Keller (EBK) quantization, which is the generalization of Bohr-Sommerfeld quantization for high-dimensional integrable systems. Therefore, we can also regard the left side of (\ref{qc2}) as phase volumes corresponding to each periods of the mirror curve,
\be\label{vol}
\mathrm{Vol}_i({\bf t}, \hbar)=\hbar C_{ij}\frac{{F^{\rm NS}({\bf t}, \hbar)}}{\partial{t_j}},\ i=1,\cdots,g.
\ee
Now the NS quantization conditions for the mirror curve are just the EBK quantization conditions,
\be\label{qc3}
\mathrm{Vol}_i({\bf t}, \hbar)=2\pi\hbar \left(n_i+\frac{1}{2}\right),\ i=1,\cdots,g.
\ee
However, the story does not end here. It is easy to see from the definition of NS free energy (\ref{NS-j}) that the phase volumes (\ref{vol}) contains poles at rational Planck constants $\hbar=2\pi p/q$. The existence of poles in volumes indicates that one must add the non-perturbative effect to obtain the exact quantization conditions. This is not unusually in quantum mechanics. For example, it is long known that even including all-order WKB expansion it is not enough to characterize the energy spectrum of quantum mechanics double-well potential. One must consider the instanton effect as well, see \cite{ZinnJustin:2004ib}. For another fascinating recent developments on the non-perturbative effects in quantum theories, in particular the resurgence theory, see \cite{Dunne:2013ada}\cite{Dunne:2014bca}. In a series of papers \cite{Krefl:2013bsa, Krefl:2014nfa, Krefl:2016svj}, Krefl used the NS quantization condition for defining a non-perturbative completion, in the cases of topological strings dual to matrix models and 4d gauge theory. The NS quantization conditions are also related with the exact WKB analysis of $N=2$ gauge theory, see \cite{Ashok:2016yxz}\cite{Kashani-Poor:2015pca}\cite{Basar:2015xna}. For the Bethe/gauge correspondence on squashed $S^5$ and $S^3$ and the relation with open topological string, see \cite{Sciarappa:2016ctj}.
\subsection{Exact Nekrasov-Shatashvili Quantization Conditions}\label{sec:exact}
The non-perturbative completion of the NS quantization conditions (\ref{qc3}) was first proposed in \cite{Wang:2015wdy} for the mirror curve of genus one, and soon be generalized to higher genus in \cite{Franco:2015rnr}. In this section, we review the exact NS quantization conditions in the notations similar with \cite{Wang:2015wdy}, but essentially the same with \cite{Franco:2015rnr}.\\
\indent First, let us introduce a modified version of the instanton part of the NS free energy,
\be
\label{NSinstmodified}
F_{\rm NS}^{\rm inst}({\bf t}, \hbar)=\sum_{j_L, j_R} \sum_{w, {\bf d} }
N^{{\bf d}}_{j_L, j_R} (-)^{{\bf B}\cdot{\bf d}}\,\frac{\sin\frac{\hbar w}{2}(2j_L+1)\sin\frac{\hbar w}{2}(2j_R+1)}{2 w^2 \sin^3\frac{\hbar w}{2}} \re^{-w {\bf d}\cdot{\bf  t}},
\ee
where we add a crucial sign determined by ${\bf B}\cdot{\bf d}$. The ${\bf B}$ field is a constant integral vector, defined up to an even lattice. It was introduced in \cite{Hatsuda:2013oxa} and satisfies the following important property: for all triples of degree ${\bf d}$, spin $j_L$ and $j_R$ such that the refined BPS invariants $N^{{\bf d}}_{j_L, j_R}(X) $ is non-vanishing, they must satisfy
\be
(-1)^{2j_L + 2 j_R-1}= (-1)^{{\bf B} \cdot {\bf d}}.
\ee
This ${\bf B}$ field is crucial for the pole cancellation in both NS and GHM quantization schemes. We shall establish its existence for arbitrary toric Calabi-Yau in Section \ref{sec:remark}.

Then we introduce the following notations:
\begin{equation}
f_i(\mathbf{t},\hbar)=\frac{\partial F^{\mathrm{inst}}_{\mathrm{NS}}(\mathbf{t},\hbar)}{\partial t_i}
\end{equation}
\begin{equation}
\widehat{f}_i(\mathbf{t},\hbar)=f_i(\mathbf{t}+\ri\pi\mathbf{B},\hbar).
\end{equation}
A hat on a function means the operation of adding the first variable with $\ri\pi\mathbf{B}$. For example,
\begin{equation}
\widehat{f}_i(\frac{2\pi\mathbf{t}}{\hbar},\frac{4\pi^2}{\hbar})=f_i(\frac{2\pi\mathbf{t}}{\hbar}+\ri\pi\mathbf{B},\frac{4\pi^2}{\hbar}).
\end{equation}
We also introduce the hat operation for variables,
\begin{equation}
\widehat{\mathbf{t}}=\mathbf{t}+\ri\pi\mathbf{B}.
\end{equation}
Note the following property:
\begin{equation}\label{eq:doublehat}
\widehat{f}_i(\widehat{\mathbf{t}},\hbar)=f_i(\mathbf{t},\hbar).
\end{equation}
For polynomial part of NS free energy, we have
\begin{equation}
f_i^{\mathrm{poly}}(\mathbf{t},\hbar)=\frac{\partial F^{\mathrm{poly}}_{\mathrm{NS}}(\mathbf{t},\hbar)}{\partial t_i}
=\frac{1}{2\hbar}a_{ijk}t^jt^k+b_i^{\mathrm{NS}}\hbar.
\end{equation}
Then, we define the $g$ phase volumes as follows:
\begin{equation}\label{eq:volumeunshiftori}
\mathrm{Vol}_j/\hbar=C_{ji}\left(f_i^{\mathrm{poly}}(\widehat{\mathbf{t}},\hbar)+\widehat{f}_i(\widehat{\mathbf{t}},\hbar)
+\widehat{f}_i(\frac{2\pi\widehat{\mathbf{t}}}{\hbar},\frac{4\pi^2}{\hbar})+\mathrm{const}\right).
\end{equation}
Since there is $\widehat{\mathbf{t}}$ in every term, we simply drop the hat on $\mathbf{t}$ for the writing convenience from now on. Then we have the non-perturbative volumes as
\begin{equation}\label{eq:volumeunshift}
\mathrm{Vol}_j/\hbar=C_{ji}\left(f_i^{\mathrm{poly}}(\mathbf{t},\hbar)+\widehat{f}_i(\mathbf{t},\hbar)
+\widehat{f}_i(\frac{2\pi\mathbf{t}}{\hbar},\frac{4\pi^2}{\hbar})+\mathrm{const}\right).
\end{equation}
One should keep in mind that every $\mathbf{t}$ from now on is actually $\widehat{\mathbf{t}}$. This notation turns out to be convenient for the GHM conjecture as well. Besides, the K\"{a}hler moduli with or without the hat do not affect the identity (\ref{check}) or our statement for the equivalence either. In fact, we believe the genuine physical K\"{a}hler moduli should always be attached with a $\mathbf{B}$ field, which is $t_i=\pi\ri B_i+\log z_i+\dots$, rather than the traditional definition $t_i=\log z_i+\dots$. When calculating the quantum spectra of the mirror curve and comparing it with the numerical methods, one should be careful that only the quantization of volumes (\ref{eq:volumeunshiftori}) result in the correct spectra of the mirror curve, if one use the traditional definition of the K\"{a}hler moduli. Or equivalently, one can stick to the volumes in (\ref{eq:volumeunshift}) and solve the spectra, but this spectra will be the mirror curve with the complex moduli having the following transformations,
\be
z_i\rightarrow(-)^{B_i}z_i,\quad\quad i=1,2,\dots,s.
\ee

The key structure of (\ref{eq:volumeunshift}) is
\begin{equation}\label{eq:structure1}
\widehat{f}_i(\mathbf{t},\hbar)
+\widehat{f}_i(\frac{2\pi\mathbf{t}}{\hbar},\frac{4\pi^2}{\hbar}),
\end{equation}
which can be proved to contains no pole for arbitrary $\hbar$. Besides, this structure naturally appears from the NS limit of non-perturbative completion of topological string free energy proposed by Lockhart and Vafa.\\
\indent The constant in (\ref{vol}) is proportional to $\pi^2/\hbar$ and can be determined analytically by the semiclassical analysis\footnote{For most geometries, the constant terms can be most easily determined by numerical method.}. For the self-S duality of the polynomial part of volumes\footnote{This was first noticed in \cite{Hatsuda:2015fxa}}, it is conjectured to have the following value
\begin{equation}\label{eq:BBCrelation}
\mathrm{const}=\frac{4\pi^2b_i^{\mathrm{NS}}}{\hbar}.
\end{equation}
We have checked this is true for all models which have been studied, including local $\mathbb{P}^2$, $\mathbb{F}_0$, $\mathbb{F}_1$, $\mathbb{F}_2$, $\mathfrak{B}_2$, $E_8$ del Pezzo, $\mathbb{C}^3/\mathbb{Z}_5$ orbifold and $SU(N)$ geometries. We will give a physical explanation on this value in section \ref{sec:LV}.
\subsection{Pole Cancellation}\label{sec:pole}
It is obvious that either $F^{\mathrm{inst}}_{\mathrm{NS}}(\mathbf{t},\hbar)$ or $F^{\mathrm{inst}}_{\mathrm{NS}}(2\pi\mathbf{t}/\hbar,4\pi^2/\hbar)$ itself has poles for rational Planck constant. In this subsection, we show that for $\hbar=2\pi p/q$ the summation (\ref{eq:structure1})
\begin{equation}
\widehat{f}_i(\mathbf{t},\hbar)
+\widehat{f}_i(\frac{2\pi\mathbf{t}}{\hbar},\frac{4\pi^2}{\hbar}).
\end{equation}
contains no pole. This structure was first proposed in \cite{Wang:2015wdy}. It was soon realized in \cite{Hatsuda:2015qzx} that this structure is in fact a self-S duality,
\be
\ba
\bf{t}\rightarrow\,&\frac{2\pi\bf{t}}{\hbar}\rightarrow\bf{t},\\
\hbar\rightarrow\,&\frac{4\pi^2}{\hbar}\rightarrow\hbar.
\ea
\ee
The pole cancellation is easy to show for general cases using the property of $\mathbf{B}$ field, see \cite{Hatsuda:2015qzx}\cite{Franco:2015rnr}\cite{Kashani-Poor:2016edc}. The key point is that both perturbative part and non-perturbative part contain $\mathbf{B}$ field. Here we present the complete calculations for each term in the quantization conditions at rational Planck constant for future use in the proof of the equivalence with GHM conjecture.\\
\indent First, assume $\hbar=2\pi\frac{p}{q}+\zeta$, then
\be
\ba
&\widehat{F}^{\rm NS}_{\rm inst}({\bf t}, \hbar) =\sum_{j_L, j_R,\bf d}\sum_{q\nmid w}
N^{{\bf d}}_{j_L, j_R}(-)^{w{\bf B}\cdot{\bf d}}\frac{\sin\frac{\hbar w}{2}(2j_L+1)\sin\frac{\hbar w}{2}(2j_R+1)}{2 w^2 \sin^3\frac{\hbar w}{2}}\bigg|_{\hbar=2\pi\frac{p}{q}} \re^{-w {\bf d}\cdot{\bf  t}}\\
&+\sum_{j_L, j_R,\bf d}\sum_{w=nq}N^{{\bf d}}_{j_L, j_R}\frac{(-)^{nq{\bf B}\cdot{\bf d}}(-)^{np(2j_L+2j_R-1)}}{2n^2q^2}\re^{-nq {\bf d}\cdot{\bf  t}}\bigg(\frac{2m_Lm_R}{w\zeta}+\frac{m_Lm_R(3-m_L^2-m_R^2)w\zeta}{12}+o(\zeta^2)\bigg),
\ea
\ee
in which $m_{L/R}=2j_{L/R}+1$. The dual NS free energy is
\be
\ba
&\widehat{F}^{\rm NS}_{\rm inst}(\frac{2\pi\bf t}{\hbar}, \frac{4\pi^2}{\hbar}) =\sum_{j_L, j_R,\bf d}\sum_{p\nmid w}
N^{{\bf d}}_{j_L, j_R}(-)^{w{\bf B}\cdot{\bf d}}\frac{\sin\frac{\hbar w}{2}(2j_L+1)\sin\frac{\hbar w}{2}(2j_R+1)}{2 w^2 \sin^3\frac{\hbar w}{2}}\bigg|_{\hbar=2\pi\frac{p}{q}} \re^{-w\frac{q}{p} {\bf d}\cdot{\bf  t}}\\
&+\sum_{j_L, j_R,\bf d}\sum_{w=np}N^{{\bf d}}_{j_L, j_R}\frac{(-)^{np{\bf B}\cdot{\bf d}}(-)^{nq(2j_L+2j_R-1)}}{2n^2p^2}\re^{-nq {\bf d}\cdot{\bf  t}}\bigg(-\frac{2m_Lm_R}{(\frac{q}{p})^2w\zeta}-\frac{m_Lm_R(1+nq\bf d\cdot\bf t)}{n\pi q}\bigg).
\ea
\ee
Then the perturbative BPS part becomes
\be\label{eq:f}
\ba
&\widehat{f}_i({\bf t}, \hbar) =\sum_{j_L, j_R,\bf d}\sum_{q\nmid w}
N^{{\bf d}}_{j_L, j_R}(-)^{w{\bf B}\cdot{\bf d}}\frac{\sin\frac{\hbar w}{2}(2j_L+1)\sin\frac{\hbar w}{2}(2j_R+1)}{2 w^2 \sin^3\frac{\hbar w}{2}}\bigg|_{\hbar=2\pi\frac{p}{q}} \re^{-w {\bf d}\cdot{\bf  t}}(-wd_i)+\\
&\sum_{j_L, j_R,\bf d}\sum_{w=nq}N^{{\bf d}}_{j_L, j_R}\frac{(-)^{nq{\bf B}\cdot{\bf d}}(-)^{np(2j_L+2j_R-1)}}{2n^2q^2}\re^{-nq {\bf d}\cdot{\bf  t}}(-d_i)\bigg(\frac{2m_Lm_R}{\zeta}+\frac{m_Lm_R(3-m_L^2-m_R^2)w^2\zeta}{12}+o(\zeta^2)\bigg).
\ea
\ee
The non-perturbative part becomes
\be\label{eq:fdual}
\ba
&\widehat{f}_i(\frac{2\pi\bf t}{\hbar}, \frac{4\pi^2}{\hbar}) =\sum_{j_L, j_R,\bf d}\sum_{p\nmid w}
N^{{\bf d}}_{j_L, j_R}(-)^{w{\bf B}\cdot{\bf d}}\frac{\sin\frac{\hbar w}{2}(2j_L+1)\sin\frac{\hbar w}{2}(2j_R+1)}{2 w^2 \sin^3\frac{\hbar w}{2}}\bigg|_{\hbar=2\pi\frac{p}{q}} \re^{-w\frac{q}{p} {\bf d}\cdot{\bf  t}}(-w\frac{q}{p}d_i)\\
&+\sum_{j_L, j_R,\bf d}\sum_{w=np}N^{{\bf d}}_{j_L, j_R}\frac{(-)^{np{\bf B}\cdot{\bf d}}(-)^{nq(2j_L+2j_R-1)}}{2n^2p^2}\re^{-nq {\bf d}\cdot{\bf  t}}(-d_i)\bigg(-\frac{2m_Lm_R}{(\frac{q}{p})^2\zeta}-\frac{m_Lm_R(1+nq\bf d\cdot\bf t)}{\pi \frac{q}{p}}\bigg).
\ea
\ee
Using the fact that
\be
(-)^{2j_L+2j_R-1}=(-)^{\bf B\cdot\bf d},
\ee
It is easy to see that the terms proportional to $\zeta^{-1}$ in $\widehat{f}_i({\bf t}, \hbar)$ and $\widehat{f}_i(\frac{2\pi\bf t}{\hbar}, \frac{4\pi^2}{\hbar})$ cancel with each other perfectly. Therefore, the total volumes (\ref{vol}) contain no pole for rational Planck constant. Note the proof does not rely on the choice of variable $\bf t$, thus under any shift ${\bf t}\rightarrow{\bf t}+\ri\pi \bf r$, the total volumes contain no pole either. We shall use this fact in the future.
\subsection{Derivation from Lockhart-Vafa Partition Function}\label{sec:LV}
Based on the localization calculation on the partition function of superconformal theories on squashed $S^5$, a non-perturbative definition of refined topological string was proposed by Lockhart and Vafa in \cite{Lockhart:2012vp}. Roughly speaking, the proposal for the non-perturbative topological string partition function take the form
\be
Z_{\mathrm{LV}}(t_i,m_j,\tau_1,\tau_2)={Z_{\text{ref}}(t_i,m_j;\tau_1,\tau_2)\over Z_{\text{ref}}(t_i/\tau_1,m_j/\tau_1; -1/\tau_1,\tau_2/\tau_1)\cdot Z_{\text{ref}}(t_i/\tau_2,m_j/\tau_2; \tau_1/\tau_2,-1/\tau_2)}
\ee
where $t_i, m_j$ are normalizable and non-normalizable K\"{a}hler classes, and $\tau_1,\tau_2$
are the two couplings of the refined topological strings with relation $\epsilon_{1,2}=2\pi\tau_{1,2}$. By carefully considering the spin structure in gauge precise and make the K\"{a}hler parameters suitable for the current context, the precise form of Lockhart-Vafa partition function should be
\be \label{eq:Z-LV}
Z_{\mathrm{LV}}(\mathbf{t},\tau_1,\tau_2)={Z_{\text{ref}}(\mathbf{t},\tau_1+1,\tau_2)\over Z_{\text{ref}}(\mathbf{t}/\tau_1,-1/\tau_1,\tau_2/\tau_1+1)\cdot Z_{\text{ref}}(\mathbf{t}/\tau_2. \tau_1/\tau_2+1,-1/\tau_2)}
\ee
Here we do not bother to distinguish the mass parameters from the true K\"{a}hler moduli. Then the non-perturbative free energy of refined topological string is given by
\be
F_{\text{LV}}({\bf{t}},\tau_1,\tau_2)=F_{\text{ref}}({\bf{t}},\tau_1+1,\tau_2)-F_{\text{ref}} \( \frac{{\bf{t}}}{\tau_1},-\frac{1}{\tau_1},\frac{\tau_2}{\tau_1}+1 \)
-F_{\text{ref}} \( \frac{{\bf{t}}}{\tau_2},\frac{\tau_1}{\tau_2}+1,-\frac{1}{\tau_2} \).
\label{eq:F-LV}
\ee
\indent In \cite{Hatsuda:2015fxa}, it was shown for local $\mathbb{P}^2$ and $\mathbb{F}_0$ that the non-perturbative volume is roughly the derivative of non-perturbative NS free energy. However, the shift of $\tau$ in the proposal of Lockhart-Vafa was omitted there, and the $\mathbf{B}$ field effect in the perturbative instanton part and non-perturbative part of quantum volumes was added by hands. Here we show that the instanton part of the volume (\ref{vol}) is in fact exactly the derivative of the NS limit of Lockhart-Vafa proposal for the refined partition function.\\
\indent The NS limit is defined as
\be
\ba
F^{\text{NS}}({\bf{t}} ,\hbar)&:=\lim_{\tau_2 \to 0} (-2\pi \tau_2)F_{\text{ref}}\({\bf{t}},\tau_1=\frac{\hbar}{2\pi}, \tau_2 \).
\ea
\ee
After some simple calculations, we observe the following elegant relations:
\be
\ba
F_{\text{ref}}({\bf{t}},\tau_1+1,\tau_2)&=F_{\text{ref}}({\bf{t}}+\ri\pi{\bf B},\tau_1,\tau_2)\\
F_{\text{ref}} \( \frac{{\bf{t}}}{\tau_1},-\frac{1}{\tau_1},\frac{\tau_2}{\tau_1}+1 \)&=F_{\text{ref}} \( \frac{{\bf{t}}}{\tau_1}+\ri\pi{\bf B},-\frac{1}{\tau_1},\frac{\tau_2}{\tau_1}\)\\
F_{\text{ref}} \( \frac{{\bf{t}}}{\tau_2},\frac{\tau_1}{\tau_2}+1,-\frac{1}{\tau_2} \)&=F_{\text{ref}} \( \frac{{\bf{t}}}{\tau_2}+\ri\pi{\bf B},\frac{\tau_1}{\tau_2},-\frac{1}{\tau_2} \).
\ea
\ee
When we take NS limit $\tau_2 \to +0$ and keep $\bf{t}$ finite, the third term on the right hand side
in \eqref{eq:F-LV} vanishes because ${\bf{t}}/\tau_2 \to \infty$. Remembering the hat operation defined in section \ref{sec:exact}, then the second term becomes
\be
\ba
&\lim_{\tau_2 \to 0}
(-2\pi \tau_2) F_{\text{ref}} \( \frac{\bf{t}}{\tau},-\frac{1}{\tau},\frac{\tau_2}{\tau}+1 \) \\
&=-\tau \sum_{j_L, j_R} \sum_{w, \bf{d}} \frac{1}{2w^2}
N^{\bf{d}}_{j_L,j_R}(-)^{w{\bf d}\cdot{\bf B}/\tau} \frac{\sin \frac{\pi w }{\tau} (2j_L+1) \sin \frac{\pi w}{\tau} (2j_R+1)}{\sin^3 \frac{\pi w}{\tau}}
\re^{-w \bf{d} \cdot \bf{t}/\tau}\\
&=\tau \widehat{F}^{\text{NS}}\( \frac{\bf{t}}{\tau},-\frac{1}{\tau}\)
\ea
\label{eq:F-NS-nppart}
\ee
Therefore the non-perturbative NS free energy can be written as
\be\label{eq:non-perturbative NS free energy}
\ba
F^{\text{NS}}_{\text{np}}({\bf{t}},\hbar)&:= \lim_{\tau_2 \to 0}(-2\pi \tau_2)F^{\text{ref}}_{\text{np}}({\bf{t}},\tau,\tau_2) \\
&=\widehat{F}^{\text{NS}}({\bf{t}},\tau)-\tau \widehat{F}^{\text{NS}}\( \frac{\bf{t}}{\tau},-\frac{1}{\tau}\)\\
&=\widehat{F}^{\text{NS}}({\bf{t}},\tau)+\tau \widehat{F}^{\text{NS}}\( \frac{\bf{t}}{\tau},\frac{1}{\tau}\).
\ea
\ee
In the last step, we assumed the analytic continuation.\\
\indent Note $\tau=\hbar/(2\pi)$, it is easy to see that
\be
\ba
f_i(\mathbf{t},\hbar)&=\frac{\partial}{\partial t_i}\widehat{F}^{\text{NS}}({\bf{t}},\tau),\\
\widehat{f}_i(\frac{2\pi\mathbf{t}}{\hbar},\frac{4\pi^2}{\hbar})&=\tau \frac{\partial}{\partial t_i}\widehat{F}^{\text{NS}}\( \frac{\bf{t}}{\tau},\frac{1}{\tau}\).
\ea
\ee
In summary, we have proved the BPS part of non-perturbative volumes naturally appears in the spirit of NS quantization by simply replacing the standard NS free energy with the non-perturbative completion proposed in \cite{Lockhart:2012vp}.\\
\indent Now let us look at the polynomial part of the non-perturbative volumes (\ref{vol}), which is
\begin{equation}\label{volpoly}
\frac{\partial F^{\mathrm{poly}}_{\mathrm{NS}}(\mathbf{t},\hbar)}{\partial t_i}+\text{const}
=\frac{1}{2\hbar}a_{ijk}t^jt^k+b_i^{\mathrm{NS}}\hbar+\frac{4\pi^2b_i^{\mathrm{NS}}}{\hbar}.
\end{equation}
We will show that this value of constant term can also be ascribed to a refinement of the polynomial part of topological string partition function. In the unrefined cases, it is long known that the polynomial part of topological string partition function for compact Calabi-Yau threefold can be written as
\be
F_{\text{unref}}^{\text(p)}={1\over 6 g_{s}^2}\int_{CY}J\wedge J\wedge J +{1\over 24}({1\over g_{s}^2}-1)\int_{CY} J\wedge c_2,
\ee
where $J$ denotes the K\"{a}hler form on the CY and $c_2$ is the second
Chern class. In the refined case where $\tau_1+\tau_2\not=0$, Lockhart-Vafa proposed the following refinement \cite{Lockhart:2012vp}:
\be\label{refinepoly}
F_{\text{ref}}^{\text{(p)}}={1\over 6 \tau_1\tau_2}\int_{CY}J\wedge J\wedge J -{1\over 24 } ({\tau_1\over \tau_2}+{\tau_2\over \tau_1} +{1\over \tau_1\tau_2}+3) \int_{CY} J\wedge c_2.
\ee
Note this formula is only formal, since for compact Calabi-Yau manifolds which make the integral well-defined, the refinement is not well understood. While for the local Calabi-Yau manifolds, the refinement is well established, but the integrals in (\ref{refinepoly}) are not well-defined due to the non-compactness. Nevertheless, we still want to make some observation from this proposal.

To write (\ref{refinepoly}) with components, we have
\be
F_{\text{ref}}^{\text{(p)}}(\mathbf{t},\epsilon_1,\epsilon_2)=-\frac{a_{ijk}t^it^jt^k}{6(2\pi)^2\tau_1\tau_2}
+b_it_i({\tau_1\over \tau_2}+{\tau_2\over \tau_1} +{1\over \tau_1\tau_2}+3).
\ee
In the unrefined limit $\epsilon_1=-\epsilon_2=g_s$, this becomes
\be\label{unrefpold}
F_{\text{unref}}^{\text{(p)}}(\mathbf{t},g_s)=\frac{a_{ijk}t^it^jt^k}{6g_s^2}
+b_it_i(1-\frac{4\pi^2}{g_s^2}),
\ee
which is just the known form of traditional topological string partition function.

In the NS limit, this becomes
\be\label{NSp}
\ba
F_{\text{NS}}^{\text{(p)}}(\mathbf{t},\hbar)&=\lim_{\tau_2 \to 0} (-2\pi \tau_2)F^{\text{ref}}\({\bf{t}},\tau_1=\frac{\hbar}{2\pi}, \tau_2 \)\\
&=\frac{a_{ijk}t^it^jt^k}{6\hbar}
-b_it_i(\hbar+\frac{4\pi^2}{\hbar}).
\ea
\ee
Although this formula may be merely formal, the self-S dual structure like $\hbar+4\pi^2/\hbar$ already appears. Compare (\ref{NSp}) with the definition of NS free energy, we request
\be\label{eq:bibins}
b_i^{\text{NS}}=-b_i.
\ee
We conjecture this is true for all toric Calabi-Yau threefolds which can engineer to gauge theories. We have verified this for all known examples including local $\mathbb{F}_0$, $\mathbb{F}_1$, $\mathbb{F}_2$, $SU(3)$ geometries with $m=0,1,2$ and $SU(4)$, $SU(5)$ geometries with $m=0$.\\
\indent Submit the above relation (\ref{eq:bibins}) to equation (\ref{NSp}), we have
\be\label{FNSp}
\ba
F_{\text{NS}}^{\text{(p)}}(\mathbf{t},\hbar)=\frac{a_{ijk}t^it^jt^k}{6\hbar}
+b_i^{\text{NS}}t_i(\hbar+\frac{4\pi^2}{\hbar}).
\ea
\ee
The derivatives of this polynomial part result in exactly the expression we need in (\ref{volpoly}). This explain the origin of the constant term in the exact NS quantization condition.

Here we emphasis that the relation (\ref{eq:bibins}) may not be true for those toric Calabi-Yau threefolds which cannot engineer to gauge theories, such as local $\mathbb{P}^2$ and $\mathbb{C}^3/\mathbb{Z}_5$ orbifold. However, final expression $F_{\text{NS}}^{\text{(p)}}(\mathbf{t},\hbar)$ (\ref{FNSp}) is expected to be correct for all toric Calabi-Yau. The above argument allows us to propose the following polynomial part of unrefined topological string on local Calabi-Yau,
\be\label{unrefpnew}
F_{\text{unref}}^{\text{(p)}}(\mathbf{t},\hbar)=\frac{1}{6\hbar^2}a_{ijk}t_it_jt_k+b_i t_i+\frac{4\pi^2}{\hbar^2}b_i^{\text{NS}}t_i.
\ee
The linear terms in this proposal which belong to genus zero free energy $F_0$ are sometimes different from the normal expectation in the compact cases (\ref{unrefpold}). This is not so peculiar, since the linear terms in $F_0$ are usually omitted in most circumstances. But it we accept this definition, we will benefit a lot for the conciseness in the expression of GHM conjecture, where we typically encounter $b_i^{\text{NS}}$ combined with the unrefined free energy. We will show this in Section \ref{sec:ghm2} and \ref{sec:genericcheck}. Besides, the new definition (\ref{unrefpnew}) is naturally identical with the traditional definition (\ref{unrefpold}) for the geometries which can engineer to gauge theories.
\subsection{Derivation from Mellin-Barnes Representation}
In \cite{Krefl:2015qva}, Krefl introduced the Mellin-Barnes representation of topological string, which serves as a non-perturbative completion of refined topological string. However, the form derived there is inconsistent with the exact NS quantization conditions for generic Planck constant. In this subsection, we will show that after some modification of topological string, the Mellin-Barnes representation can be used to derive the right non-perturbative NS quantization conditions.\\
\indent First, let us introduce the following integral representation of logarithm,
\be\label{rep1}
\log(1+z)=\frac{1}{2\pi \ri}\int_{\mathcal{C}_+}dx\frac{\Gamma^2(1+x)\Gamma(-x)}{\Gamma(2+x)}z^{1+x}
\ee
where $\mathcal{C}_+$ runs from $-\ri\infty$ to $\ri\infty$ in the usual definition. For our later purpose to compare it with the Faddeev's quantum dilogarithm, we define $\mathcal{C}_+$ run from $-\ri\infty-1_+$ to $\ri\infty-1_+$, where $1_+$ stands for infinitely closing to $-1$ from the right hand side of real axis.

The quantum dilogarithm function is defined as
\be
\prod_{n=0}^{\infty}(1+zq^{n+1/2}),  \ \ \ |q|<1,
\ee
where $q=e^{2\pi \ri(\tau+\ri\epsilon)},\,\epsilon \rightarrow 0_+$. Using (\ref{rep1}), the logarithm of quantum dilogarithm function can be represented as
\be\label{ex1}
\sum_{n=0}^{\infty}\log(1+zq^{n+1/2})\sim\frac{1}{2\pi \ri}\int_{\mathcal{C}_+}dx\frac{\Gamma^2(1+x)\Gamma(-x)}{2i\Gamma(2+x)\sin((1+x)\pi\tau)}z^{1+x}.
\ee
Using the identity
\be
\Gamma(1-z)\Gamma(z)=\frac{\pi}{\sin(\pi z)},
\ee
we have
\be
\frac{\Gamma^2(1+x)\Gamma(-x)}{2i\Gamma(2+x)\sin((1+x)\pi\tau)}z^{1+x}=\frac{\pi}{2\ri(1+x)\sin((1+x)\pi)\sin((1+x)\pi\tau)}z^{1+x}.
\ee
Thus there are two sorts of poles contributing to the integral, including perturbative poles $x_p$,
\be
1+x_p^*=n_p,\ n_p=1,2,\cdots.
\ee
and non-perturbative poles $x_{np}$
\be
1+x_{np}^*=\frac{n_p}{\tau},\ n_{np}=1,2,\cdots.
\ee
The integral over non-perturbative poles gives non-perturbative expansion of quantum dilogarithm. Finally, the non-perturbative generalization of quantum dilogarithm function is
\be\label{quantumdlog}
\prod_{n=0}^{\infty}(1+ze^{2\pi \ri\tau(n+1/2)})\rightarrow \prod_{n=0}^{\infty}(1+ze^{2\pi \ri\tau(n+1/2)})\prod_{n=0}^{\infty}(1+z^{\frac{1}{\tau}}e^{\frac{2\pi \ri}{\tau}(n+1/2)})
\ee
On the other hand, the Faddeev's quantum dilogarithm is defined as
\be
\Phi_b(z)=\exp \left(-\frac{1}{4}\int_{\mathcal{C}_+} \frac{e^{2 z x}}{\sin(x b)\sin(x/b)}\frac{dx}{x}\right).
\ee
It is easy to show that the Mellin-Barnes representation of quantum dilogarithm function gives the same function as Faddeev's quantum dilogarithm.\\

\indent In the Lockhart-Vafa proposal on non-perturbative topological string (\ref{eq:Z-LV})(\ref{eq:F-LV}), they redefine the traditional topological string partition function by inserting a $(-1)^F$. This is equivalent to shifting $\tau_1$ (or $\tau_2$) by $1$ or adding $\ri\pi \mathbf{B}$ to the K\"{a}hler class parameters.

With this modification, the refined topological string partition can be written as
\begin{small}
\be\label{ref}
Z_{\mathrm{ref}}(\mathbf{t},\tau_1,\tau_2)=\prod_{\mathbf{d},j_L,j_R}\prod_{m_L=-j_L}^{j_L}\prod_{m_R=-j_R}^{j_R}\prod_{n_1,n_2=0}^{\infty}
(1+e^{2\pi \ri (\tau_1+1)(m_L+m_R)} e^{2\pi \ri \tau_2(m_L-m_R)} e^{2\pi \ri \tau_1(n_1+\frac{1}{2})}e^{2\pi \ri \tau_2(n_2+\frac{1}{2})}\mathbf{Q}^{\mathbf{d}})^{N^{\mathbf{d}}_{j_L,j_R}}
\ee
\end{small}
Transform into Gopakumar-Vafa expansion, we have
\be
F_{\mathrm{ref}}=\sum_{\mathbf{d}}\sum_{g_L,g_R=0}^{\infty}\sum_{m=1}^{\infty} n_{\mathbf{d}}^{(g_L,g_R)}
\frac{\sin \left(\frac{\pi m (\tau_-+1)}{2}\right)^{2g_L}\sin \left(\frac{\pi m (\tau_++1)}{2}\right)^{2g_R}}{4m \sin(\pi m \tau_1)\sin(\pi m \tau_2)}(-\mathbf{Q}^{\mathbf{d}})^m
\ee
where $\tau_{\pm}=\tau_1 \pm \tau_2$. After this modification, the Mellin-Barnes representation of refined topological string \cite{Krefl:2015qva} becomes
\be
Z_{\mathrm{ref}}=\prod_{\mathbf{d}}\prod_{g_L,g_R=0}^{\infty}\left(L_{\tau_1,\tau_2}^{(g_L,g_R)}(Q^{\mathbf{d}})\right)^{\frac{(2g_L)!(2g_R)!}{(-4)^{g_L+g_R-1}}n_{\mathbf{d}}^{(g_L,g_R)}},
\ee
where
\begin{small}
\be\label{2}
L_{\tau_1,\tau_2}^{(g_L,g_R)}(Q^{\mathbf{d}})=
-\frac{(-4)^{g_L+g_R}}{8\pi \ri(2g_L)!(2g_R)!}\int_{\mathcal{C}_+} dx \frac{\sin^{2g_L}\left(\frac{\pi (\tau_-+1)(1+x)}{2}\right)\sin^{2g_R}\left(\frac{\pi (\tau_++1)(1+x)}{2}\right)\Gamma^2(1+x)\Gamma(-x)}{\sin(\pi \tau_1(1+x))\sin (\pi \tau_2(1+x)) \Gamma(2+x)}(Q^{\mathbf{d}})^{1+x}
\ee
\end{small}
Using the identity $\Gamma(1-z)\Gamma(z)=\pi/\sin(\pi z)$, it is easy to see that the integral (\ref{2}) has poles at
\be
x^*_p=n, \ \ x^*_{np}=-1+\frac{n}{\tau_{1,2}},
\ee
where $n$ are positive integers. If $\tau_1/\tau_2$ is an irrational number, then the all the poles of the integral are order one\footnote{Also requires $x^*_p \neq x^*_{np}$}. For special choices of $\tau_1,\tau_2$ such that there exist high order poles, they can be viewed as the limit of the cases with only order-one poles. In the following, we simply assume that all the poles of the integral are order one poles. In this case, for perturbative poles $x^*_p$, we have
\be
F_{\text{ref}}^{p}=\tilde{F}_{\mathrm{ref}}(\mathbf{t},\tau_1+1,\tau_2),
\ee
where the tilde means usual definition of refined topological string. For the poles at $x^*_{np}=-1+n/\tau_{1}$, one can check
\be
F_{\mathrm{ref}}^{np1}=\tilde{F}_{\mathrm{ref}}\left(\frac{\mathbf{t}}{\tau_1},\frac{1}{\tau_1}+1,\frac{\tau_2}{\tau_1}\right)=\tilde{F}_{\mathrm{ref}}\left(\frac{\mathbf{t}}{\tau_1},\frac{1}{\tau_1},\frac{\tau_2}{\tau_1}+1\right).
\ee
While for the poles at $x^*_{np}=-1+n/\tau_{2}$, one can not express the residue with usual topological string free energy. In fact, we have,
\be
F_{\mathrm{ref}}^{np2}=\sum_{\mathbf{d}}\sum_{g_L,g_R=0}^{\infty}\sum_{m=1}^{\infty} n_{\mathbf{d}}^{(g_L,g_R)}
\frac{\sin \left(\frac{\pi m (\tau_1/\tau_2-1+1/\tau_2)}{2}\right)^{2g_L}\sin \left(\frac{\pi m (\tau_1/\tau_2+1+1/\tau_2)}{2}\right)^{2g_R}}{4m \sin(\pi m \tau_1/\tau_2)\sin(\pi m/\tau_2)}(-\mathbf{Q}^{\mathbf{d}/\tau_2})^m
\ee
Note if $\tau_2<0$, these poles should be $x^*_{np}+1=-\frac{n}{\tau_2}$, then
\be
F_{\mathrm{ref}}^{np2}=\sum_{\mathbf{d}}\sum_{g_L,g_R=0}^{\infty}\sum_{m=1}^{\infty} n_{\mathbf{d}}^{(g_L,g_R)}
\frac{\sin \left(\frac{\pi m (\tau_1/\tau_2-1+1/\tau_2)}{2}\right)^{2g_L}\sin \left(\frac{\pi m (\tau_1/\tau_2+1+1/\tau_2)}{2}\right)^{2g_R}}{4m \sin(\pi m \tau_1/\tau_2)\sin(\pi m/\tau_2)}(-\mathbf{Q}^{-\mathbf{d}/\tau_2})^m
\ee

Finally, we obtain the non-perturbative refined topological sting free energy as
\be
F_{\mathrm{ref}}^{np}=F_{\mathrm{ref}}^{p}+F_{\mathrm{ref}}^{np1}+F_{\mathrm{ref}}^{np2}
\ee
This formula is different from the Lockhart Vafa partition function of refined topological string (\ref{eq:F-LV})! However, in the NS limit ($2\pi\tau_1 \rightarrow \hbar,\tau_2\rightarrow 0$), it is easy to see the refined free energy becomes
\be\label{4}
F_{\text{NS}}^{np}(\mathbf{t},\hbar)=F_{\text{NS}}(\mathbf{t}+\ri\pi\mathbf{B},\hbar)+\frac{\hbar}{2\pi}F_{\text{NS}}\left(\frac{2\pi}{\hbar}\mathbf{t}+\ri\pi\mathbf{B},\frac{4\pi^2}{\hbar}\right).
\ee
Note this exactly the same non-perturbative NS free energy we need to derived the exact NS quantization conditions in the last section, see equation (\ref{eq:non-perturbative NS free energy}).

The exact NS quantization conditions can also be derived directly via the Faddeev's quantum dilogarithm function \cite{Faddeev:1993rs}\cite{Kashaev:2011se}. To see this, since
\begin{small}
\be
\begin{split}
\frac{\partial}{\partial{t_i}}F_{\text{NS}}^{inst}(\mathbf{t}+\ri\pi\mathbf{B},\hbar)&=-\frac{1}{2}\sum_{n=0}^{\infty}\sum_{\mathbf{d},j_L,j_R}(-1)^{2(j_L+j_R)}\frac{d_iN_{j_l,j_R}^{\mathbf{d}}}{n}\frac{\sin\left(\frac{n\hbar}{2}(2j_R+1)\right)\sin\left(\frac{n\hbar}{2}(2j_L+1)\right)}{\sin^3\left(\frac{n\hbar}{2}\right)}e^{n\mathbf{d}\cdot\mathbf{t}+\ri\pi n\mathbf{d}\cdot\mathbf{B}}\\
&=\log\left( \prod_{m_L=-j_L}^{j_L}\prod_{m_L=-j_R}^{j_R}\prod_{m=0}^{\infty}(1+q^{m+1/2}e^{\ri (\hbar+2\pi) (m_L+m_R)}e^{\mathbf{d}\cdot\mathbf{t}})^{\ri(-1)^{2(j_L+j_R)}d_i N_{j_l,j_R}^{\mathbf{d}}} \right),
\end{split}
\ee
\end{small}
where $q=e^{\ri\hbar}$. Comparing it with (\ref{quantumdlog}), ($\tau=\hbar/2\pi$, $z=-e^{\ri (\hbar+2\pi) (m_L+m_R)}e^{\mathbf{d}\cdot\mathbf{t}}$), we have the non-perturbative completion
\begin{small}
\be
(1+q^{m+1/2}e^{\ri (\hbar+2\pi) (m_L+m_R)}e^{\mathbf{d}\cdot\mathbf{t}}) \rightarrow  (1+q^{m+1/2}e^{\ri (\hbar+2\pi) (m_L+m_R)}e^{\mathbf{d}\cdot\mathbf{t}}) (1+e^{\frac{4\pi^2}{\hbar}(m+1/2)}e^{\ri \left(2\pi+\frac{4\pi^2}{\hbar}\right) (m_L+m_R)}e^{\frac{2\pi}{\hbar}\mathbf{d}\cdot\mathbf{t}})
\ee
\end{small}
Again we obtain the same non-perturbative completion as in (\ref{eq:non-perturbative NS free energy}) and (\ref{4}),
\be
\frac{\partial}{\partial{t_i}}F_{\text{NS}}^{inst}(\mathbf{t}+\ri\pi\mathbf{B},\hbar) \rightarrow \frac{\partial}{\partial{t_i}}F_{\text{NS}}^{inst}(\mathbf{t}+\ri\pi\mathbf{B},\hbar)+\frac{\partial}{\partial{t_i}}F_{\text{NS}}^{inst}\left(\frac{2\pi}{\hbar}\mathbf{t}+\ri\pi\mathbf{B},\frac{4\pi^2}{\hbar}\right).
\ee
\subsection{Remarks on $\mathbf{B}$ Field}
\label{sec:remark}
${\bf B}$ field is the key of the pole cancellation in both exact NS quantization conditions \cite{Wang:2015wdy} and HMO mechanism \cite{Hatsuda:2012dt}\cite{Hatsuda:2013gj}. Mathematically, ${\bf B}$ field can be defined as a constant vector satisfying the following requirement:
for all ${\bf d}$, $j_L$ and $j_R$ such that the refined BPS invariant $N^{{\bf d}}_{j_L, j_R} $ is non-vanishing, it must have
\be\label{defB}
(-1)^{2j_L + 2 j_R-1}= (-1)^{{\bf B} \cdot {\bf d}}.
\ee
Note ${\bf B}$ field is only defined up to an even lattice. For local del Pezzo CY threefolds, the existence of such a vector was established in \cite{Hatsuda:2013oxa}. Here we give a physical explanation on the existence of ${\bf B}$ field and an effective way to calculate ${\bf B}$ field for arbitrary toric Calabi-Yau threefold.\\
\indent Let us go back to the general form of mirror curve
\be
\label{mirrorcurve}
H(\re^x, \re^p)=\sum_{i=0}^{k+2}x_i  \exp\left( \nu^{(i)}_1 x+  \nu^{(i)}_2 p\right)=0.
 \ee
When we promote it to quantum mechanics, $x,p$ become operators. Let us introduce the function
\be
V_n(x)={\psi(x+ \ri n\hbar) \over \psi (x)},
\ee
where $\psi(x)$ is a wave function in the $x$-representation. Now the mirror curve (\ref{mirrorcurve}) becomes

\be\label{quantummirror1}
\sum_{i=0}^{k+2}x_i  \exp\left( \nu^{(i)}_1 \mx+  \nu^{(i)}_2 \mathsf{p}\right)|\psi \rangle=0.
\ee
Note
\be
\ba
e^{n\mx}|\psi \rangle&=e^{nx}|\psi \rangle,\\
e^{n\mathsf{p}}|\psi \rangle&=|\psi(x+n\ri\hbar) \rangle.
\ea
\ee
For mixed terms with both $\mx$ and $\mathsf{p}$, one should be careful because the Heisenberg relation
\be
[\mx, \mathsf{p}]= \im \hbar.
\ee
To obtain the right operator equation, we need to consider the Baker-Campbell-Hausdorff formula
\be
\log(e^Xe^Y)=X+Y+\frac{1}{2}[X,Y]+\cdots.
\ee
Therefore,
\be
e^{m\mx+n\mathsf{p}}|\psi \rangle=e^{mx}q^{mn}|\psi(x+n\ri\hbar) \rangle,
\ee
where
\be
q=e^{\frac{i\hbar}{2}}.
\ee
Let us denote
\be
X=e^x,
\ee
then the quantum mirror curve (\ref{quantummirror1}) can be written as quantum operator equation
\be\label{mirrorcurve2}
\sum_{i=0}^{k+2}x_i q^{\nu^{(i)}_1\nu^{(i)}_2}X^{\nu^{(i)}_1}V_{\nu^{(i)}_2}(X)=0.
\ee
The quantum A-periods can be solved from this difference equation \cite{Aganagic:2011mi}\cite{Huang:2014eha}, which means the form of equation (\ref{mirrorcurve2}) uniquely determines the quantum A-periods.\\
\indent When one changes the Planck constant $\hbar$ to $\hbar+2\pi$, parameter $q$ becomes $-q$. Apparently, it is always possible to change the signs of complex moduli $x_i$ to keep the form of mirror curve (\ref{mirrorcurve2}) unchanged. In fact, these sign-changing of complex moduli, translating to the sign-changing of Batyrev coordinates, are exactly the effect of $\mathbf{B}$ field.\\
\indent As we know, the information of a toric Calabi-Yau threefold are totally encoded in its mirror curve. Therefore, if we shift Planck constant $\hbar$ by $2\pi$ and appropriately change the signs of complex moduli simultaneously to keep the form of quantum mirror curve, which are defined to be a $\mathbf{B}$ field adding to K\"{a}hler parameters, then all quantities including A-periods and B-periods should remain the same. To see why this definition of $\mathbf{B}$ field is exactly the same definition with (\ref{defB}), let us look at the expression of B-periods. B-periods are the linear combinations of the derivatives of NS free energy (\ref{NS-j}). The polynomial part does not concern us now, let us only pay attention to the BPS part,
\be
\sum_{j_L, j_R} \sum_{w, {\bf d} }
N^{{\bf d}}_{j_L, j_R}  \frac{\sin\frac{\hbar w}{2}(2j_L+1)\sin\frac{\hbar w}{2}(2j_R+1)}{2 w^2 \sin^3\frac{\hbar w}{2}} \re^{-w {\bf d}\cdot{\bf  t}}.
\ee
If we shift Planck constant $\hbar$ from zero to $2\pi$, obviously we add a factor to the above expression,
\be
(-1)^{w(2j_L + 2 j_R-1)}
\ee
To keep the mirror curve unchanged, we also shift the K\"{a}hler parameters with $\mathbf{B}$ field, which add the other factor to the above expression,
\be
(-1)^{w{\bf B} \cdot {\bf d}}.
\ee
As the B-periods should not change under these two operations simultaneously, it is only possible to require that for all non-vanishing refined BPS invariant $N^{{\bf d}}_{j_L, j_R} $, they must have
\be
(-1)^{2j_L + 2 j_R+1}= (-1)^{{\bf B} \cdot {\bf d}}.
\ee
This is not only the physical explanation of $\mathbf{B}$ field, but also an effective method to determine the $\mathbf{B}$ field for arbitrary toric Calabi-Yau.\\
\indent Let us look at one example. The mirror curve of $\IC^3/\IZ_5$ geometry can be written as \cite{Codesido:2015dia}\cite{Klemm:2015iya}
\be
\re^x + \re^p +z_1 z_2^3 \re^{-3x -p} + z_2  \re^{-x}+1=0.
\ee
Using the notation above, the equation
\be
\left(\re^\mx + \re^{\rm p} +z_1 z_2^3 \re^{-3\mx -\mathrm{p}} + z_2  \re^{-\mx}+1\right) |\psi \rangle=0
\ee
becomes
\be
X+ z_2 X^{-1} +1 + V_1(X)+ z_1 z_2^3 q^{3}X^{-3} V_{-1}(X)=0,
\ee
where
\be
q=\re^{\ri \hbar/2}, \qquad X=\re^x.
\ee
When one shifts $\hbar$ by $2\pi$, parameter $q$ becomes $-q$. To keep the form of quantum mirror curve, obviously one must have the following phase change for the Batyrev coordinates:
\be
z_1\rightarrow-z_1,\quad\quad z_2\rightarrow z_2.
\ee
This means for the current base choice,
\be
\mathbf{B}=(1,0),
\ee
which is exactly the value conjectured in \cite{Codesido:2015dia}.
\section{Grassi-Hatsuda-Mari\~no Quantization Scheme}
\label{sec:ghm}
In this section, we focus on the GHM quantization scheme for the mirror curve of arbitrary toric Calabi-Yau threefold. It should be noted that Grassi-Hatsuda-Mari\~no proposed the first precise form of the conjecture for the genus one cases in \cite{Grassi:2014zfa}, while for the cases with arbitrary genus, the generalization was done by Codesido-Grassi-Mari\~no in \cite{Codesido:2015dia}. Here we stick the GHM for brevity, and always refer it to the full conjecture for arbitrary toric Calabi-Yau threefold. In the following subsection, We review the GHM quantization scheme, for more details see \cite{Marino:2015ixa}\cite{Codesido:2015dia}. In the second subsection, we explicitly present the calculation of the generalized grand potential and quantum Riemann theta at rational Planck constant, which will be useful to our proof in next section. In subsection three, we consider a set of generalizations of the GHM conjecture, which is crucial to establish the equivalence with Nekrasov-Shatashvili quantization scheme.
\subsection{Review on GHM Conjecture}\label{sec:ghm1}
In this section, we briefly review the GHM conjecture. The GHM conjecture contains rich information about the mirror curve. Here we focus on the quantization conditions which is relevant with this paper. Our convention is the same with \cite{Marino:2015nla}\cite{Codesido:2015dia}.

For a mirror curve $\Sigma$ with genus $g_{\Sigma}$, there are $g_{\Sigma}$ different canonical forms for the curve,
\be
\CO_i (x,y) + \kappa_i=0, \qquad i=1, \cdots, g_{\Sigma}.
\ee
Here, $\kappa_i$ is normally a true modulus $x_i$ of $\Sigma$. The different canonical forms of the curves are related by reparametrizations and overall factors,
\be
\label{o-rel}
\CO_i +\kappa_i  = \CP_{ij} \left( \CO_j +\kappa_j\right), \qquad i,j=1, \cdots, g_{\Sigma},
\ee
where $\CP_{ij}$ is of form $\re^{ \lambda x + \mu y}$. Equivalently, we can write
\be
\label{oi-exp}
\CO_i = \CO_i^{(0)}+ \sum_{j \not=i} \kappa_j \CP_{ij}.
\ee
Perform the Weyl quantization of the operators $\CO_i(x,y)$, we obtain
$g_{\Sigma}$ different Hermitian operators $\mO_i$, $i=1, \cdots, g_{\Sigma}$,
\be
\mO_i=\mO_i^{(0)}+ \sum_{j \not=i} \kappa_j \mP_{ij}.
\ee
The operator $\mO_i^{(0)}$ is the unperturbed operator, while the moduli $\kappa_j$ encode different
perturbations. It turns out that the most interesting operator was not $\mO$, but its inverse
$\rho$. This is because $\rho$ is expected to be of trace class and
positive-definite, therefore it has a discrete, positive spectrum, and its Fredholm (or spectral) determinant is well-defined. We have

\be
\label{Xops}
\rho_i =\mO_i^{-1}, \qquad i=1, \cdots, g_\Sigma,
\ee
and
\be
\label{unp-inv}
\rho_i^{(0)}=\left( \mO_i^{(0)}\right)^{-1}, \qquad i=1, \cdots, g_{\Sigma}.
\ee
For the discussion on the eigenfunctions of $\rho$, see a recent paper \cite{Marino:2016rsq}. In order to construct the generalized spectral determinant, we need to introduce the following operators,
\be
\label{ajl}
\mA_{jl}= \rho_j^{(0)} \mP_{jl}, \quad j, l=1, \cdots, g_{\Sigma}.
\ee
Now the generalized spectral determinant is defined as
\be
\label{gsd}
\Xi_X ( {\boldsymbol \kappa}; \hbar)= {\rm det} \left( 1+\kappa_1 \mA_{j1} +\cdots+ \kappa_{g_{\Sigma}} \mA_{j g_{\Sigma}}  \right).
\ee
It is easy to prove this definition does not depend on the index $j$.

This function is an entire function on the complex moduli space, and can be expanded as follows,
\be
\label{or-exp}
\Xi_X ({\boldsymbol \kappa};\hbar)= \sum_{N_1\ge 0} \cdots \sum_{N_{g_{\Sigma}}\ge 0} Z_X(\boldsymbol{N}, \hbar) \kappa_1^{N_1} \cdots \kappa_{g_{\Sigma}}^{N_{g_{\Sigma}}},
\ee
with the convention that
\be
\label{zzeros}
Z_X(0, \cdots, 0;\hbar)=1.
\ee
$Z_X(\boldsymbol{N}, \hbar)$ are called fermionic spectral traces.

Using Fredholm-Plemelj's formula, we can also obtain the following expansion,
\be
\ba
\CJ_X (\boldsymbol{\kappa}; \hbar)&= \log\, \Xi_X ( {\boldsymbol \kappa}; \hbar)=
 -\sum_{\ell \ge 1} {(-1)^\ell \over \ell} \tr \left( \kappa_1 \mA_{11} +\cdots+ \kappa_{g_\Sigma} \mA_{1 g_\Sigma}  \right)^\ell \\
 &=-\sum_{\ell_1\ge 0} \sum_{\ell_2 \ge 0}  \cdots \sum_{\ell_{g_\Sigma} \ge 0} {1\over \ell_1+ \cdots + \ell_{g_\Sigma} } (-\kappa_1)^{\ell_1} \cdots \left(-\kappa_{g_\Sigma} \right)^{\ell_{g_\Sigma}} \sum_{W \in \CW_{\boldsymbol{\ell}} } \tr (W),
 \ea
 \ee
where
\be
\boldsymbol{\ell}=\left( \ell_1, \cdots, \ell_{g_\Sigma} \right)
\ee
and $\CW_{\boldsymbol{\ell}}$ is the set of all possible ``words" made of $\ell_i$ copies of the letters $\mA_{1i}$ defined in (\ref{ajl}).

This completes the definitions on quantum mirror curve from the quantum-mechanics side. Let us now turn to the topological string side. The total modified grand potential for CY with arbitrary-genus mirror curve is defined as
\be
\label{jtotal}
\mathsf{J}_{X}(\boldsymbol{\mu}, \boldsymbol{\xi},\hbar) = \mathsf{J}^{\rm WKB}_X (\boldsymbol{\mu}, \boldsymbol{\xi},\hbar)+ \mathsf{J}^{\rm WS}_X
(\boldsymbol{\mu},  \boldsymbol{\xi} , \hbar),
\ee
where
\be
\label{jm2}
\mathsf{J}^{\rm WKB}_X(\boldsymbol{\mu}, \boldsymbol{\xi}, \hbar)= {t_i(\hbar) \over 2 \pi}   {\partial F^{\rm NS}({\bf t}(\hbar), \hbar) \over \partial t_i}
+{\hbar^2 \over 2 \pi} {\partial \over \partial \hbar} \left(  {F^{\rm NS}({\bf t}(\hbar), \hbar) \over \hbar} \right) + {2 \pi \over \hbar} b_i t_i(\hbar) + A({\boldsymbol \xi}, \hbar).
\ee
and
\be
\label{jws}
\mathsf{J}^{\rm WS}_X(\boldsymbol{\mu}, \boldsymbol{\xi}, \hbar)=F^{\rm GV}\left( {2 \pi \over \hbar}{\bf t}(\hbar)+ \pi \ri {\bf B} , {4 \pi^2 \over \hbar} \right).
\ee
The modified grand potential has the following structure,
\be
\label{j-larget}
\mathsf{J}_{X}(\boldsymbol{\mu}, \boldsymbol{\xi},\hbar)= {1\over 12 \pi \hbar} a_{ijk} t_i(\hbar) t_j(\hbar) t_k(\hbar) + \left( {2 \pi b_i \over \hbar} + {\hbar b_i^{\rm NS} \over 2 \pi} \right) t_i(\hbar) +
\CO\left( \re^{-t_i(\hbar)}, \re^{-2 \pi t_i(\hbar)/\hbar} \right).
\ee
$A({\boldsymbol \xi}, \hbar)$ is some unknown function, which is relevant to the spectral determinant but does not affect the quantum Riemann theta function, therefore does not appear in the quantization conditions.

GHM conjecture says that the generalized spectral determinant (\ref{gsd}) is given by
%
\be
\label{our-conj}
\Xi_X({\boldsymbol \kappa}; \hbar)= \sum_{ {\bf n} \in \IZ^{g_{\Sigma}}} \exp \left( \mathsf{J}_{X}(\boldsymbol{\mu}+2 \pi \ri  {\bf n}, \boldsymbol{\xi}, \hbar) \right).
\ee
As a corollary, we have, the quantization condition for the mirror curve is given by
\be
\Xi_X({\boldsymbol \kappa}; \hbar)=0.
\ee
We can also define the quantum Riemann theta function from
\be
\label{qtf}
\Xi_X({\boldsymbol \kappa}; \hbar)= \exp\left( \mathsf{J}_{X}(\boldsymbol{\mu}, \boldsymbol{\xi}, \hbar) \right) \Theta_X({\boldsymbol \kappa}; \hbar).
\ee
Equivalently, the quantization condition for the mirror curve can be written as
\be
\Theta_X({\boldsymbol \kappa}; \hbar)=0.
\ee
\subsection{GHM Conjecture at Rational Planck Constant}\label{sec:ghm2}
In this section, we present detailed calculation of GHM with rational Planck constants. Let $\hbar=2\pi p/q$. It was proved for genus one cases in \cite{Grassi:2014zfa} that the poles cancel with each other in the total modified grand potential, albeit either $J_{\rm {WKB}}$ or $J_{\rm {WS}}$ itself has poles at rational Planck constant. The proof can be easily generalized to higher genus cases, therefore the total modified grand potential (\ref{jtotal}) contains no poles. In the following, we only show the non-singular part in $J_{\rm {WKB}}$ and $J_{\rm {WS}}$.\\
\indent The non-singular part of $J_{\rm {WS}}$ is
\be
\label{eq:JWScal}
\ba
&\mathsf{J}_{\rm WS}(\boldsymbol{\mu}, \boldsymbol{\xi}, \hbar)=F^{\rm GV}\left( {2 \pi \over \hbar}{\bf t}(\hbar)+ \pi \ri {\bf B} , {4 \pi^2 \over \hbar} \right)\\
&=\sum_{p\nmid w}d_w\left(2\pi\frac{p}{q}\right)(-)^{w\bf B\cdot \bf d}\re^{-\frac{q}{p}w\bf d\cdot\bf t}\\
&+\sum_{g=0}\sum_{\bf d}\sum_{w=np}\frac{(-)^{np\bf B\cdot\bf d}}{np}\re^{-nq\bf d\cdot\bf t}\left[
n_0^{\bf d}\left(\frac{6+2n^2\pi^2q^2+6nq\mathbf{d} \cdot \mathbf{t} +3n^2q^2({\bf d}\cdot{\bf t})^2}{24n^2q^2\pi^2}\right)+n_1^{\bf d}\right],
\ea
\ee
where
\be
d_w(\hbar)=\sum_{j_L, j_R,\bf d}N^{{\bf d}}_{j_L, j_R}\frac{2j_R+1}{w\left(2\sin\frac{2\pi^2w}{\hbar}\right)^2}
\frac{\sin\left(\frac{4\pi^2w}{\hbar}(2j_L+1)\right)}{\sin\frac{4\pi^2w}{\hbar}}.
\ee
We define the following analogies of standard free energy:
\be\label{eq:g0inst}
\widehat{G}_0^{\rm inst}=\sum_{g,\bf d}\sum_{w=nq}\frac{(-)^{np\bf B\cdot\bf d}\re^{-w\bf d\cdot\bf t}}{w^3}n_0^{\bf d},
\ee
\be
\widehat{G}_1^{\rm inst}=\sum_{g,\bf d}\sum_{w=nq}\frac{(-)^{np\bf B\cdot\bf d}\re^{-w\bf d\cdot\bf t}}{w}\left(\frac{n_0^{\bf d}}{12}+n_1^{\bf d}\right).
\ee
Note when $\hbar=2\pi$, i.e. $p=q=1$, $\widehat{G}_0^{\rm inst}$ and $\widehat{G}_1^{\rm inst}$ are just $\widehat{F}_0^{\rm inst}$ and $\widehat{F}_1^{\rm inst}$.\\
\indent Now the second part of (\ref{eq:JWScal}) can be written as
\be
\ba
&\sum_{g=0}\sum_{\bf d}\sum_{w=np}\frac{(-)^{np\bf B\cdot\bf d}}{np}\re^{-nq\bf d\cdot\bf t}\left[
n_0^{\bf d}\left(\frac{6+2n^2\pi^2q^2+6nq\mathbf{d} \cdot \mathbf{t} +3n^2q^2({\bf d}\cdot{\bf t})^2}{24n^2q^2\pi^2}\right)+n_1^{\bf d}\right]\\
=&\frac{1}{2\pi\hbar}\left(\widehat{G}_0^{\rm inst}-t_i\frac{\partial}{\partial t_i} \widehat{G}_0^{\rm inst}+\frac{t_it_j}{2}\frac{\partial^2}{\partial t_i\partial t_j}\widehat{G}_0^{\rm inst}\right)+\frac{2\pi}{\hbar}\widehat{G}_1^{\rm inst}
\ea
\ee
Now we turn to the calculation of $J_{\rm WKB}$. The nonsingular part of ${t_i(\hbar) \over 2 \pi}   {\partial F^{\rm NS,inst}({\bf t}(\hbar), \hbar) \over \partial t_i}$ is
\be
\frac{1}{2\pi}\sum_{j_L, j_R,\bf d}\sum_{q\nmid w}
N^{{\bf d}}_{j_L, j_R}\frac{\sin\frac{\hbar w}{2}(2j_L+1)\sin\frac{\hbar w}{2}(2j_R+1)}{2 w^2 \sin^3\frac{\hbar w}{2}}\bigg|_{\hbar=2\pi\frac{p}{q}} \re^{-w {\bf d}\cdot{\bf  t}}(-w\bf d\cdot\bf t).
\ee
The nonsingular part of ${\hbar^2 \over 2 \pi} {\partial \over \partial \hbar} \left(  {F^{\rm NS,inst}({\bf t}(\hbar), \hbar) \over \hbar} \right)$ is
\be
\label{eq:Jwkb2}
\ba
&\frac{1}{2\pi}\sum_{j_L, j_R,\bf d} \sum_{q\nmid w}
N^{{\bf d}}_{j_L, j_R}\frac{\re^{-w\bf d\cdot \bf t}}{2w^2}\left[\hbar^2\frac{\partial}{\partial \hbar}\left(  \frac{\sin\frac{\hbar w}{2}(2j_L+1)\sin\frac{\hbar w}{2}(2j_R+1)}{\sin^3\frac{\hbar w}{2}}\right)\right]\Bigg|_{\hbar=2\pi\frac{p}{q}}\\
&+\frac{1}{2\pi}\sum_{j_L, j_R,\bf d} \sum_{w=nq}
N^{{\bf d}}_{j_L, j_R}\frac{(-)^{n\bf B\cdot\bf d}\re^{-nq\bf d\cdot \bf t}}{2w}\left(\frac{2\pi p}{q}\right)\left(\frac{1}{12}m_Lm_R(3-m_L^2-m_R^2)\right).
\ea
\ee
We define the following analogy from $F_1^{\rm NS,inst}$:
\be
\widehat{G}_1^{\rm NS,inst}=\sum_{j_L, j_R,\bf d} \sum_{w=nq}
N^{{\bf d}}_{j_L, j_R}\frac{(-)^{n\bf B\cdot\bf d}\re^{-w\bf d\cdot \bf t}}{24w}m_Lm_R(3-m_L^2-m_R^2),
\ee
Note when $\hbar=2\pi$, i.e. $p=q=1$, $\widehat{G}_1^{\rm NS,inst}$ is just $\widehat{F}_1^{\rm NS,inst}$. Then the second term in (\ref{eq:Jwkb2}) can be written as
\be
\frac{\hbar}{2\pi}\widehat{G}_1^{\rm NS,inst}.
\ee
Therefore, the total modified grand potential is
\be
\ba
&\mathsf{J}(\boldsymbol{\mu}, \boldsymbol{\xi},\hbar)= {1\over 12 \pi \hbar} a_{ijk} t_i(\hbar) t_j(\hbar) t_k(\hbar) + \left( {2 \pi b_i \over \hbar} + {\hbar b_i^{\rm NS} \over 2 \pi} \right) t_i(\hbar)\\
&+\frac{1}{2\pi\hbar}\left(\widehat{G}_0^{\rm inst}-t_i\frac{\partial}{\partial t_i} \widehat{G}_0^{\rm inst}+\frac{t_it_j}{2}\frac{\partial^2}{\partial t_i\partial t_j}\widehat{G}_0^{\rm inst}\right)+\frac{2\pi}{\hbar}\widehat{G}_1^{\rm inst}+\frac{\hbar}{2\pi}\widehat{G}_1^{\rm NS,inst}\\
&+\frac{1}{2\pi}\sum_{j_L, j_R,\bf d}\sum_{q\nmid w}
N^{{\bf d}}_{j_L, j_R}\frac{\sin\frac{\hbar w}{2}(2j_L+1)\sin\frac{\hbar w}{2}(2j_R+1)}{2 w^2 \sin^3\frac{\hbar w}{2}}\bigg|_{\hbar=2\pi\frac{p}{q}} \re^{-w {\bf d}\cdot{\bf  t}}(-w\bf d\cdot\bf t)\\
&+\frac{1}{2\pi}\sum_{j_L, j_R,\bf d} \sum_{q\nmid w}
N^{{\bf d}}_{j_L, j_R}\frac{\re^{-w\bf d\cdot \bf t}}{2w^2}\left[\hbar^2\frac{\partial}{\partial \hbar}\left(  \frac{\sin\frac{\hbar w}{2}(2j_L+1)\sin\frac{\hbar w}{2}(2j_R+1)}{\sin^3\frac{\hbar w}{2}}\right)\right]\Bigg|_{\hbar=2\pi\frac{p}{q}}\\
&+\sum_{p\nmid w}d_w\left(2\pi\frac{p}{q}\right)(-)^{w\bf B\cdot \bf d}\re^{-\frac{q}{p}w\bf d\cdot\bf t}+ A({\boldsymbol \xi}, \hbar).
\ea
\ee
A more compact expression can be obtained if we introduce the following analogies:
\be\label{eq:g0}
\widehat{G}_0^{\rm}=\frac{1}{6}a_{ijk}t_it_jt_k+\widehat{G}_0^{\rm inst},
\ee
\be
\widehat{G}_1^{\rm}=b_it_i+\widehat{G}_1^{\rm inst}.
\ee
\be
\widehat{G}_1^{\rm NS}=b^{\rm NS}_it_i+\widehat{G}_1^{\rm NS,inst},
\ee
Then the total modified grand potential becomes
\be
\label{eq:totalJ2}
\ba
&\mathsf{J}(\boldsymbol{\mu}, \boldsymbol{\xi},\hbar)=\frac{1}{2\pi\hbar}\left(\widehat{G}_0-t_i\frac{\partial}{\partial t_i} \widehat{G}_0+\frac{t_it_j}{2}\frac{\partial^2}{\partial t_i\partial t_j}\widehat{G}_0\right)+\frac{2\pi}{\hbar}\widehat{G}_1+\frac{\hbar}{2\pi}\widehat{G}_1^{\rm NS}\\
&+\frac{1}{2\pi}\sum_{j_L, j_R,\bf d}\sum_{q\nmid w}
N^{{\bf d}}_{j_L, j_R}\frac{\sin\frac{\hbar w}{2}(2j_L+1)\sin\frac{\hbar w}{2}(2j_R+1)}{2 w^2 \sin^3\frac{\hbar w}{2}}\bigg|_{\hbar=2\pi\frac{p}{q}} \re^{-w {\bf d}\cdot{\bf  t}}(-w\bf d\cdot\bf t)\\
&+\frac{1}{2\pi}\sum_{j_L, j_R,\bf d} \sum_{q\nmid w}
N^{{\bf d}}_{j_L, j_R}\frac{\re^{-w\bf d\cdot \bf t}}{2w^2}\left[\hbar^2\frac{\partial}{\partial \hbar}\left(  \frac{\sin\frac{\hbar w}{2}(2j_L+1)\sin\frac{\hbar w}{2}(2j_R+1)}{\sin^3\frac{\hbar w}{2}}\right)\right]\Bigg|_{\hbar=2\pi\frac{p}{q}}\\
&+\sum_{p\nmid w}d_w\left(2\pi\frac{p}{q}\right)(-)^{w\bf B\cdot \bf d}\re^{-\frac{q}{p}w\bf d\cdot\bf t}+ A({\boldsymbol \xi}, \hbar).
\ea
\ee
We give some comments on (\ref{eq:totalJ2}). The third line does not appear in the generalized Riemann theta function, therefore it never affects the quantization condition or energy spectrum. Neither does $A({\boldsymbol \xi}, \hbar)$ affect the quantization condition.\\
\indent  The first term in the fourth line is the only obstruction for $\Theta({\boldsymbol \kappa}; \hbar)$ to be a genuine Riemann theta function. In fact, this term is also the main obstacle to establish the equivalence with our quantization condition for general rational Planck constant. Note for $p=1$, which is $\hbar=2\pi/k$, this term disappears. In these special values, the quantum Riemann theta function is indeed a Riemann theta function and we can establish an equivalence between GHM conjecture and NS quantization conditions.\\
\indent Now we can write down the generalized theta function for $\hbar=2\pi/k$, i.e. $p=1$, $q=k$.
\be
\label{theta-2pik}
\Theta({\boldsymbol \kappa}; \hbar=2 \pi/k)=\sum_{ {\bf n} \in \IZ^{g_{\Sigma}}} \exp\left[ \pi \ri {\bf n} \tau {\bf n} + 2\pi \ri {\bf n} \cdot \boldsymbol {\upsilon} -{ 2\pi^2\ri \over {3\hbar}} a_{ijk} C_{li} C_{mj} C_{pk} n_l n_m n_p \right].
\ee
In this equation, $\tau$ is a $g_{\Sigma}\times g_{\Sigma}$ matrix given by
\be
\label{tau-xi}
\tau_{l m}= {i\over\hbar} C_{lj} C_{ mk} {\partial^2 \widehat G_0 \over \partial t_j\partial t_k}, \qquad l,m=1, \cdots, g_{\Sigma}.
\ee
This is a generalization of the $\tau$ matrix of the mirror curve. It is a symmetric matrix satisfying
\be
{\rm Im}(\tau) >0.
\ee
In addition, the vector $\boldsymbol {\upsilon}$ appearing in (\ref{theta-2pik}) has components
\be
\label{ups}
\ba
&\upsilon_m= {C_{mj} \over 2 \pi\hbar} \left\{  {\partial^2 \widehat G_0 \over \partial t_j  \partial t_k} t_k  -
{\partial \widehat G_0 \over \partial t_j } \right\} +  C_{mj} \left( \frac{2\pi}{\hbar}b_j +\frac{\hbar}{2\pi}  b_j^{\text{NS}}\right)\\
&+\frac{C_{mj}}{2\pi}\sum_{j_L, j_R,\bf d}\sum_{k\nmid w}
N^{{\bf d}}_{j_L, j_R}\frac{\sin\frac{\hbar w}{2}(2j_L+1)\sin\frac{\hbar w}{2}(2j_R+1)}{2 w^2 \sin^3\frac{\hbar w}{2}}\re^{-w {\bf d}\cdot{\bf  t}}(-wd_j),
\ea
\ee
where $m=1, \cdots, g_{\Sigma}$. Surprisingly $\upsilon_m$ has a direct relation with $g_{\Sigma}$ non-perturbative volumes which we will give in the following section.
\subsection{Generalized GHM Conjecture}\label{sec:ghm3}
It is not clear before how the GHM conjecture and NS quantization scheme are related. The most significant difference lies in the number of the quantization conditions. In GHM quantization, there is one single quantization condition for the mirror curve of arbitrary Calabi-Yau threefold, the spectra of the quantum system lie on the divisor determined by the single quantization condition. While in NS quantization, there are $g$ quantization conditions for a mirror curve of genus $g$, and the spectra can be directly solved from the $g$ quantization conditions. An explanation of this difference was given in \cite{Codesido:2015dia}, in which the quantum spectral curve is not regarded as a quantum integrable system but a normal one-dimensional quantum mechanics with a potential. This undoubtedly loses much information of the integrability of the quantum spectral curve. After all, it is proved that for every toric Calabi-Yau, its mirror curve corresponds to an integrable system \cite{Goncharov:2011hp}\cite{Franco:2015rnr}.

In this subsection, we propose a simple method to exhaust the integrability of the quantum mirror curve by introduce a set of constant integral $\mathbf{r}$ fields in GHM conjecture. The $\mathbf{r}$ fields characterize the phase-changing of complex moduli, which results in the equivalent mirror curve but nonequivalent spectral determinants. In general, we find that there exist usually more than one non-equivalent $\mathbf{r}$ fields for a local toric Calabi-Yau to make the discrete spectra satisfy the vanishing of spectral determinants. The existence of more than one non-equivalent $\mathbf{r}$ field should not be very surprising. This in fact has been indicated in the previous study on the quantization of the mirror curve of resolved $\IC^3/\IZ_5$ orbifold. In \cite{Franco:2015rnr}, it was noticed when promoting the mirror curve to quantum mechanics, they obtained two different Baxter equations and two generalized spectral determinants,
\be
\Xi(H_2,-H_1)=0,
\ee
and
\be
\Xi(e^{\frac{6\pi\ri}{5}}H_2,e^{\frac{3\pi\ri}{5}}H_1)=0,
\ee
where $H_1,H_2$ are the Hamiltonians (Batyrev coordinates). This in fact shows that there exist at least two non-equivalent $\mathbf{r}$ fields for resolved $\IC^3/\IZ_5$ orbifold. We will have a detailed study on this model in the next section, where we will see that up to complex conjugation, there are indeed two non-equivalent $\mathbf{r}$ fields, each of which result in one of the above generalized spectral determinants.

Now we state the details of $\mathbf{r}$ field. The $\mathbf{r}$ field characterizes the phase-changing of complex moduli in the way that when one makes a transformation for the Batyrev coordinates
\be
(z_1,\dots,z_n)\rightarrow(z_1e^{r_1\pi\ri},\dots,z_ne^{r_n\pi\ri}),
\ee
equivalently, we have the following translation on the K\"{a}hler parameters:
\be
\mathbf{t}\rightarrow\mathbf{t}+\pi\ri\mathbf{r}.
\ee
This makes the effect of $\mathbf{r} $ field just like $\mathbf{B} $ field. For some specific choices of $\mathbf{r} $ fields, we have the generalized GHM conjecture as
\be
\Xi(\mathbf{t}+\pi\ri\mathbf{r},\hbar)=0.
\ee
Or equivalently,
\be
\Theta(\mathbf{t}+\pi\ri\mathbf{r},\hbar)=0.
\ee
Because the definition of the generalized spectral determinant and quantum Riemann theta function involves infinite sum, it is easy to see that different choices of $\mathbf{r} $ fields may result in the same functions. We define non-equivalent $\mathbf{r} $ fields as those which produce non-equivalent quantum Riemann theta functions (generalized spectral determinant).\\
\indent For a mirror curve $\Sigma$, we denote the number of non-equivalent $\mathbf{r}$ fields as $w_{\Sigma}$. Then we have the following conjecture: \\\newline
{ \noindent\emph{The spectra of quantum mirror curve are solved by the simultaneous equations:
\be
\bigg\{\Theta(\mathbf{t}+\ri\pi\mathbf{r}^a,\hbar)=0,\ a=1,\cdots,w_{\Sigma}.\bigg\}
\ee}}
One can view these equations as the quantization conditions of the quantum mirror curve. We believe these quantization conditions, though quite different the exact Nekrasov-Shatashvili quantization conditions, also exhaust the integrability of the mirror curve. Besides, these quantization conditions also have clear geometric meaning. Considering the solution of each equation is actually the divisor of each quantum Riemann theta function, we can also say\\\newline
{ \noindent\emph{The spectra of quantum mirror curve are the intersections of the divisors of all non-equivalent quantum Riemann theta functions.
}}\newline\\
Here we should clarify some facts on the theta divisors in GHM quantization scheme. Naively, one may suspect the generalized spectral determinant and quantum Riemann Theta function may vanish not only at the physical divisors, but also at some unphysical divisors which contain no physical spectra. But this is not the case. Because if the generalized spectral determinant defined from topological string contain more zeros than the one defined directly from quantum mechanics, the two sides of GHM conjecture will be different functions. As we mentioned before, the GHM conjecture contains more information than just the spectra. There are plenty of independent evidences showing the two sides are indeed the same function. For example, see the checks on the large $N$ limit of spectral trace in \cite{Grassi:2014zfa}\cite{Codesido:2015dia}.\footnote{We thank Marcos Mari\~no for clarifying this to us.}

For all known examples, we always have $w_{\Sigma}\geq g_{\Sigma}$. This of course raises the issue such as why more than $g_\Sigma$ divisors should intersect in the same points or whether part of all non-equivalent quantum Riemann theta functions can already determine the full spectra. These questions are still not perfectly clear because we are still in lack of a proof of the original GHM conjecture. At current stage, however, we still hope to address these issues with specific examples in the next section.

We also provide a physical explanation on our generalization. The standard integrable system, especially in the literature of NS quantization conditions, describe a relativistic quantum system with $g$ particles. For a stable system, each particle must have specific energy. For a quantum system, their energy spectrum is discrete, then there are $g$ quantization conditions. In the original version of GHM quantization scheme for higher-genus mirror curves \cite{Codesido:2015dia}, when quantizing one complex modulus, they in fact are studying the one-particle energy spectrum, and fix the energy of all other particles, which means all other particles are set as background.

However, for a $g$ particle system, there may be some symmetry, and the $\mathbf{r}^{a}$ field here characterize the symmetry. Different $\mathbf{r}^a$ fields here is only differed by some transformation in the symmetry. To calculate the physical quantities, in particular, the energy spectrum we have studied here, we should solve all the nonequivalent spectral determinants with nonequivalent $\mathbf{r}$ fields. Then we can obtain the physical energy spectrum, which is equivalent to the exact NS spectrum.
\section{Equivalence between the Two Quantization Schemes}
\label{sec:equivalence}
In this section, we study the relation between the exact Nekrasov-Shatashvili quantization conditions and generalized GHM conjecture. First, we propose a method to check the equivalence for generic Planck constant $\hbar$. To be specific, we want to test if the spectra solved by the exact NS quantization conditions lie on the divisors of each quantum theta function. This can be ascribed to a set of novel identities for the topological string partition function and NS free energy. As long as one has the refined BPS invariants, one can check the identities to arbitrary high orders and therefore verify the equivalence to arbitrary orders. Although the refined BPS invariants of toric Calabi-Yau can be calculated by the topological vertex technique to arbitrary orders, we did not find a simple way to prove the identities directly. In fact, these identities require infinite constrains among refined BPS invariants.\\
\indent Albeit the difficulty to prove the equivalence at generic Planck constant for arbitrary toric Calabi-Yau, it is possible to prove it at $\hbar=2\pi/k$ for some cases. As we mentioned before, at $\hbar=2\pi/k$ the quantum Riemann theta functions become true Riemann theta functions (with characteristic). So it is possible to directly prove the identities using the properties of Riemann theta functions.
\subsection{Generic Planck Constant $\hbarOrd\mkern-9mu{h}$}\label{sec:genericcheck}
In the original GHM conjecture, there is only one quantization condition, realizing by the vanishing of quantum Riemann theta function. Since GHM conjecture and NS quantization conditions are two quantization schemes for the same mirror curve and both should be consistent with the numerical method, naturally one expect the discrete spectra solved by the exact NS quantization conditions should lie on the theta divisor of GHM quantum Riemann theta function. That is to say the GHM conjecture has solutions at the spectra determined by the NS quantization conditions,
\begin{equation}\label{qc}
\text{Vol}_i=2\pi\hbar \left(n_i+\frac{1}{2}\right),\ i=1,\cdots,g,
\end{equation}
where the phase volume is defined as the derivatives of non-perturbative NS free energy with respect to the true K\"{a}hler moduli
\begin{small}
\begin{equation}\label{vol11}
 \begin{split}
\text{Vol}_j=\hbar C_{ji}\left(\frac{1}{2\hbar}a_{ilm}{t}_lt_m+b_i^{\text{NS}}\hbar+b_i^{\text{NS}} \frac{4\pi^2}{\hbar}+\frac{\partial }{\partial t_i}F_{\text{NS}}^{inst}(\mathbf{t}+\ri\pi\mathbf{B},\hbar)+\frac{\hbar}{2\pi}\frac{\partial }{\partial t_i}F_{\text{NS}}^{inst}\left(\frac{2\pi}{\hbar}\mathbf{t}+\ri\pi \mathbf{B},\frac{4\pi^2}{\hbar}\right)\right).
 \end{split}
 \end{equation}
 \end{small}
However, as we have mentioned, the K\"{a}hler parameters in the GHM conjecture are not the same with the K\"{a}hler parameters in NS quantization, but with a shift of a constant integral vector, which we call the $\mathbf{r}$ field.
To make the GHM quantization scheme consistent with the spectrum of the quantum mirror curve, this shift $\mathbf{r}$ should be chosen carefully. In this subsection, we shall systematically study what kind of $\bf{r}$ fields are admissible and how to classify these $\bf{r}$ fields.

In the examples studied in \cite{Grassi:2014zfa}\cite{Franco:2015rnr}, they always transform the mirror curve to a canonical form for some $x_i$, make others as arbitrary parameters and then quantize that $x_i$. This somehow makes the question difficult to study in general. Here we stick the form of quantum Riemann theta function of GHM, and focus on the shift of $\mathbf{t}$. Since the K\"{a}hler moduli $\mathbf{t}$ in NS quantization scheme are fixed, all we need to do is to study the shift of $\mathbf{t}$ between GHM and NS quantization scheme.\\

\indent The primary condition for $\mathbf{r}$ field is that it must be a $\mathbf{B}$ field, which means they share the same parity,
\be\label{r}
\mathbf{r}\equiv\mathbf{B}\,(\mathrm{Mod}\ 2\mathbb{Z}^s).
\ee
This is requested by the reality of the energy of quantum mirror curve and the consistence between the two quantization scheme as we will see later.\\
\indent The modified grand potential in GHM conjecture is given in Section \ref{sec:ghm1}, with a shift of $\mathbf{r}$ field, it becomes
\begin{equation}\label{qcsd}
\begin{split}
&J(\mathbf{t}+\ri\pi\mathbf{r},\hbar)=\frac{1}{12\pi \hbar}a_{ijk}(t_i+\ri\pi r_i)(t_j+\ri\pi r_j)(t_k+\ri\pi r_k)+\left(\frac{2\pi}{\hbar} b_i +\frac{\hbar b_i^{\text{NS}}}{2\pi}\right)(t_i+\ri\pi r_i)+A(\hbar)\\
&+\frac{t_i+\ri\pi r_i}{2\pi}\frac{\partial F_{\text{NS}}^{inst}(\mathbf{t}+\ri\pi\mathbf{r},\hbar)}{\partial t_i} +\frac{\hbar^2}{2\pi} \frac{\partial}{\partial \hbar}\left(\frac{F_{\text{NS}}^{inst}(\mathbf{t}+\ri\pi\mathbf{r},\hbar)}{\hbar}\right)+F_{\text{GV}}^{inst}\left( \frac{2\pi}{\hbar}(\mathbf{t}+\ri\pi \mathbf{r})+\pi \ri \mathbf{B},\frac{4\pi^2}{\hbar}\right).
 \end{split}
 \end{equation}
Then the quantization condition is encoded in the vanishing loci:
\be\label{def0}
\sum_{\mathbf{n}\in \mathbb{Z}^g}\exp{\left(J(t_i+2\pi \ri n_jC_{ji}+\ri\pi r_i,\hbar)\right)}=0.
\ee
Since $\mathbf{B}$ and $\mathbf{r}$ share the same parity, we have the same perturbative instanton part
\be
\frac{\partial }{\partial t_i}F_{\text{NS}}^{inst}(\mathbf{t}+\ri\pi\mathbf{B},\hbar),
\ee
in both (\ref{vol11}) and (\ref{def0}). Substitute the NS quantization condition (\ref{qc}) into the modified grand potential (\ref{qcsd}) and cancel out the above perturbative instanton part, we have
\begin{equation}\label{qcsdeff}
\begin{split}
J^{eff}(&t_i+2\pi \ri n_jC_{ji}+\ri\pi r_i,\hbar)=J^p(t_i+2\pi \ri n_jC_{ji}+\ri\pi r_i,\hbar)+F_{\text{GV}}^{inst}\left( \frac{2\pi}{\hbar}(\mathbf{t}+\ri\pi \mathbf{r})+\pi \ri \mathbf{B},\frac{4\pi^2}{\hbar}\right)\\
&+\ri n_i \pi
-\ri n_j C_{ji}\left(\frac{1}{2\hbar}a_{ilm}{t}_lt_m+b_i^{\text{NS}}\hbar+b_i^{\text{NS}} \frac{4\pi^2}{\hbar}+\frac{\hbar}{2\pi}\frac{\partial }{\partial t_i}F_{\text{NS}}^{inst}\left(\frac{2\pi}{\hbar}\mathbf{t}+\ri\pi \mathbf{B},\frac{4\pi^2}{\hbar}\right)\right), \\
 \end{split}
 \end{equation}
where
\be
J^p(t,\hbar)=\frac{1}{12\pi \hbar}a_{ijk}t_it_jt_k+\left(\frac{2\pi}{\hbar} b_i +\frac{\hbar b_i^{\text{NS}}}{2\pi}\right)t_i.
\ee
The superscript $eff$ means we have dropped those terms irrelevant to the quantum Riemann theta function. Now the quantization condition of GHM becomes
\be\label{def1}
\sum_{\mathbf{n}\in \mathbb{Z}^g}\exp{\left(J^{eff}(t_i+2\pi i n_jC_{ji}+\ri\pi r_i,\hbar)\right)}=0.
\ee
The terms $\frac{1}{2\hbar}a_{lim}{t}_lt_m$ and  $b_i^{\text{NS}}\hbar$ in the volumes cancel with the same terms in $J^p$, thus $J^{eff}$ does not contain any $t$ square term. This is a well-defined expression and we can expand equation (\ref{def1}) as a series of $\exp{\frac{2\pi}{\hbar}t_i}$ and check this equation to arbitrary orders. We find that for appropriately chosen $\mathbf{r}$ field, equation (\ref{def1}) is an identity, regardless of the K\"{a}hler parameters. Note this identity is far stronger than the requirement that the spectra of exact NS quantization conditions lie on the theta divisor of GHM.

Now we want to give a close form of (\ref{def1}). We define the unrefined topological string free energy as\footnote{Here we modified the definition of the instanton part of the unrefined free energy by adding a $\mathbf{B}$ field to the K\"ahler parameters $\mathbf{t}$, or equivalently add a 1 to one of $\tau_1,\tau_2$ in the refined topological string. We also add a factor $4\pi^2b_i^{\text{NS}}t_i$ in the genus zero part $F_0$ of the free energy. This make senses since we add this factor to both NS and unrefined topological string.  }
\be\label{unrefnew}
F_{\text{unref}}(\mathbf{t},\hbar)=\frac{1}{6\hbar^2}a_{ijk}t_it_jt_k+b_i t_i+\frac{4\pi^2}{\hbar^2}b_i^{\text{NS}}t_i +F_{\text{GV}}^{inst}(\mathbf{t}+\ri\pi \mathbf{B},\hbar).
\ee
Then
\be
F_{\text{unref}}\left(\frac{2\pi}{\hbar}\mathbf{t},\frac{4\pi^2}{\hbar}\right)=\frac{1}{12\pi\hbar}a_{ijk}t_it_jt_k+\frac{2\pi}{\hbar}b_i t_i +\frac{\hbar b_i^{\text{NS}}}{2\pi}t_i+F_{\text{GV}}^{inst}\left(\frac{2\pi}{\hbar}\mathbf{t}+\ri\pi\mathbf{B},\frac{4\pi^2}{\hbar}\right).
\ee
Finally, we have
\begin{small}
\be\label{qcsd2}
J^{eff}_X(t_i+2\pi \ri n_jC_{ji}+\ri\pi r_i,\hbar)=\sum_{i=1}^{g}n_i \pi\ri+ F_{\text{unref}}\left(\frac{2\pi}{\hbar}(\mathbf{t}+2\pi \ri \mathbf{n}\cdot\mathbf{C}+\ri\pi\mathbf{r}),\frac{4\pi^2}{\hbar}\right)
 -\ri n_jC_{ji}\frac{\hbar}{2\pi}\frac{\partial }{\partial t_i}F_{\text{NS}}\left(\frac{2\pi}{\hbar}\mathbf{t},\frac{4\pi^2}{\hbar}\right),
\ee
\end{small}
and (\ref{def1}) indicates a remarkable relation between unrefined and NS free energy of topological string
\be\label{conjecture0}
 \sum_{\mathbf{n}\in\mathbb{Z}^g}\exp\left(i\sum_{i=1}^{g}n_i \pi+F_{\text{unref}}\left(\frac{2\pi}{\hbar}(\mathbf{t}+2\pi \ri \mathbf{n}\cdot\mathbf{C}+\ri\pi\mathbf{r}),\frac{4\pi^2}{\hbar}\right)
 -\ri n_jC_{ji}\frac{\hbar}{2\pi}\frac{\partial }{\partial t_i}F_{\text{NS}}\left(\frac{2\pi}{\hbar}\mathbf{t},\frac{4\pi^2}{\hbar}\right)\right)=0,
\ee
or equivalently, via S-duality $\mathbf{t}\rightarrow 2\pi\mathbf{t}/\hbar, \hbar\rightarrow 4\pi^2/\hbar$, we have
\be\label{conjecture}
\sum_{\mathbf{n}\in\mathbb{Z}^g}\exp\left(i\sum_{i=1}^{g}n_i \pi+F_{\text{unref}}\left(\mathbf{t}+ \ri\hbar \mathbf{n}\cdot\mathbf{C}+\frac{1}{2}\ri\hbar\mathbf{r},\hbar\right)
 -\ri n_jC_{ji}\frac{\partial }{\partial t_i}F_{\text{NS}}\left(\mathbf{t},\hbar\right)\right)=0.
\ee
In the last step, we transform the non-perturbative to the perturbative. This is the most convenient form to check its validity. It is also noteworthy that to check this identity, we need not the knowledge of mirror map, since we directly expand this formula with respect to $Q_i=\exp{t_i}$. The coefficients are normally some complicated expression of $\hbar$ and refined BPS invariants. The identity holds for generic $\hbar$ owing to infinite series of constraints among the refined BPS invariants.\\
\indent After establishing the above identity, we now consider the properties of $\mathbf{r}$ field. The primary requirement for $\mathbf{r}$ field is that it must be a representative of the $\mathbf{B}$ field. However, not arbitrary representative of the $\mathbf{B}$ field can make the identity (\ref{conjecture}) hold. We want to find out all possible $\mathbf{r}$ fields which make the identity (\ref{conjecture}) hold. Apparently, because of the summation on $\mathbb{Z}^g$ in the definition of the quantum Riemann theta function, different values of $\mathbf{r}$ field could result in the same quantum Riemann theta function. \emph{We define the $\mathbf{r}$ fields as equivalent if they result in the same quantum Riemann theta function or generalized spectral determinant, and non-equivalent if they do not.}

In general, we find that there always are only finite non-equivalent $\mathbf{r}$ fields for an arbitrary local toric Calabi-Yau threefold. Denote the number of non-equivalent $\mathbf{r}$ fields for mirror curve $\Sigma$ as $w_{\Sigma}$, we find $w_{\Sigma}$ is always no less than the genus of the mirror curve. There is no preference on the nonequivalent $\mathbf{r}$ fields. For every $\mathbf{r}^a$, the identity (\ref{conjecture}) guarantees that the discrete spectra solved by the exact NS quantization conditions lie on the theta divisor of quantum Riemann theta function $\Theta(\mathbf{t}+\ri\pi\mathbf{r}^a,\hbar)$. This leads to the following equivalence between GHM and NS quantization schemes:\\\newline
\emph{For the mirror curve $\Sigma$ of an arbitrary toric Calabi-Yau threefold $X$ with K\"{a}hler moduli $\mathbf{t}$, suppose the non-perturbative phase volumes of quantum mirror curve in Nekrasov-Shatashvili quantization are denoted as $\mathrm{Vol}_i(\mathbf{t},\hbar)$, $i=1,\dots,g_{\Sigma}$, and the quantum Riemann theta function defined by Grassi-Hatsuda-Mari\~no is denoted as $\Theta(\mathbf{t},\hbar)$, then there exist a set of constant integral vectors $\mathbf{r}^a$, $a=1,\cdots,w_{\Sigma}$, such that the intersections of the theta divisors of all $w_\Sigma$ quantum Riemann theta functions $\Theta(\mathbf{t}+\ri\pi\mathbf{r}^a,\hbar)$ coincide with the spectra determined by the exact Nekrasov-Shatashvili quantization conditions.
\be\label{equiv2}
\bigg\{\Theta(\mathbf{t}+\ri\pi\mathbf{r}^a,\hbar)=0,\ a=1,\cdots,w_{\Sigma}.\bigg\} \Leftrightarrow\left\{\mathrm{Vol}_i(\mathbf{t},\hbar)=2\pi\hbar \left(n_i+\frac{1}{2}\right),\ i=1,\cdots,g_{\Sigma}.\right\}
\ee}
\newline
This indicates the quantum Riemann theta function equipped with all non-equivalent $\mathbf{r}^a$ fields can exhaust the full quantum integrability of mirror curve just like the exact NS quantization conditions.

Let us say a few more things on the non-equivalent $\mathbf{r}$ fields. As we will see in the next section with specific examples, the number of non-equivalent $\mathbf{r}$ fields $w_\Sigma$ are normally larger than the genus of mirror curve $g$. Since the theta divisors are in moduli space $\mathbb{C}^g$, it is quite nontrivial that more than $g$ different theta should intersect on the same discrete points. Actually, this is just as nontrivial as the identity (\ref{conjecture}) holds for all non-equivalent $\mathbf{r}$ fields. One may suspect whether the $w_\Sigma$ non-equivalent $\mathbf{r}$ fields are indeed non-equivalent, as we will show in the next section with some examples, the identities with non-equivalent $\mathbf{r}$ fields indeed may result in different constraints for the refined BPS invariants.
\subsection{Proof at $\hbarOrd\mkern-9mu{h}=2\pi/k$}\label{sec:eq2}
The identity (\ref{conjecture}) seems rather difficult to prove for generic Planck constants. However, at $\hbar=2\pi/k$, the problems will be significantly simplified. As we have seen in equation (\ref{theta-2pik}), the quantum Riemann theta function becomes a genuine Riemann theta function at $\hbar=2\pi/k$. In this section, as a main step to prove the identity (\ref{conjecture}) at $\hbar=2\pi/k$, we establish the following relation between the non-perturbative volumes in NS quantization and the elliptic parameters in the Riemann theta function (\ref{theta-2pik}),
\be\label{eq:relation}
\upsilon_m\sim\frac{\mathrm{Vol}_i}{2\pi\hbar}.
\ee
$\sim$ means roughly, which we will amplify later. Read from equation (\ref{ups}), we have
\be
\label{vm}
\ba
&\upsilon_m= {C_{mj} \over 2 \pi\hbar} \left\{  {\partial^2 \widehat G_0 \over \partial t_j  \partial t_k} t_k  -
{\partial \widehat G_0 \over \partial t_j } \right\} +  C_{mj} \left( \frac{2\pi}{\hbar}b_j +\frac{\hbar}{2\pi}  b_j^{\text{NS}}\right)\\
&+\frac{C_{mj}}{2\pi}\sum_{j_L, j_R,\bf d}\sum_{k\nmid w}
N^{{\bf d}}_{j_L, j_R}\frac{\sin\frac{\hbar w}{2}(2j_L+1)\sin\frac{\hbar w}{2}(2j_R+1)}{2 w^2 \sin^3\frac{\hbar w}{2}}\re^{-w {\bf d}\cdot{\bf  t}}(-wd_j),
\ea
\ee
where $m=1, \cdots, g_\Sigma$. Note $\upsilon_m$ contains $b_j$ while the volumes do not.

As we have emphasized in the last subsection, there is a shift between the K\"{a}hler parameters in $\upsilon_m$ and those in volumes. Denote the K\"{a}hler parameters in $\upsilon_m$ as $t_i$, then the K\"{a}hler parameters in volumes becomes $ t_i-\ri\pi r_i$. From (\ref{eq:f}), we can see after pole cancellation, the remaining part of $\widehat{f}_i({\widehat{\bf t}}, \hbar)$ becomes
\be
\sum_{j_L, j_R,\bf d}\sum_{k\nmid w}
N^{{\bf d}}_{j_L, j_R}\frac{\sin\frac{\hbar w}{2}(2j_L+1)\sin\frac{\hbar w}{2}(2j_R+1)}{2 w^2 \sin^3\frac{\hbar w}{2}}\re^{-w {\bf d}\cdot{\bf  t}}(-wd_i),
\ee
in which we have used the property
\be
\label{rprop}
(-1)^{2j_L + 2 j_R-1}= (-1)^{{\bf r} \cdot {\bf d}},\quad a=1,\cdots,w_{\Sigma}.
\ee
From (\ref{eq:fdual}), we can see the first line disappears for $p=1$. Thus the remaining part of $\widehat{f}_i(\frac{2\pi\widehat{\bf t}}{\hbar}, \frac{4\pi^2}{\hbar})$ becomes
\be\label{eq:1}
\sum_{j_L, j_R,\bf d}\sum_{w=np}N^{{\bf d}}_{j_L, j_R}\frac{(-)^{np{\bf B}\cdot{\bf d}}}{2n^2p^2}\re^{-nq {\bf d}\cdot{\bf  t}}(-d_i)\bigg(-\frac{m_Lm_R(1+nq{\bf d}\cdot({\bf t}-\ri\pi r))}{\pi \frac{q}{p}}\bigg)
\ee
Remind our previous definition (\ref{eq:g0inst}), then (\ref{eq:1}) can be written as
\be
\frac{1}{\hbar}\left(t_j\frac{\partial^2}{\partial t_j\partial t_i}\widehat{G}_0^{\rm inst}-\frac{\partial}{\partial t_i}\widehat{G}_0^{\rm inst}-\ri\pi r_j\frac{\partial^2}{\partial t_j\partial t_i}\widehat{G}_0^{\rm inst}\right)
\ee
Combining the polynomial part, we have
\be\label{eq:Voli1}
\ba
\mathrm{Vol}_m/\hbar=&C_{mi}\Bigg(\frac{1}{2\hbar}a_{ijk}(t_j-\pi\ri r_j)(t_k-\pi\ri r_k)+b_i^{\mathrm{NS}}\hbar+\frac{1}{\hbar}\left(t_j\frac{\partial^2}{\partial t_j\partial t_i}\widehat{G}_0^{\rm inst}-\frac{\partial}{\partial t_i}\widehat{G}_0^{\rm inst}-\ri\pi r_j\frac{\partial^2}{\partial t_j\partial t_i}\widehat{G}_0^{\rm inst}\right)\\
&+b_i^{\text{NS}} \frac{4\pi^2}{\hbar}+\sum_{j_L, j_R,\bf d}\sum_{k\nmid w}
N^{{\bf d}}_{j_L, j_R}\frac{\sin\frac{\hbar w}{2}(2j_L+1)\sin\frac{\hbar w}{2}(2j_R+1)}{2 w^2 \sin^3\frac{\hbar w}{2}}\re^{-w {\bf d}\cdot{\bf  t}}(-wd_i)\Bigg).
\ea
\ee
Remind our previous definition (\ref{eq:g0}), then (\ref{eq:Voli1}) can be written as
\be\label{eq:Voli2}
\ba
\mathrm{Vol}_m/\hbar=&\,C_{mi}\Bigg(-\frac{a_{ijk}r_jr_k\pi^2}{2\hbar}+b_i^{\mathrm{NS}}\hbar+b_i^{\text{NS}} \frac{4\pi^2}{\hbar}+\frac{1}{\hbar}\left(t_j\frac{\partial^2}{\partial t_j\partial t_i}\widehat{G}_0-\frac{\partial}{\partial t_i}\widehat{G}_0-\ri\pi r_j\frac{\partial^2}{\partial t_j\partial t_i}\widehat{G}_0\right)\\
&+\sum_{j_L, j_R,\bf d}\sum_{k\nmid w}
N^{{\bf d}}_{j_L, j_R}\frac{\sin\frac{\hbar w}{2}(2j_L+1)\sin\frac{\hbar w}{2}(2j_R+1)}{2 w^2 \sin^3\frac{\hbar w}{2}}\re^{-w {\bf d}\cdot{\bf  t}}(-wd_i)\Bigg).
\ea
\ee
Now we can see the most terms in the above volumes have correspondences in ($\ref{vm}$). Compare (\ref{eq:Voli2}) with (\ref{vm}), using also the NS quantization conditions
\be
\mathrm{Vol}_m=(N_m+\frac{1}{2})2\pi\hbar,
\ee
we have the following relation
\be\label{vm1}
\upsilon_m=N_m+\frac{1}{2}+C_{mi}\left(\frac{2\pi}{\hbar}(b_i-b_i^{\text NS})+\frac{\pi}{4\hbar}a_{ijk}r_jr_k+\frac{\ri r_j}{2\hbar}\frac{\partial^2}{\partial t_j\partial t_i}\widehat{G}_0\right).
\ee
Now to prove the equivalence between the two quantization scheme at $\hbar=2\pi/k$, is just to prove the theta function
\be
\label{theta1-2pik}
\Theta(\mathbf{t}; \hbar=2 \pi/k)=\sum_{ {\bf n} \in \IZ^{g_\Sigma}} \exp\left[ \pi \ri {\bf n} \tau {\bf n} + 2\pi \ri {\bf n} \cdot \boldsymbol {\upsilon} -{ 2\pi^2\ri \over {3\hbar}} a_{ijk} C_{li} C_{mj} C_{pk} n_l n_m n_p \right]
\ee
vanishes with $\boldsymbol {\upsilon}$ defined in (\ref{vm1}) and $\tau$ defined as
\be
\label{tau-xi1}
\tau_{l m}= {i\over\hbar} C_{lj} C_{ mk} {\partial^2 \widehat G_0 \over \partial t_j\partial t_k}, \qquad l,m=1, \cdots, g_\Sigma.
\ee
For all known examples, the cubic terms in (\ref{theta1-2pik}) can reduce to quadratic or linear term. Let us assume this is true for general local toric Calabi-Yau. Then expression (\ref{theta1-2pik}) becomes a Riemann theta function with characteristics. In general, Riemann theta function with characteristics $\boldsymbol{\alpha}$, $\boldsymbol{\beta}$ is defined by
\be  \vartheta \left[\begin{array}{cc} \boldsymbol{\alpha}\\ \boldsymbol{\beta}
 \end{array}\right] \left({\boldsymbol{z}}, \Omega \right)= \sum_{ {\boldsymbol{n}} \in \IZ^g }e^{  \pi\ri  ({\boldsymbol{n}}  + {\boldsymbol{\alpha}})  \Omega
({\boldsymbol{n}}  + {\boldsymbol{\alpha}}) +2 \pi \ri\left( {\boldsymbol{n}} + {\boldsymbol{\alpha}} \right)\cdot ({\boldsymbol{z}}  + {\boldsymbol{\beta}})}.
\ee
See more details of Riemann theta function in Appendix A. The Riemann theta functions with half-period characteristics ($2\boldsymbol{\alpha}\in\mathbb{Z}^g$, $2\boldsymbol{\beta}\in\mathbb{Z}^g$) have the following property: if
\be
4\boldsymbol{\alpha}\cdot\boldsymbol{\beta}=1\ (\mathrm{Mod}\ 2),
\ee
then $\boldsymbol{z}=0$ is its zero point. With this fine property, we make the following statement: If the function (\ref{theta1-2pik}) can be written as the following Riemann theta function with characteristics,
\be  \vartheta \left[\begin{array}{cc} \boldsymbol{\alpha}\\ \boldsymbol{\beta}
 \end{array}\right] \left({\boldsymbol{0}}, \tau+S \right)= \sum_{ {\boldsymbol{n}} \in \IZ^g }e^{  \pi\ri  ({\boldsymbol{n}}  + {\boldsymbol{\alpha}})  (\tau+S)
({\boldsymbol{n}}  + {\boldsymbol{\alpha}}) +2 \pi \ri\left( {\boldsymbol{n}} + {\boldsymbol{\alpha}} \right)\cdot  {\boldsymbol{\beta}}}
\ee
such that $2\boldsymbol{\alpha}\in\mathbb{Z}^g$, $2\boldsymbol{\beta}\in\mathbb{Z}^g$ and
\be\label{4ab}
4\boldsymbol{\alpha}\cdot\boldsymbol{\beta}=1\ (\mathrm{Mod}\ 2),
\ee
then the quantum Riemann theta function (\ref{theta1-2pik}) vanishes at the spectra determined by the exact NS quantization conditions. This is a sufficient condition to prove the equivalence between the two quantization schemes at $\hbar=2\pi/k$.

Let us now examine when the function (\ref{theta1-2pik}) can be written as the above Riemann theta function with characteristics. Compare the exponential, we have
\be\label{5}
\ba
\pi \ri {\bf n} \tau {\bf n} + 2\pi \ri n_m & \left(\frac{1}{2}  +C_{mi}\left(k(b_i-b_i^{\text NS})+\frac{k}{8}a_{ijk}r_jr_k+\frac{\ri k r_j}{4\pi}\frac{\partial^2}{\partial t_j\partial t_i}\widehat{G}_0\right)\right) -{ 2\pi^2\ri \over {3\hbar}} a_{ijk} C_{li} C_{mj} C_{pk} n_l n_m n_p\\
\simeq & \,\pi\ri  \({\boldsymbol{n}}  + {\boldsymbol{\alpha}}\)\cdot  \(\tau+S\)\cdot
\({\boldsymbol{n}}  + {\boldsymbol{\alpha}}\) +2 \pi \ri\left( {\boldsymbol{n}} + {\boldsymbol{\alpha}} \right)\cdot  {\boldsymbol{\beta}}\\
\simeq & \,\pi \ri \, {\bf n}\cdot \tau\cdot {\bf n}+\pi \ri\,  {\bf n}\cdot S\cdot {\bf n}+2\pi\ri\, {\bf n}\cdot\left((\tau+S)\cdot\boldsymbol{\alpha}+\boldsymbol{\beta}\right).
\ea
\ee
where $\simeq$ means up to $2\pi\ri$. Let us look at the terms that are not constants. The quadratic term ${\bf n} \tau {\bf n}$ cancels out. Remind the definition of $\tau$, we observe that only when
\be\label{rca}
\mathbf{r}=2C\cdot\boldsymbol{\alpha},
\ee
then term ${\bf n}\cdot\tau\cdot\boldsymbol{\alpha}$ can exactly cancels with the term
\be
\frac{\ri k r_j}{4\pi}\frac{\partial^2}{\partial t_j\partial t_i}\widehat{G}_0.
\ee
The remaining terms are just constants. $S$ usually comes from the reduction of the cubic terms in quantum Riemann theta functions.\\
\indent With the above discussion, we sketch a possible proof for the equivalence at $\hbar=2\pi/k$ for a general toric Calabi-Yau. Suppose we have known a $\mathbf{r}$ field for a toric Calabi-Yau $X$, we first check if there exists an $\boldsymbol{\alpha}$, ($2\boldsymbol{\alpha}\in\mathbb{Z}^g$) such that the criterion (\ref{rca}) is satisfied. Then we need to check if the cubic terms in quantum Riemann theta function (\ref{theta1-2pik}) can reduce to quadratic or linear terms. Third, we should use the equation (\ref{5}) to identify the $\boldsymbol{\beta}$. If $2\boldsymbol{\beta}\in\mathbb{Z}^g$, and condition (\ref{4ab}) is satisfied, then we rigorously proved for this $\mathbf{r}$ field, the identity (\ref{conjecture}) holds. This may look quite cumbersome, but in the next section, following this approach we will prove that for resolved $\IC^3/\IZ_5$ orbifold, identity (\ref{conjecture}) indeed holds for $\mathbf{r}=(1,-2)$ at $\hbar=2\pi/k$.

The proof here does not need the knowledge of refined BPS invariants. We call these cases $(X,\mathbf{r})$ as trivial cases. One should be warned that the $\mathbf{r}$ fields satisfying the criterion $\mathbf{r}=2C\cdot\boldsymbol{\alpha}$ are in fact not common. And the trivial cases are even rare. Normally, at $\hbar=2\pi/k$, even though the quantum Riemann theta function becomes a genuine Riemann theta function and only genus zero BPS invariants are relevant, the identity (\ref{conjecture}) is still nontrivial, and one can obtain some constraints on the genus-zero BPS invariants of the Calabi-Yau.

Let us say a few more things on the cubic $n$ terms in (\ref{theta1-2pik}). We conjecture these terms can always reduce to some quadratic or linear terms. But so far we do not have a proof on this. For local toric Calabi-Yau threefolds, the coefficients $a_{ijk} C_{li} C_{mj} C_{pk}$ can be interpreted as intersection numbers for appropriate divisors, therefore are integers by geometric meaning.\footnote{Note however $a_{ijk}$ themselves are normally not integers, see for example \cite{Klemm:2015iya}.} This is a necessary condition for the cubic terms reducing to quadratic or linear terms. Then for all known examples, we can always use the fact that for an integer $n$,
\be
n^3\equiv n\ (\mathrm{Mod }6),
\ee
and
\be
n^2\equiv n\ (\mathrm{Mod }2),
\ee
to reduce the cubic terms like $n_i^3$ to $n_i$ and $n_i^2n_j$ to $n_in_j$. However, in general, there may be cubic terms like $n_in_jn_k$ with $i\neq j\neq k$ for mirror curves with high genus and a large number of mass parameters. There seems no simple way to reduce such terms to quadratic or linear terms. We conjecture that by appropriately choosing basis for the moduli space, such terms will not appear and thus (\ref{theta1-2pik}) can still be equivalent to a genuine Riemann theta function. Indeed, we find this is true for all $SU(N)$ geometries with $m=0$. See our proof in Section \ref{sec:ex3}.
\subsection{Remarks on the Equivalence}\label{sec:rem}
In this section, we give some further clarification to make the equivalence (\ref{equiv2}) precise. From the definitions of generalization spectral determinant and quantum Riemann theta function in Section \ref{sec:ghm1}, it is obvious that they have a manifest symmetry $\mathbf{t}\rightarrow\mathbf{t}+2\pi\ri\mathbf{n}\cdot\mathbf{C}$. However, this symmetry is missing in the NS quantization scheme. One consequence of this difference is that the theta divisors are bi-periodic while the NS spectra are only singly periodic. Let us look at a simple example which is local $\mathbb{P}^2$ at $\hbar=2\pi$ \footnote{We thank the referee for pointing out this example.}. In this case, the quantum theta function has been calculated in \cite{Grassi:2014zfa},
 \be\label{p2theta}
    \Theta_{\mathbb{P}^2}(\mu, 2 \pi)= \vartheta_3\left( v, {9 \tau \over 4}\right),
    \ee
    with
    \be
 \label{taup2}
 \tau={2 \ri \over \pi} \partial_t^2 \widehat F_0(t),
 \ee
 \be
 \label{xip2}
 \xi= {3\over 4 \pi^2} \left( t \partial_t^2  \widehat F_0(t) -\partial_t  \widehat F_0(t)\right),
 \ee
 and
    \be
    v= \xi -\frac{3}{8}.
    \ee
Since this is a standard Jacobi theta function, the zeros can be written as
\be\label{p2ghm}
\xi -\frac{3}{8}=n+\frac{1}{2}+\left(m+\frac{1}{2}\right)\frac{9\tau}{4},\quad\quad n,m\in\mathbb{Z}.
\ee
On the other hand, the NS quantization condition in this case can be read from our previous calculation (\ref{eq:Voli2}), which is
\be\label{p2ns}
\xi -\frac{3}{8}=n+\frac{1}{2}+\frac{9\tau}{8}.
\ee
Apparently, the former equation has more solutions than the latter one. To conciliate this seeming disagreement, we need to consider the meaning of symmetry $\mathbf{t}\rightarrow\mathbf{t}+2\pi\ri\mathbf{n}\cdot\mathbf{C}$ in the NS quantization approach. The mirror curve remains the same under this translation, since the complex moduli only have a phase change of multiple $2\pi\ri$. For example, for local $\mathbb{P}^2$, $r=3$. Then under the translation $t\rightarrow t+6\pi\ri$, we have $z\rightarrow z\re^{-6\pi\ri}$ and $E\rightarrow E+2\pi\ri$. Obviously, the mirror curve
\be
\re^x+\re^p+\re^{-x-p}+z^{-1/3}=0
\ee
or quantum mechanics
\be
\re^x+\re^p+\re^{-x-p}=\re^E
\ee
remains the same. Therefore, it is natural to expect this symmetry should be kept in some way in the NS quantization scheme. In fact, we find that combining with this symmetry, the missing $\tau$ periodicity will emerge in the NS quantization conditions and Equation (\ref{p2ns}) will become exactly the same with Equation (\ref{p2ghm}). To see this, let
\be
t'=t+6\pi\ri m,\quad\quad m\in\mathbb{Z}.
\ee
Using the formulas in Equation (\ref{taup2}) and (\ref{xip2}), we have
\be
\tau(t')=\tau(t)-4m,\quad\quad m\in\mathbb{Z},
\ee
and
\be
\xi(t')=\xi(t)+\frac{9}{4}m\tau(t')+\frac{9m^2}{2}.
\ee
Suppose $t$ satisfy the NS quantization condition (\ref{p2ns}),
\be
\xi(t) -\frac{3}{8}=n+\frac{1}{2}+\frac{9\tau(t)}{8},
\ee
then
\be
\xi(t') -\frac{3}{8}=n+\frac{1}{2}+\left(m+\frac{1}{2}\right)\frac{9\tau(t')}{4}+\frac{9m(m+1)}{2}.
\ee
Since $m,n$ run over all integers, $9m(m+1)/2$ are integers and can be absorbed into $n$. Now we can see the NS quantization condition with symmetry $t\rightarrow t+6\pi\ri$ becomes exactly the same with the zeros of Jacobi theta function (\ref{p2ghm}).

For arbitrary local toric Calabi-Yau and generic Planck constant, we expect the similar phenomenon. The hidden symmetry $\mathbf{t}\rightarrow\mathbf{t}+2\pi\ri\mathbf{n}\cdot\mathbf{C}$ in the NS quantization conditions is also a request to keep the identity (\ref{conjecture0}) invariant under integral shift of $\mathbf{n}$. Therefore, to make the equivalence (\ref{equiv2}) precise, one can either reduce the symmetry in the GHM side or enlarge the symmetry in the NS side. Practically, since the physical quantity -- energy should be real, we usually reduce some symmetry in the quantum Riemann theta function, which is just the $\tau$ periodicity when it becomes a genuine Riemann theta function. In other word, the symmetry $E\rightarrow E+2\pi\ri$ is gauged.

Let us make a further remark on the Equation (\ref{p2ns}). Note there is a half period of $\tau$ on the right side. It is easy to make it disappear and restore the original form of Bohr-Sommerfeld quantization condition by using the variable $\widehat{\mathbf{t}}$ or equivalently the notation $E_{\mathrm{eff}}$ in \cite{Grassi:2014zfa}. In this case,
\be
\widehat{t}=t+3\pi\ri,
\ee
and
\be
\widehat{t}=3E_{\mathrm{eff}}.
\ee
Then from Equation (\ref{p2ns}) we have
\be
{1\over 2} \widehat{t}^2 -\pi^2 - 3 \partial_t  \widehat{F}_0^{\rm inst}(\widehat{t})+ 3\widehat{t} \partial^2_t  \widehat{F}_0^{\rm inst}(\widehat{t})=4 \pi^2 \left( n+{1\over 2} \right), \qquad  n=0,1,2,\cdots
\ee
and
\be
\label{Eeffp2}
{9\over 2} E_{\rm eff}^2 -\pi^2 - 3 \partial_t  \widehat{F}_0^{\rm inst}(\widehat{t})+ 9 E_{\rm eff} \partial^2_t  \widehat{F}_0^{\rm inst}(\widehat{t})=4 \pi^2 \left( n+{1\over 2} \right), \qquad  n=0,1,2,\cdots
\ee
The latter is just the quantization condition given in \cite{Grassi:2014zfa}.

One may also be concerned whether the exact NS quantization conditions such as Equation (\ref{Eeffp2}) have non-real solutions for energy. Although we don't have a rigorous proof for general cases yet, we expect the NS quantization conditions should have only real energy solutions. For local $\mathbb{P}^2$ at $\hbar=2\pi$, there are strong evidences to support this. Let us assume the quantization condition (\ref{p2ns}) has non-real solutions, then these solutions also belong to the solutions of theta function (\ref{p2ghm}), therefore are the roots of spectral determinant
\be
\Xi(\mu,2\pi)=\re^{J(\mu,2\pi)}\Theta(\mu,2\pi).
\ee
On the other hand, the GHM conjecture says this expression exactly equals the spectral determinant of local $\mathbb{P}^2$ quantum mechanics
\be
\Xi(\mu,2\pi)=\prod_{n=0}^{\infty}\left(1+\re^{\mu-E_n}\right).
\ee
Since we have a Hermitian Hamiltonian, the energy eigenvalues $E_n$ must be real. Therefore, if quantization condition (\ref{p2ns}) has more solutions other than the real spectrum $E_n$, then the GHM conjecture and the equality would fail. However, there already are plenty of independent checks and partial proofs to show that the equality should indeed holds, such as in the orbifold expansion and conifold expansion, see \cite{Marino:2015nla} for example. Thus the correctness of the GHM conjecture would guarantee the NS quantization condition (\ref{p2ns}) only has real solutions.\footnote{We thank Marcos Marino for this argument.} This argument is easy to extend to $\hbar=2\pi/k$, $k=1,2,3...$
\section{Examples}
\label{sec:examples}
In this section, we elaborate our theory with some examples. For local del Pezzo surfaces, the equivalence between the two quantization scheme has been clarified in \cite{Wang:2015wdy} with local $\mathbb{P}^2$, $\mathbb{F}_0$, $\mathbb{F}_1$ and $\mathbb{F}_2$. The method there are not very useful for the higher genus cases. Here we will look at the equivalence in a more unified perspective. For the higher genus cases, the most simple example is resolved $\IC^3/\IZ_5$ orbifold, which has mirror curve of genus two and can be reduced from the $SU(3)$ CY with Chern-Simons number $m=2$. We will have a detailed study on this model in subsection two. This model is quite typical in that for one among the three $\mathbf{r}$ fields, the equivalence can be rigorously proved at $\hbar=2\pi/k$, which means no specific data of invariants are needed, while the other two $\mathbf{r}$ fields are conjugate and result in nontrivial constraints among the refined BPS invariants even at $\hbar=2\pi/k$. After this, we study the $SU(N)$ geometries in general and propose an effective method to generate all non-equivalent $\mathbf{r}$ fields for these models. Then we will have a detailed study on the equivalence for $SU(3)$ geometries with $m=0,1,2$ and $SU(4)$, $SU(5)$ geometries with $m=0$.
\subsection{Local Del Pezzo Surfaces}\label{sec:ex1}
Toric almost del Pezzo surfaces are given by
reflexive polyhedra in two dimensions \cite{Batyrev:1994hm}, see all sixteen reflexive polyhedra in Figure 1. The first four surfaces are just $\mathbb{P}^2$, $\mathbb{F}_0$, $\mathbb{F}_1$ and $\mathbb{F}_2$. The fifth and sixth are blown-up surfaces $\mathfrak{B}_2(\mathbb{P}^2)$ and $\mathfrak{B}_1(\mathbb{F}_2)$. The anti-canonical class is only semi-positive
if there is a point on one edge of the toric diagram,
otherwise positive and ample. In particular the polyhedra 1,2,3,5,6
correspond to toric del Pezzo surfaces. See \cite{Huang:2013yta}\cite{Huang:2014nwa} for more details of the quantum geometry of local del Pezzo surfaces. Note also polyhedra 2,3,4 can engineer to $SU(2)$ $N=2$ gauge theories with Chern-Simons number $m=0,1,2$.\\

\begin{figure}[h!]
\begin{center}
\includegraphics[angle=0,width=1.0\textwidth]{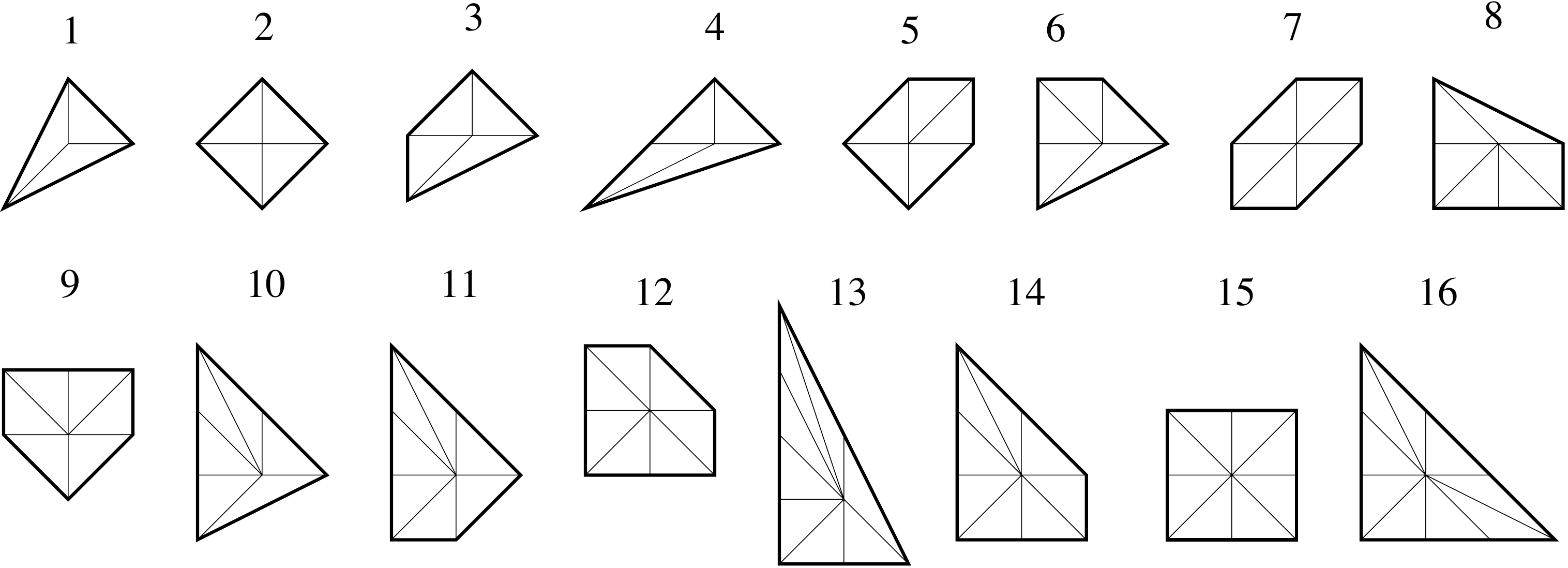}
\begin{quote}
\caption{These are the 16 reflexive polyhedra $\Delta$ in two dimensions. Polyhedron $k$ is dual to
polyhedron $17-k$ for $k=1,\ldots,6$. The polyhedra $7,\ldots,10$
are self-dual.
\vspace{-1.2cm}} \label{poly}
\end{quote}
\end{center}
\end{figure}
\indent The genus of the mirror curve of a local toric Calabi-Yau is just the number of inferior integral points of its toric diagram. Therefore, it is clear that all local almost del Pezzo surfaces has mirror curve of genus one. This makes the question easier in many occasions, especially in GHM quantization scheme. Note in genus one cases, the numbers of quantization conditions are the same between NS quantization and the original version of GHM conjecture. This make is possible to directly compare two quantization scheme.\\
\indent Now we review the method in \cite{Wang:2015wdy} to check to the equivalence between quantization schemes. The key is to written down a non-perturbative volume from GHM conjecture and compare it with the non-perturbative volume from exact NS quantization conditions. Note for genus higher than one, it is not clear how to written down the non-perturbative volumes from GHM conjecture, therefore this method is only valid in genus one. we focus on the local $\mathbb{P}^2$ here. For more details and the local $\mathbb{F}_0$, $\mathbb{F}_1$ and $\mathbb{F}_2$, see \cite{Wang:2015wdy}.\\
\indent The quantum mirror curve of local $\mathbb{P}^2$ is described as the quantum mechanics with the following Hamiltonian,
\begin{eqnarray} \label{localP2}
\exp{(\hat{H})} = \exp{(\hat{x})} + \exp{(\hat{p})} + \exp{(-\hat{x}-\hat{p})} ,
\end{eqnarray}
where the $\hat{x}, \hat{p}$ operators satisfy the canonical commutation relation $[\hat{x},\hat{p}]=i\hbar$. In Nekrasov-Shatashvili quantization scheme, the discrete energy spectrum satisfies the following exact quantization condition,
\begin{eqnarray}  \label{BScondition}
\textrm{vol}_{\rm p}(E, \hbar) + \textrm{vol}_{\rm np}(E, \hbar) = 2\pi \hbar(n+\frac{1}{2}),
\end{eqnarray}
where $n=0,1,2, \cdots$ is the discrete energy level, and the formulas for perturbative and non-perturbative parts
$\textrm{vol}_{\rm p}(E, \hbar)$ and $\textrm{vol}_{\rm np}(E, \hbar)$ are

\begin{eqnarray} \label{volu}
\textrm{vol}_{\rm p}(E,\hbar ) &=& \frac{t^2-\pi^2}{2} -\frac{\hbar^2}{8}+\hbar   f_{NS} (t , \hbar),
 \nonumber \\
\textrm{vol}_{\rm np}(E,\hbar ) &=& \hbar   f_{NS} (\frac{2\pi t}{\hbar} , \frac{4\pi^2}{\hbar}).
\end{eqnarray}
Here the quantum A-period $t$ is $t (E,\hbar)  \sim  - rE$ for large $E$ and can be expressed as a power series of $e^{-rE}$. For the local $\mathbb{P}^2$ model we have $r= 3$, and the first few terms are
\begin{eqnarray} \label{deformedA}
t (E,\hbar) =  -3E+ 3 \left(q + \frac{1}{q}\right) e^{-3E}
+ 3\left(  q^2 + \frac{1}{q^2} + \frac{7}{2} \left(q + \frac{1}{q}\right)  + 6 \right) e^{-6E} +\mathcal{O}(e^{-9E}) ,
\end{eqnarray}
in which $q=e^{\frac{i\hbar}{2}}$. The function $f_{NS} (t , \hbar)$ denotes the derivative of world-sheet instanton part of the topological string amplitude in the Nekrasov-Shatashvili limit with respect to the quantum A-period $t$, and can be expressed in terms of the refined Gopakumar-Vafa $n^{d}_{j_L,j_R}$ as
\begin{eqnarray} \label{fNS}
f_{NS} (t , \hbar) = - \frac{r}{2} \sum_{j_L,j_R} \sum_{w,d=1}^{\infty}   (-1)^{2j_L+2j_R+rwd}   n^{d}_{j_L,j_R}    e^{wdt}    \cdot \frac{d}{w} \cdot
\frac{\sin \frac{w\hbar(2j_R+1)}{2} \sin \frac{w\hbar(2j_L+1)}{2}}{ \sin^3 (\frac{w\hbar}{2})}.
\end{eqnarray}
To compare the above quantization condition with GHM quantization from the quantum Riemann theta function, we follow \cite{Grassi:2014zfa} and define some functions as
\begin{eqnarray}
 f &:= &  \frac{ n^{d}_{j_L,j_R}}{4w} (-1)^{2j_L+2j_R+rwd} Q^{wd}   \cdot
  \frac{(2j_R+1) \sin [4x w (2j_L+1) ]}{\sin^2 (2x w)\sin(4x w ) },
 \nonumber \\
 f_c (n) & := &   \sum_{j_L,j_R} \sum_{w,d=1}^{\infty} [ \cos (2r(2n + 1) x wd )
    - \cos(2r x wd) ]\cdot f ,
  \nonumber \\
   f_s (n) & := & \sum_{j_L,j_R} \sum_{w,d=1}^{\infty}   [ \sin (2r (2n + 1) x wd )
   - (2n+1) \sin (2rx wd) ] \cdot f,
\end{eqnarray}
where we denote $x=\pi^2/\hbar$ and $Q=\exp{(2\pi t/\hbar )}$. Besides the same perturbative volume, it was shown in \cite{Grassi:2014zfa} that the GHM quantization condition gives the following non-perturbative phase volume
\begin{eqnarray} \label{GHMvol}
 \textrm{vol}_{\rm np}^{\rm GHM}(E,\hbar) = - 2 \hbar  \sum_{j_L,j_R} \sum_{w,d=1}^{\infty}  f\cdot \sin( 2r x wd )    +\lambda    , ~~~~~  \end{eqnarray}
where $\lambda := \sum_{k=1}^{\infty} c_k(x) Q^k $  is the correction term and is determined by the following equation
\begin{eqnarray}
\sum_{n=0}^{\infty}  \sin \left[3n(n + 1)(2n + 1)x + f_s(n) + (n + \frac{1}{2})\lambda \right]  \cdot  Q^{\frac{3n(n + 1)}{2}} (-1)^n e^{ f_c(n)} =0.
\end{eqnarray}
The correction terms can be solved perturbatively by expansion of the equation in powers of $Q$.  Using the data of refined Gopakumar-Vafa invariants, we can easily check that the formula for non-perturbative phase volume $\textrm{vol}_{\rm np}(E,\hbar)$ in (\ref{volu}) is equal to the GHM formula (\ref{GHMvol}) for the first few order coefficients. We can check the equality in the convenient variables $x$ and $Q$ here, and shall not need the formula for quantum A-period (\ref{deformedA}). The checks are performed successfully up to order $Q^7$ in \cite{Wang:2015wdy}. We can also show the equality $\textrm{vol}_{\rm np} = \textrm{vol}_{\rm np}^{\rm GHM} $ for the special cases when $\hbar=2\pi/k$. For these special cases, one can check $f_s(n)=0$, so the correction term $\lambda$ vanishes.

It should be emphasized that the check on $\textrm{vol}_{\rm np} = \textrm{vol}_{\rm np}^{\rm GHM} $ here is in fact the same check on the identity (\ref{check}) specializing to local $\mathbb{P}^2$ with only one non-equivalent $r$ field as 3. In fact, it is easy to see all equivalent $r$ fields for local $\mathbb{P}^2$ are $6k-3$, $k\in\mathbb{Z}$. However, to check the identity (\ref{check}) is more convenient and more general, while the method here is difficult to apply to the higher-genus cases. The main reason is that the concept of phase volumes are obscure in the GHM quantization scheme for the mirror curve of higher genus. Moreover, the correction term $\lambda$ in (\ref{GHMvol}) can be generalized to the higher-genus cases, which will be crucial for us to check whether there are extra solutions besides the NS spectra for all quantum Riemann theta functions. We will see this in the following subsections.

Now let us look at all del Pezzo cases in an unified way. For an arbitrary local del Pezzo surface, there is one true modulus $\tilde{u}$ and possibly several mass parameters $\xi_i$. In section \ref{sec:remark}, we have proposed a method to determine the $\bf{B}$ field. Our statement on the equivalence for all genus-one cases is as follows: \emph{For an arbitrary local del Pezzo surface, there exist one and only one non-equivalent $\mathbf{r}$ field such that the identity (\ref{check}) holds. The unique $\bf{r}$ field is determined by the following conditions:\\\newline
\noindent The true modulus $\tilde{u}$ becomes its opposite, while the phases of mass parameters keep zero.\newline\\}
First, we need to show why the $\mathbf{r}$ field determined in this way is a $\mathbf{B}$ field, after all, the $\mathbf{B}$ field is defined in a rather different way in Section \ref{sec:remark}. Let us look at the mirror curve of an arbitrary local del Pezzo surface, suppose the origin in the fan is corresponding to $x_0$, then we have
\be
H(\re^x, \re^p)=\sum_{i=1}^{k+3}x_i  \exp\left( \nu^{(i)}_1 x+  \nu^{(i)}_2 p\right)+x_0=0,
\ee
where $k$ is the number of the mass parameters. For all but the eighth and thirteenth polyhedra in Fig \ref{poly}, there are terms $\re^x$ and $\re^p$ in the mirror curve. To absorb the $x_0$ into the other terms, one can introduce the shifts $x\rightarrow x-\log x_0$ and $p\rightarrow p-\log x_0$ to fix the relative sign between $\re^x$ and $\re^p$ and $1$. Suppose the terms $\re^x$ and $\re^p$ correspond to $x_1$ and $x_2$, then the mirror curve becomes
\be
H(\re^x, \re^p)=1+\re^x+\re^p+\sum_{i=3}^{k+3}x_ix_0^{\nu^{(i)}_1+\nu^{(i)}_2-1}  \exp\left( \nu^{(i)}_1 x+  \nu^{(i)}_2 p\right)=0.
\ee
Remember our procedure to obtain the $\mathbf{B}$ field in Section \ref{sec:remark}, after promoting the above curve to operator equation and change $\hbar$ from zero to $2\pi$, one need to add a factor of
\be
(-1)^{\nu^{(i)}_1\nu^{(i)}_2}
\ee
to each term in the mirror curve. On the other hand, to obtain the $\mathbf{r}$ field, we change the sign of $x_0$ and keep the others the same. Then we add a factor of
\be
(-1)^{\nu^{(i)}_1+\nu^{(i)}_2-1}
\ee
to each term. Since in the fan of arbitrary del Pezzo surface, all vectors but the origin have at least one coordinate as $\pm1$, then obviously
\be
(-1)^{\nu^{(i)}_1\nu^{(i)}_2}=(-1)^{\nu^{(i)}_1+\nu^{(i)}_2-1}.
\ee
This concludes that our rule here to determine the $\mathbf{r}$ field is consistent with our general definition of the $\mathbf{B}$ field.

For the remaining eighth and thirteenth fans which do not contain both vector $(1,0)$ and $(0,1)$, one can choose to fix the coefficients of terms $\re^{-x}$ and $\re^{-p}$ in the mirror curve and introduce the shifts $x\rightarrow x+\log x_0$ and $p\rightarrow p+\log x_0$ to absorb the $x_0$. The consistent check follows almost the same procedure.

\begin{figure}[h]
\begin{center}
\includegraphics[scale=0.5]{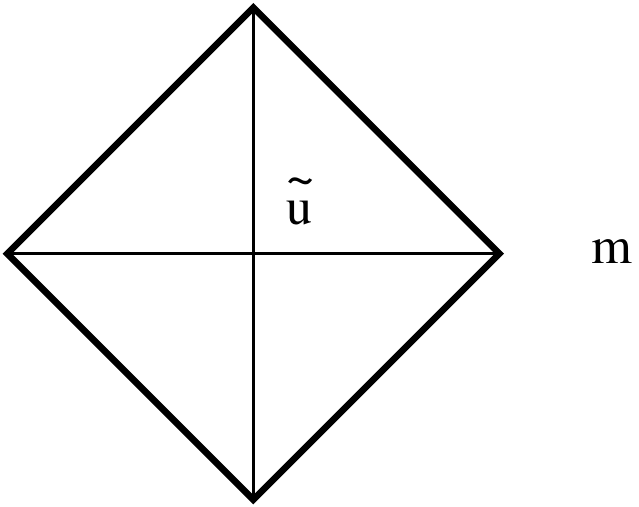}
\caption{\label{fig:poly2} Polyhedron 2 depicting the toric geometry $\mathbb{F}_0$.}
\end{center}
\end{figure}

Now let us amplify the above rule for local $\mathbb{F}_0$. All toric data and vector choice follows \cite{Huang:2014eha}. The Mori cone is depicted in Fig.~\ref{fig:poly2}
\begin{equation}
 \label{dataf0}
 \begin{array}{cc|rrr|rrl}
    \multicolumn{5}{c}{\nu_i }    &l^{(1)}& l^{(2)}&\\
    D_u    &&     1&     0&   0&         -2&  -2&       \\
    D_1    &&     1&     1&   0&         1&   0&        \\
    D_2    &&     1&     0&   1&         0&   1&        \\
    D_3    &&     1&    -1&   0&         1&   0&        \\
    D_4    &&     1&     0&   -1&        0&   1&       \\
  \end{array} \ .
\end{equation}
From the toric data we find the Batyrev coordinates,
\begin{equation}\label{eq:f0mpara}
z_1=\frac{m}{\tilde{u}^2}\, ,\quad z_2 = \frac{1}{\tilde{u}^2} \, .
\end{equation}
After setting $u=\frac{1}{\tilde{u}^2}$ the mirror curve is given by
\begin{equation}
  -1 + e^x + e^p + m u\, e^{-x} + u\, e^{-p} = 0 \, .
  \label{eqn:P1tP1curve}
\end{equation}
Since there is no mixed term with both $x$ and $p$, obviously from the method in section \ref{sec:remark}, the $\mathbf{B}$ field of local $\mathbb{F}_0$ should be $(0,0)$. From the above rule one, the phase of $\tilde{u}$ should minus $\pi\ri$, while the phase of $m$ keep unchanged. Therefore, the phases of both $z_1$ and $z_2$ should add $2\pi\ri$, which means the $\mathbf{r}$ field should be $(2,2)$. In the notation of \cite{Wang:2015wdy}, this is in fact just
\be
\mathbf{r}=r\mathbf{c},
\ee
where $r=2$, $\mathbf{c}=(1,1)$. This formula is also true for local $\mathbb{F}_1$ and $\mathbb{F}_2$ listed there. Besides, it is easy to see that all $\mathbf{r}$ fields equivalent with $(2,2)$, which can be obtained by the $2\pi\ri$ shift of $\tilde{u}$, are generated by $(2,2)+k(4,4)$, $k\in\mathbb{Z}$. The procedure here should be able to apply to all del Pezzo surfaces.
\subsection{Resolved $\IC^3/\IZ_5$ Orbifold}\label{sec:ex2}
The resolved $\mathbb{C}^3/\mathbb{Z}_5$ orbifold is the simplest local toric Calabi-Yau with genus-two mirror curve. It has two true complex moduli and no mass parameter. This model has been extensively studied in \cite{Klemm:2015iya}\cite{Codesido:2015dia}\cite{Franco:2015rnr}. In this subsection, we amplify our theory with this example, determine all the $\mathbf{r}$ fields and check the identities (\ref{conjecture}). We find for resolved $\mathbb{C}^3/\mathbb{Z}_5$, there are three non-equivalent $\mathbf{r}$ fields. For the $\mathbf{r}$ field $(-1,2)$, which is a trivial case, we give a rigorous proof on the identity at $\hbar=2\pi/k$. For the other two $\mathbf{r}$ fields which are conjugate, we show the identities give nontrivial constraints among the refined BPS invariants. We also study whether two of the three non-equivalent $\mathbf{r}$ fields can already determine the full NS spectra.\\
\begin{equation}\label{c3z5charge}
\begin{array}{c|rrr|rr}
    \multicolumn{5}{c}{v_i}    $Q_1$&  Q_2 \\
    x_0    &     0&     0&  \ \ 1\,&         -3&  1  \\
    x_1    &     1&     0&   \ \ 1\, &         1&     -2\\
    x_2    &     2&     0&   \ \ 1\, &         0&     1  \\
    x_3    &     0&    1&   \ \ 1\, &         1&   0\\
    x_4    &     -1&    -1&   \ \ 1\, &         1&     0\\
  \end{array}
\end{equation}

\begin{figure}[htbp]
\centering\includegraphics[width=2.5in]{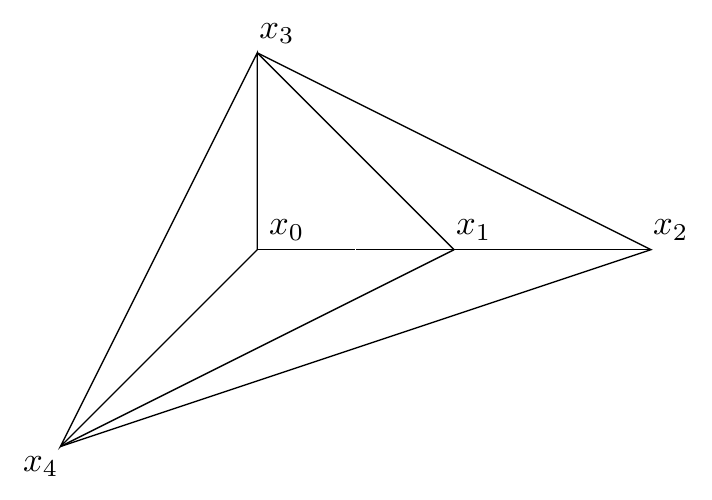}
\caption{Fan diagram of resolved $\cz$ model.}\label{fan:c3z5}
\end{figure}

Resolved $\mathbb{C}^3/\mathbb{Z}_5$ can be obtained by taking the limit $x_5=0$ in the $SU(3)$ geometry with $m=2$. The toric data of this model is listed in (\ref{c3z5charge})\footnote{One should not mix the charges $Q_i$ here and exponential of K\"{a}hler parameter $Q_i=e^{-t_i}$}. The fan diagram is illustrated in Figure \ref{fan:c3z5}. From the toric data, we can see there are two Batyrev coordinates,
\begin{equation}
z_1=\frac{x_1x_3x_4}{x_0^3}, \ \ \ z_2=\frac{x_0 x_2}{x_1^2}.
\end{equation}
The true moduli of this model are $x_0,x_1$ and the $C$ matrix is
\be
C=\left(\begin{array}{cc}-3 & 1 \\1 & -2\end{array}\right).
\ee
The genus zero free energy of this geometry is \cite{Klemm:2015iya}\footnote{In this and the following subsections, we use the notation $Q_i=\re^{t_i}$ for our convenience.}
\be
F_0=\frac{1}{15}t_1^3+\frac{1}{10}t_1^2t_2+\frac{3}{10}t_1t_2^2+\frac{3}{10}t_2^3+3 Q_1-2 Q_2-{45 \over 8} Q_1^2 +4 Q_1 Q_2 -
{Q_2^2 \over 4}+\mathcal O(Q_i^3).
\ee
The genus one free energy in NS limit is
\be
F^{\rm NS}_1=-\frac{1}{12}t_1-\frac{1}{8}t_2-{7 Q_1 \over 8} +{Q_2 \over 6} + {129 Q_1^2 \over 16} -{5 Q_1 Q_2 \over 6}+{Q_2^2 \over 12}+ \mathcal O(Q_i^3),
\ee
and the genus one unrefined free energy is
\be
F^{\rm GV}_1=\frac{2}{15}t_1+\frac{3}{20}t_2+ {Q_1 \over 4} -{Q_2\over 6}-{3Q_1^2 \over 8} +{Q_1 Q_2 \over 3}-{Q_2^2 \over 12}+ \mathcal O(Q_i^3).
\ee
To check the identities (\ref{conjecture}), we need to find the correct $\mathbf{r}$ fields. The most direct method is that we could set some special value of $t_i,\hbar$ in (\ref{conjecture}), and scan all the integral vectors $\mathbf{r}(=\mathbf{B} \mod 2)$ in a region to see if the identity (\ref{conjecture}) holds. For the current case, we scan from $-5$ to $5$, and the following $\mathbf{r}$ fields make the identity (\ref{conjecture}) holds,
\be\label{rc3z5}
\begin{split}
&(-5,-4),(-5,0),(-5,4),(-3,-4),(-3,0),(-3,2),(-1,-4),(-1,-2),\\
&(-1,2),(1,-2),(1,2),(1,4),(3,-2),(3,0),(3,4),(5,-4),(5,0),(5,4).
\end{split}
\ee
Many of the above $\mathbf{r}$ fields may result in the same spectral determinant, which are defined to be equivalent. In fact, all equivalent $\mathbf{r}$ fields are generated by the shift,
\be\label{c3z5:symmetry}
\mathbf{r} \rightarrow \mathbf{r} +2\mathbf{n}\cdot C, \ \ \mathbf{n}\in \mathbb{Z}^{g}.
\ee
Under this symmetry, we obtain three classes of non-equivalent $\mathbf{r}$ fields. $\mathbf{r}=(-3,2)$ is equivalent to
$$
(-5,-4),(-3,2),(-1,-2),(1,4),(3,0),(5,-4),\dots.
$$
$\mathbf{r}=(-3,0)$ is equivalent to
$$
(-5,4),(-3,0),(-1,-4),(1,2),(3,-2),(5,4),\dots.
$$
$\mathbf{r}=(-1,2)$ is equivalent to
$$
(-5,0),(-3,-4),(-1,2),(1,-2),(3,4),(5,0),\dots.
$$
The list (\ref{rc3z5}) are exactly the union of these three classes of $\mathbf{r}$ fields. Our conjecture for the equivalence between GHM quantization scheme and NS quantization scheme states, choosing arbitrarily one $\mathbf{r}$ field from each of the above three classes, suppose they are denoted as $\mathbf{r}_1, \mathbf{r}_2,\mathbf{r}_3$, then the intersection points of the theta divisors of all three quantum Riemann theta function $\Theta(\mathbf{t}+\ri\pi\mathbf{r}^a)$, $a=1,2,3$ coincide with the spectra determined by the exact NS quantization conditions. This in guaranteed by the identities (\ref{conjecture}).

Note the first two classes are complex conjugated, while the third class is self-conjugated. In fact, the third $\mathbf{r}$ field $(1,-2)$ is very special in that it can be expressed as
\be
r_i=2\alpha_jC_{ij},
\ee
where
\be
\boldsymbol{\alpha}=\left(0,\frac{1}{2}\right).
\ee
This fact is crucial for us to prove the identity (\ref{conjecture}) for $\mathbf{r}=(1,-2)$ at $\hbar=2\pi/k$.

To verify our conjecture, it is easy to expand the (\ref{conjecture}) as a series of $e^{t_i}$, and then check it order by order. For all three nonequivalent $\mathbf{r}$ fields, we have check the three identities up to degree $2d_1+d_2 <14$. So we conclude that all the solution solved from the exact NS quantization conditions lie on the divisor of each of the three nonequivalent spectral determinants.

Now we follow the approach in Section \ref{sec:eq2} to prove for $\mathbf{r}=(1,-2)$, the identity (\ref{conjecture}) holds at $\hbar=2\pi/k$. The cubic terms in (\ref{theta1-2pik}) for resolved $\mathbb{C}^3/\mathbb{Z}_5$ is
\be
-2k\pi\ri\left(\frac{3n_1^3}{2}-\frac{3n_1^2n_2}{2}+\frac{n_1n_2^2}{2}+\frac{4n_2^3}{3}\right).
\ee
Using the fact that for an integer $n$,
\be
n^3\equiv n\ (\mathrm{Mod }6),
\ee
and
\be
n^2\equiv n\ (\mathrm{Mod }2),
\ee
we can reduce the above cubic terms as linear and quadratic terms,
\be
-2k\pi\ri\left(\frac{3n_1}{2}-n_1n_2+\frac{4n_2}{3}\right).
\ee
After carefully checking every term, we find that the quantum Riemann theta function at $\hbar=2\pi/k$ (\ref{theta1-2pik}) for resolved $\mathbb{C}^3/\mathbb{Z}_5$ can be written as a Riemann theta function with characteristics,
\be
\vartheta \left[\begin{array}{cc} \boldsymbol{\alpha}\\ \boldsymbol{\beta}
 \end{array}\right] \left({\boldsymbol{0}}, \tau+S \right),
\ee
in which
\be
\boldsymbol{\alpha}=\left(0,\frac{1}{2}\right),\quad\quad\boldsymbol{\beta}=\left(\frac{1}{2}-\frac{3k}{2},\frac{1}{2}\right),
\ee
and
\be
S=\left(\begin{array}{cc} 0 & k\\ k & 0\\
 \end{array}\right).
\ee
Since
\be
4\boldsymbol{\alpha}\cdot\boldsymbol{\beta}=1,
\ee
the last criterion (\ref{rca}) is satisfied. Therefore, zero lies on the divisor of the quantum Riemann theta function at $\hbar=2\pi/k$ (\ref{theta1-2pik}). This concludes the proof that for $\mathbf{r}=(1,-2)$ the identity (\ref{conjecture}) holds at $\hbar=2\pi/k$.

The equivalence between the two quantization schemes and the identities (\ref{conjecture}) are highly nontrivial. In the following we explicitly show how the identities (\ref{conjecture}) result in constraints among refined BPS invariants. To see this, let us suppose we know the degrees $\mathbf{d}$ and spin contents $(j_L,j_R)$ of all non-zero BPS invariants but do not know the BPS numbers $N^{\mathbf{d}}_{j_L,j_R}$, the identities will require a series of constraints on the BPS numbers, or even fix some of the invariants at the first few orders.

Now suppose we do not know the exact nonzero BPS numbers, and try to expand (\ref{conjecture}) with respect to $e^{2\pi/\hbar t_i}$. Then for $\mathbf{r}=(1,-2)$, we have
$$
0=e^{-\frac{2 \ri\pi ^2}{\hbar}} \left(-1+e^{\frac{4 \ri\pi ^2}{\hbar}}\right)e^{2\pi/\hbar t_1} (N^{(0,1)}_{0,1}+1)+\cdots.
$$
Thus we have $N^{(0,1)}_{0,1}=-1$. Continue to expand to higher orders, we obtain\footnote{For convenience, we absorb the $(-1)^{2j_L+2j_R}$ term into $N^{\mathbf{d}}_{j_L,j_R}$.}

$$
N^{(0,2)}_{0,{5\over 2}}=1;N^{(0,3)}_{0,3}=-1;N^{(0,3)}_{{1\over 2},{9\over 2}}=-1;N^{(1,1)}_{0,1}=-1;N^{(1,1)}_{0,1}=-1;N^{(1,0)}_{0,{1\over 2}}=0 \text{ or } 1,\cdots.
$$

For $\mathbf{r}=(1,2)$ and $\mathbf{r}=(-1,-2)$, since they are complex conjugate, they have the same constraints on the refined BPS invariants. For the first few order we have expanded, the two identities result in different constraints on the invariants from the $\mathbf{r}=(1,-2)$ case.
\be
\begin{split}
&N^{(1,0)}_{0,{1\over 2}}=1;N^{(1,1)}_{0,0}=-1;N^{(0,3)}_{{1\over 2},{9\over 2}}=-1;N^{(0,3)}_{0,3}=-1;N^{(2,1)}_{0,1}=-1;N^{(1,2)}_{0,{5\over 2}}=\frac{4}{15} (N^{(0,1)}_{0,1})^2-\frac{2}{5}N^{(0,1)}_{0,1}+\frac{1}{3};\\
&N^{(1,2)}_{0,{3\over 2}}=\frac{1}{10} (N^{(0,1)}_{0,1})^2-\frac{9}{10} N^{(0,1)}_{0,1};N^{(0,1)}_{0,1}=0 \text{ or } -1;\cdots.
\end{split}
\ee
Note if we requires $N^{\mathbf{d}}_{j_L,j_R}$ to be integer, then we can fix $N^{(0,1)}_{0,1}=-1$.

From the constraints on the BPS invariants, we can see the that the equivalence between the two quantization schemes are highly nontrivial. As there are infinity number of constraints and infinity number of BPS invariants, it is not clear whether we can solve all the BPS invariants from the identities.

Finally, we want to check whether all three nonequivalent theta divisors have extra intersection points besides the spectra determined by the exact NS quantization conditions, and whether part of the three nonequivalent theta divisors can already determine the full NS spectra and without extra intersections. To study these questions, let us go back to the original picture of GHM quantization, where the divisors of quantum Riemann theta function are hypersurfaces in the moduli space and the discrete spectra of exact NS quantization conditions lie on the divisors. Consider those points on the divisors but do not coincide with the NS spectra, apparently their K\"{a}hler parameters $\mathbf{t}$  still satisfy the vanishing of spectral determinant and quantum Riemann theta function, but may result in some corrections to $(n_i+1/2)2\pi\hbar$ when submitted to the non-perturbative volumes,
\be\label{lambdadef}
\text{Vol}_i=2\pi \hbar \left(n_i+\frac{1}{2}+\lambda_i\left(e^{\frac{2\pi}{\hbar} \mathbf{t}}\right)\right),
\ee
where
\be
\lambda_i\left(e^{\frac{2\pi}{\hbar} \mathbf{t}}\right)=c_i^{0} +\sum_{n=1}^{\infty}\sum_jc_{i,j}^{n}e^{n\frac{2\pi}{\hbar}t_j},
\ee
with unknown complex coefficients $c_i^{0}=c_i^{0}(\hbar),c_{i,j}^{n}=c_{i,j}^{n}(\hbar)$. This is in fact the higher dimensional generalization of the $\lambda$ correction for genus one cases considered in \cite{Grassi:2014zfa}\cite{Wang:2015wdy} and our Section \ref{sec:ex1}. Every point on the quantum theta is attached with a set of correction functions $\{\lambda_i(\exp{\frac{2\pi}{\hbar} \mathbf{t}})\},i=1,\cdots,g$ which characterize its deviation from the NS spectra. Substitute equation ($\ref{lambdadef}$) to the exponential of the spectral determinant just like what we did in (\ref{qcsdeff}) and cancel out the perturbative part, one can regard the vanishing of the resulting expression as the constraint equation for the corrections. Then the corrections $\lambda_i$ can be solved order by order with respect to $\exp{\frac{2\pi}{\hbar} \mathbf{t}}$ (at least to some constraints). In general, for a theta divisor with one $\mathbf{r}$ field and dimension larger than zero, the coefficients in $\lambda_i$ can vary smoothly within the constraints while the corresponding points run all through the divisor. If one want to study the intersection loci of two theta divisors with two nonequivalent $\mathbf{r}$ fields, one should solve the constraint equations simultaneously. For some theta divisors with some nonequivalent $\mathbf{r}$ fields, if the simultaneous constraint equations require all corrections $\lambda_i$ to be zero, then their intersection points coincide with the NS spectra without extra points. Practically, one should solve the simultaneous constraints among the coefficients in $\lambda_i$ order by order.


For illustration, here we list the coefficients up to order two for resolved $\IC^3/\IZ_5$ orbifold. For $\mathbf{r}=(-1,2)$ the coefficients for the first few orders are

\be\begin{split}
&c_2^0\in \mathbb{Z},\\
&c_{2,1}^{1}=\frac{2}{\pi}\cos[\pi(2\pi/\hbar-c_{1}^0)]\sin(\pi c_1^0),\\
&c_{2,2}^{1}=0,\\
&c_{2,2}^{2}=0,\\
&c_{2,1}^2=\frac{1}{2 \pi }[4 \pi  c_{1,1}^1 \cos (\frac{2 \pi  (\pi -c_1^0 \hbar )}{\hbar
   })-6 \sin (\frac{4 \pi ^2}{\hbar }-2 \pi  c_1^0)-4 \sin
   (\frac{8 \pi ^2}{\hbar }-2 \pi  c_1^0)\\
   &\ \ \   -\sin (\frac{4 \pi
   (\pi -c_1^0 \hbar )}{\hbar })+6 \sin (2 \pi  c_1^0)+7
   \sin (\frac{4 \pi ^2}{\hbar })+4 \sin (\frac{8 \pi ^2}{\hbar
   })],\\
&c_{1,2}^{1}=\frac{1}{\pi}\cos[\pi(2\pi/\hbar-c_1^0)]\sin(\pi c_1^0) \sec(4\pi^2/h-2\pi c_1^0) , \\
&\ \ \ \ \ \ \vdots\\
\end{split}
\ee

For $\mathbf{r}=(1,2)$ the coefficients for the first few orders are
\be\begin{split}
&c_1^0\in \mathbb{Z},\\
&c_{2,1}^{1}=\frac{2\ri}{\pi}\cos(2\pi^2/\hbar)\left(1-e^{2\ri\pi c_2^0}\right),\\
&c_{1,1}^{1}=0,\\
&c_{1,1}^{2}=0,\\
&c_{1,2}^{1}=\frac{1}{\pi}\sin(\pi c_2^0) e^{-\ri\pi c_2^0},  \\
&c_{1,2}^{2}=\frac{1}{2\pi}\left(\sin(2\pi c_2^0) +2\pi c_{2,2}^1 \right) e^{2\ri\pi c_2^0},\\
&\ \ \ \ \ \ \vdots\\
\end{split}
\ee

For $\mathbf{r}=(-1,-2)$, the coefficients for the first few orders are
\be\begin{split}
&c_1^0\in \mathbb{Z},\\
&c_{2,1}^{1}=-\frac{2\ri}{\pi}\cos(2\pi^2/\hbar)\left(1-e^{-2\ri\pi c_2^0}\right),\\
&c_{1,1}^{1}=0,\\
&c_{1,1}^{2}=0,\\
&c_{1,2}^{1}=\frac{1}{\pi}\sin(\pi c_2^0) e^{\ri\pi c_2^0},  \\
&c_{1,2}^{2}=\frac{1}{2\pi}\left(\sin(2\pi c_2^0) +2\pi c_{2,2}^1 \right) e^{-2\ri\pi c_2^0},\\
&\ \ \ \ \ \ \vdots\\
\end{split}
\ee
Let all the three ${\lambda_i\left(\exp{\frac{2\pi}{\hbar}\mathbf{t}}\right)}$ equal to each other, we find that all the coefficients are requested to vanish to any high orders. We have checked this up to order $d=d_1+d_2\leqslant6$.

Note that even though all the three theta divisors intersect at the NS spectra, it does not rule out the possibility that only two of the three theta divisors can already determine the full NS spectra. In fact, we can choose two divisors as $\mathbf{r}=(-1,2),(1,2)$ or $\mathbf{r}=(-1,2),(-1,-2)$. From the above method, we can show that order by order that these pairs of $\mathbf{r}$ can already solve all the coefficients to be zero. One may wonder about the pair $\mathbf{r}=(1,2),(-1,-2)$, which are complex conjugated to each other. However, these two $\mathbf{r}$ fields do not fix all the coefficients $c_{i,j}^n$. Since the number of equations is infinite, we do not have a proper way to state whether this two $\mathbf{r}$ field is enough to fix all the non-perturbative corrections to be zero.

\subsection{$SU(N)$ Geometries}\label{sec:ex3}
In this subsection, we consider the general features of toric CY geometries $SU(N)$ with arbitrary Chern-Simons number $m$ and leave the detailed study on $SU(3)$, $SU(4)$ and $SU(5)$ geometries in the following subsections. We give the unified description on charge vectors, fans and mirror curves, and determine the $\mathbf{B}$ field from the curve. We also propose a general method to generate all $(N-1)^2$ non-equivalent $\mathbf{r}$ fields. Although we do not prove our method of generating $\mathbf{r}$ fields, we amass considerable evidence.\\
\indent $SU(N)$ geometries are also called $X_{N-1}$ geometries. They are of physical interest because they can engineer to $SU(N)$ supersymmetric gauge theories \cite{Iqbal:2003zz}. In mathematics, they can be realized as $A_{N-1}$ fibration over $\mathbb{P}^1$, or resolutions of cone over the Sasaki-Einstein manifold $Y^{p,q}$ \cite{Brini:2008rh}. The mirror of $SU(N)$ CY is a Riemann surface with genus $N-1$. Besides the rank of corresponding gauge group, there is the other parameter $m=0,\dots,N-1$, which labels the Chern-Simons number of the gauge theories from geometric engineering. For example, $SU(N)$ geometries with $m=0$ are related to periodic Toda system of $N$ particles \cite{Hatsuda:2015qzx}. $SU(N)$ geometries with $m=N$ can be realized as $\mathbb{C}^3/2N$ orbifold \cite{Klemm:2015iya}\cite{Iqbal:2002ep}.\\
\indent Now we introduce the following set of charge vectors in $\mathbb{C}^{N+3}$ for $SU(N)$ Calabi-Yau with Chern-Simons number $m$,
\be\label{suncharge}
\ba
Q_1&=(0,0,1,-2,1,0,0,&&0,0,0,0,\cdots,0,0,0)\\
 Q_2&=(0,0,0,1,-2,1,0,&&0,0,0,0,\cdots,0,0,0)\\
 Q_{3}&=(0,0,0,0,1,-2,1,&&0,0,0,0,\cdots,0,0,0)\\
  &\ \ \ \ \ \  &&\vdots\\
Q_{N-1}&=(0,0,0,0,0,0,0,0,&&0,0,0,\cdots,1,-2,1)\\
Q_{N}&=(1,1,0,0,\cdots,0,0,&&-1,0,\cdots,0,-1,0)\\
& &&\downarrow\\
& && (m+4)\text{th component of } Q_{N} \text{ is } -1\\
\ea
\ee
for $m=N$, $Q_{N}=(1,1,0,0,\cdots,0,0,-2)$.

The corresponding fan is
\be
\label{sunfan}
\ba
v_0&=(0,-1,1), \\
v_1&=(m-N+2, 1, 1), \\
v_{i+2}&=(i-N+1, 0, 1), \qquad i=0, \cdots, N.
\ea
\ee
This determines the mirror curve to be
\be\label{eq:mirrorcurve}
x_0\re^{-p}+x_1\re^{(m-N+2)x+p}+\sum_{i=0}^Nx_{i+2}\re^{(-N+i+1)x}=0.
\ee
The genus of mirror curve can be read from the number of inner points of fan diagram, which is $N-1$ for $SU(N)$ geometries. The corresponding Batyrev coordinates for the moduli space are\footnote{Here we focus on the $m<N$ cases. The $m=N$ cases are slightly different due to the charge choice, but they are essentially the same and can analyzed in a similar way.}
\be\label{eq:Batyrev}
\ba
z_i&={x_{i+1} x_{i+3} \over x_{i+2}^2}, \qquad i=1, \cdots, N-1, \\
z_N&={x_0 x_1 \over x_{m+3} x_{N+1}}.
\ea
\ee
When we quantize the mirror curve (\ref{eq:mirrorcurve}), it is easy to see that after changing $\hbar$ from zero to $2\pi$, only $x_1$ need a phase change of $2\pi \ri(m-N+2)$ to keep the form of the curve. Combining the Batyrev coordinates relations (\ref{eq:Batyrev}), we can immediately write down the $\mathbf{B}$ field,
\be\label{eq:Bfield}
\mathbf{B}=(0,0,\dots,0, N-m).
\ee
Note the value of $\mathbf{B}$ field relies on the choice of bases and is only defined up to a even lattice.\\
\indent Now we turn to the discussion of $\mathbf{r}$ field. From the very beginning, $\mathbf{r}$ fields are only representatives of $\mathbf{B}$ field, which means
\be
\mathbf{r}=\mathbf{B}\,(\mathrm{Mod}\ 2\mathbb{Z}^N).
\ee
There is no known method to determine the values of all non-equivalent $\mathbf{r}$ fields for an general toric Calabi-Yau threefold. Practically, one can assume the identity (\ref{conjecture}) and obtain $\mathbf{r}$ fields from the first few orders, then perform the check to higher order. This is of course not an ideal method and can only be done one by one Calabi-Yau. However, for $SU(N)$ CY, we introduce the following rule to generate all $(N-1)^2$ non-equivalent $\mathbf{r}$ fields. Although this is not rigorously proved, it successfully produces the right $\mathbf{r}$ fields for all $SU(N)$ geometries we have studied, including $SU(3)$ geometries with $m=0,1,2$ and $SU(4),SU(5)$ geometry with $m=0$. Besides, as a large quantity of toric CY can be realized as the reduction of $SU(N)$ geometries, such as resolved $\mathbb{C}^3/\mathbb{Z}_5$ orbifold, this method turns out to be quite useful.\\
\indent In $SU(N)$ geometries, the complex parameters $x_i$, $i=0, \cdots, N+2$, give a redundant parametrization of the moduli space, and some of them can be set to one. Here we always set
\be
\ba
x_0&=1,\\
x_1&=1.
\ea
\ee
Now the $N$ Batyrev coordinates can be written as
\be\label{eq:Batyrev1}
\ba
z_i&={x_{i+1} x_{i+3} \over x_{i+2}^2}, \qquad i=1, \cdots, N-1, \\
z_N&={1 \over x_{m+3} x_{N+1}}.
\ea
\ee
Among these $N$ parameters, there are only $g_{\Sigma}=N-1$ true moduli of the geometry, which are $z_i$, $i=1, \cdots, N-1$, and in addition a mass parameter $z_N$.\\
\indent The $\mathbf{r}$ fields characterize the phase of Batyrev coordinates $z_i$, which may be translated to the phase of $x_i$. In the cases of $SU(N)$ geometry, it is indeed much clearer to study $\mathbf{r}$ fields from the aspect of moduli $x_i$. As we discussed before, $\mathbf{r}$ fields are also equivalence classes like $\mathbf{B}$ fields, but more obscure. The value of two $\mathbf{r}$ fields can be different, but may still produce the same quantum Riemann theta function, and therefore are equivalent by definition. For $SU(N)$ geometries with Chern-Simons number $m$, we have the following claim:\\\newline
{ \noindent\emph{1, All equivalent $\mathbf{r}$ fields are generated by the $2\pi i$ phase change of $x_i,i=3, \cdots, N+1$;\\}
\emph{2, All non-equivalent $\mathbf{r}$ fields are generated by $g^2$ pairs of phases of $(x_2,x_{N+2})$, which are determined by the Diamond rule, where $g=N-1$.}}\newline\\
\indent Now we explain the Diamond rule. On the two-dimensional lattice $\mathbb{Z}^2$, draw a square with $(g,0)$, $(-g,0)$, $(0,g)$ and $(0,-g)$ as its four vertexes. Then divide this square as $g^2$ squares in the obvious way. Denote the coordinates of the center of each small squares as $(a_i,b_i)$, $i=1,2,\dots,g^2$. Here $(a_i,b_i)$ are the integer pairs such that $|a_i|+|b_i|\leq g-1$ and $a_i+b_i\equiv g-1\,(\mathrm{Mod}\ 2)$. Then we obtain the $g^2$ pairs of the phase of $(x_2,x_{N+2})$ as
\be
(a_i,b_i)\pi \ri,\quad\quad i=1,2,\dots,g^2.
\ee
Note the fact
\be\label{abn}
a_i+b_i\equiv N\,(\mathrm{Mod}\ 2),\quad i=1,\dots,g^2.
\ee

Now we turn to the phases of $x_i,i=3, \cdots, N+1$. Albeit the freedom of phase changing $2\pi\ri$, the relative phases between $x_i,i=3, \cdots, N+1$ are fixed. We find that there are only two possible fundamental choices for the phases of $x_i,i=3, \cdots, N+1$ which can make the identities (\ref{conjecture}) hold: either
\be
(1,0,1,0,1,0,\dots)\pi\ri
\ee
or
\be
(0,1,0,1,0,1,\dots)\pi\ri.
\ee
From the relation (\ref{eq:Batyrev1}), it is easy to see both choices are consistent with the $\mathbf{B}$ field (\ref{eq:Bfield}) components $B_i=0$ for $i=2,\dots,N-2$. We leave the discussion on $B_1$, $B_{N-1}$ and $B_N$ later. Note there are two other phase choices, $(0,0,0,\dots)\pi\ri$ and $(1,1,1,\dots)\pi\ri$,  also consistent with these $\mathbf{B}$-field components, but they do not result in the correct $\mathbf{r}$ fields. For example, the choice $(0,0,0,\dots)\pi\ri$, which means no sign-changing for the any true modulus, can be ruled out by the fact that Riemann theta function with characteristics $\boldsymbol{\alpha}=\boldsymbol{\beta}=\boldsymbol{0}$ in general does not have real constant roots. The phase choice $(1,1,1,\dots)\pi\ri$ can be ruled out by similar arguments.

To determine which is consistent with a specific pair of the phase of $(x_2,x_{N+2})$, one should keep in mind the Mod two equivalence between $\mathbf{r}$ field and $\mathbf{B}$ field. As the $\mathbf{B}$ field for $SU(N)$ CY with arbitrary $m$ are obtained in (\ref{eq:Bfield}), for Batyrev coordinate $z_{N-1}$, we require
\be
r_{N-1}=0\,(\mathrm{Mod}\ 2).
\ee
On the other hand, from equation (\ref{eq:Batyrev1}), we know
\be\label{constraint2}
r_{N-1}=\theta_{N}+b_i\,(\mathrm{Mod}\ 2),
\ee
where $\theta_{i}\pi\ri$ is the phase of $x_{i}$, $i=3,\dots,N+1$. This determines the odd-even property of $x_{N}$ and therefore which of two fundamental choices is consistent with the phase pair of $(x_2,x_{N+2})$. Since $z_1$ and $z_N$ are related to $x_2$ and $x_{N+2}$, we still need to show the above rules are consistent with the $\mathbf{B}$-field components for $z_1$ and $z_{N}$. The Batyrev relation for $z_1$ is
\be
z_1=\frac{x_2x_4}{x_3^2}.
\ee
Using the fact (\ref{abn}) and the property that for either fundamental choice,
\be
\theta_i\equiv\theta_N+(4-N)\,(\mathrm{Mod}\ 2),\quad i=3,\dots,N+1,
\ee
we have
\be
\ba
B_1&\equiv a_i+\theta_4\,(\mathrm{Mod}\ 2)\\
&\equiv a_i+\theta_N+(4-N)\,(\mathrm{Mod}\ 2)\\
&\equiv \theta_N+b_i\,(\mathrm{Mod}\ 2)\\
&\equiv r_N\,(\mathrm{Mod}\ 2)\\
&\equiv 0\,(\mathrm{Mod}\ 2)
\ea
\ee
This show that for all phase pairs $(a_i,b_i)$ with the corresponding fundamental choice, $B_1$ is consistent with the prerequisite on the $\mathbf{B}$ field (\ref{eq:Bfield}). Now let us perform the similar check on $B_N$. Using the Batyrev relation for $z_N$ in (\ref{eq:Batyrev1}), we have
\be
\ba
B_N&\equiv \theta_{m+3}+\theta_{N+1}\,(\mathrm{Mod}\ 2)\\
&\equiv \theta_N+(m+3-N)+\theta_N+1\,(\mathrm{Mod}\ 2)\\
&\equiv N-m\,(\mathrm{Mod}\ 2),
\ea
\ee
which is exactly the $B_N$ in (\ref{eq:Bfield}). Note this consistent check applies to all pairs $(a_i,b_i)$.

Let us say a few more things about the diamond rule. Apparently, from (\ref{constraint2}) we can see the fundamental choice of any two adjacent small squares should be different. Besides, it is easy to prove that a pair of conjugate $\mathbf{r}$ fields are central symmetric with respect to the original point in the square. It should be noted this elegant rule is closely related to the charge and vector choice in (\ref{suncharge},\ref{sunfan}). If one constructs a toric Calabi-Yau with a random fan, which is acceptable, there may not be such a clear rule on the nonequivalent $\mathbf{r}$ fields. We also emphasize that the diamond it itself is only relevant to the fan vectors, or equivalently complex moduli $x_i$, while the value of $\mathbf{B}$ field and $\mathbf{r}$ fields also rely on the charge choices. Different charge choices may result in different Batyrev relations, and therefore different values of $\mathbf{B}$ field and $\mathbf{r}$ fields. Another important thing is that once we know all nonequivalent $\mathbf{r}$ fields of $SU(N)$ geometries with arbitrary $m$, we can easily written down all nonequivalent $\mathbf{r}$ fields for the toric Calabi-Yau which are the reduction of $SU(N)$ geometries. These already includes a large number of interesting toric Calabi-Yau, such as resolved $\IC^3/\IZ_5$ orbifold.
\begin{figure}[htbp]
\centering\includegraphics[width=2.0in]{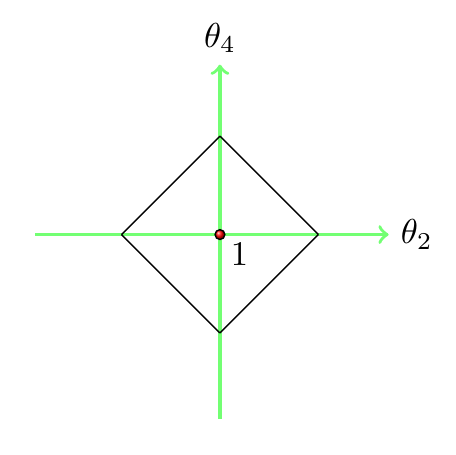}
\caption{Diamond rule for $SU(2)$ geometries}\label{diamond1}
\end{figure}

In the following subsections, we will amplify the diamond rule in $SU(3)$, $SU(4)$ and $SU(5)$ geometries. But first, let us see why the diamond rule in genus one is consistent with the rule to determine the $\mathbf{r}$ field for the del Pezzo surfaces in Section \ref{sec:ex1}. For local $\mathbb{F}_0$, $\mathbb{F}_1$ and $\mathbb{F}_2$, which are $SU(2)$ geometries with $m=0,1,2$, the diamond is just a square (Fig \ref{diamond1}), and the phase change $(\theta_2\pi\ri,\theta_4\pi\ri)$ of the two moduli $x_2,x_4$ is characterized by the origin, which means the phases of $(x_2,x_4)$ do not change. The true modulus here is $x_3$. To make the $\mathbf{r}$ field consistent with the $\mathbf{B}$ field, the fundamental choice of the phase of $x_3$ is $\pi\ri$, which means the true moduli $x_i$ should become its opposite. Remember we have assumed $x_0,x_1$ are constants in the first place, to summary we have, for $SU(2)$ geometries the true modulus $x_3$ becomes its opposite and other moduli $x_0,x_1,x_2,x_4$ remain the same. This is exactly the rule we has proposed for local del Pezzo in Section \ref{sec:ex1}.

Now let us turn to another issue. As we have mentioned in Section \ref{sec:ghm2}, it is not obvious the quantum Riemann theta function at $\hbar=2\pi/k$ must be a genuine Riemann theta function, due to the cubic term in (\ref{theta-2pik}). However, here we can prove for all $SU(N)$ geometries with $m=0$, the quantum Riemann theta function at $\hbar=2\pi/k$ is indeed a genuine Riemann theta function. This is owing to the nice structure of the coefficients of cubic terms in partition function and the $C$ matrix of $SU(N)$ geometries. The perturbative part of the NS free energy can be obtained in many approaches, see for example \cite{Iqbal:2003zz}\cite{Iqbal:2007ii}\cite{Huang:2014eha}. Here only the cubic terms concern us, which are
\be
\label{sunpertns}
\ba
\hbar F^{\rm NS}({\mathbf{t}}, \hbar)={1\over 6} \sum_{1\le l <n\le N-1} (t_l + \cdots + t_n)^3 +{t_N\over 2 N}  \sum_{1\le l <n\le N-1} (t_l + \cdots + t_n)^2+\cdots\\
\ea
\ee
And the $C$ matrix is
\be\label{sunc}
C=\left(\begin{array}{ccccccc}  -2&      1  & & & & & 0\\
   1&      -2 &1& & & & 0\\
    &        1& -2& & & & 0\\
     & & &\ddots & & & 0\\
     & & & & -2 & 1 & 0\\
     & & & & 1 & -2 & 0\end{array}\right)_{(N-1)\times N}.
\ee
The relevant cubic term in (\ref{theta-2pik}) is
\be
{ 2\pi^2\ri \over {3\hbar}}a_{ijk} C_{il} C_{jm} C_{k p} n_l n_m n_p.
\ee
At $\hbar=2\pi/k$, this becomes
\be
\frac{\pi\ri k}{3}a_{ijk} C_{il} C_{jm} C_{k p} n_l n_m n_p.
\ee
We need to prove this cubic term is effectively only quadratic. The $a_{ijk}$ coming from the second term in (\ref{sunpertns}) only produces quadratic terms of $n$, therefore we just need to deal with the $a_{ijk}$ coming from the first term in (\ref{sunpertns}). Noticing the structure of $a_{ijk}$ and $C$ matrix in (\ref{sunc}), we introduce the following variables,
\be
\ba
b_0&=-n_1,\\
b_i&=n_i-n_{i+1},\quad i=1,\dots,N-1,\\
b_N&=n_N.
\ea
\ee
Then for the $a_{ijk}$ coming from the first term in (\ref{sunpertns}), we have
\be
a_{ijk} C_{il} C_{jm} C_{k p} n_l n_m n_p=\sum_{i=0}^{N-2}\sum_{j=i+2}^{N}(b_i-b_j)^3.
\ee
Since $b_i$ and $k$ are integers, and
\be
n^3\equiv n\,(\text{Mod}\ 6),
\ee
then
\be
\ba
&\frac{\pi\ri k}{3}a_{ijk} C_{il} C_{jm} C_{k p} n_l n_m n_p\\
=&\frac{\pi\ri k}{3}\sum_{i=0}^{N-2}\sum_{j=i+2}^{N}(b_i-b_j)^3\\
\equiv&\frac{\pi\ri k}{3}\sum_{i=0}^{N-2}\sum_{j=i+2}^{N}(b_i-b_j),\,(\text{Mod}\ 2\pi\ri).
\ea
\ee
Since in theta function, the terms on the exponent are only defined up to $2\pi\ri$. Therefore, these cubic term are effectively linear terms. This concludes the proof that for all $SU(N)$ geometries with $m=0$, the quantum Riemann theta function at $\hbar=2\pi/k$ is indeed a genuine Riemann theta function. Using the perturbative part of topological string partition function of $SU(N)$ geometries with other $m$ \cite{Iqbal:2002ep}\cite{Iqbal:2003zz}, it should also be easy to extend the proof here to those cases.
\subsection{$SU(3)$ Geometries with $m=0,1,2$}\label{sec:ex4}
\subsubsection{$SU(3)$ Geometries with $m=0$}
\begin{figure}[htbp]
\centering\includegraphics[width=2.5in]{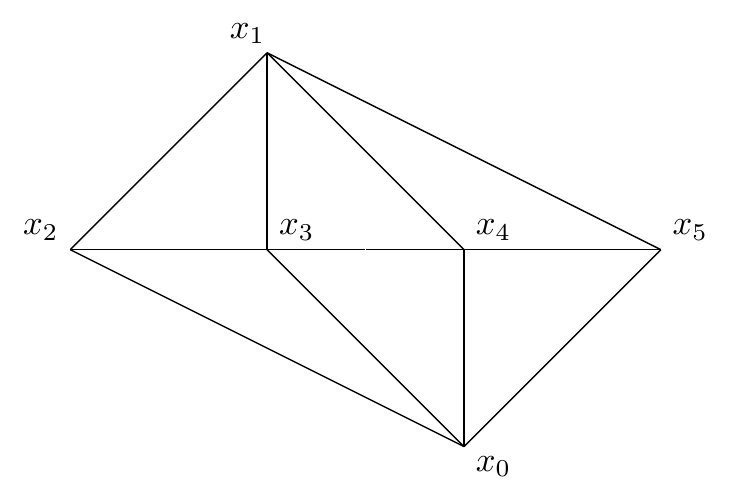}
\caption{Fan diagram of $SU(3)$ Geometry with $m=0$. Here $t_B=t_1+t_2+t_3.$}\label{fan:su3m0}
\end{figure}

The toric data for $SU(3)$ geometry with $m=0$ is as follows,\footnote{The charge choice here is not the same with the general choice in the last subsection. In fact the $Q_3$ here is $Q_1+Q_2+Q_3$ there. As we have mentioned, the different charge choices only affect the values of $\mathbf{B}$ field and $\mathbf{r}$ fields, but not the diamond rule.}
\begin{equation}
\begin{array}{c|crr|rrrr|}
    \multicolumn{5}{c}{v_i}    $Q_1$&  Q_2 & Q_3\\
    x_0    &     1&     -1&   0&           0&      0  &1\\
    x_1    &     1&     1&   -1&         0&      0  &1 \\
    x_2    &     1&     0&   -2&          1&      0  & -1 \\
    x_3    &     1&    0&   -1&          -2&      1  &0  \\
    x_4    &     1&    0&   0&            1&      -2 &0\\
     x_5    &     1&    0&   1&           0&        1& -1\\
  \end{array}
\end{equation}
The fan diagram of this geometry is illustrated in Figure \ref{fan:su3m0}.
Read from the data, the mirror curve is
\be\label{su3m0}
z_1e^{2x}+e^x+z_2e^{-x}+e^p+z_1 z_2 z_3 e^{x-p}+1=0.
\ee
This curve corresponds to set $x_4=1,x_3=e^x,x_1=e^p$ in the Batyrev's method. Here,
\be\label{Batyrevsu2m0}
z_1=\frac{x_2 x_4}{x_3^2},\ \ \ z_2=\frac{x_3 x_5}{x_4^2},\ \ \ z_3=\frac{x_0x_1}{x_2x_5}.
\ee
For $x,p$ as complex variables, this is a Riemann surface of genus two.
The true moduli here are $x_3,x_4$, therefore the $C$ matrix is
\be
C=\left(\begin{array}{ccc} -2&      1  &0 \\  1&      -2 &0\\ \end{array}\right).
\ee
Using the general result in (\ref{eq:Bfield}) or read from the refined BPS invariants, the $\mathbf{B}$ field is
\be
\mathbf{B}=(0,0,1).
\ee
We can express the Riemann surface (\ref{su3m0}) in standard form by changing the variables,
\be\label{su3m0:sf}
y^2=\frac{1}{4}(z_1 x^3+x^2+x+z_2)^2-z_1 z_2 z_3 x^3.
\ee
The discriminant (\ref{su3m0:sf}) is
\begin{dmath}
\Delta=1-8 \left(z_1+z_2\right)+16 z_1^2+68 z_2 z_1+16 z_2^2-144 z_2 z_1^2-144 z_2^2 z_1-z_2 z_3 z_1+270 z_2^2 z_1^2+36 z_2 z_3 z_1^2+36 z_2^2 z_3 z_1+216 z_2^2 z_1^3+216 z_2^3 z_1^2-513 z_2^2 z_3 z_1^2-972 z_2^3 z_1^3+540 z_2^2 z_3 z_1^3+540 z_2^3 z_3 z_1^2-27 z_2^2 z_3^2 z_1^3-243 z_2^3 z_3 z_1^3-27 z_2^3 z_3^2 z_1^2+729 z_1^4 z_2^4+1215 z_1^3 z_3^2 z_2^3-2187 z_1^4 z_2^4 z_3+2187 z_1^4 z_2^4 z_3^2-729 z_1^4 z_2^4 z_3^3.
\end{dmath}
The discriminant is very important for computing the genus one free energy.

We define $\theta_i=z_i\frac{\partial}{\partial z_i}$.  The classical periods are the solutions of Picard-Fuchs system. The solutions to such a systems can be obtained from the derivatives of a fundamental period $\omega_0$.  A-periods are the first order derivatives of the fundamental period $\partial_i \omega_0$, while B-periods are the linear combinations of the second order derivatives $\omega_{ij}=\partial_i\partial_j \omega_0$, where the coefficients of the linear combination can be fixed by the Picard-Fuchs functions. The Picard-Fuchs operators for $SU(3)$ geometry with $m=0$ are
\be
\begin{split}
\mathcal L_1&=(\theta_1-\theta_3)(\theta_1-2\theta_2)-z_1(-2\theta_1+\theta_2-1)(-2\theta_1+\theta_2),\\
\mathcal L_2&=(\theta_2-\theta_3)(\theta_2-2\theta_1)-z_2(-2\theta_2+\theta_1-1)(-2\theta_2+\theta_1),\\
\mathcal L_3&=\theta_3^2-z_3(\theta_1-\theta_3)(\theta_2-\theta_3),\\
\mathcal L_4&=\theta_3^2-z_1 z_2 z_3(\theta_1-2\theta_2)(\theta_2-2\theta_1).\\
\end{split}
\ee
Using Picard-Fuchs operators and the known refined BPS invariants, we have\footnote{The terms $b^{\mathrm{NS,GV}}_{m} \log(m)$ in the genus one expansion of free energy could not be fixed by the known refined BPS invariants, where $m$ represents the mass parameters. This is because these terms do not contribute to the BPS part of the free energy. Fortunately, these terms do not contribute to the zeros of spectral determinant, because these terms decouple with $n_i$. They do not contribute to the phase volumes in NS quantization either, since after the derivatives with respect to the true moduli, these terms vanish. Therefore we can simply set $b^{\mathrm{NS,GV}}_{m}=0.$}
\be
\begin{split}
\frac{\partial}{\partial t_1}F_0&=\omega_{11}+\frac{1}{2}\omega_{22}+\omega_{12}+\frac{2}{3}\omega_{13}+\frac{1}{3}\omega_{23},\\
\frac{\partial}{\partial t_2}F_0&=\frac{1}{2}\omega_{11}+\omega_{22}+\omega_{12}+\frac{1}{3}\omega_{13}+\frac{2}{3}\omega_{23}.\\
\end{split}
\ee
Define $Q_i=e^{t_i}$, the genus zero free energy is
\be
F_0=\frac{1}{3}t_1^3+\frac{1}{3}t_2^3+\frac{1}{2}(t_1^2t_2+t_2^2t_1)+\frac{1}{3}t_3(t_1^2+t_2^2+t_1t_2)+F_0^{inst}(t_i),
\ee
where
\be
F_0^{inst}(t_i)=2Q_1+2Q_2+2Q_3+\frac{1}{4}Q_1^2+\frac{1}{4}Q_2^2+\frac{1}{4}Q_3^2+2Q_1Q_2+2Q_1Q_3+4Q_2Q_3+\mathcal{O}(Q_i^3).
\ee
The genus one free energy can be obtained as
\be
F_1^{\mathrm{NS}}=\frac{1}{24}\log(\Delta z_1^{-4}z_2^{-4})=-\frac{1}{6}t_1-\frac{1}{6}t_2-\frac{1}{6}Q_1-\frac{1}{6}Q_2-\frac{1}{12}Q_1^2-\frac{1}{12}Q_2^2-\frac{1}{6}Q_1Q_2+\mathcal{O}(Q_i^3),
\ee
\be
F_1^{\mathrm{GV}}=\frac{1}{12}\log(\Delta z_1^{8}z_2^{8})+\frac{1}{2}\log(|G_{ij}|)=\frac{1}{6}t_1+\frac{1}{6}t_2+\frac{1}{6}Q_1+\frac{1}{6}Q_2+\frac{1}{12}Q_1^2+\frac{1}{12}Q_2^2+\frac{1}{6}Q_1Q_2+\mathcal{O}(Q_i^3),
\ee
where $G_{ij}$ is defined in \cite{Klemm:2015iya} as
\be
G_{ij}=\frac{\partial t_i}{\partial z_j}.
\ee

Using the above ingredients and the refined BPS invariants in the Appendix, we can now check the identities (\ref{check}). Scan all possible $\mathbf{r}$ fields with every components ranging from $-10$ to $10$, we finally obtain four non-equivalent $\mathbf{r}$ fields,
$$
\mathbf{r}=(-2,2,-1),(0,-2,1),(2,-2,-1),(-2,0,1).
$$
These four $\mathbf{r}$ fields can be divided as two groups $(-2,2,-1),(-2,0,1)$ and $(2,-2,-1),(0,-2,1)$ via complex conjugation, which means if $t_i\in \mathbb{R}$, then the spectral determinant determined by $(-2,2,-1)$ and $(-2,0,1)$ can transform to each other by complex conjugation. After obtaining the $\mathbf{r}$ fields, we can check the identity (\ref{conjecture}) analytically, order by order. Indeed we have verified the identities for these three $\mathbf{r}$ fields up to order $d=d_1+d_2+d_B\leqslant  6$.

\begin{figure}[htbp]
\centering\includegraphics[width=2.5in]{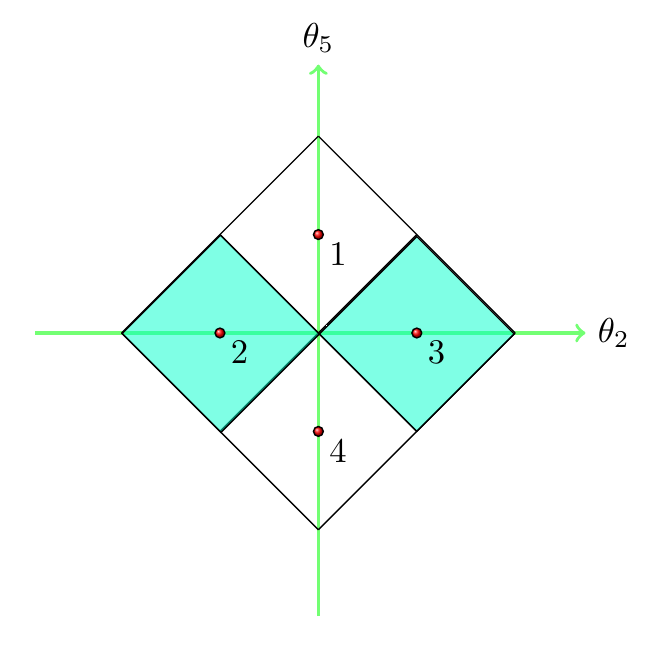}
\caption{Diamond rule for $SU(3)$ geometries}\label{diamond2}
\end{figure}

Now let us see how these four $\mathbf{r}$ fields can are generated by the Diamond rule. The Diamond for $SU(3)$ geometry is depicted in Fig \ref{diamond2}.
The coordinates of the four points in the centers of four small squares represent the phase of $x_2$ and $x_5$. Based on the diamond rule in Section \ref{sec:ex3} and the Batyrev relation (\ref{Batyrevsu2m0}) the four $\mathbf{r}$ fields can be generated by the following phases $\theta_i$ of $x_i$,
\begin{equation}
\begin{array}{|c|cccc|}\hline
    \mathbf{r} & \theta_2 & \theta_3 & \theta_4 & \theta_5\\\hline
    (-2,2,-1) & 0 & 1 & 0 & 1\\
    (0,-2,1) & -1 & 0 & 1 & 0\\
    (2,-2,-1) & 1 & 0 & 1 & 0\\
    (-2, 0, 1) & 0 & 1 & 0 & -1\\\hline
  \end{array}
\end{equation}
The two small squares with shadow correspond to the fundamental choice of the phases of $(x_3,x_4)$ as $(0,1)\pi$, while the blank small squares correspond to the fundamental choice of the phases as $(1,0)\pi$.

The identities for these four non-equivalent $\mathbf{r}$ fields are all non-trivial. For example, suppose we know the degrees $\mathbf{d}$ and spin contents $(j_L,j_R)$ of all non-zero refined BPS invariants but do not know the exact BPS numbers $N^{\mathbf{d}}_{j_L,j_R}$, the identities (\ref{check}) can result in a series of constraints on the BPS numbers. For $\mathbf{r}=(-2,2,-1),(-2,0,1)$, we obtain the following constraints from the first few orders of the identity (\ref{check}),
\be
\begin{split}
&N^{(1,0,0)}_{0,{1\over 2}}=1;N^{(0,0,1)}_{0,{0}}=-1;N^{(1,1,0)}_{0,{1\over 2}}=N^{(0,1,0)}_{0,{1\over 2}};N^{(1,1,1)}_{0,{1}}=-N^{(0,1,0)}_{0,{1\over 2}};N^{(1,1,1)}_{0,{0}}=-N^{(0,1,0)}_{0,{1\over 2}};\\
&N^{(2,0,1)}_{0,{2}}=-1;N^{(1,0,1)}_{0,{1}}=-1;N^{(0,1,0)}_{0,{1\over 2}}=0 \text{ or } 1;\cdots.
\end{split}
\ee
For $\mathbf{r}=(2,-2,-1),(0,-2,1)$, we have
\be
\begin{split}
&N^{0,1,0}_{0,\frac{1}{2}}=1;N^{0,0,1}_{0,0}=-1;N^{1,1,0}_{0,\frac{1}{2}}=N^{1,0,0}_{0,\frac{1}{2}};N^{0,2,1}_{0,2}=-1;N^{0,1,1}_{0,1}=-1;N^{1,1,1}_{0,1}=-N^{1,0,0}_{0,\frac{1}{2}};\\
&N^{1,1,1}_{0,0}=-N^{1,0,0}_{0,\frac{1}{2}};N^{1,0,0}_{0,\frac{1}{2}}=0\text{ or } 1;\cdots.
\end{split}
\ee

Finally, we can also check whether par of the four spectral determinants with the four different $\mathbf{r}$ field can already result in the spectra determined by the exact NS quantization conditions. Following the method in resolved $\cz$ model,  we find that, for any two $\mathbf{r}$ fields selecting from the two groups respectively, solve their non-perturbative corrections $\lambda$ simultaneously, then the resulting coefficients of $\lambda$ are always zero to any order. We have check this for $d_1+d_2+d_3\leqslant6$. To illustrate this phenomenon explicitly, we list the coefficients $c_{i,j}^n$ of $\lambda_i$ computed up to degree $d=d_1+d_2+d_3\leqslant2$.

For $\mathbf{r}=(0,-2,1)$,
\be\label{su3cons1}
\begin{split}
&c_{2}^0 \in \mathbb{Z},\\
&c_{2,3}^1=0,\\
&c_{2,2}^1=0,\\
&c_{2,1}^1=-\frac{\ri }{2
   \pi }e^{-2 \ri\pi  c_{1}^0} (-1+e^{2 \ri\pi  c_{1}^0}),\\
&c_{2,3}^2=0,\\
&c_{2,2}^2=0,\\
&c_{1,3}^1=0,\\
&c_{1,2}^1=-\frac{\ri}{2 \pi } (-1+e^{2 \ri\pi c_{1}^0}),\\
&c_{2,1}^2=\frac{1}{4 \pi }e^{-4 \ri\pi  c_{1}^0} (\ri-\ri e^{4 \ri\pi  c_{1}^0}+4 e^{2 \ri
   \pi  c_{1}^0} \pi c_{1,1}^1),\\
 &\text{with all other $c_{i,j}^1,c_{i,j}^2$ arbitrary(, up to $d=2)$.}
\end{split}
\ee
For $\mathbf{r}=(2,-2,-1)$ which is complex conjugated with $(0,-2,1)$, the $c_{i,j}^n$ is obviously complex conjugate to the $c_{i,j}^n$ of $\mathbf{r}=(0,-2,1)$ (in form). For simplicity, we do not list them here.

For $\mathbf{r}=(-2,2,-1)$,
\be\label{su3cons2}
\begin{split}
&c_{1}^0 \in \mathbb{Z},\\
&c_{1,3}^1=1,\\
&c_{1,2}^1=-\frac{\ri}{2\pi}(-1+e^{2\ri\pi c_{2}^0}),\\
&c_{1,1}^1=0,\\
&c_{1,3}^2=0,\\
&c^{1}_{2,3}=0,\\
&c_{1,2}^2=\frac{1}{4 \pi }(\ri-\ri e^{4 \ri\pi c_{2}^0}+4 e^{2 \ri\pi  c_{2}^0} \pi )
   c_{2,2}^1,\\
&c_{2,1}^1=-\frac{\ri}{2
   \pi } e^{-2 \ri\pi  c_{2}^0} (-1+e^{2 \ri\pi c_{2}^0}),\\
&c_{1,1}^2=0,\\
 &\text{with all other $c_{i,j}^1,c_{i,j}^2$ arbitrary(, up to $d=2)$.}
\end{split}
\ee
with (\ref{su3cons1}) and (\ref{su3cons2}), we can see that all $c_{i,j}^1,c_{i,j}^2$ are zero (except that the constant term which is an arbitrary integer can be absorbed to be zero).
\subsubsection{$SU(3)$ Geometries with $m=1$}
\begin{figure}[htbp]
\centering\includegraphics[width=2.5in]{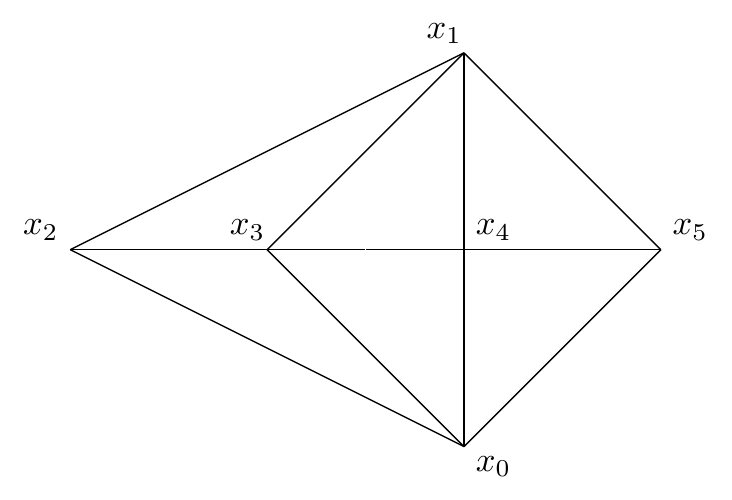}
\caption{Fan diagram of $SU(3)$ Geometry with $m=1$}\label{fan:su3m1}
\end{figure}

The toric data for $SU(3)$ geometry with $m=1$ is as follows,
\begin{equation}
\begin{array}{c|crr|rrrr|}
    \multicolumn{5}{c}{v_i}    $Q_1$&  Q_2 & Q_3\\
    x_0    &     1&     -1&   0&           0&      0  &1\\
    x_1    &     1&     1&   0&         0&      0  &1 \\
    x_2    &     1&     0&   -2&          1&      0  & 0 \\
    x_3    &     1&    0&   -1&          -2&      1  &0  \\
    x_4    &     1&    0&   0&            1&      -2 &-2\\
     x_5    &     1&    0&   1&           0&        1& 0\\
  \end{array}
\end{equation}
The fan diagram of this geometry is illustrated in Figure \ref{fan:su3m1}.
Read from the toric data, the mirror curve is
\be
z_1e^{2x}+e^x+z_2e^{-x}+e^p+z_3 e^{-p}+1=0.
\ee
This curve corresponds to setting $x_4=1,x_3=e^x,x_1=e^p$ in the Batyrev's method. Here,
\be\label{Batyrevsu3m1}
z_1=\frac{x_2 x_4}{x_3^2},\ \ \ z_2=\frac{x_3 x_5}{x_4^2},\ \ \ z_3=\frac{x_0x_1}{x_4^2}.
\ee
The true moduli here are $x_3,x_4$, therefore the $C$ matrix are
\be
C=\left(\begin{array}{ccc}-2&      1  &0  \\  1&      -2 &-2\\ \end{array}\right).
\ee
From the refined BPS invariants, or the general formula (\ref{eq:Bfield}), we see the $\mathbf{B}$ field is
\be
\mathbf{B}=(0,0,0).
\ee
Write the mirror curve into the standard form of a Riemann surface, we have
\be
y^2=\frac{1}{4}(z_1 x^3+x^2+x+z_2)^2-z_3 x^2.
\ee
The discriminant is
\begin{dmath}
\Delta=1-8 \left(z_1+z_2+z_3\right)+16 z_1^2+68 z_2 z_1+64 z_3 z_1+16 z_2^2+16 z_3^2-32 z_2 z_3-144 z_2 z_1^2-192 z_3 z_1^2-144 z_2^2 z_1-128 z_3^2 z_1+240 z_2 z_3 z_1+270 z_1^2 z_2^2+768 z_1^2 z_3^2+216 z_2^2 z_1^3+216 z_2^3 z_1^2-1024 z_3^3 z_1^2+2304 z_2 z_3^2 z_1^2-1512 z_2^2 z_3 z_1^2+2592 z_1^3 z_2^2 z_3-972 z_1^3 z_2^3+0+729 z_1^4 z_2^4.
\end{dmath}
The Picard-Fuchs operators which determine the classical periods of the geometry are
\be
\begin{split}
\mathcal L_1&=\theta_1(\theta_1-2\theta_2-2\theta_3)-z_1(-2\theta_1+\theta_2-1)(-2\theta_1+\theta_2),\\
\mathcal L_2&=\theta_2(-2\theta_1+\theta_2)-z_2(\theta_1-2\theta_2-2\theta_3-1)(\theta_1-2\theta_2-2\theta_3),\\
\mathcal L_3&=\theta_3^2-z_3(\theta_1-2\theta_2-2\theta_3-1)(\theta_1-2\theta_2-2\theta_3),\\
\mathcal L_4&=\theta_1\theta_2\theta_3^2-z_1 z_2 z_3(-2\theta_1+\theta_2)(\theta_1-2\theta_2-2\theta_3-2)(\theta_1-2\theta_2-2\theta_3-1)(\theta_1-2\theta_2-2\theta_3).\\
\end{split}
\ee
Using the Picard-Fuchs operators and the known refined BPS invariants, we have
\be
\begin{split}
\frac{\partial}{\partial t_1}F_0&=\frac{4}{9}\omega_{11}-\frac{2}{9}\omega_{22}-\frac{2}{9}\omega_{12}+\frac{2}{3}\omega_{13}+\frac{1}{3}\omega_{23},\\
\frac{\partial}{\partial t_2}F_0&=-\frac{1}{9}\omega_{11}-\frac{4}{9}\omega_{22}-\frac{4}{9}\omega_{12}+\frac{1}{3}\omega_{13}+\frac{2}{3}\omega_{23},\\
\frac{\partial}{\partial t_3}F_0&=-\frac{1}{3}\omega_{11}+\frac{1}{3}\omega_{22}+\frac{1}{3}\omega_{12}.\\
\end{split}
\ee
Then the genus zero free energy is
\be
F_0=\frac{4}{27}(t_1^3-t_2^3)+\frac{1}{3}t_3(t_1^2+t_2^2+t_1t_2)-\frac{1}{9}t_1t_2(t_1+2t_2)+F_0^{inst}(t_i),
\ee
where
\be
F_0^{inst}(t_i)=2Q_1+2Q_2+2Q_3+\frac{1}{4}Q_1^2+\frac{1}{4}Q_2^2+\frac{1}{4}Q_3^2+2Q_1Q_2+2Q_1Q_3+4Q_2Q_3+\mathcal{O}(Q_i^3).
\ee
The genus one free energy is
\be
\begin{split}
F_1^{\mathrm{NS}}=\frac{1}{24}\log(\Delta z_2^{4}z_3^{-6})=&\frac{1}{6}t_2-\frac{1}{4}t_3-\frac{1}{6}Q_1-\frac{1}{6}Q_2-\frac{1}{6}Q_3-\frac{1}{12}Q_1^2-\frac{1}{12}Q_2^2-\frac{1}{12}Q_3^2\\&-\frac{1}{6}Q_1Q_2-\frac{1}{6}Q_1Q_3-\frac{7}{3}Q_2Q_3+\mathcal{O}(Q_i^3),\\
\end{split}
\ee
and
\be
\begin{split}
F_1^{\mathrm{GV}}=\frac{1}{12}\log(\Delta z_1^{6}z_2^{4}z_3^9)+&\frac{1}{2}\log(|G_{ij}|)=-\frac{1}{6}t_2+\frac{1}{4}t_3+\frac{1}{6}Q_1+\frac{1}{6}Q_2+\frac{1}{6}Q_3+\frac{1}{12}Q_1^2\\&+\frac{1}{12}Q_2^2+\frac{1}{12}Q_3^2+\frac{1}{6}Q_1Q_2+\frac{1}{6}Q_1Q_3+\frac{1}{3}Q_2Q_3+\mathcal{O}(Q_i^3).\\
\end{split}
\ee

With all the ingredients, we now want to check our identities. For this model, we find there are four non-equivalent $\mathbf{r}$ fields
$$
\mathbf{r}=(-2,2,0),(-2,0,0),(0,-2,-2),(2,-2,-2).
$$
Combining the Batyrev relations (\ref{Batyrevsu3m1}), these $\mathbf{r}$ fields can be generated by the Diamond \ref{diamond2} as follows,
\begin{equation}
\begin{array}{|c|cccc|}\hline
    \mathbf{r} & \theta_2 & \theta_3 & \theta_4 & \theta_5\\\hline
    (-2,2,0) & 0 & 1 & 0 & 1\\
    (-2,0,0) & 0 & 1 & 0 & -1\\
    (0,-2,-2) & -1 & 0 & 1 & 0\\
    (2,-2,-2) & 1 & 0 & 1 & 0\\\hline
  \end{array}
\end{equation}
The two small squares with shadow correspond to the fundamental choice of the phases of $(x_3,x_4)$ as $(0,1)\pi$, while the two blank small squares correspond to the fundamental choice of the phases as $(1,0)\pi$. This is the same with $m=0$ case.

We have checked the identities (\ref{conjecture}) for these $\mathbf{r}$ fields up to $d=d_1+d_2+d_3\leqslant6$.  The $\mathbf{r}$ fields in this model can be divided into two groups $(-2,2,0),(-2,0,0)$ and $(0,-2,-2),(2,-2,-2)$ via complex conjugation. Selecting one $\mathbf{r}$ field from each group, the vanishing of their corresponding quantum Riemann theta functions solves the non-perturbative corrections $\lambda$ to be zero, which means the intersection points are identical to the spectra determined by the exact NS quantization conditions and do not have extra points. The vanishing of all four quantum Riemann theta function with all four $\mathbf{r}$ fields of course also is equivalent to the spectra from the exact NS quantization conditions.
\subsubsection{$SU(3)$ Geometries with $m=2$}
\begin{figure}[htbp]
\centering\includegraphics[width=2.5in]{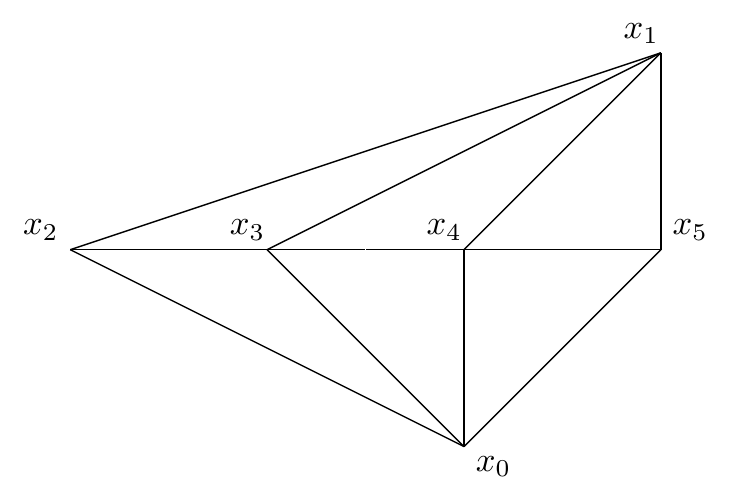}
\caption{Fan diagram $SU(3)$ Geometry with $m=2$}\label{fan:su3m2}
\end{figure}

The toric data for $SU(3)$ geometry with $m=2$ is as follows,
\begin{equation}
\begin{array}{c|crr|rrrr|}
    \multicolumn{5}{c}{v_i}    $Q_1$&  Q_2 & Q_3\\
    x_0    &     1&     -1&   0&           0&      0  &1\\
    x_1    &     1&     1&   1&         0&      0  &1 \\
    x_2    &     1&     0&   -2&          1&      0  & 0 \\
    x_3    &     1&    0&   -1&          -2&      1  &0  \\
    x_4    &     1&    0&   0&            1&      -2 &-1\\
     x_5    &     1&    0&   1&           0&        1& -1\\
  \end{array}
\end{equation}
The fan diagram of this geometry is illustrated in Figure \ref{fan:su3m2}.

Read from the toric data, the mirror curve is
\be
z_1e^{2x}+e^x+z_2e^{-x}+e^p+z_2 z_3 e^{-x-p}+1=0.
\ee
This curve correspond to setting $x_4=1,x_3=e^x,x_1=e^p$ in the Batyrev's method. Here,
\be\label{Batyrevsu3m2}
z_1=\frac{x_2 x_4}{x_3^2},\ \ \ z_2=\frac{x_3 x_5}{x_4^2},\ \ \ z_3=\frac{x_0x_1}{x_4x_5}.
\ee
The true moduli here are $x_3,x_4$, therefore the $C$ matrix are
$$C=\left(\begin{array}{ccc}  -2&      1  &0\\  1&      -2 &-1\\ \end{array}\right).$$
From the refined BPS invariants, or general formula (\ref{eq:Bfield}), we can see the $\mathbf{B}$ field is
$$
\mathbf{B}=(0,0,1).
$$
The standard form of the mirror curve is
\be
y^2=\frac{1}{4}(z_1 x^3+x^2+x+z_2)^2-z_2 z_3 x.
\ee
We immediately get the discriminant as
\begin{dmath}
\Delta=1-8 z_1-8 z_2-z_3+16 z_1^2+68 z_2 z_1+8 z_3 z_1+16 z_2^2+36 z_2 z_3-144 z_2 z_1^2-16 z_3 z_1^2-144 z_2^2 z_1-301 z_2 z_3 z_1-27 z_2 z_3^2+270 z_2^2 z_1^2+660 z_2 z_3 z_1^2+225 z_2 z_3^2 z_1+12 z_2^2 z_3 z_1+216 z_2^2 z_1^3+216 z_2^3 z_1^2-500 z_2 z_3^2 z_1^2+405 z_2^2 z_3 z_1^2-972 z_2^3 z_1^3-2700 z_2^2 z_3 z_1^3-375 z_2^2 z_3^2 z_1^2+5625 z_2^2 z_3^2 z_1^3+2025 z_2^3 z_3 z_1^3+729 z_1^4 z_2^4-3125 z_1^3 z_2^2 z_3^3.
\end{dmath}
The Picard-Fuchs operators read from the charges are
\be
\begin{split}
\mathcal L_1&=\theta_1(\theta_1-2\theta_2-\theta_3)-z_1(-2\theta_1+\theta_2-1)(-2\theta_1+\theta_2),\\
\mathcal L_2&=(\theta_2-\theta_3)(-2\theta_1+\theta_2)-z_2(\theta_1-2\theta_2-\theta_3-1)(\theta_1-2\theta_2-\theta_3),\\
\mathcal L_3&=\theta_3^2-z_3(\theta_1-2\theta_2-\theta_3)(\theta_2-\theta_3),\\
\mathcal L_4&=\theta_1\theta_3^2-z_1 z_2 z_3(\theta_1-2\theta_2-\theta_3-1)(\theta_1-2\theta_2-\theta_3)(-2\theta_1+\theta_2).\\
\end{split}
\ee
Using the Picard-Fuchs operators and the known refined BPS invariants, we have
\be
\begin{split}
\frac{\partial}{\partial t_1}F_0&=\frac{8}{9}\omega_{11}+\frac{1}{18}\omega_{22}+\frac{5}{9}\omega_{12}+\frac{2}{3}\omega_{13}+\frac{1}{3}\omega_{23},\\
\frac{\partial}{\partial t_2}F_0&=\frac{5}{18}\omega_{11}+\frac{1}{9}\omega_{22}+\frac{1}{9}\omega_{12}+\frac{2}{3}\omega_{13}+\frac{2}{3}\omega_{23},\\
\frac{\partial}{\partial t_3}F_0&=\frac{1}{3}\omega_{11}+\frac{1}{3}\omega_{22}+\frac{1}{3}\omega_{12}.\\
\end{split}
\ee
Then the genus zero free energy is
\be
F_0=\frac{1}{27}(8t_1^3+t_2^3)+\frac{1}{3}t_3(t_1^2+t_2^2+t_1t_2)+\frac{1}{18}t_1t_2(5t_1+t_2)+F_0^{inst}(t_i),
\ee
where
\be
F_0^{inst}(t_i)=2Q_1+2Q_2-Q_3+\frac{1}{4}Q_1^2+\frac{1}{4}Q_2^2-\frac{1}{8}Q_3^2+2Q_1Q_2-3Q_2Q_3+\mathcal{O}(Q_i^3).
\ee
The genus one free energy is
\be
\begin{split}
F_1^{\mathrm{NS}}=\frac{1}{24}\log(\Delta z_2^{4}z_3^{-12})=&\frac{1}{6}t_2-\frac{1}{2}t_3-\frac{1}{6}Q_1-\frac{1}{6}Q_2-\frac{1}{24}Q_3-\frac{1}{12}Q_1^2-\frac{1}{12}Q_2^2-\frac{1}{48}Q_3^2\\&-\frac{1}{6}Q_1Q_2+\frac{7}{8}Q_2Q_3+\mathcal{O}(Q_i^3).\\
\end{split}
\ee
\be
\begin{split}
F_1^{\mathrm{GV}}=\frac{1}{12}\log(\Delta z_1^{6}z_2^{4}z_3^9)+&\frac{1}{2}\log(|G_{ij}|)=-\frac{1}{6}t_2+\frac{1}{2}t_3+\frac{1}{6}Q_1+\frac{1}{6}Q_2-\frac{1}{12}Q_3+\frac{1}{12}Q_1^2\\&+\frac{1}{12}Q_2^2-\frac{1}{24}Q_3^2+\frac{1}{6}Q_1Q_2-\frac{1}{4}Q_2Q_3+\mathcal{O}(Q_i^3).\\
\end{split}
\ee
We find there are four non-equivalent $\mathbf{r}$ fields for this model
$$
\mathbf{r}=(-2,2,-1),(-2,0,1),(2,-2,-1),(0,-2,-1).
$$
Combining the Batyrev relations (\ref{Batyrevsu3m2}), these $\mathbf{r}$ fields can be generated by the Diamond (\ref{diamond2}) as follows,
\begin{equation}
\begin{array}{|c|cccc|}\hline
    \mathbf{r} & \theta_2 & \theta_3 & \theta_4 & \theta_5\\\hline
    (-2,2,-1) & 0 & 1 & 0 & 1\\
    (-2,0,1) & 0 & 1 & 0 & -1\\
    (0,-2,-1) & 1 & 0 & 1 & 0\\
    (2,-2,-1) & -1 & 0 & 1 & 0\\\hline
  \end{array}
\end{equation}
The two small squares with shadow correspond to the fundamental choice of the phases of $(x_3,x_4)$ as $(0,1)\pi$, while the two blank small squares correspond to the fundamental choice of the phases as $(1,0)\pi$. This is the same with $m=0,1$ cases.

These four non-equivalent $\mathbf{r}$ fields are divided into two groups $(-2,2,-1)$, $(-2,0,1)$ and $(2,-2,-1)$, $(0,-2,-1)$ via complex conjugation. The identities (\ref{conjecture}) for these $\mathbf{r}$ fields are also check analytically up to $d=d_1+d_2+d_3\leqslant6$. Selecting one $\mathbf{r}$ field from each group, the vanishing of their corresponding quantum Riemann theta functions solves the non-perturbative corrections $\lambda$ to be zero, which means the intersection points are identical to the spectra determined by the exact NS quantization conditions and do not have extra points.

As we have mentioned, the resolved $\mathbb{C}^3/\mathbb{Z}_5$ orbifold can be obtained from the $SU(3)$ geometry with $m=0$ by taking the limit $x_5=0$. Compare the fan diagram, it is easy to see the Batyrev coordinates $(z_1,z_2)$ of $\mathbb{C}^3/\mathbb{Z}_5$ are $z_2z_3,z_1$ with the Batyrev coordinates of $SU(3)$ geometry with $m=0$. Here we show how the three $\mathbf{r}$ fields of $\mathbb{C}^3/\mathbb{Z}_5$ are obtained from the four $\mathbf{r}$ fields of this $SU(3)$ geometry under the reduction. As the $\mathbf{r}$ field are actually the phases of the Batyrev coordinates, the transformation of Batyrev coordinates should translate to the transformation of $\mathbf{r}$ field. Therefore, we have
\be
\ba
(-2,2,-1) & \quad\rightarrow\quad(1,-2),\\
    (-2,0,1) &  \quad\rightarrow\quad(1,-2),\\
    (0,-2,-1) &  \quad\rightarrow\quad(-3,0),\\
    (2,-2,-1) &  \quad\rightarrow\quad(-3,2),
\ea
\ee
which are just the three nonequivalent $\mathbf{r}$ fields of $\mathbb{C}^3/\mathbb{Z}_5$ obtained in Section \ref{sec:ex2}. In fact, this is a general phenomenon. Under the reduction of toric Calabi-Yau which means to take some mass parameters to be zero, the $\mathbf{r}$ fields of the reduced CY can be obtained from the $\mathbf{r}$ fields of the original CY. This is consistent with taking the reduction in the identities (\ref{conjecture}).
\subsection{$SU(4)$ Geometries with $m=0$}\label{sec:ex5}
\begin{figure}[htbp]
\centering\includegraphics[width=3.5in]{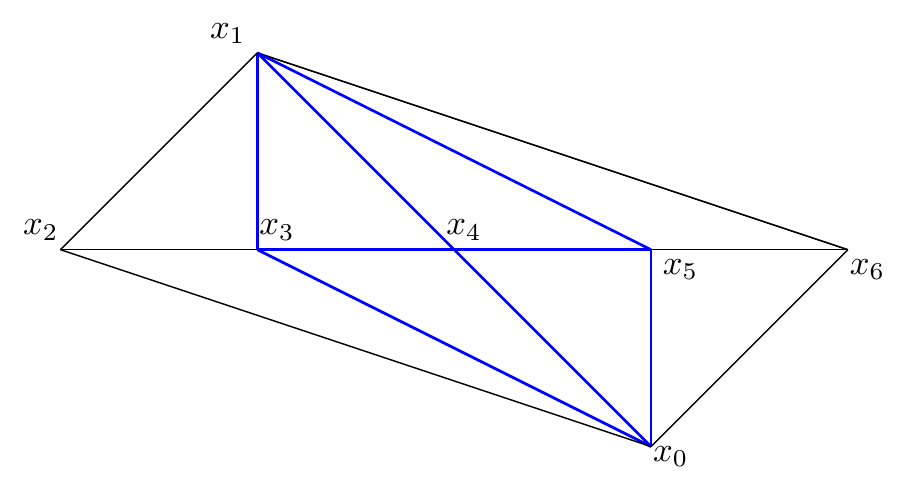}
\caption{Fan diagram of $SU(4)$ Geometry with $m=0$. There is a $\mathbb{F}_0=\mathbb{P}_1\times\mathbb{P}_1$ in this geometry (blue Fan). We choose the base of this $\mathbb{F}_0$ corresponding to $t_B$ in the refined BPS invariants in (\ref{appendix:invariants}). The charge of the fiber of $\mathbb{F}_0$ corresponding to $(v_1-v_4)+(v_0-v_4)=0.$ Then $z_B=\frac{x_1x_0}{x_4^2}=z_1z_2^2z_3z_4$, so that $t_B=t_1+2t_2+t_3+t_4.$}\label{fan:su4m0}
\end{figure}

The toric data for $SU(4)$ geometry with $m=0$ is as follows,
\begin{equation}
\begin{array}{c|crr|rrrrr|}
    \multicolumn{5}{c}{v_i}    $Q_1$&  Q_2 & Q_3 &Q_4\\
    x_0    &     1&     -1&   0&          0&      0  &0 &1\\
    x_1    &     1&     1&   -2&         0&      0  &0&1 \\
    x_2    &     1&     0&   -3&        1&      0  & 0 &-1\\
    x_3    &     1&    0&   -2&         -2&      1  &0 &0 \\
    x_4    &     1&    0&   -1&            1&      -2 &1&0\\
    x_5    &     1&    0&   0&           0&        1& -2&0\\
    x_6    &     1&    0&   1&           0&        0& 1&-1\\
  \end{array}
\end{equation}
The fan diagram of this geometry is illustrated in Figure \ref{fan:su4m0}.

\begin{figure}[htbp]
\centering\includegraphics[width=3in]{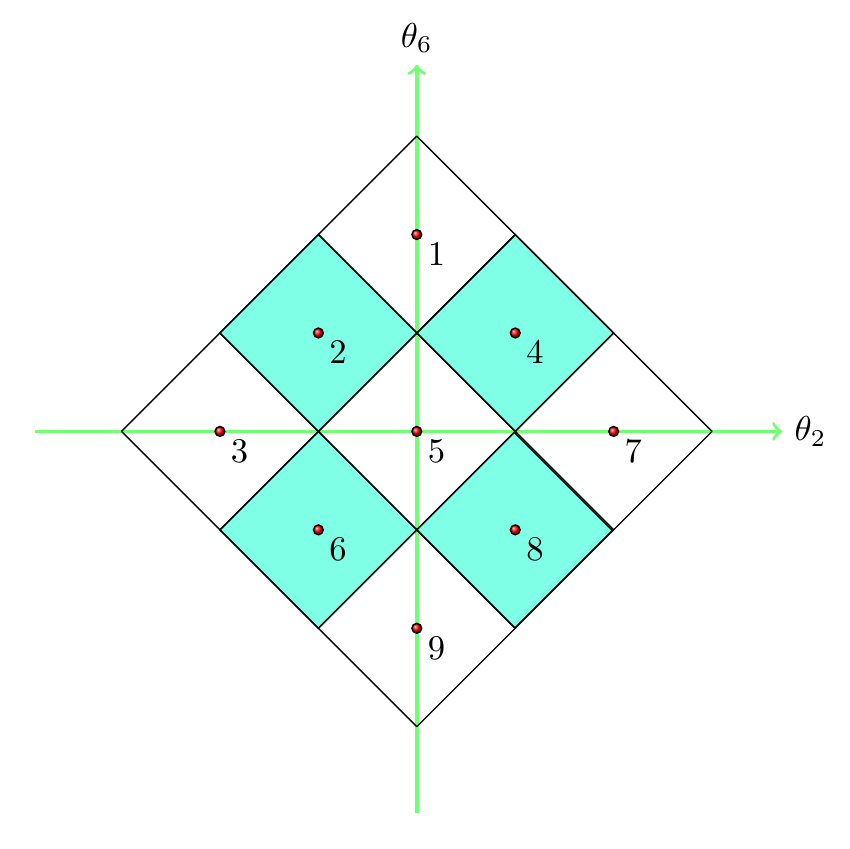}
\caption{Diamond rule for $SU(4)$ geometries. The number in each box stands for the sequence number $a$ of the each $\mathbf{r}^a$ field.}\label{dianomd3}
\end{figure}

Read from the toric data, the mirror curve is
$$\label{a4}
1+e^{x}+e^{p}+z_1e^{-x}+z_2e^{2x}+z_2^2z_3e^{3x}+z_1z_2^2z_3z_4e^{2x-p}=0.
$$
The Batyrev coordinates are
\be
z_1=\frac{x_2 x_4}{x_3^2},\ \ \ z_2=\frac{x_3 x_5}{x_4^2},\ \ \ z_3=\frac{x_4x_6}{x_5^2}\ \ \ z_4=\frac{x_0x_1}{x_2x_6}.
\ee
The true moduli here are $x_3,x_4,x_5$, therefore the $C$ matrix are
$$C=\left(\begin{array}{cccc}  -2&      1  &0 &0\\   1&      -2 &1&0\\  0&        1& -2&0\\ \end{array}\right)$$
The standard form of the mirror curve (\ref{a4}) is
$$
y^2=\frac{1}{4}(z_1+x+x^2+z_2x^3+z_2^2z_3x^4)^2-z_1z_2^2z_3z_4x^4.
$$
Then the discriminant is
\begin{dseries}
\begin{math}
\Delta=1-8 \left(z_1+z_2+z_3\right)+16 z_1^2+68 z_2 z_1+64 z_3 z_1+16 z_2^2+16 z_3^2+68 z_2 z_3-144 z_2 z_1^2-128 z_3 z_1^2-144 z_2^2 z_1-128 z_3^2 z_1-576 z_2 z_3 z_1-144 z_2 z_3^2-144 z_2^2 z_3+270 z_2^2 z_1^2+256 z_3^2 z_1^2+1216 z_2 z_3 z_1^2+1216 z_2 z_3^2 z_1+1276 z_2^2 z_3 z_1+270 z_2^2 z_3^2+216 z_2^2 z_1^3+216 z_2^3 z_1^2-2560 z_2 z_3^2 z_1^2-2328 z_2^2 z_3 z_1^2-2328 z_2^2 z_3^2 z_1+48 z_2^3 z_3 z_1-8 z_2^2 z_3 z_4 z_1+216 z_2^2 z_3^3+216 z_2^3 z_3^2-972 z_2^3 z_1^3-2016 z_2^2 z_3 z_1^3+3648 z_2^2 z_3^2 z_1^2-2340 z_2^3 z_3 z_1^2+64 z_2^2 z_3 z_4 z_1^2-2016 z_2^2 z_3^3 z_1-2340 z_2^3 z_3^2 z_1+64 z_2^2 z_3^2 z_4 z_1-32 z_2^3 z_3 z_4 z_1-972 z_2^3 z_3^3+5632 z_2^2 z_3^2 z_1^3+9504 z_2^3 z_3 z_1^3-192 z_2^2 z_3 z_4 z_1^3+5632 z_2^2 z_3^3 z_1^2+11968 z_2^3 z_3^2 z_1^2-512 z_2^2 z_3^2 z_4 z_1^2+240 z_2^3 z_3 z_4 z_1^2+9504 z_2^3 z_3^3 z_1-192 z_2^2 z_3^3 z_4 z_1+240 z_2^3 z_3^2 z_4 z_1+729 z_1^4 z_2^4+729 z_3^4 z_2^4+324 z_1 z_3^3 z_2^4+3030 z_1^2 z_3^2 z_2^4+324 z_1^3 z_3 z_2^4-26112 z_1^2 z_3^3 z_2^3-26112 z_1^3 z_3^2 z_2^3-1792 z_1^2 z_3^2 z_4 z_2^3-4096 z_1^3 z_3^3 z_2^2+1536 z_1^2 z_3^3 z_4+\dots.
\end{math}
\end{dseries}
The Picard-Fuchs operators that cancels the A,B-periods are
\be
\begin{split}
\mathcal L_1&=\left(\theta _1-2 \theta _2+\theta _3\right) \left(\theta _1-\theta _4\right)-z_1\left(-2 \theta _1+\theta _2-1\right) \left(\theta _2-2 \theta
   _1\right), \\
\mathcal L_2&=\left(-2 \theta _1+\theta _2\right) \left(\theta _2-2 \theta _3\right)-z_2 \left(-1+\theta _1-2 \theta _2+\theta _3\right) \left(\theta _1-2 \theta
   _2+\theta _3\right),\\
\mathcal L_3&=\left(\theta _1-2 \theta _2+\theta _3\right) \left(\theta _3-\theta _4\right)-z_3 \left(-1+\theta _2-2 \theta _3\right) \left(\theta _2-2 \theta
   _3\right),\\
\mathcal L_4&=\theta _4^2-z_4 \left(\theta _1-\theta _4\right) \left(\theta _3-\theta _4\right),\\
\mathcal L_5&=\theta _4^2-z_1 z_2 z_3 z_4 \left(-2 \theta _1+\theta _2\right) \left(\theta _2-2 \theta _3\right).\\
\end{split}
\ee
Solving from these operators, we have the following A-periods,
\be
\begin{split}
\frac{\partial}{\partial t_1}F_0&=\frac{3 \omega _{11}}{2}+2 \omega _{12}+\omega _{13}+\frac{3 \omega
   _{14}}{4}+\omega _{22}+\omega _{23}+\frac{\omega _{24}}{2}+\frac{\omega
   _{33}}{2}+\frac{\omega _{34}}{4},\\
\frac{\partial}{\partial t_2}F_0&=\omega _{11}+2 \omega _{12}+\omega _{13}+\frac{\omega _{14}}{2}+2 \omega _{22}+2 \omega _{23}+\omega _{24}+\omega _{33}+\frac{\omega _{34}}{2},\\
\frac{\partial}{\partial t_3}F_0&=\frac{\omega _{11}}{2}+\omega _{12}+\omega _{13}+\frac{\omega _{14}}{4}+\omega _{22}+2 \omega _{23}+\frac{\omega _{24}}{2}+\frac{3 \omega
   _{33}}{2}+\frac{3 \omega _{34}}{4}.\\
\end{split}
\ee
Using the known BPS invariants from refined topological vertex, we obtain the genus zero free energy of the refined topological string as
\begin{dseries}
\begin{math}
F_0=\frac{t_1^3}{2}+t_2 t_1^2+\frac{1}{2} t_3 t_1^2+\frac{3}{8} t_4 t_1^2+t_2^2
   t_1+\frac{1}{2} t_3^2 t_1+t_2 t_3 t_1+\frac{1}{2} t_2 t_4 t_1+\frac{1}{4} t_3
   t_4 t_1+\frac{2 t_2^3}{3}+\frac{t_3^3}{2}+t_2 t_3^2+t_2^2 t_3+\frac{1}{2}
   t_2^2 t_4+\frac{3}{8} t_3^2 t_4+\frac{1}{2} t_2 t_3 t_4+2 Q_1+2 Q_2+2 Q_3+\frac{Q_1^2}{4}+2 Q_2 Q_1+\frac{Q_2^2}{4}+\frac{Q_3^2}{4}+2 Q_2 Q_3+\mathcal O(Q_i^3).
\end{math}
\end{dseries}
The genus one free energy is
\begin{dseries}
\begin{math}
F_1^{\mathrm{GV}}=\frac{t_1}{4}+\frac{t_2}{3}+\frac{t_3}{4}+\frac{Q_1}{6}+\frac{Q_2}{6}+\frac{Q_3}{6}+\frac{Q_1^2}{12}+\frac{Q_2 Q_1}{6}+\frac{Q_2^2}{12}+\frac{Q_2 Q_3}{6}+\mathcal O(Q_i^3),
\end{math}
\end{dseries}
and
\begin{dseries}
\begin{math}
F_1^{\mathrm{NS}}=-\frac{t_1}{4}-\frac{t_2}{3}-\frac{t_3}{4}-\frac{Q_1}{6}-\frac{Q_2}{6}-\frac{Q_3}{6}-\frac{Q_1^2}{12}-\frac{Q_2 Q_1}{6}-\frac{Q_2^2}{12}-\frac{Q_2 Q_3}{6}+\mathcal O(Q_i^3).
\end{math}
\end{dseries}
From the refined BPS invariants, or general formula (\ref{eq:Bfield}), we can see the $\mathbf{B}$ field is
$$
\mathbf{B}=(0,0,0,0).
$$
For this model, we find there are nine classes of non-equivalent $\mathbf{r}$ fields for which the identities (\ref{conjecture}) holds,
\be
\begin{split}
&\ \cycle{1}:(-2,2,0,-2),\ \cycle{2}:(-2,2,0,0),\cycle{3}:(0,-2,2,2),\cycle{4}:(0,2,0,-2),\ \cycle{5}:(-2,0,2,0),\\
&\ \cycle{6}:(0,-2,0,2),\ \cycle{7}:(0,0,2,-2),\ \cycle{8}:(0,2,-2,0),\ \cycle{9}:(-2,0,0,2).\\
\end{split}
\ee
The Diamond for $SU(4)$ geometry is depicted in Fig \ref{dianomd3}.
We do not bother to present the exact phases of each $x_i$ for every $\mathbf{r}$ field here, since this has been shown many times in the $SU(3)$ geometries in the last subsection. Here we simply point out that the each $\mathbf{r}$ field with sequence number $n$ comes from the point with index $n$ in the Diamond. The four small squares with shadow correspond to the fundamental choice of the phases of $(x_3,x_4,x_5)$ as $(0,1,0)\pi$, while the five blank small squares correspond to the fundamental choice of the phases as $(1,0,1)\pi$.

We have checked analytically that all the nine $\mathbf{r}$ fields together can determine the non-perturbative correction $\lambda_i$ to be 0 for $d=d_1+d_2+d_3+d_B \leqslant 6$. From the Diamond rule, the $\mathbf{r}$ fields on the edge can be divided into 3 groups $\{\cycle{1},\cycle{9}\}$, $\{\cycle{3},\cycle{7}\}$ and $\{\cycle{2},\cycle{4},\cycle{6},\cycle{8}\}$. We can check that by selecting one $\mathbf{r}$ field from each group, the three $\mathbf{r}$ fields can already determine the non-perturbative correction $\lambda_i$ to be zero. The diamond in the middle ${\cycle{5}}$ has many free parameters in its non-perturbative correction and is less constraining.
\subsection{$SU(5)$ Geometries with $m=0$}\label{sec:ex6}
\begin{figure}[htbp]
\centering\includegraphics[width=3.5in]{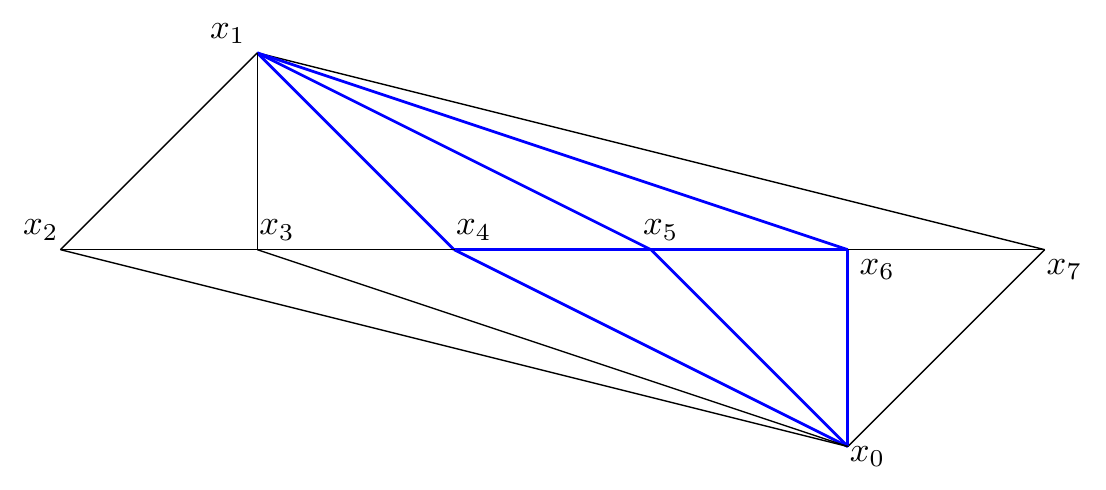}
\caption{Fan diagram of $SU(5)$ Geometry with $m=0$. There is a $\mathbb{F}_1$ inside this geometry (blue Fan). We choose the base of this $\mathbb{F}_1$ corresponding to $t_B$ in the refined BPS invariants in (\ref{appendix:invariants}). The charge of the fiber K\"{a}hler parameter $t_B$ of corresponding to $(v_1-v_5)+(v_0-v_5)-(v_4-v_5)=0.$ Then $z_B=\frac{x_0x_1}{x_4x_5}=z_1z_2^2z_3^2z_4z_5$, so that $t_B=t_1+2t_2+2t_3+t_4+t_5.$}\label{fan:su5m0}
\end{figure}

The toric data for $SU(5)$ geometry with $m=1$ is as follows,
\begin{equation}
\begin{array}{c|crr|rrrrrr|}
    \multicolumn{5}{c}{v_i}    $Q_1$&  Q_2 & Q_3 &Q_4&Q_5\\
    x_0    &     1&     -1&   0&           0&      0  &0 &0&1\\
    x_1    &     1&     1&   -3&       0&      0  &0&0&1 \\
    x_2    &     1&     0&   -4&       1&      0  & 0 &0&-1\\
    x_3    &     1&    0&   -3&        -2&      1  &0 &0&0 \\
    x_4    &     1&    0&   -2&           1&      -2 &1&0&0\\
    x_5    &     1&    0&   -1&          0&        1& -2&1&0\\
    x_6    &     1&    0&   0&          0&        0& 1&-2&0\\
    x_7    &     1&    0&   1&          0&        0& 0&1&-1\\
  \end{array}
\end{equation}
The fan diagram of this geometry is illustrated in Figure \ref{fan:su5m0}.

Read from the toric data, the mirror curve is
$$\label{a5}
1+e^{x}+e^{p}+z_3e^{-x}+z_4e^{2x}+z_2z_3^2e^{-2x}+z_1z_2^2z_3^3e^{-3x}+z_1z_2^2z_3^3z_4z_5e^{-x-p}=0.
$$
The Batyrev coordinates are
\be\label{batyrevsu4}
z_1=\frac{x_2 x_4}{x_3^2},\ \ \ z_2=\frac{x_3 x_5}{x_4^2},\ \ \ z_3=\frac{x_4x_6}{x_5^2}\ \ \ z_4=\frac{x_5x_7}{x_6^2}\ \ \ z_5=\frac{x_0x_1}{x_2x_7}.
\ee
The true moduli here are $x_3,x_4,x_5,x_6$, therefore the $C$ matrix is
$$C=\left(\begin{array}{ccccc}  -2&      1  &0 &0&0 \\    1&      -2 &1&0&0\\   0&        1& -2&1&0\\   0&        0& 1&-2&0\\ \end{array}\right)$$
The standard form of the mirror curve (\ref{a4}) is
$$
y^2=\frac{1}{4}(z_1+x+x^2+z_2x^3+z_2^2z_3x^4)^2-z_1z_2^2z_3z_4x^4.
$$
The discriminant is
\begin{small}
\begin{dseries}
\begin{math}
\Delta=1-8 \left(z_1+z_2+z_3+z_4\right)+16 z_1^2+68 z_2 z_1+64 z_3 z_1+64 \
z_4 z_1+16 z_2^2+16 z_3^2+16 z_4^2+68 z_2 z_3+64 z_2 z_4+68 z_3 \
z_4-144 z_2 z_1^2-128 z_3 z_1^2-128 z_4 z_1^2-144 z_2^2 z_1-128 z_3^2 \
z_1-128 z_4^2 z_1-576 z_2 z_3 z_1-544 z_2 z_4 z_1-544 z_3 z_4 z_1-144 \
z_2 z_3^2-128 z_2 z_4^2-144 z_3 z_4^2-144 z_2^2 z_3-128 z_2^2 z_4-144 \
z_3^2 z_4-576 z_2 z_3 z_4+270 z_2^2 z_1^2+256 z_3^2 z_1^2+256 z_4^2 \
z_1^2+1216 z_2 z_3 z_1^2+1152 z_2 z_4 z_1^2+1088 z_3 z_4 z_1^2+1216 \
z_2 z_3^2 z_1+1088 z_2 z_4^2 z_1+1152 z_3 z_4^2 z_1+1276 z_2^2 z_3 \
z_1+1152 z_2^2 z_4 z_1+1152 z_3^2 z_4 z_1+4880 z_2 z_3 z_4 z_1+270 \
z_2^2 z_3^2+256 z_2^2 z_4^2+270 z_3^2 z_4^2+1216 z_2 z_3 z_4^2+1276 \
z_2 z_3^2 z_4+1216 z_2^2 z_3 z_4+216 z_2^2 z_1^3+216 z_2^3 z_1^2-2560 \
z_2 z_3^2 z_1^2-2304 z_2 z_4^2 z_1^2-2304 z_3 z_4^2 z_1^2-2328 z_2^2 \
z_3 z_1^2-2160 z_2^2 z_4 z_1^2-2304 z_3^2 z_4 z_1^2-10304 z_2 z_3 z_4 \
z_1^2-2328 z_2^2 z_3^2 z_1-2304 z_2^2 z_4^2 z_1-2160 z_3^2 z_4^2 \
z_1-10304 z_2 z_3 z_4^2 z_1+48 z_2^3 z_3 z_1-10784 z_2 z_3^2 z_4 \
z_1-10784 z_2^2 z_3 z_4 z_1+216 z_2^2 z_3^3+216 z_3^2 z_4^3+216 z_2^3 \
z_3^2+216 z_3^3 z_4^2-2328 z_2 z_3^2 z_4^2-2560 z_2^2 z_3 z_4^2+48 \
z_2 z_3^3 z_4-2328 z_2^2 z_3^2 z_4+\cdots.
\end{math}
\end{dseries}
\end{small}
The Picard-Fuchs operators that cancels the A,B-periods are
\be
\begin{split}
\mathcal L_1&=\left(\theta _1-2 \theta _2+\theta _3\right) \left(\theta _1-\theta _5\right)-z_1 \left(-1-2 \theta _1+\theta _2\right) \left(-2 \theta
   _1+\theta _2\right), \\
\mathcal L_2&=\left(-2 \theta _1+\theta _2\right) \left(\theta _2-2 \theta _3+\theta _4\right)-z_2 \left(-1+\theta _1-2 \theta _2+\theta _3\right) \left(\theta _1-2
   \theta _2+\theta _3\right),\\
\mathcal L_3&=\left(\theta _1-2 \theta _2+\theta _3\right) \left(\theta _3-2 \theta _4\right)-z_3 \left(-1+\theta _2-2 \theta _3+\theta _4\right) \left(\theta _2-2
   \theta _3+\theta _4\right),\\
\mathcal L_4&=\left(\theta _2-2 \theta _3+\theta _4\right) \left(\theta _4-\theta _5\right)-z_4 \left(-1+\theta _3-2 \theta _4\right) \left(\theta _3-2 \theta
   _4\right),\\
\mathcal L_5&=\theta _5^2-z_5 \left(\theta _1-\theta _5\right) \left(\theta _4-\theta _5\right),\\
\mathcal L_6&=\theta _5^2-z_1 z_2 z_3 z_4 z_5 \left(-2 \theta _1+\theta _2\right) \left(\theta _3-2 \theta _4\right).\\
\end{split}
\ee
Solving from these Picard-Fuchs operators, we have the following relations with the fundamental period
\small{\be
\begin{split}
\frac{\partial}{\partial t_1}F_0&=2 \omega _{11}+3 \omega _{12}+2 \omega _{13}+\omega _{14}+\frac{4 \omega
   _{15}}{5}+\frac{3 \omega _{22}}{2}+2 \omega _{23}+\omega _{24}+\frac{3 \omega
   _{25}}{5}+\omega _{33}+\omega _{34}+\frac{2 \omega _{35}}{5}+\frac{\omega
   _{44}}{2}+\frac{\omega _{45}}{5},\\
\frac{\partial}{\partial t_2}F_0&=\frac{3 \omega _{11}}{2}+3 \omega _{12}+2 \omega _{13}+\omega _{14}+\frac{3
   \omega _{15}}{5}+3 \omega _{22}+4 \omega _{23}+2 \omega _{24}+\frac{6 \omega
   _{25}}{5}+2 \omega _{33}+2 \omega _{34}+\frac{4 \omega _{35}}{5}+\omega
   _{44}+\frac{2 \omega _{45}}{5},\\
\frac{\partial}{\partial t_3}F_0&=\omega _{11}+2 \omega _{12}+2 \omega _{13}+\omega _{14}+\frac{2 \omega
   _{15}}{5}+2 \omega _{22}+4 \omega _{23}+2 \omega _{24}+\frac{4 \omega
   _{25}}{5}+3 \omega _{33}+3 \omega _{34}+\frac{6 \omega _{35}}{5}+\frac{3
   \omega _{44}}{2}+\frac{3 \omega _{45}}{5},\\
   \frac{\partial}{\partial t_4}F_0&=\frac{\omega _{11}}{2}+\omega _{12}+\omega _{13}+\omega _{14}+\frac{\omega
   _{15}}{5}+\omega _{22}+2 \omega _{23}+2 \omega _{24}+\frac{2 \omega
   _{25}}{5}+\frac{3 \omega _{33}}{2}+3 \omega _{34}+\frac{3 \omega _{35}}{5}+2
   \omega _{44}+\frac{4 \omega _{45}}{5},\\
\end{split}
\ee}
Using the known BPS invariants from refined topological vertex, we have the following genus zero free energy of the refined topological string.
\begin{dseries}
\begin{math}
F_0=\frac{2 t_1^3}{3}+\frac{3}{2} t_2 t_1^2+t_3 t_1^2+\frac{1}{2} t_4
   t_1^2+\frac{2}{5} t_5 t_1^2+\frac{3}{2} t_2^2 t_1+t_3^2 t_1+\frac{1}{2} t_4^2
   t_1+2 t_2 t_3 t_1+t_2 t_4 t_1+t_3 t_4 t_1+\frac{3}{5} t_2 t_5 t_1+\frac{2}{5}
   t_3 t_5 t_1+\frac{1}{5} t_4 t_5 t_1+t_2^3+t_3^3+\frac{2 t_4^3}{3}+2 t_2
   t_3^2+t_2 t_4^2+\frac{3}{2} t_3 t_4^2+2 t_2^2 t_3+t_2^2 t_4+\frac{3}{2} t_3^2
   t_4+2 t_2 t_3 t_4+\frac{3}{5} t_2^2 t_5+\frac{3}{5} t_3^2 t_5+\frac{2}{5}
   t_4^2 t_5+\frac{4}{5} t_2 t_3 t_5+\frac{2}{5} t_2 t_4 t_5+\frac{3}{5} t_3 t_4
   t_5+2 Q_1+2 Q_2+2 Q_3+2 Q_4+\frac{Q_1^2}{4}+2 Q_2 Q_1+\frac{Q_2^2}{4}+\frac{Q_3^2}{4}+\frac{Q_4^2}{4}+2 Q_2
   Q_3+2 Q_3 Q_4+\mathcal O(Q_i^3).
\end{math}
\end{dseries}
The genus one free energy is
\begin{small}
\begin{dmath}
F_1^{\mathrm{GV}}=\frac{t_1}{3}+\frac{t_2}{2}+\frac{t_3}{2}+\frac{t_4}{3}+\frac{Q_1}{6}+\frac{Q_2}{6}+\frac{Q_3}{6}+\frac{Q_4}{6}+\frac{Q_1^2}{12}+\frac{Q_2
   Q_1}{6}+\frac{Q_2^2}{12}+\frac{Q_3^2}{12}+\frac{Q_4^2}{12}+\frac{Q_2
   Q_3}{6}+\frac{Q_3 Q_4}{6}+\mathcal O(Q_i^3),
\end{dmath}
\end{small}
and
\begin{small}
\begin{dmath}
F_1^{\mathrm{NS}}=-\frac{t_1}{3}-\frac{t_2}{2}-\frac{t_3}{2}-\frac{t_4}{3}-\frac{Q_1}{6}-\frac{Q_2}{6}-\frac{Q_3}{6}-\frac{Q_4}{6}-\frac{Q_1^2}{12}-\frac{Q_2
   Q_1}{6}-\frac{Q_2^2}{12}-\frac{Q_3^2}{12}-\frac{Q_4^2}{12}-\frac{Q_2
   Q_3}{6}-\frac{Q_3 Q_4}{6}+\mathcal O(Q_i^3).
\end{dmath}
\end{small}
From the refined BPS invariants, or general formula (\ref{eq:Bfield}), we can see the $\mathbf{B}$ field is
$$
\mathbf{B}=(0,0,0,0,1).
$$
For this model, we find there are sixteen classes of non-equivalent $\mathbf{r}$ fields for which the identities (\ref{conjecture}) hold,
\be
\begin{split}
&\cycle{1}:(0,0,0,2,-3),\ \cycle{2}:(0,0,2,-2,-1),\ \cycle{3}:(0,2,-2,0,1),\ \cycle{4}:(2,-2,0,0,3),\\
&\cycle{5}:(0,0,2,0,-3),\ \cycle{6}:(0,2,-2,2,-1),\ \cycle{7}:(2,0,-2,0,1),\ \cycle{8}:(0,-2,0,0,3),\\
&\cycle{9}:(0,2,0,0,-3),\ \cycle{10}:(2,-2,2,0,-1),\ \cycle{11}:(0,-2,0,2,1),\ \cycle{12}:(0,0,-2,0,3),\\
&\cycle{13}:(2,0,0,0,-3),\ \cycle{14}:(-2,2,0,0,-1),\ \cycle{15}:(0,-2,2,0,1),\ \cycle{16}:(0,0,-2,2,3).\\
\end{split}
\ee
\begin{figure}[htbp]
\centering\includegraphics[width=3.5in]{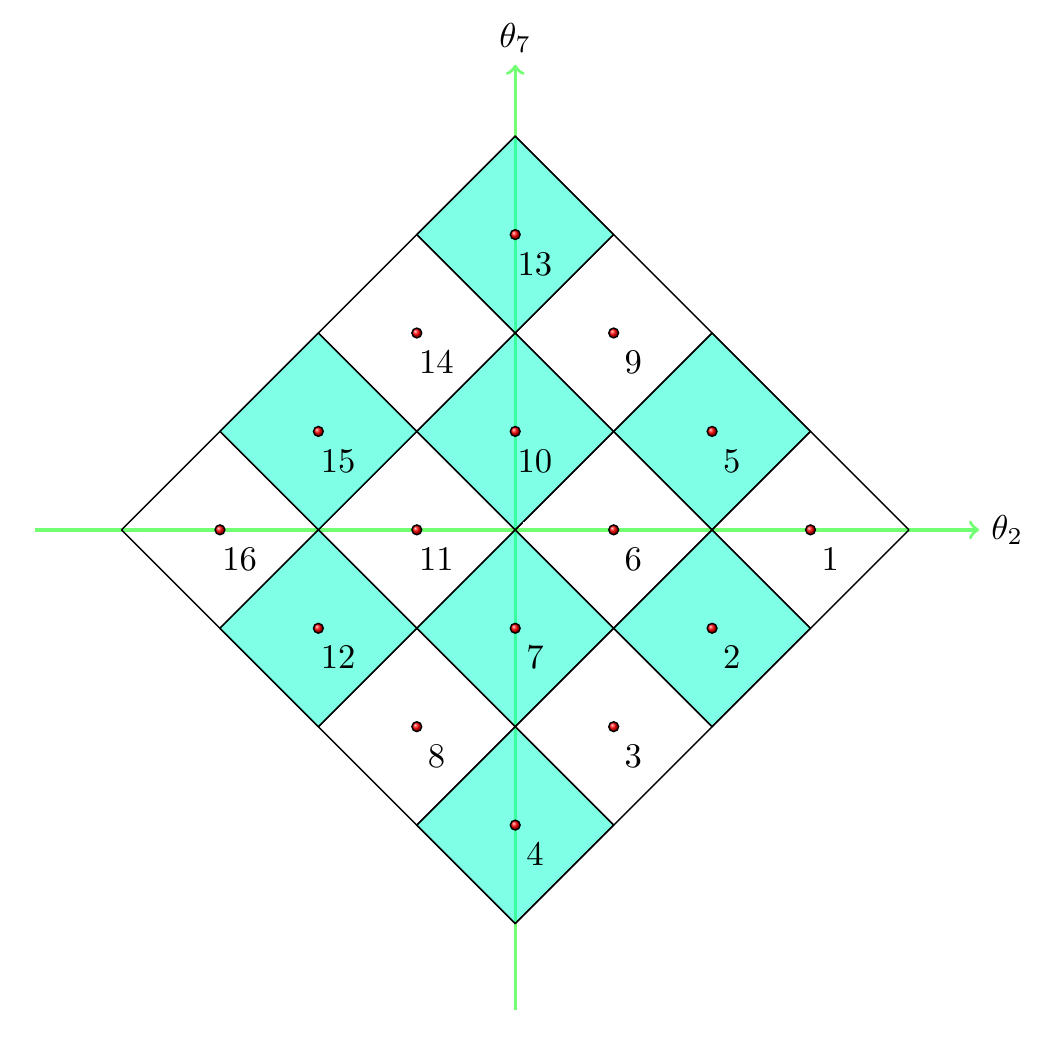}
\caption{Diamond rule for $SU(5)$ geometries. The number in each box stands for the sequence number $a$ of the each $\mathbf{r}^a$ field.}
\label{diamond4}
\end{figure}
The Diamond for $SU(4)$ geometry is depicted in Fig \ref{diamond4}. Each of the sixteen $\mathbf{r}$ fields with sequence number $a$ comes from the point with index $a$ in the Diamond. The eight small squares with shadow correspond to the fundamental choice of the phases of $(x_3,x_4,x_5,x_6)$ as $(0,1,0,1)\pi$, while the five blank small squares correspond to the fundamental choice of the phases as $(1,0,1,0)\pi$.

We have checked analytically that all the sixteen $\mathbf{r}$ fields together can solve the non-perturbative corrections $\lambda_i$ to be zero for $d=d_1+d_2+d_3+d_4+d_B \leqslant 6$. From the Diamond rule, the $\mathbf{r}$ fields on the edge can be divided into four groups $\{\cycle{1},\cycle{16}\}$, $\{\cycle{4},\cycle{13}\}$, $\{\cycle{3},\cycle{8},\cycle{9},\cycle{14}\}$ and $\{\cycle{2},\cycle{5},\cycle{12},\cycle{15}\}$. We can check that by selecting one $\mathbf{r}$ field from each group, the four $\mathbf{r}$ fields can already solve the non-perturbative corrections $\lambda_i$ to be zero.

We propose the following empirical rule to classify the $\mathbf{r}$ fields for $SU(N)$ geometries. We divide the $\mathbf{r}$ fields corresponding to the outermost layer of small squares into $g$ groups in a central symmetric way: Each group contains the following center points of the small squares,
\be
\ba
\mathrm{Group}\ 1\ :\,&(0,g-1),\ (0,1-g);\\
\mathrm{Group}\ 2\ :\,&(1,g-2),\ (-1,g-2),\ (1,2-g),\ (-1,2-g);\\
&\cdots\\
\mathrm{Group}\ n\ :\,&(n-1,g-n),\ (1-n,g-n),\ (n-1,n-g),\ (1-n,n-g);\\
&\cdots\\
\mathrm{Group}\ g-1\ :\,&(g-2,1),\ (2-g,1),\ (g-2,-1),\ (2-g,-1);\\
\mathrm{Group}\  g\ :\,&(g-1,0),\ (1-g,0).
\ea
\ee
We conjecture that choosing one $\mathbf{r}$ field from each group, then the resulting $g$ quantum Riemann theta functions can already determine the full Nekrasov-Shatashvili spectra without extra intersection points.
\section{Outlook}
\label{sec:outlook}
In this paper, by introducing the concept of $\mathbf{r}$ fields, we establish the precise relation between Nekrasov-Shatashvili quantization scheme and Grassi-Hatsuda-Mari\~no quantization scheme for the mirror curve of arbitrary toric Calabi-Yau threefold. We find a set of novel identities which guarantee the equivalence. We provide a physical explanation and an effective way to determine the previously mysterious $\mathbf{B}$ fields for arbitrary toric Calabi-Yau threefold. By improving the existing methods, we give two derivations of exact NS quantization conditions from Lockhart-Vafa partition function. We extend the GHM conjectures and check their equivalence with exact NS quantization for plenty of local Calabi-Yau and rigorously prove the equivalence for some models at special Planck constants. We also propose a general method to written down all non-equivalent $\mathbf{r}$ fields for $SU(N)$ geometries.\\
\indent Although abundant with physical arguments and nontrivial tests, it should be remembered that both exact NS quantization conditions and GHM quantization conditions are still conjectures which are well-defined in mathematics, and in lack of proofs for even the simplest model like local $\mathbb{P}^2$ at generic Planck constant. We regard our equivalence-check in this paper as strong support to both of them. As we can see in the paper, the equivalence between two quantization schemes are highly nontrivial and requires infinite constraints among the refined Gopakumar-Vafa invariants. Therefore, it is plausible that the rigorous proof of the equivalence should emerge after we have a deeper understanding on non-perturbative topological string. Although we do not seek a general proof in the current paper, we point out that both NS and GHM conjecture are directly related to the Lockhart-Vafa partition function of non-perturbative topological string, one in its NS limit, the other in its unrefined limit. It is shown in \cite{Aganagic:2003qj} that when one considers the branes in refined topological string theory, their wave-functions satisfy a multi-time version of the Schr\"{o}dinger equation. In NS limit, the time dependence vanishes and time-independent Schr\"{o}dinger equation is exactly the equation of quantum mirror curve. Therefore, one would like to engage the possibility that there exists a more general conjecture concerning the full refined topological string and the Schr\"{o}dinger equation depending on multiple times. And the different limits of this unknown grand conjecture may lead to NS and GHM conjectures. We expect the structure of infinite sum of exponential will remain in the grand conjecture, since this actually indicates the inclusion of branes \cite{Hollands:2009ar,Dijkgraaf:2002ac}. Besides, the form of Nekrasov-Okounkov partition function defined in \cite{Nekrasov:2003rj} should also give some hints on the subject. We hope to address this generalization in the future.\\
\indent Let us say a few more things on the quantization of mirror curve. The different quantization schemes are in fact the results of different non-perturbative completions of topological string. It is quite natural that one perturbative theory, either field theory or string theory, may have many different non-perturbative completions. For example, it was considered in \cite{Grassi:2014cla} to use Borel resummation to study the asymptotic series of topological string partition function. However it turns out that in many cases like $\mathbb{P}^2$ the result is different from the completion defined in GHM conjectures. Therefore, it should be worthwhile to quest for a more general non-perturbative completion of refined topological string which includes the various features of current approaches, such as modular invariance, branes and resummation. \\
\indent Although we have established the relation between NS and GHM quantization schemes, many questions remains. The most important one is to find an effective method to determine all non-equivalent $\mathbf{r}$ fields merely with the knowledge of the toric data. We have achieved this for $SU(N)$ geometries and their reductions. Another direction is to complete our proof at $\hbar=2\pi/k$ for all toric Calabi-Yau. As we have shown, at $\hbar=2\pi/k$ only genus zero invariants remain, which are easy to calculate in B model. Our proof in the text only apply to more restricted cases with $\mathbf{r}=2C\cdot \boldsymbol{\alpha}$ and other conditions. It would be valuable to give a proof of the equivalence for all local toric Calabi-Yau at $\hbar=2\pi/k$. This will involves the issue like whether the quantum Riemann theta functions are always genuine Riemann theta functions at Planck constant $\hbar=2\pi/k$. We conjecture this to be true but it is not very obvious for general toric Calabi-Yau with higher genus mirror curve and a large number of mass parameters. One obstacle is that unlike the compact cases there is no instinctive definition of the coefficients of cubics in the local cases. Nevertheless, this is indeed true for all known examples and all $SU(N)$ geometries with $m=0$. Besides, as we have noticed in the local del Pezzo cases and all $SU(N)$ geometries, the $b_i$ and $b_i^{\rm NS}$ in free energy are closed related. Although this is only one nontrivial constrain from the infinite list determined by the identity (\ref{check}), it would be nice to single out an elegant relation for $b_i$ and $b_i^{\rm NS}$, as our conjecture $b_i+b_i^{\rm NS}=0$ for those $SU(N)$ geometries.\\
\indent All through this paper, we study the local toric Calabi-Yau, which is noncompact. It is commonly believed that only on local Calabi-Yau, topological string can have a well-defined refinement. However, it was recently indicated in \cite{Huang:2015sta} that compact Calabi-Yau, in particular elliptic $\mathbb{P}^2$ may also have the refinement. It would be intriguing to see whether there exist identities like (\ref{check}) for compact Calabi-Yau. If so, the identities may give us more constraints among the enumerative invariants.\\
\indent Another intriguing issue is that because the number of non-equivalent $\mathbf{r}$ fields are typically larger than the genus of mirror curve, one need to explain why more than $g$ divisors of quantum theta function should intersect at the same discrete points. Although part of the coincidence results from the complex conjugation, the complete explanation is still lacking. Physically, the divisors for non-equivalent $\mathbf{r}$ fields describe the same mirror curve, but they indeed have different loci. Besides, one may also ask whether part of the $w_{\Sigma}$ divisors can already determine all the spectra. We have studied this for every model in the paper but only obtain partial answer. It seems to be a general phenomenon that with appropriate choose, only $g_{\Sigma}$ nonequivalent divisors can already determine the full spectra and have no extra intersection point. The methods there should be useful to study this question for arbitrary toric Calabi-Yau in the future.
\section*{Acknowledgements}
We thank Alba Grassi, Wei Gu, Babak Haghighat, Qingyuan Jiang, Marcos Mari\~no, Bin Xu, Jie Yang and especially Albrecht Klemm for discussion. MH is supported by the ``Young Thousand People" plan by the Central Organization Department in China,  Natural Science Foundation of China and CAS Center for Excellence in Particle Physics(CCEPP).
\appendix

\section{Riemann Theta Function}
\label{sec:Theta}
In this section, we collect some basic properties of Riemann theta function. Our notation follows \cite{mumford1,mumford2}. See also the chapter VI of \cite{farkas}.\\
\indent Siegel upper half-space is the multi-dimensional analog of the Poincare upper half-plane and is defined as
\be
\mathbb{H}_n=\{F\in M(g,\mathbb{C})|F=F^T ~\textrm{and}~\mathrm{Im} F>0\},
\ee
which is the set of symmetric matrices whose imaginary part is positive definite.\\
\indent Let $\boldsymbol{z}\in\mathbb{C}^g$, and $\Omega\in\mathbb{H}_g$, Riemann theta function is defined by
\be \label{thetadef}  \vartheta \left({\boldsymbol{z}}, \Omega \right)= \sum_{ {\boldsymbol{n}} \in \IZ^g }e^{ \pi\ri {\boldsymbol{n}}\cdot  \Omega\cdot
{\boldsymbol{n}}  +2 \pi \ri {\boldsymbol{n}}   \cdot{\boldsymbol{z}}}.
\ee
Riemann theta function is the multi-dimensional generalization of Jacobi theta function and shares many fine properties. For example, it is easy to prove the following transformation property, for all vector $\boldsymbol{n},\boldsymbol{m}\in\mathbb{Z}^g$,
\be\label{thetatrans}
\vartheta \left(\boldsymbol{z}+\boldsymbol{n}+\boldsymbol{m}\Omega, \Omega \right)=e^{2\pi\ri\left(-\boldsymbol{m}\boldsymbol{z}-\frac{1}{2}\boldsymbol{m}\cdot\Omega\cdot\boldsymbol{m}\right)}\vartheta \left({\boldsymbol{z}}, \Omega \right).
\ee
Also, Riemann theta function is a even function,
\be \label{thetaeven}  \vartheta \left({\boldsymbol{z}}, \Omega \right)=\vartheta \left({\boldsymbol{-z}}, \Omega \right).
\ee
Let $\alpha,\beta\in\mathbb{R}^g$, Riemann theta function with characteristics $\boldsymbol{\alpha}$, $\boldsymbol{\beta}$ is defined by
\be \label{thetachardef}  \vartheta \left[\begin{array}{cc} \boldsymbol{\alpha}\\ \boldsymbol{\beta}
 \end{array}\right] \left({\boldsymbol{z}}, \Omega \right)= \sum_{ {\boldsymbol{n}} \in \IZ^g }e^{  \pi\ri  ({\boldsymbol{n}}  + {\boldsymbol{\alpha}})  \Omega
({\boldsymbol{n}}  + {\boldsymbol{\alpha}}) +2 \pi \ri\left( {\boldsymbol{n}} + {\boldsymbol{\alpha}} \right)\cdot ({\boldsymbol{z}}  + {\boldsymbol{\beta}})}.
\ee
It is a translation of the standard Riemann theta function multiplied by an exponential factor:
\be \label{thetachartrans}  \vartheta \left[\begin{array}{cc} \boldsymbol{\alpha}\\ \boldsymbol{\beta}
 \end{array}\right] \left({\boldsymbol{z}}, \Omega \right)=e^{   \pi\ri  {\boldsymbol{\alpha}}\cdot \Omega
\cdot{\boldsymbol{\alpha}} +2 \pi \ri {\boldsymbol{\alpha}}\cdot ({\boldsymbol{z}}  + {\boldsymbol{\beta}}) } \vartheta\left({\boldsymbol{z}+\Omega\boldsymbol{\alpha}+\boldsymbol{\beta}}, \Omega \right).
\ee
Characteristics whose elements are either $0$ or $1/2$ are called half-period characteristics. For given $\Omega$, there are $2^g$ g-dimensional Riemann theta functions with half-period characteristics.\\
\indent At genus one, the four one dimensional Riemann theta functions are just the well-known four Jacobi theta functions,
\be
\ba
\theta_1(z,\tau)&=-\vartheta \left[\begin{array}{cc} \tiny\frac{1}{2}\\ \tiny\frac{1}{2}
 \end{array}\right] \left({z}, \tau \right),\\
 \theta_2(z,\tau)&=\vartheta \left[\begin{array}{cc} \tiny\frac{1}{2}\\ 0
 \end{array}\right] \left({z}, \tau \right),\\
 \theta_3(z,\tau)&=\vartheta \left[\begin{array}{cc} 0\\ 0
 \end{array}\right] \left({z}, \tau \right),\\
 \theta_4(z,\tau)&=\vartheta \left[\begin{array}{cc} 0\\ \tiny\frac{1}{2}
 \end{array}\right] \left({z}, \tau \right).\\
\ea
\ee
Another useful transformation law is
\be
\vartheta \left[\begin{array}{cc} \boldsymbol{\alpha}+\boldsymbol{n}\\ \boldsymbol{\beta}+\boldsymbol{m}
 \end{array}\right] \left({\boldsymbol{z}}, \Omega \right)=e^{   2\pi\ri\boldsymbol{\alpha}\cdot\boldsymbol{m} } \vartheta \left[\begin{array}{cc} \boldsymbol{\alpha}\\ \boldsymbol{\beta}
 \end{array}\right] \left({\boldsymbol{z}}, \Omega \right).
\ee
Because this property, the elements of $\boldsymbol{\alpha}$ and $\boldsymbol{\beta}$ are usually restricted to $[0,1)$, without loss of generality.\\
It is also easy to prove for Riemann theta functions with half-period characteristics,
\be \label{thetaformula}  \vartheta \left[\begin{array}{cc} \boldsymbol{\alpha}\\ \boldsymbol{\beta}
 \end{array}\right] \left(-{\boldsymbol{z}}, \Omega \right)=(-1)^{4\boldsymbol{\alpha}\cdot\boldsymbol{\beta}}\vartheta \left[\begin{array}{cc} \boldsymbol{\alpha}\\ \boldsymbol{\beta}
 \end{array}\right] \left({\boldsymbol{z}}, \Omega \right).
\ee
Immediately we obtain the following corollary: if
\be
4\boldsymbol{\alpha}\cdot\boldsymbol{\beta}=1\ (\mathrm{Mod}\ 2),
\ee
then $\boldsymbol{z}=0$ is a zero point of the Riemann theta functions with half-period characteristics. This fact is crucial for us to prove the equivalence at $\hbar=2\pi/k$.
\section{Refined Topological Vertex}
In this section, we introduce some basic facts of $SU(N)$ geometry for self-consistence and fix the conventions. This kind of geometries have been studied in many papers, e.g.  \cite{Iqbal:2003zz}.

\begin{figure}[htbp]
\centering\includegraphics[width=3.5in]{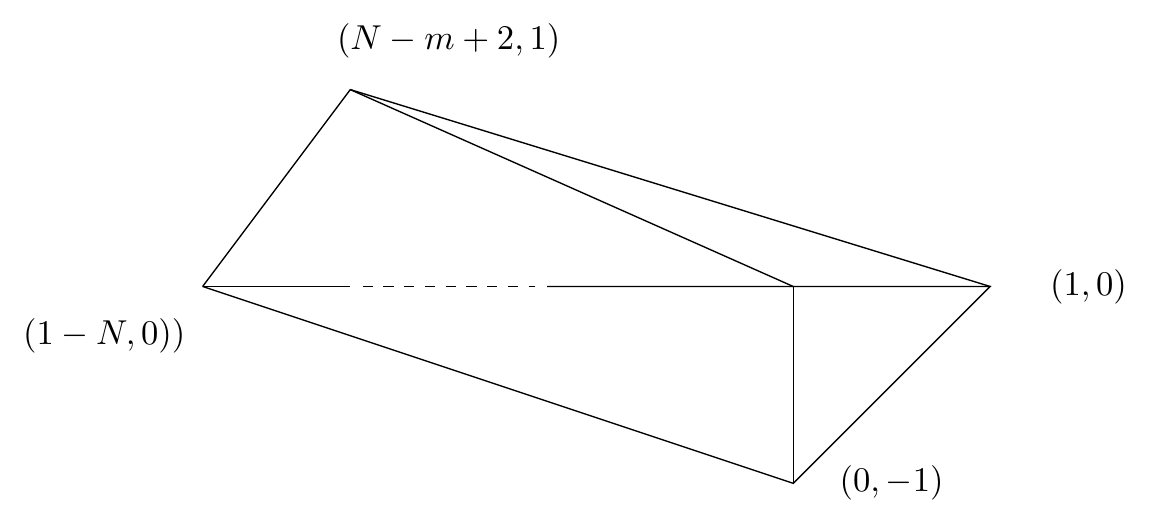}
\caption{Fan diagram}\label{genericfan}
\end{figure}

The fan diagram for $SU(N)$ geometries with Chern-Simons number $m$ is depicted in Fig \ref{genericfan}. This geometry is $A_{N-1}$ fibration over $\mathbb{P}^1$ (with another canonical line bundle added). There are $N+1$ types of the fibration, which lead to $N+1$ different geometries labelled by $m$, $(m=0,1,\cdots,N)$. The vectors of the fan diagram are
\be
\bs
v_1&=(1,-1,0),\\
v_2&=(1,1,m+2-N),\\
v_{i+3}&=(1,0,-N+1+i),\ \ i=0,1,\cdots, N.\\
\end{split}
\ee
The charges for $m=0,\cdots,N-1$ are
\begin{alignat*}{3}
Q_1&=(0,0,1,-2,1,0,0,&&0,0,0,0,\cdots,0,0,0),\\
 Q_2&=(0,0,0,1,-2,1,0,&&0,0,0,0,\cdots,0,0,0),\\
 Q_{3}&=(0,0,0,0,1,-2,1,&&0,0,0,0,\cdots,0,0,0),\\
  &\ \ \ \ \ \  &&\vdots\\
 Q_{N-1}&=(0,0,0,0,0,0,0,0,&&0,0,0,\cdots,1,-2,1),\\
Q_{N}&=(1,1,0,0,\cdots,0,-&&1,0,\cdots,0,0,0,\cdots,0,-1,0).\\
& &&\downarrow\\
& && (m+4)\text{th component of } Q_{N} \text{ have another } -1.\\
\end{alignat*}

For $m=N$, $Q_{N}=(1,1,0,0,\cdots,0,0,-2)$. The rank of all charges is $N+3$.\footnote{The charges $Q_1,\cdots,Q_{N-1}$ are related to the fiber $t_{F_i}$ and $Q_N$ is related to one of the $t_{b_i}$. We will fix the notation in the specific models.}

\begin{figure}[htbp]
\centering\includegraphics[width=7.5in]{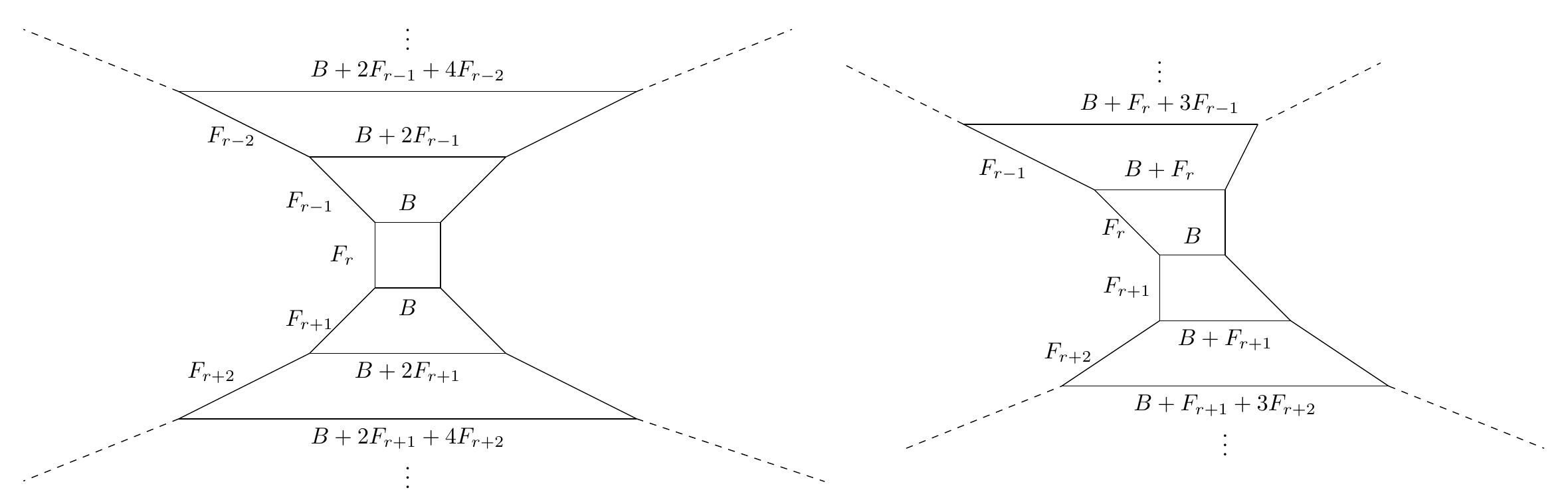}
\caption{The web diagram of $SU(N)$ geometry. It is a combination of even(left) or odd(right) Hirzebruch surface. From the divisors of Hirzebruch surface, we can easily read the relation with each base K\"{a}hler parameter $t_{b_i}$ with $t_B$ and $t_{F_i}$}
\label{webdiagram}
\end{figure}

As shown in \cite{Iqbal:2003zz}, the base of the $SU(N)$ geometries is a combination of many Hirzebruch surfaces, and we can easily write down the K\"{a}hler parameters of each edge of the web diagram, which is dual to the fan diagram.  The web diagram is illustrated in Fig. \ref{webdiagram}.

From Fig. \ref{webdiagram}, we can see the K\"{a}hler parameters $t_{b_i}$ are related with the K\"{a}hler parameters of base $t_B$ and fiber $t_{F_i}$ by

for $N+m=2r+1$,
\begin{eqnarray}\label{t1}
t_{b_{r+1}}&=&t_{B}\,,\\ \nonumber
t_{b_{r+1-i}}&=&t_{B}+\sum_{j=1}^{i}(2j-1)t_{F_{r+1-j}}\,,\,\,i=1,\cdots,r\,,\\ \nonumber
t_{b_{r+1+i}}&=&t_{B}+\sum_{j=1}^{i}(2j-1)t_{F_{r+j}}\,,\,\,i=1,\cdots,N-r-1\,,\\ \nonumber
\end{eqnarray}
and for $N+m=2r$,
\begin{eqnarray}\label{t2}
t_{b_{r}}&=&t_{b_{r+1}}=t_{B} \,,\\ \nonumber
t_{b_{r-i}}&=&t_{B}+\sum_{j=1}^{i}2j\,t_{F_{r-j}}\,,i=1,\cdots,r-1\\ \nonumber
t_{b_{r+1+i}}&=&t_{B}+\sum_{j=1}^{i}2j\,t_{F_{r+j}}\,,i=1,\cdots,N-r-1\,,\nonumber
\label{k2}
\end{eqnarray}

The partition function of topological string on toric Calabi-Yau can be computed via refined topological vertex \cite{Iqbal:2007ii}\footnote{One can also refer to \cite{Bao:2013pwa} for a good introduction on refined topological vertex}.  For $SU(N)$ geometries, the refined partition function has been computed in \cite{Taki:2007dh}.
\be\label{pf}
Z^{SU(N),m}=\sum_{\mu_1,\cdots,\mu_N}\prod_{a=1}^{N}\left[(-Q_{b,a})^{|\mu_a|}f_{\mu_a}(t,q)^{n_a}\right]K_{\mu_1\cdots\mu_N}(Q_{F,a},t,q)K_{\mu_N^{t}\cdots\mu_1^{t}}(Q_{F,a}^{t},q,t),
\ee
where the framing factor are given by $n_a=-(N+m-2a+1)$, $\mu$ is the partitions of a number, and
\be
K_{\mu_1\cdots\mu_N}(Q_{F,a},t,q)=\prod_{a=1}^N\left[q^{\frac{||\mu_a||^2}{2}}\tilde{Z}_{\mu_a}(t,q) \right]\times \prod_{1\leqslant a<b \leqslant N}\prod_{i,j=1}^{\infty}\frac{1}{1-Q_{ab}t^{-\mu_{a,i}^t+j}q^{-\mu_{b,j}+i-1}},
\ee
\be
\tilde{Z}_{\mu}(t,q)=\prod_{(i,j)\in \mu} (1-t^{\mu_{j}^t-i+1}q^{\mu_i-j})^{-1},
\ee
\be
Q_{ab}=\prod_{i=a}^{b-1}Q_{i},
\ee
\be
|\mu_a|=\sum_{i\in \mu_a} \mu_{a,i},\ \ ||\mu_a||=\sum_{i\in \mu_a} \mu_{a,i}^2.
\ee
Using the notation (\ref{t1}) (\ref{t2}), the product term with $Q_{b,a}$ can be write as
\be\nonumber
\prod_{i=1}^{N}Q_{b,a}^{|\mu_a|}
=
\begin{cases}
Q_{B}^{\sum_{i=1}^{N}l_{i}}
\,\prod_{i=1}^{\frac{N+m-1}{2}}Q_{F_{i}}^{(N+m-2i)(l_{1}+\cdots l_{i})}
\prod_{i=\frac{N+m-1}{2}+1}^{N-1}Q_{F_{i}}^{(2i-N-m)(l_{i+1}+\cdots l_{N})}\,,\,\,N+m=\mbox{odd}\,\\ \nonumber
Q_{B}^{\sum_{i=1}^{N}l_{i}}
\,\prod_{i=1}^{\frac{N+m}{2}-1}Q_{F_{i}}^{(N+m-2i)(l_{1}+\cdots l_{i})}
\prod_{i=\frac{N+m}{2}+1}^{N-1}Q_{F_{i}}^{(2i-N-m)(l_{i+1}+\cdots l_{N})}\,,\,\,N+m=\mbox{even}\,.
\end{cases}
\label{product}
\ee
where $l_a=|\mu_a|$.

With the knowledge of refine free energy, we can extract the refined BPS invariants by comparing (\ref{pf}) with the form of Gopakumar-Vafa formula. A standard symbolic computation will cost a lot of computing time. However, we can extract the invariants by setting some arbitrary numbers to $q,t$ and regarding the invariants as a unknown number. Then the problem becomes to solve a linear system, which will be much easier and saves a lot of computing time. This method was used in \cite{Wang:2014ega} to extract the BPS invariants of local $\mathbb{P}^1 \times \mathbb{P}^1 $ up to degree $d=d_1+d_2$ to 18. The refined BPS invariants of each geometry we have studied have been listed in the appendix for the first few degrees.
\section{Mirror Maps}
For completeness, we list the classical mirror maps of the $SU(N)$ geometries studied in this paper. Here we follow the traditional definition of mirror map without an imaginary phase attached. The A,B-periods satisfy the Picard-Fuchs operators, which are defined as
\be
\mathcal{L}_a=\prod_{Q^a_i>0}(\frac{\partial}{\partial x_i})^{Q^a_{i}}-\prod_{Q^a_i<0}(\frac{\partial}{\partial x_i})^{Q^a_{i}}.
\ee
In the models we have computed, we choose $Q^a$ to be the charge vectors of each model, and the summation of each charges vectors $Q^{N+1}=Q_1+Q_2+\cdots+Q_{N}$ as an additional charge.

Then the fundamental period is
\be
\omega_0(z,\rho)=\sum_{i_1,\cdots,i_N=1}^{\infty}\frac{\prod_{k=1}^{N+3}\Gamma(1+\sum_{a=1}^N Q^{a}_k\rho_a)}{\prod_{k=1}^{N+3}\Gamma(1+\sum_{a=1}^N Q^{a}_k(\rho_a+i_a))}\prod_{a=1}^{N}z_a^{\rho_a+i_a},
\ee
where $\Gamma(x)$ is gamma function. Then the notations in the Section \ref{sec:examples} are
\be
t_i=\omega_i=\frac{\partial}{\partial \rho_i}\omega_0(z,\rho)|_{\rho\rightarrow 0},
\ee
\be
\omega_{ij}=\frac{\partial^2}{\partial \rho_i\partial \rho_j}\omega_0(z,\rho)|_{\rho\rightarrow 0}.
\ee
\subsection*{$SU(3)$ geometry with $m=0$}
\begin{dgroup*}
\bdm
t_1=\log \left(z_1\right)+2 z_1-z_2+3 z_1^2-\frac{3 z_2^2}{2}+\frac{20 z_1^3}{3}-2 z_2 z_1^2+z_2^2 z_1-\frac{10 z_2^3}{3}+\frac{35 z_1^4}{2}-8 z_2 z_1^3-4 z_2 z_3 z_1^2+4 z_2^3 z_1+2 z_2^2 z_3 z_1-\frac{35 z_2^4}{4}+\frac{252 z_1^5}{5}-30 z_2 z_1^4-24 z_2 z_3 z_1^3+15 z_2^4 z_1+12 z_2^3 z_3 z_1-\frac{126 z_2^5}{5},
\edm
\bdm
t_2=\log \left(z_2\right)+2 z_2-z_1+3 z_2^2-\frac{3 z_1^2}{2}-\frac{10 z_1^3}{3}+z_2 z_1^2-2 z_2^2 z_1+\frac{20 z_2^3}{3}-\frac{35 z_1^4}{4}+4 z_2 z_1^3+2 z_2 z_3 z_1^2-8 z_2^3 z_1-4 z_2^2 z_3 z_1+\frac{35 z_2^4}{2}-\frac{126 z_1^5}{5}+15 z_2 z_1^4+12 z_2 z_3 z_1^3-30 z_2^4 z_1-24 z_2^3 z_3 z_1+\frac{252 z_2^5}{5},
\edm
\bdm
t_3=\log \left(z_3\right).
\edm
\end{dgroup*}
\subsection*{$SU(3)$ geometry with $m=1$}
\begin{dgroup*}
\bdm
t_1=\log \left(z_1\right)+2 z_1-z_2-z_3+3 z_1^2-\frac{3 z_2^2}{2}-\frac{3 z_3^2}{2}-6 z_2 z_3+\frac{20 z_1^3}{3}-2 z_2 z_1^2+6 z_3 z_1^2+z_2^2 z_1-\frac{10 z_2^3}{3}-\frac{10 z_3^3}{3}-30 z_2 z_3^2-30 z_2^2 z_3+\frac{35 z_1^4}{2}-8 z_2 z_1^3+40 z_3 z_1^3+4 z_2^3 z_1+12 z_2^2 z_3 z_1-\frac{35 z_2^4}{4}-\frac{35 z_3^4}{4}-140 z_2 z_3^3-315 z_2^2 z_3^2-140 z_2^3 z_3+\frac{252 z_1^5}{5}-30 z_2 z_1^4+210 z_3 z_1^4+15 z_2^4 z_1+90 z_2^2 z_3^2 z_1+120 z_2^3 z_3 z_1-\frac{126 z_2^5}{5}-\frac{126 z_3^5}{5}-630 z_2 z_3^4-2520 z_2^2 z_3^3-2520 z_2^3 z_3^2-630 z_2^4 z_3,
\edm
\bdm
t_2=\log \left(z_2\right)-z_1+2 z_2+2 z_3-\frac{3 z_1^2}{2}+3 z_2^2+3 z_3^2+12 z_2 z_3-\frac{10 z_1^3}{3}+z_2 z_1^2-3 z_3 z_1^2-2 z_2^2 z_1+\frac{20 z_2^3}{3}+\frac{20 z_3^3}{3}+60 z_2 z_3^2+60 z_2^2 z_3-\frac{35 z_1^4}{4}+4 z_2 z_1^3-20 z_3 z_1^3-8 z_2^3 z_1-24 z_2^2 z_3 z_1+\frac{35 z_2^4}{2}+\frac{35 z_3^4}{2}+280 z_2 z_3^3+630 z_2^2 z_3^2+280 z_2^3 z_3-\frac{126 z_1^5}{5}+15 z_2 z_1^4-105 z_3 z_1^4-30 z_2^4 z_1-180 z_2^2 z_3^2 z_1-240 z_2^3 z_3 z_1+\frac{252 z_2^5}{5}+\frac{252 z_3^5}{5}+1260 z_2 z_3^4+5040 z_2^2 z_3^3+5040 z_2^3 z_3^2+1260 z_2^4 z_3,
\edm
\bdm
t_3=\log \left(z_3\right)+2 z_2+2 z_3+3 z_2^2+12 z_3 z_2+3 z_3^2+\frac{20 z_2^3}{3}-2 z_1 z_2^2+60 z_3 z_2^2+60 z_3^2 z_2+\frac{20 z_3^3}{3}+\frac{35 z_2^4}{2}-8 z_1 z_2^3+280 z_3 z_2^3+630 z_3^2 z_2^2-24 z_1 z_3 z_2^2+280 z_3^3 z_2+\frac{35 z_3^4}{2}+\frac{252 z_2^5}{5}-30 z_1 z_2^4+1260 z_3 z_2^4+5040 z_3^2 z_2^3-240 z_1 z_3 z_2^3+5040 z_3^3 z_2^2-180 z_1 z_3^2 z_2^2+1260 z_3^4 z_2+\frac{252 z_3^5}{5},
\edm
\begin{dsuspend}
with
\end{dsuspend}
\end{dgroup*}
\bdm
2t_1+4t_2-3t_3=2\log \left(z_1\right)+4\log \left(z_2\right)-3\log \left(z_3\right).
\edm
\subsection*{$SU(3)$ geometry with $m=2$}
\begin{dgroup*}
\bdm
t_1=\log \left(z_1\right)+2 z_1-z_2+3 z_1^2-\frac{3 z_2^2}{2}+2 z_2 z_3+\frac{20 z_1^3}{3}-2 z_2 z_1^2+z_2^2 z_1-\frac{10 z_2^3}{3}+12 z_2^2 z_3+\frac{35 z_1^4}{2}-8 z_2 z_1^3+4 z_2^3 z_1-6 z_2^2 z_3 z_1-\frac{35 z_2^4}{4}-15 z_2^2 z_3^2+60 z_2^3 z_3+\frac{252 z_1^5}{5}-30 z_2 z_1^4-8 z_2 z_3 z_1^3+15 z_2^4 z_1+6 z_2^2 z_3^2 z_1-60 z_2^3 z_3 z_1-\frac{126 z_2^5}{5}-210 z_2^3 z_3^2+280 z_2^4 z_3,
\edm
\bdm
t_2=\log \left(z_2\right)+2 z_2-z_1-\frac{3 z_1^2}{2}+3 z_2^2-4 z_2 z_3-\frac{10 z_1^3}{3}+z_2 z_1^2-2 z_2^2 z_1+\frac{20 z_2^3}{3}-24 z_2^2 z_3-\frac{35 z_1^4}{4}+4 z_2 z_1^3-8 z_2^3 z_1+12 z_2^2 z_3 z_1+\frac{35 z_2^4}{2}+30 z_2^2 z_3^2-120 z_2^3 z_3-\frac{126 z_1^5}{5}+15 z_2 z_1^4+4 z_2 z_3 z_1^3-30 z_2^4 z_1-12 z_2^2 z_3^2 z_1+120 z_2^3 z_3 z_1+\frac{252 z_2^5}{5}+420 z_2^3 z_3^2-560 z_2^4 z_3,
\edm
\bdm
t_3=\log \left(z_3\right)+z_2+\frac{3 z_2^2}{2}-2 z_2 z_3+\frac{10 z_2^3}{3}-z_1 z_2^2-12 z_3 z_2^2+\frac{35 z_2^4}{4}-4 z_1 z_2^3-60 z_3 z_2^3+15 z_3^2 z_2^2+6 z_1 z_3 z_2^2+\frac{126 z_2^5}{5}-15 z_1 z_2^4-280 z_3 z_2^4+210 z_3^2 z_2^3+60 z_1 z_3 z_2^3-6 z_1 z_3^2 z_2^2,
\edm
\begin{dsuspend}
with
\end{dsuspend}
\end{dgroup*}
\bdm
t_1+2t_2-3t_3=\log \left(z_1\right)+2\log \left(z_2\right)-3\log \left(z_3\right).
\edm
\subsection*{$SU(4)$ geometry with $m=0$}
\begin{dgroup*}
\bdm
t_1=\log \left(z_1\right)+2 z_1-z_2+3 z_1^2-\frac{3 z_2^2}{2}+\frac{20 z_1^3}{3}-2 z_2 z_1^2+z_2^2 z_1-\frac{10 z_2^3}{3}+z_2^2 z_3+\frac{35 z_1^4}{2}-8 z_2 z_1^3+4 z_2^3 z_1-z_2^2 z_3 z_1-\frac{35 z_2^4}{4}+4 z_2^3 z_3+\frac{252 z_1^5}{5}-30 z_2 z_1^4+15 z_2^4 z_1-6 z_2^3 z_3 z_1-z_2^2 z_3 z_4 z_1-\frac{126 z_2^5}{5}+15 z_2^4 z_3,
\edm
\bdm
t_2=\log \left(z_2\right)-z_1+2 z_2-z_3-\frac{3 z_1^2}{2}+3 z_2^2-\frac{3 z_3^2}{2}-\frac{10 z_1^3}{3}+z_2 z_1^2-2 z_2^2 z_1+\frac{20 z_2^3}{3}-\frac{10 z_3^3}{3}+z_2 z_3^2-2 z_2^2 z_3-\frac{35 z_1^4}{4}+4 z_2 z_1^3-8 z_2^3 z_1+2 z_2^2 z_3 z_1+\frac{35 z_2^4}{2}-\frac{35 z_3^4}{4}+4 z_2 z_3^3-8 z_2^3 z_3-\frac{126 z_1^5}{5}+15 z_2 z_1^4-30 z_2^4 z_1+12 z_2^3 z_3 z_1+2 z_2^2 z_3 z_4 z_1+\frac{252 z_2^5}{5}-\frac{126 z_3^5}{5}+15 z_2 z_3^4-30 z_2^4 z_3,
\edm
\bdm
t_3=\log \left(z_3\right)+2 z_3-z_2+3 z_3^2-\frac{3 z_2^2}{2}-\frac{10 z_2^3}{3}+z_1 z_2^2+z_3 z_2^2-2 z_3^2 z_2+\frac{20 z_3^3}{3}-\frac{35 z_2^4}{4}+4 z_1 z_2^3+4 z_3 z_2^3-z_1 z_3 z_2^2-8 z_3^3 z_2+\frac{35 z_3^4}{2}-\frac{126 z_2^5}{5}+15 z_1 z_2^4+15 z_3 z_2^4-6 z_1 z_3 z_2^3-z_1 z_3 z_4 z_2^2-30 z_3^4 z_2+\frac{252 z_3^5}{5},
\edm
\bdm
t_4=\log \left(z_4\right).
\edm
\end{dgroup*}
\subsection*{$SU(5)$ geometry with $m=0$}
\begin{dgroup*}
\bdm
t_1=\log \left(z_1\right)+2 z_1-z_2+3 z_1^2-\frac{3 z_2^2}{2}+\frac{20 z_1^3}{3}-2 z_2 z_1^2+z_2^2 z_1-\frac{10 z_2^3}{3}+z_2^2 z_3+\frac{35 z_1^4}{2}-8 z_2 z_1^3+4 z_2^3 z_1-z_2^2 z_3 z_1-\frac{35 z_2^4}{4}+4 z_2^3 z_3+\frac{252 z_1^5}{5}-30 z_2 z_1^4+15 z_2^4 z_1-6 z_2^3 z_3 z_1-\frac{126 z_2^5}{5}+15 z_2^4 z_3,
\edm
\bdm
t_2=\log \left(z_2\right)-z_1+2 z_2-z_3-\frac{3 z_1^2}{2}+3 z_2^2-\frac{3 z_3^2}{2}-\frac{10 z_1^3}{3}+z_2 z_1^2-2 z_2^2 z_1+\frac{20 z_2^3}{3}-\frac{10 z_3^3}{3}+z_2 z_3^2-2 z_2^2 z_3+z_3^2 z_4-\frac{35 z_1^4}{4}+4 z_2 z_1^3-8 z_2^3 z_1+2 z_2^2 z_3 z_1+\frac{35 z_2^4}{2}-\frac{35 z_3^4}{4}+4 z_2 z_3^3-8 z_2^3 z_3+4 z_3^3 z_4-z_2 z_3^2 z_4-\frac{126 z_1^5}{5}+15 z_2 z_1^4-30 z_2^4 z_1+12 z_2^3 z_3 z_1+\frac{252 z_2^5}{5}-\frac{126 z_3^5}{5}+15 z_2 z_3^4-30 z_2^4 z_3+15 z_3^4 z_4-6 z_2 z_3^3 z_4,
\edm
\bdm
t_3=\log \left(z_3\right)-z_2+2 z_3-z_4-\frac{3 z_2^2}{2}+3 z_3^2-\frac{3 z_4^2}{2}-\frac{10 z_2^3}{3}+z_1 z_2^2+z_3 z_2^2-2 z_3^2 z_2+\frac{20 z_3^3}{3}-\frac{10 z_4^3}{3}+z_3 z_4^2-2 z_3^2 z_4-\frac{35 z_2^4}{4}+4 z_1 z_2^3+4 z_3 z_2^3-z_1 z_3 z_2^2-8 z_3^3 z_2+2 z_3^2 z_4 z_2+\frac{35 z_3^4}{2}-\frac{35 z_4^4}{4}+4 z_3 z_4^3-8 z_3^3 z_4-\frac{126 z_2^5}{5}+15 z_1 z_2^4+15 z_3 z_2^4-6 z_1 z_3 z_2^3-30 z_3^4 z_2+12 z_3^3 z_4 z_2+\frac{252 z_3^5}{5}-\frac{126 z_4^5}{5}+15 z_3 z_4^4-30 z_3^4 z_4,
\edm
\bdm
t_4=\log \left(z_4\right)+2 z_4-z_3+3 z_4^2-\frac{3 z_3^2}{2}-\frac{10 z_3^3}{3}+z_2 z_3^2+z_4 z_3^2-2 z_4^2 z_3+\frac{20 z_4^3}{3}-\frac{35 z_3^4}{4}+4 z_2 z_3^3+4 z_4 z_3^3-z_2 z_4 z_3^2-8 z_4^3 z_3+\frac{35 z_4^4}{2}-\frac{126 z_3^5}{5}+15 z_2 z_3^4+15 z_4 z_3^4-6 z_2 z_4 z_3^3-30 z_4^4 z_3+\frac{252 z_4^5}{5},
\edm
\bdm
t_5=\log \left(z_5\right).
\edm
\end{dgroup*}
\newpage\section{Refined Gopakumar-Vafa Invariants}\label{appendix:invariants}
We list the refined BPS invariants from refined topological vertex in this section. The refined BPS invariants $N^{\mathbf{d}}_{j_L,j_R}$ defined in this paper differ by a factor $(-1)^{2j_L+2j_R+1}$ with the conventional refined BPS invariants which are always positive. Here we list the data with the later definition.
\subsection*{Resolved $\IC^3/\IZ_5$ Orbifold}
\begin{small}
\begin{spacing}{1.3}
\begin{longtable}{| p{0.01\textwidth} | p{\textwidth} |} \hline  $d$&$\sum_{d_1+d_2=d} \sum_{j_L,j_R} \oplus N^{\mathbf{d}}_{j_L,j_R}(j_L,j_R)_{d_1,d_2} $\\*[0.1cm] \hline 1 & \begin{math}(0,\frac{1}{2})_{0,1}\oplus(0,1)_{1,0}\end{math}\\*[0.1cm] \hline2 & \begin{math}(0,0)_{1,1}\oplus(0,1)_{1,1}\oplus(0,\frac{5}{2})_{2,0}\end{math}\\*[0.1cm] \hline3 & \begin{math}(0,1)_{1,2}\oplus(0,\frac{3}{2})_{2,1}\oplus(0,\frac{5}{2})_{2,1}\oplus(0,3)_{3,0}\oplus(\frac{1}{2},\frac{9}{2})_{3,0}\end{math}\\*[0.1cm] \hline4 & \begin{math}(0,2)_{1,3}\oplus(0,\frac{1}{2})_{2,2}\oplus(0,\frac{3}{2})_{2,2}\oplus(0,\frac{5}{2})_{2,2}\oplus(0,2)_{3,1}\oplus2(0,3)_{3,1}\oplus(0,4)_{3,1}\oplus(\frac{1}{2},\frac{7}{2})_{3,1}\oplus(\frac{1}{2},\frac{9}{2})_{3,1}\oplus(0,\frac{5}{2})_{4,0}\oplus(0,\frac{9}{2})_{4,0}\oplus(0,\frac{13}{2})_{4,0}\oplus(\frac{1}{2},4)_{4,0}\oplus(\frac{1}{2},5)_{4,0}\oplus(\frac{1}{2},6)_{4,0}\oplus(1,\frac{11}{2})_{4,0}\oplus(\frac{3}{2},7)_{4,0}\end{math}\\*[0.1cm] \hline5 & \begin{math}(0,3)_{1,4}\oplus(0,\frac{1}{2})_{2,3}\oplus(0,\frac{3}{2})_{2,3}\oplus(0,\frac{5}{2})_{2,3}\oplus(0,1)_{3,2}\oplus2(0,2)_{3,2}\oplus3(0,3)_{3,2}\oplus(0,4)_{3,2}\oplus(\frac{1}{2},\frac{5}{2})_{3,2}\oplus(\frac{1}{2},\frac{7}{2})_{3,2}\oplus(\frac{1}{2},\frac{9}{2})_{3,2}\oplus(0,\frac{3}{2})_{4,1}\oplus2(0,\frac{5}{2})_{4,1}\oplus3(0,\frac{7}{2})_{4,1}\oplus3(0,\frac{9}{2})_{4,1}\oplus2(0,\frac{11}{2})_{4,1}\oplus(0,\frac{13}{2})_{4,1}\oplus(\frac{1}{2},3)_{4,1}\oplus3(\frac{1}{2},4)_{4,1}\oplus4(\frac{1}{2},5)_{4,1}\oplus2(\frac{1}{2},6)_{4,1}\oplus(1,\frac{9}{2})_{4,1}\oplus2(1,\frac{11}{2})_{4,1}\oplus(1,\frac{13}{2})_{4,1}\oplus(\frac{3}{2},6)_{4,1}\oplus(\frac{3}{2},7)_{4,1}\oplus(0,1)_{5,0}\oplus(0,3)_{5,0}\oplus(0,4)_{5,0}\oplus2(0,5)_{5,0}\oplus2(0,6)_{5,0}\oplus2(0,7)_{5,0}\oplus(0,8)_{5,0}\oplus(\frac{1}{2},\frac{5}{2})_{5,0}\oplus(\frac{1}{2},\frac{7}{2})_{5,0}\oplus2(\frac{1}{2},\frac{9}{2})_{5,0}\oplus2(\frac{1}{2},\frac{11}{2})_{5,0}\oplus3(\frac{1}{2},\frac{13}{2})_{5,0}\oplus2(\frac{1}{2},\frac{15}{2})_{5,0}\oplus(\frac{1}{2},\frac{17}{2})_{5,0}\oplus(1,4)_{5,0}\oplus(1,5)_{5,0}\oplus2(1,6)_{5,0}\oplus2(1,7)_{5,0}\oplus2(1,8)_{5,0}\oplus(1,9)_{5,0}\oplus(\frac{3}{2},\frac{11}{2})_{5,0}\oplus(\frac{3}{2},\frac{13}{2})_{5,0}\oplus2(\frac{3}{2},\frac{15}{2})_{5,0}\oplus(\frac{3}{2},\frac{17}{2})_{5,0}\oplus(\frac{3}{2},\frac{19}{2})_{5,0}\oplus(2,7)_{5,0}\oplus(2,8)_{5,0}\oplus(2,9)_{5,0}\oplus(\frac{5}{2},\frac{17}{2})_{5,0}\oplus(3,10)_{5,0}\end{math}\\*[0.1cm] \hline6 & \begin{math}(0,4)_{1,5}\oplus(0,\frac{3}{2})_{2,4}\oplus2(0,\frac{5}{2})_{2,4}\oplus(0,\frac{7}{2})_{2,4}\oplus(0,0)_{3,3}\oplus2(0,1)_{3,3}\oplus3(0,2)_{3,3}\oplus3(0,3)_{3,3}\oplus(0,4)_{3,3}\oplus(\frac{1}{2},\frac{3}{2})_{3,3}\oplus(\frac{1}{2},\frac{5}{2})_{3,3}\oplus(\frac{1}{2},\frac{7}{2})_{3,3}\oplus(\frac{1}{2},\frac{9}{2})_{3,3}\oplus(0,\frac{1}{2})_{4,2}\oplus2(0,\frac{3}{2})_{4,2}\oplus6(0,\frac{5}{2})_{4,2}\oplus6(0,\frac{7}{2})_{4,2}\oplus7(0,\frac{9}{2})_{4,2}\oplus2(0,\frac{11}{2})_{4,2}\oplus(0,\frac{13}{2})_{4,2}\oplus(\frac{1}{2},2)_{4,2}\oplus3(\frac{1}{2},3)_{4,2}\oplus7(\frac{1}{2},4)_{4,2}\oplus6(\frac{1}{2},5)_{4,2}\oplus3(\frac{1}{2},6)_{4,2}\oplus(1,\frac{7}{2})_{4,2}\oplus2(1,\frac{9}{2})_{4,2}\oplus3(1,\frac{11}{2})_{4,2}\oplus(1,\frac{13}{2})_{4,2}\oplus(\frac{3}{2},5)_{4,2}\oplus(\frac{3}{2},6)_{4,2}\oplus(\frac{3}{2},7)_{4,2}\oplus(0,0)_{5,1}\oplus2(0,1)_{5,1}\oplus3(0,2)_{5,1}\oplus5(0,3)_{5,1}\oplus7(0,4)_{5,1}\oplus9(0,5)_{5,1}\oplus9(0,6)_{5,1}\oplus7(0,7)_{5,1}\oplus3(0,8)_{5,1}\oplus(\frac{1}{2},\frac{3}{2})_{5,1}\oplus3(\frac{1}{2},\frac{5}{2})_{5,1}\oplus6(\frac{1}{2},\frac{7}{2})_{5,1}\oplus9(\frac{1}{2},\frac{9}{2})_{5,1}\oplus12(\frac{1}{2},\frac{11}{2})_{5,1}\oplus12(\frac{1}{2},\frac{13}{2})_{5,1}\oplus7(\frac{1}{2},\frac{15}{2})_{5,1}\oplus2(\frac{1}{2},\frac{17}{2})_{5,1}\oplus(1,3)_{5,1}\oplus3(1,4)_{5,1}\oplus6(1,5)_{5,1}\oplus9(1,6)_{5,1}\oplus9(1,7)_{5,1}\oplus6(1,8)_{5,1}\oplus2(1,9)_{5,1}\oplus(\frac{3}{2},\frac{9}{2})_{5,1}\oplus3(\frac{3}{2},\frac{11}{2})_{5,1}\oplus6(\frac{3}{2},\frac{13}{2})_{5,1}\oplus7(\frac{3}{2},\frac{15}{2})_{5,1}\oplus4(\frac{3}{2},\frac{17}{2})_{5,1}\oplus(\frac{3}{2},\frac{19}{2})_{5,1}\oplus(2,6)_{5,1}\oplus3(2,7)_{5,1}\oplus4(2,8)_{5,1}\oplus2(2,9)_{5,1}\oplus(\frac{5}{2},\frac{15}{2})_{5,1}\oplus2(\frac{5}{2},\frac{17}{2})_{5,1}\oplus(\frac{5}{2},\frac{19}{2})_{5,1}\oplus(3,9)_{5,1}\oplus(3,10)_{5,1}\oplus(0,\frac{1}{2})_{6,0}\oplus(0,\frac{3}{2})_{6,0}\oplus3(0,\frac{5}{2})_{6,0}\oplus2(0,\frac{7}{2})_{6,0}\oplus6(0,\frac{9}{2})_{6,0}\oplus4(0,\frac{11}{2})_{6,0}\oplus8(0,\frac{13}{2})_{6,0}\oplus5(0,\frac{15}{2})_{6,0}\oplus7(0,\frac{17}{2})_{6,0}\oplus2(0,\frac{19}{2})_{6,0}\oplus2(0,\frac{21}{2})_{6,0}\oplus(\frac{1}{2},1)_{6,0}\oplus2(\frac{1}{2},2)_{6,0}\oplus3(\frac{1}{2},3)_{6,0}\oplus5(\frac{1}{2},4)_{6,0}\oplus6(\frac{1}{2},5)_{6,0}\oplus9(\frac{1}{2},6)_{6,0}\oplus9(\frac{1}{2},7)_{6,0}\oplus10(\frac{1}{2},8)_{6,0}\oplus7(\frac{1}{2},9)_{6,0}\oplus5(\frac{1}{2},10)_{6,0}\oplus(\frac{1}{2},11)_{6,0}\oplus(\frac{1}{2},12)_{6,0}\oplus(1,\frac{3}{2})_{6,0}\oplus(1,\frac{5}{2})_{6,0}\oplus3(1,\frac{7}{2})_{6,0}\oplus3(1,\frac{9}{2})_{6,0}\oplus7(1,\frac{11}{2})_{6,0}\oplus7(1,\frac{13}{2})_{6,0}\oplus11(1,\frac{15}{2})_{6,0}\oplus9(1,\frac{17}{2})_{6,0}\oplus9(1,\frac{19}{2})_{6,0}\oplus4(1,\frac{21}{2})_{6,0}\oplus2(1,\frac{23}{2})_{6,0}\oplus(\frac{3}{2},3)_{6,0}\oplus(\frac{3}{2},4)_{6,0}\oplus3(\frac{3}{2},5)_{6,0}\oplus4(\frac{3}{2},6)_{6,0}\oplus7(\frac{3}{2},7)_{6,0}\oplus7(\frac{3}{2},8)_{6,0}\oplus10(\frac{3}{2},9)_{6,0}\oplus6(\frac{3}{2},10)_{6,0}\oplus4(\frac{3}{2},11)_{6,0}\oplus(2,\frac{9}{2})_{6,0}\oplus(2,\frac{11}{2})_{6,0}\oplus3(2,\frac{13}{2})_{6,0}\oplus4(2,\frac{15}{2})_{6,0}\oplus7(2,\frac{17}{2})_{6,0}\oplus6(2,\frac{19}{2})_{6,0}\oplus6(2,\frac{21}{2})_{6,0}\oplus2(2,\frac{23}{2})_{6,0}\oplus(2,\frac{25}{2})_{6,0}\oplus(\frac{5}{2},6)_{6,0}\oplus(\frac{5}{2},7)_{6,0}\oplus3(\frac{5}{2},8)_{6,0}\oplus3(\frac{5}{2},9)_{6,0}\oplus5(\frac{5}{2},10)_{6,0}\oplus3(\frac{5}{2},11)_{6,0}\oplus2(\frac{5}{2},12)_{6,0}\oplus(3,\frac{15}{2})_{6,0}\oplus(3,\frac{17}{2})_{6,0}\oplus3(3,\frac{19}{2})_{6,0}\oplus3(3,\frac{21}{2})_{6,0}\oplus3(3,\frac{23}{2})_{6,0}\oplus(3,\frac{25}{2})_{6,0}\oplus(\frac{7}{2},9)_{6,0}\oplus(\frac{7}{2},10)_{6,0}\oplus2(\frac{7}{2},11)_{6,0}\oplus(\frac{7}{2},12)_{6,0}\oplus(\frac{7}{2},13)_{6,0}\oplus(4,\frac{21}{2})_{6,0}\oplus(4,\frac{23}{2})_{6,0}\oplus(4,\frac{25}{2})_{6,0}\oplus(\frac{9}{2},12)_{6,0}\oplus(5,\frac{27}{2})_{6,0}\end{math}\\*[0.1cm] \hline\caption{The refined BPS invariants of resolved $\cz$ model. We list the invariants up to degree $d=d_1+d_2=6$. This invariant is achieved by the relation with $SU(3),m=2$ model by $(N_{j_L,j_R}^{d_1,d_2})_{\cz}=(N_{j_L,j_R}^{d_2,d_1,d_1})_{SU(3),m=2}$}\end{longtable}
\end{spacing}
\end{small}
\newpage\subsection*{$SU(3)$ Geometry with $m=0$}
\begin{spacing}{1.3}
\begin{longtable}{| p{0.01\textwidth} | p{\textwidth} |} \hline  $d$&$\sum_{d_1+d_2+d_B=d} \sum_{j_L,j_R} \oplus N^{\mathbf{d}}_{j_L,j_R}(j_L,j_R)_{d_1,d_2,d_B} $\\*[0.1cm] \hline 1 & \begin{math}(0,0)_{0,0,1}\oplus(0,\frac{1}{2})_{0,1,0}\oplus(0,\frac{1}{2})_{1,0,0}\end{math}\\*[0.1cm] \hline2 & \begin{math}(0,1)_{0,1,1}\oplus(0,1)_{1,0,1}\oplus(0,\frac{1}{2})_{1,1,0}\end{math}\\*[0.1cm] \hline3 & \begin{math}(0,2)_{0,2,1}\oplus(0,1)_{1,1,1}\oplus(0,0)_{1,1,1}\oplus(0,2)_{2,0,1}\end{math}\\*[0.1cm] \hline4 & \begin{math}(0,\frac{5}{2})_{0,2,2}\oplus(0,3)_{0,3,1}\oplus(0,2)_{1,2,1}\oplus(0,1)_{1,2,1}\oplus(0,\frac{5}{2})_{2,0,2}\oplus(0,2)_{2,1,1}\oplus(0,1)_{2,1,1}\oplus(0,3)_{3,0,1}\end{math}\\*[0.1cm] \hline5 & \begin{math}(\frac{1}{2},4)_{0,3,2}\oplus(0,\frac{7}{2})_{0,3,2}\oplus(0,\frac{5}{2})_{0,3,2}\oplus(0,4)_{0,4,1}\oplus(0,\frac{5}{2})_{1,2,2}\oplus(0,\frac{3}{2})_{1,2,2}\oplus(0,3)_{1,3,1}\oplus(0,2)_{1,3,1}\oplus(0,\frac{5}{2})_{2,1,2}\oplus(0,\frac{3}{2})_{2,1,2}\oplus(0,2)_{2,2,1}\oplus(0,1)_{2,2,1}\oplus(0,0)_{2,2,1}\oplus(\frac{1}{2},4)_{3,0,2}\oplus(0,\frac{7}{2})_{3,0,2}\oplus(0,\frac{5}{2})_{3,0,2}\oplus(0,3)_{3,1,1}\oplus(0,2)_{3,1,1}\oplus(0,4)_{4,0,1}\end{math}\\*[0.1cm] \hline6 & \begin{math}(\frac{1}{2},\frac{9}{2})_{0,3,3}\oplus(0,3)_{0,3,3}\oplus(1,\frac{11}{2})_{0,4,2}\oplus(\frac{1}{2},5)_{0,4,2}\oplus2(0,\frac{9}{2})_{0,4,2}\oplus(\frac{1}{2},4)_{0,4,2}\oplus(0,\frac{7}{2})_{0,4,2}\oplus(0,\frac{5}{2})_{0,4,2}\oplus(0,5)_{0,5,1}\oplus(\frac{1}{2},4)_{1,3,2}\oplus2(0,\frac{7}{2})_{1,3,2}\oplus(\frac{1}{2},3)_{1,3,2}\oplus3(0,\frac{5}{2})_{1,3,2}\oplus(0,\frac{3}{2})_{1,3,2}\oplus(0,4)_{1,4,1}\oplus(0,3)_{1,4,1}\oplus(0,\frac{7}{2})_{2,2,2}\oplus2(0,\frac{5}{2})_{2,2,2}\oplus2(0,\frac{3}{2})_{2,2,2}\oplus2(0,\frac{1}{2})_{2,2,2}\oplus(0,3)_{2,3,1}\oplus(0,2)_{2,3,1}\oplus(0,1)_{2,3,1}\oplus(\frac{1}{2},\frac{9}{2})_{3,0,3}\oplus(0,3)_{3,0,3}\oplus(\frac{1}{2},4)_{3,1,2}\oplus2(0,\frac{7}{2})_{3,1,2}\oplus(\frac{1}{2},3)_{3,1,2}\oplus3(0,\frac{5}{2})_{3,1,2}\oplus(0,\frac{3}{2})_{3,1,2}\oplus(0,3)_{3,2,1}\oplus(0,2)_{3,2,1}\oplus(0,1)_{3,2,1}\oplus(1,\frac{11}{2})_{4,0,2}\oplus(\frac{1}{2},5)_{4,0,2}\oplus2(0,\frac{9}{2})_{4,0,2}\oplus(\frac{1}{2},4)_{4,0,2}\oplus(0,\frac{7}{2})_{4,0,2}\oplus(0,\frac{5}{2})_{4,0,2}\oplus(0,4)_{4,1,1}\oplus(0,3)_{4,1,1}\oplus(0,5)_{5,0,1}\end{math}\\*[0.1cm] \hline7 & \begin{math}(\frac{3}{2},\frac{13}{2})_{0,4,3}\oplus(1,6)_{0,4,3}\oplus(0,6)_{0,4,3}\oplus2(\frac{1}{2},\frac{11}{2})_{0,4,3}\oplus(1,5)_{0,4,3}\oplus(0,5)_{0,4,3}\oplus2(\frac{1}{2},\frac{9}{2})_{0,4,3}\oplus2(0,4)_{0,4,3}\oplus(\frac{1}{2},\frac{7}{2})_{0,4,3}\oplus(0,3)_{0,4,3}\oplus(0,2)_{0,4,3}\oplus(\frac{3}{2},7)_{0,5,2}\oplus(1,\frac{13}{2})_{0,5,2}\oplus2(\frac{1}{2},6)_{0,5,2}\oplus(1,\frac{11}{2})_{0,5,2}\oplus2(0,\frac{11}{2})_{0,5,2}\oplus(\frac{1}{2},5)_{0,5,2}\oplus2(0,\frac{9}{2})_{0,5,2}\oplus(\frac{1}{2},4)_{0,5,2}\oplus(0,\frac{7}{2})_{0,5,2}\oplus(0,\frac{5}{2})_{0,5,2}\oplus(0,6)_{0,6,1}\oplus(\frac{1}{2},\frac{9}{2})_{1,3,3}\oplus(0,4)_{1,3,3}\oplus(\frac{1}{2},\frac{7}{2})_{1,3,3}\oplus2(0,3)_{1,3,3}\oplus(0,2)_{1,3,3}\oplus(1,\frac{11}{2})_{1,4,2}\oplus2(\frac{1}{2},5)_{1,4,2}\oplus(1,\frac{9}{2})_{1,4,2}\oplus3(0,\frac{9}{2})_{1,4,2}\oplus3(\frac{1}{2},4)_{1,4,2}\oplus5(0,\frac{7}{2})_{1,4,2}\oplus(\frac{1}{2},3)_{1,4,2}\oplus3(0,\frac{5}{2})_{1,4,2}\oplus(0,\frac{3}{2})_{1,4,2}\oplus(0,5)_{1,5,1}\oplus(0,4)_{1,5,1}\oplus(0,4)_{2,2,3}\oplus(0,3)_{2,2,3}\oplus(0,2)_{2,2,3}\oplus(0,1)_{2,2,3}\oplus(0,0)_{2,2,3}\oplus(0,\frac{9}{2})_{2,3,2}\oplus(\frac{1}{2},4)_{2,3,2}\oplus3(0,\frac{7}{2})_{2,3,2}\oplus(\frac{1}{2},3)_{2,3,2}\oplus5(0,\frac{5}{2})_{2,3,2}\oplus(\frac{1}{2},2)_{2,3,2}\oplus4(0,\frac{3}{2})_{2,3,2}\oplus2(0,\frac{1}{2})_{2,3,2}\oplus(0,4)_{2,4,1}\oplus(0,3)_{2,4,1}\oplus(0,2)_{2,4,1}\oplus(\frac{1}{2},\frac{9}{2})_{3,1,3}\oplus(0,4)_{3,1,3}\oplus(\frac{1}{2},\frac{7}{2})_{3,1,3}\oplus2(0,3)_{3,1,3}\oplus(0,2)_{3,1,3}\oplus(0,\frac{9}{2})_{3,2,2}\oplus(\frac{1}{2},4)_{3,2,2}\oplus3(0,\frac{7}{2})_{3,2,2}\oplus(\frac{1}{2},3)_{3,2,2}\oplus5(0,\frac{5}{2})_{3,2,2}\oplus(\frac{1}{2},2)_{3,2,2}\oplus4(0,\frac{3}{2})_{3,2,2}\oplus2(0,\frac{1}{2})_{3,2,2}\oplus(0,3)_{3,3,1}\oplus(0,2)_{3,3,1}\oplus(0,1)_{3,3,1}\oplus(0,0)_{3,3,1}\oplus(\frac{3}{2},\frac{13}{2})_{4,0,3}\oplus(1,6)_{4,0,3}\oplus(0,6)_{4,0,3}\oplus2(\frac{1}{2},\frac{11}{2})_{4,0,3}\oplus(1,5)_{4,0,3}\oplus(0,5)_{4,0,3}\oplus2(\frac{1}{2},\frac{9}{2})_{4,0,3}\oplus2(0,4)_{4,0,3}\oplus(\frac{1}{2},\frac{7}{2})_{4,0,3}\oplus(0,3)_{4,0,3}\oplus(0,2)_{4,0,3}\oplus(1,\frac{11}{2})_{4,1,2}\oplus2(\frac{1}{2},5)_{4,1,2}\oplus(1,\frac{9}{2})_{4,1,2}\oplus3(0,\frac{9}{2})_{4,1,2}\oplus3(\frac{1}{2},4)_{4,1,2}\oplus5(0,\frac{7}{2})_{4,1,2}\oplus(\frac{1}{2},3)_{4,1,2}\oplus3(0,\frac{5}{2})_{4,1,2}\oplus(0,\frac{3}{2})_{4,1,2}\oplus(0,4)_{4,2,1}\oplus(0,3)_{4,2,1}\oplus(0,2)_{4,2,1}\oplus(\frac{3}{2},7)_{5,0,2}\oplus(1,\frac{13}{2})_{5,0,2}\oplus2(\frac{1}{2},6)_{5,0,2}\oplus(1,\frac{11}{2})_{5,0,2}\oplus2(0,\frac{11}{2})_{5,0,2}\oplus(\frac{1}{2},5)_{5,0,2}\oplus2(0,\frac{9}{2})_{5,0,2}\oplus(\frac{1}{2},4)_{5,0,2}\oplus(0,\frac{7}{2})_{5,0,2}\oplus(0,\frac{5}{2})_{5,0,2}\oplus(0,5)_{5,1,1}\oplus(0,4)_{5,1,1}\oplus(0,6)_{6,0,1}\end{math}\\*[0.1cm] \hline\caption{The refined BPS invariants of $SU(3),m=0$ model. We list the invariants up to degree $d=d_1+d_2+d_B=7$. From the charge, we can see the symmetry $N_{j_L,j_R}^{d_1,d_2,d_B}=N_{j_L,j_R}^{d_2,d_1,d_B}$.}\end{longtable}
\end{spacing}
\newpage\subsection*{$SU(3)$ Geometry with $m=1$}
\begin{small}
\begin{spacing}{1.3}
\begin{longtable}{| p{0.01\textwidth} | p{\textwidth} |} \hline  $d$&$\sum_{d_1+d_2+d_3=d} \sum_{j_L,j_R} \oplus N^{\mathbf{d}}_{j_L,j_R}(j_L,j_R)_{d_1,d_2,d_3} $\\*[0.1cm] \hline 1 & \begin{math}(0,\frac{1}{2})_{0,0,1}\oplus(0,\frac{1}{2})_{0,1,0}\oplus(0,\frac{1}{2})_{1,0,0}\end{math}\\*[0.1cm] \hline2 & \begin{math}(0,\frac{3}{2})_{0,1,1}\oplus(0,\frac{1}{2})_{1,0,1}\oplus(0,\frac{1}{2})_{1,1,0}\end{math}\\*[0.1cm] \hline3 & \begin{math}(0,\frac{5}{2})_{0,1,2}\oplus(0,\frac{5}{2})_{0,2,1}\oplus(0,\frac{3}{2})_{1,1,1}\oplus(0,\frac{1}{2})_{1,1,1}\oplus(0,\frac{3}{2})_{2,0,1}\end{math}\\*[0.1cm] \hline4 & \begin{math}(0,\frac{7}{2})_{0,1,3}\oplus(\frac{1}{2},4)_{0,2,2}\oplus(0,\frac{7}{2})_{0,2,2}\oplus(0,\frac{5}{2})_{0,2,2}\oplus(0,\frac{7}{2})_{0,3,1}\oplus(0,\frac{5}{2})_{1,1,2}\oplus(0,\frac{3}{2})_{1,1,2}\oplus(0,\frac{5}{2})_{1,2,1}\oplus(0,\frac{3}{2})_{1,2,1}\oplus(0,\frac{3}{2})_{2,1,1}\oplus(0,\frac{1}{2})_{2,1,1}\oplus(0,\frac{5}{2})_{3,0,1}\end{math}\\*[0.1cm] \hline5 & \begin{math}(0,\frac{9}{2})_{0,1,4}\oplus(1,\frac{11}{2})_{0,2,3}\oplus(\frac{1}{2},5)_{0,2,3}\oplus2(0,\frac{9}{2})_{0,2,3}\oplus(\frac{1}{2},4)_{0,2,3}\oplus(0,\frac{7}{2})_{0,2,3}\oplus(0,\frac{5}{2})_{0,2,3}\oplus(1,\frac{11}{2})_{0,3,2}\oplus(\frac{1}{2},5)_{0,3,2}\oplus2(0,\frac{9}{2})_{0,3,2}\oplus(\frac{1}{2},4)_{0,3,2}\oplus(0,\frac{7}{2})_{0,3,2}\oplus(0,\frac{5}{2})_{0,3,2}\oplus(0,\frac{9}{2})_{0,4,1}\oplus(0,\frac{7}{2})_{1,1,3}\oplus(0,\frac{5}{2})_{1,1,3}\oplus(\frac{1}{2},4)_{1,2,2}\oplus2(0,\frac{7}{2})_{1,2,2}\oplus(\frac{1}{2},3)_{1,2,2}\oplus3(0,\frac{5}{2})_{1,2,2}\oplus(0,\frac{3}{2})_{1,2,2}\oplus(0,\frac{7}{2})_{1,3,1}\oplus(0,\frac{5}{2})_{1,3,1}\oplus(0,\frac{5}{2})_{2,1,2}\oplus(0,\frac{3}{2})_{2,1,2}\oplus(0,\frac{1}{2})_{2,1,2}\oplus(0,\frac{5}{2})_{2,2,1}\oplus(0,\frac{3}{2})_{2,2,1}\oplus(0,\frac{1}{2})_{2,2,1}\oplus(0,\frac{5}{2})_{3,0,2}\oplus(0,\frac{5}{2})_{3,1,1}\oplus(0,\frac{3}{2})_{3,1,1}\oplus(0,\frac{7}{2})_{4,0,1}\end{math}\\*[0.1cm] \hline6 & \begin{math}(0,\frac{11}{2})_{0,1,5}\oplus(\frac{3}{2},7)_{0,2,4}\oplus(1,\frac{13}{2})_{0,2,4}\oplus2(\frac{1}{2},6)_{0,2,4}\oplus(1,\frac{11}{2})_{0,2,4}\oplus2(0,\frac{11}{2})_{0,2,4}\oplus(\frac{1}{2},5)_{0,2,4}\oplus2(0,\frac{9}{2})_{0,2,4}\oplus(\frac{1}{2},4)_{0,2,4}\oplus(0,\frac{7}{2})_{0,2,4}\oplus(0,\frac{5}{2})_{0,2,4}\oplus(2,\frac{15}{2})_{0,3,3}\oplus(\frac{3}{2},7)_{0,3,3}\oplus(\frac{1}{2},7)_{0,3,3}\oplus3(1,\frac{13}{2})_{0,3,3}\oplus(\frac{3}{2},6)_{0,3,3}\oplus3(\frac{1}{2},6)_{0,3,3}\oplus2(1,\frac{11}{2})_{0,3,3}\oplus4(0,\frac{11}{2})_{0,3,3}\oplus3(\frac{1}{2},5)_{0,3,3}\oplus(1,\frac{9}{2})_{0,3,3}\oplus3(0,\frac{9}{2})_{0,3,3}\oplus2(\frac{1}{2},4)_{0,3,3}\oplus3(0,\frac{7}{2})_{0,3,3}\oplus(\frac{1}{2},3)_{0,3,3}\oplus(0,\frac{5}{2})_{0,3,3}\oplus(0,\frac{3}{2})_{0,3,3}\oplus(\frac{3}{2},7)_{0,4,2}\oplus(1,\frac{13}{2})_{0,4,2}\oplus2(\frac{1}{2},6)_{0,4,2}\oplus(1,\frac{11}{2})_{0,4,2}\oplus2(0,\frac{11}{2})_{0,4,2}\oplus(\frac{1}{2},5)_{0,4,2}\oplus2(0,\frac{9}{2})_{0,4,2}\oplus(\frac{1}{2},4)_{0,4,2}\oplus(0,\frac{7}{2})_{0,4,2}\oplus(0,\frac{5}{2})_{0,4,2}\oplus(0,\frac{11}{2})_{0,5,1}\oplus(0,\frac{9}{2})_{1,1,4}\oplus(0,\frac{7}{2})_{1,1,4}\oplus(1,\frac{11}{2})_{1,2,3}\oplus2(\frac{1}{2},5)_{1,2,3}\oplus(1,\frac{9}{2})_{1,2,3}\oplus3(0,\frac{9}{2})_{1,2,3}\oplus3(\frac{1}{2},4)_{1,2,3}\oplus5(0,\frac{7}{2})_{1,2,3}\oplus(\frac{1}{2},3)_{1,2,3}\oplus3(0,\frac{5}{2})_{1,2,3}\oplus(0,\frac{3}{2})_{1,2,3}\oplus(1,\frac{11}{2})_{1,3,2}\oplus2(\frac{1}{2},5)_{1,3,2}\oplus(1,\frac{9}{2})_{1,3,2}\oplus3(0,\frac{9}{2})_{1,3,2}\oplus3(\frac{1}{2},4)_{1,3,2}\oplus5(0,\frac{7}{2})_{1,3,2}\oplus(\frac{1}{2},3)_{1,3,2}\oplus3(0,\frac{5}{2})_{1,3,2}\oplus(0,\frac{3}{2})_{1,3,2}\oplus(0,\frac{9}{2})_{1,4,1}\oplus(0,\frac{7}{2})_{1,4,1}\oplus(0,\frac{7}{2})_{2,1,3}\oplus(0,\frac{5}{2})_{2,1,3}\oplus(0,\frac{3}{2})_{2,1,3}\oplus(\frac{1}{2},4)_{2,2,2}\oplus2(0,\frac{7}{2})_{2,2,2}\oplus(\frac{1}{2},3)_{2,2,2}\oplus4(0,\frac{5}{2})_{2,2,2}\oplus(\frac{1}{2},2)_{2,2,2}\oplus3(0,\frac{3}{2})_{2,2,2}\oplus(0,\frac{1}{2})_{2,2,2}\oplus(0,\frac{7}{2})_{2,3,1}\oplus(0,\frac{5}{2})_{2,3,1}\oplus(0,\frac{3}{2})_{2,3,1}\oplus(0,\frac{7}{2})_{3,1,2}\oplus2(0,\frac{5}{2})_{3,1,2}\oplus2(0,\frac{3}{2})_{3,1,2}\oplus(0,\frac{1}{2})_{3,1,2}\oplus(0,\frac{5}{2})_{3,2,1}\oplus(0,\frac{3}{2})_{3,2,1}\oplus(0,\frac{1}{2})_{3,2,1}\oplus(\frac{1}{2},4)_{4,0,2}\oplus(0,\frac{7}{2})_{4,0,2}\oplus(0,\frac{5}{2})_{4,0,2}\oplus(0,\frac{7}{2})_{4,1,1}\oplus(0,\frac{5}{2})_{4,1,1}\oplus(0,\frac{9}{2})_{5,0,1}\end{math}\\*[0.1cm] \hline7 & \begin{math}(0,\frac{13}{2})_{0,1,6}\oplus(2,\frac{17}{2})_{0,2,5}\oplus(\frac{3}{2},8)_{0,2,5}\oplus2(1,\frac{15}{2})_{0,2,5}\oplus(\frac{3}{2},7)_{0,2,5}\oplus2(\frac{1}{2},7)_{0,2,5}\oplus(1,\frac{13}{2})_{0,2,5}\oplus3(0,\frac{13}{2})_{0,2,5}\oplus2(\frac{1}{2},6)_{0,2,5}\oplus(1,\frac{11}{2})_{0,2,5}\oplus2(0,\frac{11}{2})_{0,2,5}\oplus(\frac{1}{2},5)_{0,2,5}\oplus2(0,\frac{9}{2})_{0,2,5}\oplus(\frac{1}{2},4)_{0,2,5}\oplus(0,\frac{7}{2})_{0,2,5}\oplus(0,\frac{5}{2})_{0,2,5}\oplus(3,\frac{19}{2})_{0,3,4}\oplus(\frac{5}{2},9)_{0,3,4}\oplus(\frac{3}{2},9)_{0,3,4}\oplus3(2,\frac{17}{2})_{0,3,4}\oplus(\frac{5}{2},8)_{0,3,4}\oplus(1,\frac{17}{2})_{0,3,4}\oplus4(\frac{3}{2},8)_{0,3,4}\oplus2(2,\frac{15}{2})_{0,3,4}\oplus(0,\frac{17}{2})_{0,3,4}\oplus2(\frac{1}{2},8)_{0,3,4}\oplus7(1,\frac{15}{2})_{0,3,4}\oplus4(\frac{3}{2},7)_{0,3,4}\oplus(2,\frac{13}{2})_{0,3,4}\oplus(0,\frac{15}{2})_{0,3,4}\oplus7(\frac{1}{2},7)_{0,3,4}\oplus6(1,\frac{13}{2})_{0,3,4}\oplus2(\frac{3}{2},6)_{0,3,4}\oplus7(0,\frac{13}{2})_{0,3,4}\oplus8(\frac{1}{2},6)_{0,3,4}\oplus5(1,\frac{11}{2})_{0,3,4}\oplus(\frac{3}{2},5)_{0,3,4}\oplus6(0,\frac{11}{2})_{0,3,4}\oplus6(\frac{1}{2},5)_{0,3,4}\oplus2(1,\frac{9}{2})_{0,3,4}\oplus7(0,\frac{9}{2})_{0,3,4}\oplus4(\frac{1}{2},4)_{0,3,4}\oplus(1,\frac{7}{2})_{0,3,4}\oplus4(0,\frac{7}{2})_{0,3,4}\oplus2(\frac{1}{2},3)_{0,3,4}\oplus3(0,\frac{5}{2})_{0,3,4}\oplus(\frac{1}{2},2)_{0,3,4}\oplus(0,\frac{3}{2})_{0,3,4}\oplus(0,\frac{1}{2})_{0,3,4}\oplus(3,\frac{19}{2})_{0,4,3}\oplus(\frac{5}{2},9)_{0,4,3}\oplus(\frac{3}{2},9)_{0,4,3}\oplus3(2,\frac{17}{2})_{0,4,3}\oplus(\frac{5}{2},8)_{0,4,3}\oplus(1,\frac{17}{2})_{0,4,3}\oplus4(\frac{3}{2},8)_{0,4,3}\oplus2(2,\frac{15}{2})_{0,4,3}\oplus(0,\frac{17}{2})_{0,4,3}\oplus2(\frac{1}{2},8)_{0,4,3}\oplus7(1,\frac{15}{2})_{0,4,3}\oplus4(\frac{3}{2},7)_{0,4,3}\oplus(2,\frac{13}{2})_{0,4,3}\oplus(0,\frac{15}{2})_{0,4,3}\oplus7(\frac{1}{2},7)_{0,4,3}\oplus6(1,\frac{13}{2})_{0,4,3}\oplus2(\frac{3}{2},6)_{0,4,3}\oplus7(0,\frac{13}{2})_{0,4,3}\oplus8(\frac{1}{2},6)_{0,4,3}\oplus5(1,\frac{11}{2})_{0,4,3}\oplus(\frac{3}{2},5)_{0,4,3}\oplus6(0,\frac{11}{2})_{0,4,3}\oplus6(\frac{1}{2},5)_{0,4,3}\oplus2(1,\frac{9}{2})_{0,4,3}\oplus7(0,\frac{9}{2})_{0,4,3}\oplus4(\frac{1}{2},4)_{0,4,3}\oplus(1,\frac{7}{2})_{0,4,3}\oplus4(0,\frac{7}{2})_{0,4,3}\oplus2(\frac{1}{2},3)_{0,4,3}\oplus3(0,\frac{5}{2})_{0,4,3}\oplus(\frac{1}{2},2)_{0,4,3}\oplus(0,\frac{3}{2})_{0,4,3}\oplus(0,\frac{1}{2})_{0,4,3}\oplus(2,\frac{17}{2})_{0,5,2}\oplus(\frac{3}{2},8)_{0,5,2}\oplus2(1,\frac{15}{2})_{0,5,2}\oplus(\frac{3}{2},7)_{0,5,2}\oplus2(\frac{1}{2},7)_{0,5,2}\oplus(1,\frac{13}{2})_{0,5,2}\oplus3(0,\frac{13}{2})_{0,5,2}\oplus2(\frac{1}{2},6)_{0,5,2}\oplus(1,\frac{11}{2})_{0,5,2}\oplus2(0,\frac{11}{2})_{0,5,2}\oplus(\frac{1}{2},5)_{0,5,2}\oplus2(0,\frac{9}{2})_{0,5,2}\oplus(\frac{1}{2},4)_{0,5,2}\oplus(0,\frac{7}{2})_{0,5,2}\oplus(0,\frac{5}{2})_{0,5,2}\oplus(0,\frac{13}{2})_{0,6,1}\oplus(0,\frac{11}{2})_{1,1,5}\oplus(0,\frac{9}{2})_{1,1,5}\oplus(\frac{3}{2},7)_{1,2,4}\oplus2(1,\frac{13}{2})_{1,2,4}\oplus(\frac{3}{2},6)_{1,2,4}\oplus3(\frac{1}{2},6)_{1,2,4}\oplus3(1,\frac{11}{2})_{1,2,4}\oplus4(0,\frac{11}{2})_{1,2,4}\oplus5(\frac{1}{2},5)_{1,2,4}\oplus(1,\frac{9}{2})_{1,2,4}\oplus7(0,\frac{9}{2})_{1,2,4}\oplus3(\frac{1}{2},4)_{1,2,4}\oplus5(0,\frac{7}{2})_{1,2,4}\oplus(\frac{1}{2},3)_{1,2,4}\oplus3(0,\frac{5}{2})_{1,2,4}\oplus(0,\frac{3}{2})_{1,2,4}\oplus(2,\frac{15}{2})_{1,3,3}\oplus2(\frac{3}{2},7)_{1,3,3}\oplus(2,\frac{13}{2})_{1,3,3}\oplus(\frac{1}{2},7)_{1,3,3}\oplus5(1,\frac{13}{2})_{1,3,3}\oplus3(\frac{3}{2},6)_{1,3,3}\oplus(0,\frac{13}{2})_{1,3,3}\oplus7(\frac{1}{2},6)_{1,3,3}\oplus8(1,\frac{11}{2})_{1,3,3}\oplus(\frac{3}{2},5)_{1,3,3}\oplus8(0,\frac{11}{2})_{1,3,3}\oplus12(\frac{1}{2},5)_{1,3,3}\oplus4(1,\frac{9}{2})_{1,3,3}\oplus13(0,\frac{9}{2})_{1,3,3}\oplus9(\frac{1}{2},4)_{1,3,3}\oplus(1,\frac{7}{2})_{1,3,3}\oplus11(0,\frac{7}{2})_{1,3,3}\oplus4(\frac{1}{2},3)_{1,3,3}\oplus7(0,\frac{5}{2})_{1,3,3}\oplus(\frac{1}{2},2)_{1,3,3}\oplus3(0,\frac{3}{2})_{1,3,3}\oplus(0,\frac{1}{2})_{1,3,3}\oplus(\frac{3}{2},7)_{1,4,2}\oplus2(1,\frac{13}{2})_{1,4,2}\oplus(\frac{3}{2},6)_{1,4,2}\oplus3(\frac{1}{2},6)_{1,4,2}\oplus3(1,\frac{11}{2})_{1,4,2}\oplus4(0,\frac{11}{2})_{1,4,2}\oplus5(\frac{1}{2},5)_{1,4,2}\oplus(1,\frac{9}{2})_{1,4,2}\oplus7(0,\frac{9}{2})_{1,4,2}\oplus3(\frac{1}{2},4)_{1,4,2}\oplus5(0,\frac{7}{2})_{1,4,2}\oplus(\frac{1}{2},3)_{1,4,2}\oplus3(0,\frac{5}{2})_{1,4,2}\oplus(0,\frac{3}{2})_{1,4,2}\oplus(0,\frac{11}{2})_{1,5,1}\oplus(0,\frac{9}{2})_{1,5,1}\oplus(0,\frac{9}{2})_{2,1,4}\oplus(0,\frac{7}{2})_{2,1,4}\oplus(0,\frac{5}{2})_{2,1,4}\oplus(1,\frac{11}{2})_{2,2,3}\oplus2(\frac{1}{2},5)_{2,2,3}\oplus(1,\frac{9}{2})_{2,2,3}\oplus4(0,\frac{9}{2})_{2,2,3}\oplus4(\frac{1}{2},4)_{2,2,3}\oplus(1,\frac{7}{2})_{2,2,3}\oplus7(0,\frac{7}{2})_{2,2,3}\oplus3(\frac{1}{2},3)_{2,2,3}\oplus8(0,\frac{5}{2})_{2,2,3}\oplus(\frac{1}{2},2)_{2,2,3}\oplus3(0,\frac{3}{2})_{2,2,3}\oplus(0,\frac{1}{2})_{2,2,3}\oplus(1,\frac{11}{2})_{2,3,2}\oplus2(\frac{1}{2},5)_{2,3,2}\oplus(1,\frac{9}{2})_{2,3,2}\oplus4(0,\frac{9}{2})_{2,3,2}\oplus4(\frac{1}{2},4)_{2,3,2}\oplus(1,\frac{7}{2})_{2,3,2}\oplus7(0,\frac{7}{2})_{2,3,2}\oplus3(\frac{1}{2},3)_{2,3,2}\oplus8(0,\frac{5}{2})_{2,3,2}\oplus(\frac{1}{2},2)_{2,3,2}\oplus3(0,\frac{3}{2})_{2,3,2}\oplus(0,\frac{1}{2})_{2,3,2}\oplus(0,\frac{9}{2})_{2,4,1}\oplus(0,\frac{7}{2})_{2,4,1}\oplus(0,\frac{5}{2})_{2,4,1}\oplus(0,\frac{9}{2})_{3,1,3}\oplus2(0,\frac{7}{2})_{3,1,3}\oplus2(0,\frac{5}{2})_{3,1,3}\oplus2(0,\frac{3}{2})_{3,1,3}\oplus2(0,\frac{1}{2})_{3,1,3}\oplus(0,\frac{9}{2})_{3,2,2}\oplus(\frac{1}{2},4)_{3,2,2}\oplus3(0,\frac{7}{2})_{3,2,2}\oplus(\frac{1}{2},3)_{3,2,2}\oplus5(0,\frac{5}{2})_{3,2,2}\oplus(\frac{1}{2},2)_{3,2,2}\oplus5(0,\frac{3}{2})_{3,2,2}\oplus(\frac{1}{2},1)_{3,2,2}\oplus4(0,\frac{1}{2})_{3,2,2}\oplus(0,\frac{7}{2})_{3,3,1}\oplus(0,\frac{5}{2})_{3,3,1}\oplus(0,\frac{3}{2})_{3,3,1}\oplus(0,\frac{1}{2})_{3,3,1}\oplus(0,\frac{7}{2})_{4,0,3}\oplus(0,\frac{9}{2})_{4,1,2}\oplus(\frac{1}{2},4)_{4,1,2}\oplus3(0,\frac{7}{2})_{4,1,2}\oplus(\frac{1}{2},3)_{4,1,2}\oplus4(0,\frac{5}{2})_{4,1,2}\oplus2(0,\frac{3}{2})_{4,1,2}\oplus(0,\frac{7}{2})_{4,2,1}\oplus(0,\frac{5}{2})_{4,2,1}\oplus(0,\frac{3}{2})_{4,2,1}\oplus(1,\frac{11}{2})_{5,0,2}\oplus(\frac{1}{2},5)_{5,0,2}\oplus2(0,\frac{9}{2})_{5,0,2}\oplus(\frac{1}{2},4)_{5,0,2}\oplus(0,\frac{7}{2})_{5,0,2}\oplus(0,\frac{5}{2})_{5,0,2}\oplus(0,\frac{9}{2})_{5,1,1}\oplus(0,\frac{7}{2})_{5,1,1}\oplus(0,\frac{11}{2})_{6,0,1}\end{math}\\*[0.1cm] \hline\caption{The refined BPS invariants of $SU(3),m=1$ model. We list the invariants up to degree $d=d_1+d_2+d_3=7$. From the charge, we can see the symmetry $N_{j_L,j_R}^{0,d_2,d_3}=N_{j_L,j_R}^{0,d_3,d_2}$.}\end{longtable}
\end{spacing}
\end{small}
\newpage\subsection*{$SU(3)$ Geometry with $m=2$}
\begin{small}
\begin{spacing}{1.3}
\begin{longtable}{| p{0.01\textwidth} | p{\textwidth} |} \hline  $d$&$\sum_{d_1+d_2+d_3=d} \sum_{j_L,j_R} \oplus N^{\mathbf{d}}_{j_L,j_R}(j_L,j_R)_{d_1,d_2,d_3} $\\*[0.1cm] \hline 1 & \begin{math}(0,0)_{0,0,1}\oplus(0,\frac{1}{2})_{0,1,0}\oplus(0,\frac{1}{2})_{1,0,0}\end{math}\\*[0.1cm] \hline2 & \begin{math}(0,1)_{0,1,1}\oplus(0,\frac{1}{2})_{1,1,0}\end{math}\\*[0.1cm] \hline3 & \begin{math}(0,2)_{0,2,1}\oplus(0,0)_{1,1,1}\oplus(0,1)_{1,1,1}\end{math}\\*[0.1cm] \hline4 & \begin{math}(0,\frac{5}{2})_{0,2,2}\oplus(0,3)_{0,3,1}\oplus(0,1)_{1,2,1}\oplus(0,2)_{1,2,1}\oplus(0,1)_{2,1,1}\end{math}\\*[0.1cm] \hline5 & \begin{math}(0,\frac{5}{2})_{0,3,2}\oplus(0,\frac{7}{2})_{0,3,2}\oplus(\frac{1}{2},4)_{0,3,2}\oplus(0,4)_{0,4,1}\oplus(0,\frac{3}{2})_{1,2,2}\oplus(0,\frac{5}{2})_{1,2,2}\oplus(0,2)_{1,3,1}\oplus(0,3)_{1,3,1}\oplus(0,0)_{2,2,1}\oplus(0,1)_{2,2,1}\oplus(0,2)_{2,2,1}\oplus(0,2)_{3,1,1}\end{math}\\*[0.1cm] \hline6 & \begin{math}(0,3)_{0,3,3}\oplus(\frac{1}{2},\frac{9}{2})_{0,3,3}\oplus(0,\frac{5}{2})_{0,4,2}\oplus(0,\frac{7}{2})_{0,4,2}\oplus2(0,\frac{9}{2})_{0,4,2}\oplus(\frac{1}{2},4)_{0,4,2}\oplus(\frac{1}{2},5)_{0,4,2}\oplus(1,\frac{11}{2})_{0,4,2}\oplus(0,5)_{0,5,1}\oplus(0,\frac{3}{2})_{1,3,2}\oplus3(0,\frac{5}{2})_{1,3,2}\oplus2(0,\frac{7}{2})_{1,3,2}\oplus(\frac{1}{2},3)_{1,3,2}\oplus(\frac{1}{2},4)_{1,3,2}\oplus(0,3)_{1,4,1}\oplus(0,4)_{1,4,1}\oplus(0,\frac{1}{2})_{2,2,2}\oplus(0,\frac{3}{2})_{2,2,2}\oplus(0,\frac{5}{2})_{2,2,2}\oplus(0,1)_{2,3,1}\oplus(0,2)_{2,3,1}\oplus(0,3)_{2,3,1}\oplus(0,1)_{3,2,1}\oplus(0,2)_{3,2,1}\oplus(0,3)_{4,1,1}\end{math}\\*[0.1cm] \hline7 & \begin{math}(0,2)_{0,4,3}\oplus(0,3)_{0,4,3}\oplus2(0,4)_{0,4,3}\oplus(0,5)_{0,4,3}\oplus(0,6)_{0,4,3}\oplus(\frac{1}{2},\frac{7}{2})_{0,4,3}\oplus2(\frac{1}{2},\frac{9}{2})_{0,4,3}\oplus2(\frac{1}{2},\frac{11}{2})_{0,4,3}\oplus(1,5)_{0,4,3}\oplus(1,6)_{0,4,3}\oplus(\frac{3}{2},\frac{13}{2})_{0,4,3}\oplus(0,\frac{5}{2})_{0,5,2}\oplus(0,\frac{7}{2})_{0,5,2}\oplus2(0,\frac{9}{2})_{0,5,2}\oplus2(0,\frac{11}{2})_{0,5,2}\oplus(\frac{1}{2},4)_{0,5,2}\oplus(\frac{1}{2},5)_{0,5,2}\oplus2(\frac{1}{2},6)_{0,5,2}\oplus(1,\frac{11}{2})_{0,5,2}\oplus(1,\frac{13}{2})_{0,5,2}\oplus(\frac{3}{2},7)_{0,5,2}\oplus(0,6)_{0,6,1}\oplus(0,2)_{1,3,3}\oplus2(0,3)_{1,3,3}\oplus(0,4)_{1,3,3}\oplus(\frac{1}{2},\frac{7}{2})_{1,3,3}\oplus(\frac{1}{2},\frac{9}{2})_{1,3,3}\oplus(0,\frac{3}{2})_{1,4,2}\oplus3(0,\frac{5}{2})_{1,4,2}\oplus5(0,\frac{7}{2})_{1,4,2}\oplus3(0,\frac{9}{2})_{1,4,2}\oplus(\frac{1}{2},3)_{1,4,2}\oplus3(\frac{1}{2},4)_{1,4,2}\oplus2(\frac{1}{2},5)_{1,4,2}\oplus(1,\frac{9}{2})_{1,4,2}\oplus(1,\frac{11}{2})_{1,4,2}\oplus(0,4)_{1,5,1}\oplus(0,5)_{1,5,1}\oplus(0,\frac{1}{2})_{2,3,2}\oplus3(0,\frac{3}{2})_{2,3,2}\oplus4(0,\frac{5}{2})_{2,3,2}\oplus2(0,\frac{7}{2})_{2,3,2}\oplus(\frac{1}{2},2)_{2,3,2}\oplus(\frac{1}{2},3)_{2,3,2}\oplus(\frac{1}{2},4)_{2,3,2}\oplus(0,2)_{2,4,1}\oplus(0,3)_{2,4,1}\oplus(0,4)_{2,4,1}\oplus(0,\frac{1}{2})_{3,2,2}\oplus(0,\frac{3}{2})_{3,2,2}\oplus(0,\frac{5}{2})_{3,2,2}\oplus(0,0)_{3,3,1}\oplus(0,1)_{3,3,1}\oplus(0,2)_{3,3,1}\oplus(0,3)_{3,3,1}\oplus(0,2)_{4,2,1}\oplus(0,3)_{4,2,1}\oplus(0,4)_{5,1,1}\end{math}\\*[0.1cm] \hline8 & \begin{math}(0,\frac{5}{2})_{0,4,4}\oplus(0,\frac{9}{2})_{0,4,4}\oplus(0,\frac{13}{2})_{0,4,4}\oplus(\frac{1}{2},4)_{0,4,4}\oplus(\frac{1}{2},5)_{0,4,4}\oplus(\frac{1}{2},6)_{0,4,4}\oplus(1,\frac{11}{2})_{0,4,4}\oplus(\frac{3}{2},7)_{0,4,4}\oplus(0,1)_{0,5,3}\oplus(0,2)_{0,5,3}\oplus3(0,3)_{0,5,3}\oplus3(0,4)_{0,5,3}\oplus5(0,5)_{0,5,3}\oplus3(0,6)_{0,5,3}\oplus2(0,7)_{0,5,3}\oplus(\frac{1}{2},\frac{5}{2})_{0,5,3}\oplus2(\frac{1}{2},\frac{7}{2})_{0,5,3}\oplus4(\frac{1}{2},\frac{9}{2})_{0,5,3}\oplus5(\frac{1}{2},\frac{11}{2})_{0,5,3}\oplus5(\frac{1}{2},\frac{13}{2})_{0,5,3}\oplus(\frac{1}{2},\frac{15}{2})_{0,5,3}\oplus(1,4)_{0,5,3}\oplus2(1,5)_{0,5,3}\oplus4(1,6)_{0,5,3}\oplus3(1,7)_{0,5,3}\oplus(1,8)_{0,5,3}\oplus(\frac{3}{2},\frac{11}{2})_{0,5,3}\oplus2(\frac{3}{2},\frac{13}{2})_{0,5,3}\oplus3(\frac{3}{2},\frac{15}{2})_{0,5,3}\oplus(2,7)_{0,5,3}\oplus(2,8)_{0,5,3}\oplus(\frac{5}{2},\frac{17}{2})_{0,5,3}\oplus(0,\frac{5}{2})_{0,6,2}\oplus(0,\frac{7}{2})_{0,6,2}\oplus2(0,\frac{9}{2})_{0,6,2}\oplus2(0,\frac{11}{2})_{0,6,2}\oplus3(0,\frac{13}{2})_{0,6,2}\oplus(\frac{1}{2},4)_{0,6,2}\oplus(\frac{1}{2},5)_{0,6,2}\oplus2(\frac{1}{2},6)_{0,6,2}\oplus2(\frac{1}{2},7)_{0,6,2}\oplus(1,\frac{11}{2})_{0,6,2}\oplus(1,\frac{13}{2})_{0,6,2}\oplus2(1,\frac{15}{2})_{0,6,2}\oplus(\frac{3}{2},7)_{0,6,2}\oplus(\frac{3}{2},8)_{0,6,2}\oplus(2,\frac{17}{2})_{0,6,2}\oplus(0,7)_{0,7,1}\oplus(0,1)_{1,4,3}\oplus3(0,2)_{1,4,3}\oplus6(0,3)_{1,4,3}\oplus7(0,4)_{1,4,3}\oplus4(0,5)_{1,4,3}\oplus(0,6)_{1,4,3}\oplus(\frac{1}{2},\frac{5}{2})_{1,4,3}\oplus4(\frac{1}{2},\frac{7}{2})_{1,4,3}\oplus7(\frac{1}{2},\frac{9}{2})_{1,4,3}\oplus4(\frac{1}{2},\frac{11}{2})_{1,4,3}\oplus(1,4)_{1,4,3}\oplus3(1,5)_{1,4,3}\oplus2(1,6)_{1,4,3}\oplus(\frac{3}{2},\frac{11}{2})_{1,4,3}\oplus(\frac{3}{2},\frac{13}{2})_{1,4,3}\oplus(0,\frac{3}{2})_{1,5,2}\oplus3(0,\frac{5}{2})_{1,5,2}\oplus5(0,\frac{7}{2})_{1,5,2}\oplus7(0,\frac{9}{2})_{1,5,2}\oplus4(0,\frac{11}{2})_{1,5,2}\oplus(\frac{1}{2},3)_{1,5,2}\oplus3(\frac{1}{2},4)_{1,5,2}\oplus5(\frac{1}{2},5)_{1,5,2}\oplus3(\frac{1}{2},6)_{1,5,2}\oplus(1,\frac{9}{2})_{1,5,2}\oplus3(1,\frac{11}{2})_{1,5,2}\oplus2(1,\frac{13}{2})_{1,5,2}\oplus(\frac{3}{2},6)_{1,5,2}\oplus(\frac{3}{2},7)_{1,5,2}\oplus(0,5)_{1,6,1}\oplus(0,6)_{1,6,1}\oplus(0,1)_{2,3,3}\oplus2(0,2)_{2,3,3}\oplus3(0,3)_{2,3,3}\oplus(0,4)_{2,3,3}\oplus(\frac{1}{2},\frac{5}{2})_{2,3,3}\oplus(\frac{1}{2},\frac{7}{2})_{2,3,3}\oplus(\frac{1}{2},\frac{9}{2})_{2,3,3}\oplus(0,\frac{1}{2})_{2,4,2}\oplus3(0,\frac{3}{2})_{2,4,2}\oplus8(0,\frac{5}{2})_{2,4,2}\oplus7(0,\frac{7}{2})_{2,4,2}\oplus4(0,\frac{9}{2})_{2,4,2}\oplus(\frac{1}{2},2)_{2,4,2}\oplus3(\frac{1}{2},3)_{2,4,2}\oplus4(\frac{1}{2},4)_{2,4,2}\oplus2(\frac{1}{2},5)_{2,4,2}\oplus(1,\frac{7}{2})_{2,4,2}\oplus(1,\frac{9}{2})_{2,4,2}\oplus(1,\frac{11}{2})_{2,4,2}\oplus(0,3)_{2,5,1}\oplus(0,4)_{2,5,1}\oplus(0,5)_{2,5,1}\oplus3(0,\frac{1}{2})_{3,3,2}\oplus4(0,\frac{3}{2})_{3,3,2}\oplus4(0,\frac{5}{2})_{3,3,2}\oplus2(0,\frac{7}{2})_{3,3,2}\oplus(\frac{1}{2},1)_{3,3,2}\oplus(\frac{1}{2},2)_{3,3,2}\oplus(\frac{1}{2},3)_{3,3,2}\oplus(\frac{1}{2},4)_{3,3,2}\oplus(0,1)_{3,4,1}\oplus(0,2)_{3,4,1}\oplus(0,3)_{3,4,1}\oplus(0,4)_{3,4,1}\oplus(0,\frac{3}{2})_{4,2,2}\oplus2(0,\frac{5}{2})_{4,2,2}\oplus(0,\frac{7}{2})_{4,2,2}\oplus(0,1)_{4,3,1}\oplus(0,2)_{4,3,1}\oplus(0,3)_{4,3,1}\oplus(0,3)_{5,2,1}\oplus(0,4)_{5,2,1}\oplus(0,5)_{6,1,1}\end{math}\\*[0.1cm] \hline\caption{The refined BPS invariants of $SU(3),m=2$ model. We compute the invariants up to degree $d=d_1+d_2+d_3=13$. We list this invariants up to degree $d=8$. }\end{longtable}
\end{spacing}
\end{small}
\newpage\subsection*{$SU(4)$ Geometry with $m=0$}
\begin{footnotesize}
\begin{spacing}{1.3}
\begin{longtable}{| p{0.01\textwidth} | p{\textwidth} |} \hline  $d$&$\sum_{d_1+d_2+d_3+d_B=d} \sum_{j_L,j_R} \oplus N^{\mathbf{d}}_{j_L,j_R}(j_L,j_R)_{d_1,d_2,d_3,d_B} $\\*[0.1cm] \hline 1 & \begin{math}(0,\frac{1}{2})_{0,0,0,1}\oplus(0,\frac{1}{2})_{0,0,1,0}\oplus(0,\frac{1}{2})_{0,1,0,0}\oplus(0,\frac{1}{2})_{1,0,0,0}\end{math}\\*[0.1cm] \hline2 & \begin{math}(0,\frac{1}{2})_{0,0,1,1}\oplus(0,\frac{3}{2})_{0,1,0,1}\oplus(0,\frac{1}{2})_{0,1,1,0}\oplus(0,\frac{1}{2})_{1,0,0,1}\oplus(0,\frac{1}{2})_{1,1,0,0}\end{math}\\*[0.1cm] \hline3 & \begin{math}(0,\frac{3}{2})_{0,0,2,1}\oplus(0,\frac{5}{2})_{0,1,0,2}\oplus(0,\frac{3}{2})_{0,1,1,1}\oplus(0,\frac{1}{2})_{0,1,1,1}\oplus(0,\frac{5}{2})_{0,2,0,1}\oplus(0,\frac{3}{2})_{1,1,0,1}\oplus(0,\frac{1}{2})_{1,1,0,1}\oplus(0,\frac{1}{2})_{1,1,1,0}\oplus(0,\frac{3}{2})_{2,0,0,1}\end{math}\\*[0.1cm] \hline4 & \begin{math}(0,\frac{5}{2})_{0,0,3,1}\oplus(0,\frac{7}{2})_{0,1,0,3}\oplus(0,\frac{5}{2})_{0,1,1,2}\oplus(0,\frac{3}{2})_{0,1,1,2}\oplus(0,\frac{3}{2})_{0,1,2,1}\oplus(0,\frac{1}{2})_{0,1,2,1}\oplus(\frac{1}{2},4)_{0,2,0,2}\oplus(0,\frac{7}{2})_{0,2,0,2}\oplus(0,\frac{5}{2})_{0,2,0,2}\oplus(0,\frac{5}{2})_{0,2,1,1}\oplus(0,\frac{3}{2})_{0,2,1,1}\oplus(0,\frac{7}{2})_{0,3,0,1}\oplus(0,\frac{5}{2})_{1,1,0,2}\oplus(0,\frac{3}{2})_{1,1,0,2}\oplus(0,\frac{3}{2})_{1,1,1,1}\oplus2(0,\frac{1}{2})_{1,1,1,1}\oplus(0,\frac{5}{2})_{1,2,0,1}\oplus(0,\frac{3}{2})_{1,2,0,1}\oplus(0,\frac{3}{2})_{2,1,0,1}\oplus(0,\frac{1}{2})_{2,1,0,1}\oplus(0,\frac{5}{2})_{3,0,0,1}\end{math}\\*[0.1cm] \hline5 & \begin{math}(0,\frac{5}{2})_{0,0,3,2}\oplus(0,\frac{7}{2})_{0,0,4,1}\oplus(0,\frac{9}{2})_{0,1,0,4}\oplus(0,\frac{7}{2})_{0,1,1,3}\oplus(0,\frac{5}{2})_{0,1,1,3}\oplus(0,\frac{5}{2})_{0,1,2,2}\oplus(0,\frac{3}{2})_{0,1,2,2}\oplus(0,\frac{1}{2})_{0,1,2,2}\oplus(0,\frac{5}{2})_{0,1,3,1}\oplus(0,\frac{3}{2})_{0,1,3,1}\oplus(1,\frac{11}{2})_{0,2,0,3}\oplus(\frac{1}{2},5)_{0,2,0,3}\oplus2(0,\frac{9}{2})_{0,2,0,3}\oplus(\frac{1}{2},4)_{0,2,0,3}\oplus(0,\frac{7}{2})_{0,2,0,3}\oplus(0,\frac{5}{2})_{0,2,0,3}\oplus(\frac{1}{2},4)_{0,2,1,2}\oplus2(0,\frac{7}{2})_{0,2,1,2}\oplus(\frac{1}{2},3)_{0,2,1,2}\oplus3(0,\frac{5}{2})_{0,2,1,2}\oplus(0,\frac{3}{2})_{0,2,1,2}\oplus(0,\frac{5}{2})_{0,2,2,1}\oplus(0,\frac{3}{2})_{0,2,2,1}\oplus(0,\frac{1}{2})_{0,2,2,1}\oplus(1,\frac{11}{2})_{0,3,0,2}\oplus(\frac{1}{2},5)_{0,3,0,2}\oplus2(0,\frac{9}{2})_{0,3,0,2}\oplus(\frac{1}{2},4)_{0,3,0,2}\oplus(0,\frac{7}{2})_{0,3,0,2}\oplus(0,\frac{5}{2})_{0,3,0,2}\oplus(0,\frac{7}{2})_{0,3,1,1}\oplus(0,\frac{5}{2})_{0,3,1,1}\oplus(0,\frac{9}{2})_{0,4,0,1}\oplus(0,\frac{7}{2})_{1,1,0,3}\oplus(0,\frac{5}{2})_{1,1,0,3}\oplus(0,\frac{5}{2})_{1,1,1,2}\oplus2(0,\frac{3}{2})_{1,1,1,2}\oplus(0,\frac{1}{2})_{1,1,1,2}\oplus(0,\frac{3}{2})_{1,1,2,1}\oplus(0,\frac{1}{2})_{1,1,2,1}\oplus(\frac{1}{2},4)_{1,2,0,2}\oplus2(0,\frac{7}{2})_{1,2,0,2}\oplus(\frac{1}{2},3)_{1,2,0,2}\oplus3(0,\frac{5}{2})_{1,2,0,2}\oplus(0,\frac{3}{2})_{1,2,0,2}\oplus(0,\frac{5}{2})_{1,2,1,1}\oplus2(0,\frac{3}{2})_{1,2,1,1}\oplus(0,\frac{1}{2})_{1,2,1,1}\oplus(0,\frac{7}{2})_{1,3,0,1}\oplus(0,\frac{5}{2})_{1,3,0,1}\oplus(0,\frac{5}{2})_{2,1,0,2}\oplus(0,\frac{3}{2})_{2,1,0,2}\oplus(0,\frac{1}{2})_{2,1,0,2}\oplus(0,\frac{3}{2})_{2,1,1,1}\oplus(0,\frac{1}{2})_{2,1,1,1}\oplus(0,\frac{5}{2})_{2,2,0,1}\oplus(0,\frac{3}{2})_{2,2,0,1}\oplus(0,\frac{1}{2})_{2,2,0,1}\oplus(0,\frac{5}{2})_{3,0,0,2}\oplus(0,\frac{5}{2})_{3,1,0,1}\oplus(0,\frac{3}{2})_{3,1,0,1}\oplus(0,\frac{7}{2})_{4,0,0,1}\end{math}\\*[0.1cm] \hline6 & \begin{math}(\frac{1}{2},4)_{0,0,4,2}\oplus(0,\frac{7}{2})_{0,0,4,2}\oplus(0,\frac{5}{2})_{0,0,4,2}\oplus(0,\frac{9}{2})_{0,0,5,1}\oplus(0,\frac{11}{2})_{0,1,0,5}\oplus(0,\frac{9}{2})_{0,1,1,4}\oplus(0,\frac{7}{2})_{0,1,1,4}\oplus(0,\frac{7}{2})_{0,1,2,3}\oplus(0,\frac{5}{2})_{0,1,2,3}\oplus(0,\frac{3}{2})_{0,1,2,3}\oplus(0,\frac{7}{2})_{0,1,3,2}\oplus2(0,\frac{5}{2})_{0,1,3,2}\oplus2(0,\frac{3}{2})_{0,1,3,2}\oplus(0,\frac{1}{2})_{0,1,3,2}\oplus(0,\frac{7}{2})_{0,1,4,1}\oplus(0,\frac{5}{2})_{0,1,4,1}\oplus(\frac{3}{2},7)_{0,2,0,4}\oplus(1,\frac{13}{2})_{0,2,0,4}\oplus2(\frac{1}{2},6)_{0,2,0,4}\oplus(1,\frac{11}{2})_{0,2,0,4}\oplus2(0,\frac{11}{2})_{0,2,0,4}\oplus(\frac{1}{2},5)_{0,2,0,4}\oplus2(0,\frac{9}{2})_{0,2,0,4}\oplus(\frac{1}{2},4)_{0,2,0,4}\oplus(0,\frac{7}{2})_{0,2,0,4}\oplus(0,\frac{5}{2})_{0,2,0,4}\oplus(1,\frac{11}{2})_{0,2,1,3}\oplus2(\frac{1}{2},5)_{0,2,1,3}\oplus(1,\frac{9}{2})_{0,2,1,3}\oplus3(0,\frac{9}{2})_{0,2,1,3}\oplus3(\frac{1}{2},4)_{0,2,1,3}\oplus5(0,\frac{7}{2})_{0,2,1,3}\oplus(\frac{1}{2},3)_{0,2,1,3}\oplus3(0,\frac{5}{2})_{0,2,1,3}\oplus(0,\frac{3}{2})_{0,2,1,3}\oplus(\frac{1}{2},4)_{0,2,2,2}\oplus2(0,\frac{7}{2})_{0,2,2,2}\oplus(\frac{1}{2},3)_{0,2,2,2}\oplus4(0,\frac{5}{2})_{0,2,2,2}\oplus(\frac{1}{2},2)_{0,2,2,2}\oplus3(0,\frac{3}{2})_{0,2,2,2}\oplus(0,\frac{1}{2})_{0,2,2,2}\oplus(0,\frac{5}{2})_{0,2,3,1}\oplus(0,\frac{3}{2})_{0,2,3,1}\oplus(0,\frac{1}{2})_{0,2,3,1}\oplus(2,\frac{15}{2})_{0,3,0,3}\oplus(\frac{3}{2},7)_{0,3,0,3}\oplus(\frac{1}{2},7)_{0,3,0,3}\oplus3(1,\frac{13}{2})_{0,3,0,3}\oplus(\frac{3}{2},6)_{0,3,0,3}\oplus3(\frac{1}{2},6)_{0,3,0,3}\oplus2(1,\frac{11}{2})_{0,3,0,3}\oplus4(0,\frac{11}{2})_{0,3,0,3}\oplus3(\frac{1}{2},5)_{0,3,0,3}\oplus(1,\frac{9}{2})_{0,3,0,3}\oplus3(0,\frac{9}{2})_{0,3,0,3}\oplus2(\frac{1}{2},4)_{0,3,0,3}\oplus3(0,\frac{7}{2})_{0,3,0,3}\oplus(\frac{1}{2},3)_{0,3,0,3}\oplus(0,\frac{5}{2})_{0,3,0,3}\oplus(0,\frac{3}{2})_{0,3,0,3}\oplus(1,\frac{11}{2})_{0,3,1,2}\oplus2(\frac{1}{2},5)_{0,3,1,2}\oplus(1,\frac{9}{2})_{0,3,1,2}\oplus3(0,\frac{9}{2})_{0,3,1,2}\oplus3(\frac{1}{2},4)_{0,3,1,2}\oplus5(0,\frac{7}{2})_{0,3,1,2}\oplus(\frac{1}{2},3)_{0,3,1,2}\oplus3(0,\frac{5}{2})_{0,3,1,2}\oplus(0,\frac{3}{2})_{0,3,1,2}\oplus(0,\frac{7}{2})_{0,3,2,1}\oplus(0,\frac{5}{2})_{0,3,2,1}\oplus(0,\frac{3}{2})_{0,3,2,1}\oplus(\frac{3}{2},7)_{0,4,0,2}\oplus(1,\frac{13}{2})_{0,4,0,2}\oplus2(\frac{1}{2},6)_{0,4,0,2}\oplus(1,\frac{11}{2})_{0,4,0,2}\oplus2(0,\frac{11}{2})_{0,4,0,2}\oplus(\frac{1}{2},5)_{0,4,0,2}\oplus2(0,\frac{9}{2})_{0,4,0,2}\oplus(\frac{1}{2},4)_{0,4,0,2}\oplus(0,\frac{7}{2})_{0,4,0,2}\oplus(0,\frac{5}{2})_{0,4,0,2}\oplus(0,\frac{9}{2})_{0,4,1,1}\oplus(0,\frac{7}{2})_{0,4,1,1}\oplus(0,\frac{11}{2})_{0,5,0,1}\oplus(0,\frac{9}{2})_{1,1,0,4}\oplus(0,\frac{7}{2})_{1,1,0,4}\oplus(0,\frac{7}{2})_{1,1,1,3}\oplus2(0,\frac{5}{2})_{1,1,1,3}\oplus(0,\frac{3}{2})_{1,1,1,3}\oplus(0,\frac{5}{2})_{1,1,2,2}\oplus2(0,\frac{3}{2})_{1,1,2,2}\oplus2(0,\frac{1}{2})_{1,1,2,2}\oplus(0,\frac{5}{2})_{1,1,3,1}\oplus(0,\frac{3}{2})_{1,1,3,1}\oplus(1,\frac{11}{2})_{1,2,0,3}\oplus2(\frac{1}{2},5)_{1,2,0,3}\oplus(1,\frac{9}{2})_{1,2,0,3}\oplus3(0,\frac{9}{2})_{1,2,0,3}\oplus3(\frac{1}{2},4)_{1,2,0,3}\oplus5(0,\frac{7}{2})_{1,2,0,3}\oplus(\frac{1}{2},3)_{1,2,0,3}\oplus3(0,\frac{5}{2})_{1,2,0,3}\oplus(0,\frac{3}{2})_{1,2,0,3}\oplus(\frac{1}{2},4)_{1,2,1,2}\oplus3(0,\frac{7}{2})_{1,2,1,2}\oplus2(\frac{1}{2},3)_{1,2,1,2}\oplus7(0,\frac{5}{2})_{1,2,1,2}\oplus(\frac{1}{2},2)_{1,2,1,2}\oplus5(0,\frac{3}{2})_{1,2,1,2}\oplus(0,\frac{1}{2})_{1,2,1,2}\oplus(0,\frac{5}{2})_{1,2,2,1}\oplus2(0,\frac{3}{2})_{1,2,2,1}\oplus2(0,\frac{1}{2})_{1,2,2,1}\oplus(1,\frac{11}{2})_{1,3,0,2}\oplus2(\frac{1}{2},5)_{1,3,0,2}\oplus(1,\frac{9}{2})_{1,3,0,2}\oplus3(0,\frac{9}{2})_{1,3,0,2}\oplus3(\frac{1}{2},4)_{1,3,0,2}\oplus5(0,\frac{7}{2})_{1,3,0,2}\oplus(\frac{1}{2},3)_{1,3,0,2}\oplus3(0,\frac{5}{2})_{1,3,0,2}\oplus(0,\frac{3}{2})_{1,3,0,2}\oplus(0,\frac{7}{2})_{1,3,1,1}\oplus2(0,\frac{5}{2})_{1,3,1,1}\oplus(0,\frac{3}{2})_{1,3,1,1}\oplus(0,\frac{9}{2})_{1,4,0,1}\oplus(0,\frac{7}{2})_{1,4,0,1}\oplus(0,\frac{7}{2})_{2,1,0,3}\oplus(0,\frac{5}{2})_{2,1,0,3}\oplus(0,\frac{3}{2})_{2,1,0,3}\oplus(0,\frac{5}{2})_{2,1,1,2}\oplus2(0,\frac{3}{2})_{2,1,1,2}\oplus2(0,\frac{1}{2})_{2,1,1,2}\oplus(\frac{1}{2},4)_{2,2,0,2}\oplus2(0,\frac{7}{2})_{2,2,0,2}\oplus(\frac{1}{2},3)_{2,2,0,2}\oplus4(0,\frac{5}{2})_{2,2,0,2}\oplus(\frac{1}{2},2)_{2,2,0,2}\oplus3(0,\frac{3}{2})_{2,2,0,2}\oplus(0,\frac{1}{2})_{2,2,0,2}\oplus(0,\frac{5}{2})_{2,2,1,1}\oplus2(0,\frac{3}{2})_{2,2,1,1}\oplus2(0,\frac{1}{2})_{2,2,1,1}\oplus(0,\frac{7}{2})_{2,3,0,1}\oplus(0,\frac{5}{2})_{2,3,0,1}\oplus(0,\frac{3}{2})_{2,3,0,1}\oplus(0,\frac{7}{2})_{3,1,0,2}\oplus2(0,\frac{5}{2})_{3,1,0,2}\oplus2(0,\frac{3}{2})_{3,1,0,2}\oplus(0,\frac{1}{2})_{3,1,0,2}\oplus(0,\frac{5}{2})_{3,1,1,1}\oplus(0,\frac{3}{2})_{3,1,1,1}\oplus(0,\frac{5}{2})_{3,2,0,1}\oplus(0,\frac{3}{2})_{3,2,0,1}\oplus(0,\frac{1}{2})_{3,2,0,1}\oplus(\frac{1}{2},4)_{4,0,0,2}\oplus(0,\frac{7}{2})_{4,0,0,2}\oplus(0,\frac{5}{2})_{4,0,0,2}\oplus(0,\frac{7}{2})_{4,1,0,1}\oplus(0,\frac{5}{2})_{4,1,0,1}\oplus(0,\frac{9}{2})_{5,0,0,1}\end{math}\\*[0.1cm] \hline\caption{The refined BPS invariants of $SU(4),m=0$ model. We list the invariants up to degree $d=d_1+d_2+d_3+d_B=6$. From the charge, we can see the symmetry $N_{j_L,j_R}^{d_1,d_2,d_3,d_B}=N_{j_L,j_R}^{d_3,d_2,d_1,d_B}$.}\end{longtable}
\end{spacing}
\end{footnotesize}
\newpage\subsection*{$SU(5)$ Geometry with $m=0$}
\begin{small}
\begin{spacing}{1.3}
\begin{longtable}{| p{0.01\textwidth} | p{\textwidth} |} \hline  $d$&$\sum_{d_1+d_2+d_3+d_4+d_B=d} \sum_{j_L,j_R} \oplus N^{\mathbf{d}}_{j_L,j_R}(j_L,j_R)_{d_1,d_2,d_3,d_4,d_B} $\\*[0.1cm] \hline 1 & \begin{math}(0,0)_{0,0,0,0,1}\oplus(0,\frac{1}{2})_{0,0,0,1,0}\oplus(0,\frac{1}{2})_{0,0,1,0,0}\oplus(0,\frac{1}{2})_{0,1,0,0,0}\oplus(0,\frac{1}{2})_{1,0,0,0,0}\end{math}\\*[0.1cm] \hline2 & \begin{math}(0,1)_{0,0,1,0,1}\oplus(0,\frac{1}{2})_{0,0,1,1,0}\oplus(0,1)_{0,1,0,0,1}\oplus(0,\frac{1}{2})_{0,1,1,0,0}\oplus(0,\frac{1}{2})_{1,1,0,0,0}\end{math}\\*[0.1cm] \hline3 & \begin{math}(0,1)_{0,0,1,1,1}\oplus(0,0)_{0,0,1,1,1}\oplus(0,2)_{0,0,2,0,1}\oplus(0,1)_{0,1,1,0,1}\oplus(0,0)_{0,1,1,0,1}\oplus(0,\frac{1}{2})_{0,1,1,1,0}\oplus(0,2)_{0,2,0,0,1}\oplus(0,1)_{1,1,0,0,1}\oplus(0,0)_{1,1,0,0,1}\oplus(0,\frac{1}{2})_{1,1,1,0,0}\end{math}\\*[0.1cm] \hline4 & \begin{math}(0,1)_{0,0,1,2,1}\oplus(0,\frac{5}{2})_{0,0,2,0,2}\oplus(0,2)_{0,0,2,1,1}\oplus(0,1)_{0,0,2,1,1}\oplus(0,3)_{0,0,3,0,1}\oplus(0,1)_{0,1,1,1,1}\oplus(0,0)_{0,1,1,1,1}\oplus(0,2)_{0,1,2,0,1}\oplus(0,1)_{0,1,2,0,1}\oplus(0,\frac{5}{2})_{0,2,0,0,2}\oplus(0,2)_{0,2,1,0,1}\oplus(0,1)_{0,2,1,0,1}\oplus(0,3)_{0,3,0,0,1}\oplus(0,1)_{1,1,1,0,1}\oplus(0,0)_{1,1,1,0,1}\oplus(0,\frac{1}{2})_{1,1,1,1,0}\oplus(0,2)_{1,2,0,0,1}\oplus(0,1)_{1,2,0,0,1}\oplus(0,1)_{2,1,0,0,1}\end{math}\\*[0.1cm] \hline5 & \begin{math}(0,2)_{0,0,1,3,1}\oplus(0,\frac{5}{2})_{0,0,2,1,2}\oplus(0,\frac{3}{2})_{0,0,2,1,2}\oplus(0,2)_{0,0,2,2,1}\oplus(0,1)_{0,0,2,2,1}\oplus(0,0)_{0,0,2,2,1}\oplus(\frac{1}{2},4)_{0,0,3,0,2}\oplus(0,\frac{7}{2})_{0,0,3,0,2}\oplus(0,\frac{5}{2})_{0,0,3,0,2}\oplus(0,3)_{0,0,3,1,1}\oplus(0,2)_{0,0,3,1,1}\oplus(0,4)_{0,0,4,0,1}\oplus(0,\frac{5}{2})_{0,1,2,0,2}\oplus(0,\frac{3}{2})_{0,1,2,0,2}\oplus(0,2)_{0,1,2,1,1}\oplus2(0,1)_{0,1,2,1,1}\oplus(0,0)_{0,1,2,1,1}\oplus(0,3)_{0,1,3,0,1}\oplus(0,2)_{0,1,3,0,1}\oplus(0,\frac{5}{2})_{0,2,1,0,2}\oplus(0,\frac{3}{2})_{0,2,1,0,2}\oplus(0,2)_{0,2,1,1,1}\oplus(0,1)_{0,2,1,1,1}\oplus(0,2)_{0,2,2,0,1}\oplus(0,1)_{0,2,2,0,1}\oplus(0,0)_{0,2,2,0,1}\oplus(\frac{1}{2},4)_{0,3,0,0,2}\oplus(0,\frac{7}{2})_{0,3,0,0,2}\oplus(0,\frac{5}{2})_{0,3,0,0,2}\oplus(0,3)_{0,3,1,0,1}\oplus(0,2)_{0,3,1,0,1}\oplus(0,4)_{0,4,0,0,1}\oplus(0,1)_{1,1,1,1,1}\oplus(0,0)_{1,1,1,1,1}\oplus(0,2)_{1,1,2,0,1}\oplus(0,1)_{1,1,2,0,1}\oplus(0,\frac{5}{2})_{1,2,0,0,2}\oplus(0,\frac{3}{2})_{1,2,0,0,2}\oplus(0,2)_{1,2,1,0,1}\oplus2(0,1)_{1,2,1,0,1}\oplus(0,0)_{1,2,1,0,1}\oplus(0,3)_{1,3,0,0,1}\oplus(0,2)_{1,3,0,0,1}\oplus(0,2)_{2,2,0,0,1}\oplus(0,1)_{2,2,0,0,1}\oplus(0,0)_{2,2,0,0,1}\oplus(0,2)_{3,1,0,0,1}\end{math}\\*[0.1cm] \hline6 & \begin{math}(0,3)_{0,0,1,4,1}\oplus(0,\frac{5}{2})_{0,0,2,2,2}\oplus(0,\frac{3}{2})_{0,0,2,2,2}\oplus(0,\frac{1}{2})_{0,0,2,2,2}\oplus(0,2)_{0,0,2,3,1}\oplus(0,1)_{0,0,2,3,1}\oplus(\frac{1}{2},\frac{9}{2})_{0,0,3,0,3}\oplus(0,3)_{0,0,3,0,3}\oplus(\frac{1}{2},4)_{0,0,3,1,2}\oplus2(0,\frac{7}{2})_{0,0,3,1,2}\oplus(\frac{1}{2},3)_{0,0,3,1,2}\oplus3(0,\frac{5}{2})_{0,0,3,1,2}\oplus(0,\frac{3}{2})_{0,0,3,1,2}\oplus(0,3)_{0,0,3,2,1}\oplus(0,2)_{0,0,3,2,1}\oplus(0,1)_{0,0,3,2,1}\oplus(1,\frac{11}{2})_{0,0,4,0,2}\oplus(\frac{1}{2},5)_{0,0,4,0,2}\oplus2(0,\frac{9}{2})_{0,0,4,0,2}\oplus(\frac{1}{2},4)_{0,0,4,0,2}\oplus(0,\frac{7}{2})_{0,0,4,0,2}\oplus(0,\frac{5}{2})_{0,0,4,0,2}\oplus(0,4)_{0,0,4,1,1}\oplus(0,3)_{0,0,4,1,1}\oplus(0,5)_{0,0,5,0,1}\oplus(0,\frac{5}{2})_{0,1,2,1,2}\oplus2(0,\frac{3}{2})_{0,1,2,1,2}\oplus(0,\frac{1}{2})_{0,1,2,1,2}\oplus(0,2)_{0,1,2,2,1}\oplus2(0,1)_{0,1,2,2,1}\oplus(0,0)_{0,1,2,2,1}\oplus(\frac{1}{2},4)_{0,1,3,0,2}\oplus2(0,\frac{7}{2})_{0,1,3,0,2}\oplus(\frac{1}{2},3)_{0,1,3,0,2}\oplus3(0,\frac{5}{2})_{0,1,3,0,2}\oplus(0,\frac{3}{2})_{0,1,3,0,2}\oplus(0,3)_{0,1,3,1,1}\oplus2(0,2)_{0,1,3,1,1}\oplus(0,1)_{0,1,3,1,1}\oplus(0,4)_{0,1,4,0,1}\oplus(0,3)_{0,1,4,0,1}\oplus(0,\frac{5}{2})_{0,2,1,1,2}\oplus(0,\frac{3}{2})_{0,2,1,1,2}\oplus(0,\frac{7}{2})_{0,2,2,0,2}\oplus2(0,\frac{5}{2})_{0,2,2,0,2}\oplus2(0,\frac{3}{2})_{0,2,2,0,2}\oplus2(0,\frac{1}{2})_{0,2,2,0,2}\oplus(0,2)_{0,2,2,1,1}\oplus2(0,1)_{0,2,2,1,1}\oplus(0,0)_{0,2,2,1,1}\oplus(0,3)_{0,2,3,0,1}\oplus(0,2)_{0,2,3,0,1}\oplus(0,1)_{0,2,3,0,1}\oplus(\frac{1}{2},\frac{9}{2})_{0,3,0,0,3}\oplus(0,3)_{0,3,0,0,3}\oplus(\frac{1}{2},4)_{0,3,1,0,2}\oplus2(0,\frac{7}{2})_{0,3,1,0,2}\oplus(\frac{1}{2},3)_{0,3,1,0,2}\oplus3(0,\frac{5}{2})_{0,3,1,0,2}\oplus(0,\frac{3}{2})_{0,3,1,0,2}\oplus(0,3)_{0,3,1,1,1}\oplus(0,2)_{0,3,1,1,1}\oplus(0,3)_{0,3,2,0,1}\oplus(0,2)_{0,3,2,0,1}\oplus(0,1)_{0,3,2,0,1}\oplus(1,\frac{11}{2})_{0,4,0,0,2}\oplus(\frac{1}{2},5)_{0,4,0,0,2}\oplus2(0,\frac{9}{2})_{0,4,0,0,2}\oplus(\frac{1}{2},4)_{0,4,0,0,2}\oplus(0,\frac{7}{2})_{0,4,0,0,2}\oplus(0,\frac{5}{2})_{0,4,0,0,2}\oplus(0,4)_{0,4,1,0,1}\oplus(0,3)_{0,4,1,0,1}\oplus(0,5)_{0,5,0,0,1}\oplus(0,\frac{5}{2})_{1,1,2,0,2}\oplus(0,\frac{3}{2})_{1,1,2,0,2}\oplus(0,2)_{1,1,2,1,1}\oplus2(0,1)_{1,1,2,1,1}\oplus(0,0)_{1,1,2,1,1}\oplus(0,3)_{1,1,3,0,1}\oplus(0,2)_{1,1,3,0,1}\oplus(0,\frac{5}{2})_{1,2,1,0,2}\oplus2(0,\frac{3}{2})_{1,2,1,0,2}\oplus(0,\frac{1}{2})_{1,2,1,0,2}\oplus(0,2)_{1,2,1,1,1}\oplus2(0,1)_{1,2,1,1,1}\oplus(0,0)_{1,2,1,1,1}\oplus(0,2)_{1,2,2,0,1}\oplus2(0,1)_{1,2,2,0,1}\oplus(0,0)_{1,2,2,0,1}\oplus(\frac{1}{2},4)_{1,3,0,0,2}\oplus2(0,\frac{7}{2})_{1,3,0,0,2}\oplus(\frac{1}{2},3)_{1,3,0,0,2}\oplus3(0,\frac{5}{2})_{1,3,0,0,2}\oplus(0,\frac{3}{2})_{1,3,0,0,2}\oplus(0,3)_{1,3,1,0,1}\oplus2(0,2)_{1,3,1,0,1}\oplus(0,1)_{1,3,1,0,1}\oplus(0,4)_{1,4,0,0,1}\oplus(0,3)_{1,4,0,0,1}\oplus(0,\frac{5}{2})_{2,2,0,0,2}\oplus(0,\frac{3}{2})_{2,2,0,0,2}\oplus(0,\frac{1}{2})_{2,2,0,0,2}\oplus(0,2)_{2,2,1,0,1}\oplus2(0,1)_{2,2,1,0,1}\oplus(0,0)_{2,2,1,0,1}\oplus(0,3)_{2,3,0,0,1}\oplus(0,2)_{2,3,0,0,1}\oplus(0,1)_{2,3,0,0,1}\oplus(0,2)_{3,2,0,0,1}\oplus(0,1)_{3,2,0,0,1}\oplus(0,3)_{4,1,0,0,1}\end{math}\\*[0.1cm] \hline\caption{The refined BPS invariants of $SU(5),m=0$ model. We list the invariants up to degree $d=d_1+d_2+d_3+d_4+d_B=6$. From the charge, we can see the symmetry $N_{j_L,j_R}^{d_1,d_2,d_3,d_4,d_B}=N_{j_L,j_R}^{d_4,d_3,d_2,d_1,d_B}$.}\end{longtable}
\end{spacing}
\end{small}


\newpage
\addcontentsline{toc}{section}{References}
\begin{thebibliography}{150}
\bibitem{Lockhart:2012vp}
  G.~Lockhart and C.~Vafa,
  ``Superconformal Partition Functions and Non-perturbative Topological Strings,''
  arXiv:1210.5909 [hep-th].



\bibitem{Dijkgraaf:2002fc}
  R.~Dijkgraaf and C.~Vafa,
  ``Matrix models, topological strings, and supersymmetric gauge theories,''
  Nucl.\ Phys.\ B {\bf 644}, 3 (2002)
  [hep-th/0206255].



\bibitem{Marino:2008ya}
  M.~Marino,
  ``Non-perturbative effects and non-perturbative definitions in matrix models and topological strings,''
  JHEP {\bf 0812}, 114 (2008)
  [arXiv:0805.3033 [hep-th]].


\bibitem{Marino:2009jd}
  M.~Marino and P.~Putrov,
  ``Exact Results in ABJM Theory from Topological Strings,''
  JHEP {\bf 1006}, 011 (2010)
  [arXiv:0912.3074 [hep-th]].



\bibitem{Kapustin:2009kz}
  A.~Kapustin, B.~Willett and I.~Yaakov,
  ``Exact Results for Wilson Loops in Superconformal Chern-Simons Theories with Matter,''
  JHEP {\bf 1003}, 089 (2010)
  [arXiv:0909.4559 [hep-th]].



\bibitem{Hatsuda:2013oxa}
  Y.~Hatsuda, M.~Marino, S.~Moriyama and K.~Okuyama,
  ``Non-perturbative effects and the refined topological string,''
  JHEP {\bf 1409}, 168 (2014)
  [arXiv:1306.1734 [hep-th]].



\bibitem{Gopakumar:1998ki}
  R.~Gopakumar and C.~Vafa,
  ``On the gauge theory / geometry correspondence,''
  Adv.\ Theor.\ Math.\ Phys.\  {\bf 3}, 1415 (1999)
  [hep-th/9811131].



\bibitem{Aganagic:2002qg}
  M.~Aganagic, M.~Marino and C.~Vafa,
  ``All loop topological string amplitudes from Chern-Simons theory,''
  Commun.\ Math.\ Phys.\  {\bf 247}, 467 (2004)
  [hep-th/0206164].



\bibitem{Krefl:2015vna}
  D.~Krefl and R.~L.~Mkrtchyan,
  ``Exact Chern-Simons / Topological String duality,''
  JHEP {\bf 1510}, 045 (2015)
  [arXiv:1506.03907 [hep-th]].



\bibitem{Katz:1996fh}
  S.~H.~Katz, A.~Klemm and C.~Vafa,
  ``Geometric engineering of quantum field theories,''
  Nucl.\ Phys.\ B {\bf 497}, 173 (1997)
  [hep-th/9609239].



\bibitem{Iqbal:2003zz}
  A.~Iqbal and A.~K.~Kashani-Poor,
  ``SU(N) geometries and topological string amplitudes,''
  Adv.\ Theor.\ Math.\ Phys.\  {\bf 10}, 1 (2006)
  [hep-th/0306032].



\bibitem{Gopakumar:1998jq}
  R.~Gopakumar and C.~Vafa,
  ``M theory and topological strings. 2.,''
  hep-th/9812127.



\bibitem{Katz:1999xq}
  S.~H.~Katz, A.~Klemm and C.~Vafa,
  ``M theory, topological strings and spinning black holes,''
  Adv.\ Theor.\ Math.\ Phys.\  {\bf 3}, 1445 (1999)
  [hep-th/9910181].



\bibitem{Dijkgraaf:2006um}
  R.~Dijkgraaf, C.~Vafa and E.~Verlinde,
  ``M-theory and a topological string duality,''
  hep-th/0602087.



\bibitem{Aganagic:2003qj}
  M.~Aganagic, R.~Dijkgraaf, A.~Klemm, M.~Marino and C.~Vafa,
  ``Topological strings and integrable hierarchies,''
  Commun.\ Math.\ Phys.\  {\bf 261}, 451 (2006)
  [hep-th/0312085].



\bibitem{Aganagic:2011mi}
  M.~Aganagic, M.~C.~N.~Cheng, R.~Dijkgraaf, D.~Krefl and C.~Vafa,
  ``Quantum Geometry of Refined Topological Strings,''
  JHEP {\bf 1211}, 019 (2012)
  [arXiv:1105.0630 [hep-th]].



\bibitem{Cheng:2010yw}
  M.~C.~N.~Cheng, R.~Dijkgraaf and C.~Vafa,
  ``Non-Perturbative Topological Strings And Conformal Blocks,''
  JHEP {\bf 1109}, 022 (2011)
  [arXiv:1010.4573 [hep-th]].



\bibitem{Ooguri:2004zv}
  H.~Ooguri, A.~Strominger and C.~Vafa,
  ``Black hole attractors and the topological string,''
  Phys.\ Rev.\ D {\bf 70}, 106007 (2004)
  [hep-th/0405146].




\bibitem{Aganagic:2004js}
  M.~Aganagic, H.~Ooguri, N.~Saulina and C.~Vafa,
  ``Black holes, q-deformed 2d Yang-Mills, and non-perturbative topological strings,''
  Nucl.\ Phys.\ B {\bf 715}, 304 (2005)
  [hep-th/0411280].



\bibitem{Santamaria:2013rua}
  R.~Couso-Santamar¨ªa, J.~D.~Edelstein, R.~Schiappa and M.~Vonk,
  ``Resurgent Transseries and the Holomorphic Anomaly,''
  Annales Henri Poincare {\bf 17}, no. 2, 331 (2016)
  [arXiv:1308.1695 [hep-th]].



\bibitem{Couso-Santamaria:2014iia}
  R.~Couso-Santamar¨ªa, J.~D.~Edelstein, R.~Schiappa and M.~Vonk,
  ``Resurgent Transseries and the Holomorphic Anomaly: Non-perturbative Closed Strings in Local ${\mathbb{C}\mathbb{P}^2}$,''
  Commun.\ Math.\ Phys.\  {\bf 338}, no. 1, 285 (2015)
  [arXiv:1407.4821 [hep-th]].



\bibitem{Hatsuda:2015owa}
  Y.~Hatsuda and K.~Okuyama,
  ``Resummations and Non-Perturbative Corrections,''
  JHEP {\bf 1509}, 051 (2015)
  [arXiv:1505.07460 [hep-th]].



\bibitem{Krefl:2015qva}
  D.~Krefl,
  ``Mellin-Barnes Representation of the Topological String,''
  arXiv:1508.04219 [hep-th].



\bibitem{Gukov:2011qp}
  S.~Gukov and P.~Sulkowski,
  ``A-polynomial, B-model, and Quantization,''
  JHEP {\bf 1202}, 070 (2012)
  [arXiv:1108.0002 [hep-th]].



\bibitem{Aganagic:2012jb}
  M.~Aganagic and C.~Vafa,
  ``Large N Duality, Mirror Symmetry, and a Q-deformed A-polynomial for Knots,''
  arXiv:1204.4709 [hep-th].



\bibitem{Dijkgraaf:2008fh}
  R.~Dijkgraaf, L.~Hollands and P.~Sulkowski,
  ``Quantum Curves and D-Modules,''
  JHEP {\bf 0911}, 047 (2009)
  [arXiv:0810.4157 [hep-th]].



\bibitem{Hatsuda:2016mdw}
  Y.~Hatsuda, H.~Katsura and Y.~Tachikawa,
  ``Hofstadter's Butterfly in Quantum Geometry,''
  arXiv:1606.01894 [hep-th].



\bibitem{Nekrasov:2009rc}
  N.~A.~Nekrasov and S.~L.~Shatashvili,
  ``Quantization of Integrable Systems and Four Dimensional Gauge Theories,''
  arXiv:0908.4052 [hep-th].



\bibitem{Wang:2015wdy}
  X.~Wang, G.~Zhang and M.~x.~Huang,
  ``New Exact Quantization Condition for Toric Calabi-Yau Geometries,''
  Phys.\ Rev.\ Lett.\  {\bf 115}, 121601 (2015)
  [arXiv:1505.05360 [hep-th]].



\bibitem{Hatsuda:2015qzx}
  Y.~Hatsuda and M.~Marino,
  ``Exact quantization conditions for the relativistic Toda lattice,''
  arXiv:1511.02860 [hep-th].




\bibitem{Franco:2015rnr}
  S.~Franco, Y.~Hatsuda and M.~Marino,
  ``Exact quantization conditions for cluster integrable systems,''
  arXiv:1512.03061 [hep-th].



\bibitem{Goncharov:2011hp}
  A.~B.~Goncharov and R.~Kenyon,
  ``Dimers and cluster integrable systems,''
  arXiv:1107.5588 [math.AG].



\bibitem{Hatsuda:2015fxa}
  Y.~Hatsuda,
  ``Comments on Exact Quantization Conditions and Non-Perturbative Topological Strings,''
  arXiv:1507.04799 [hep-th].



\bibitem{Kashani-Poor:2016edc}
  A.~K.~Kashani-Poor,
  ``Quantization condition from exact WKB for difference equations,''
  arXiv:1604.01690 [hep-th].



\bibitem{Krefl:2016svj}
  D.~Krefl,
  ``Non-Perturbative Quantum Geometry III,''
  arXiv:1605.00182 [hep-th].



\bibitem{Grassi:2014zfa}
  A.~Grassi, Y.~Hatsuda and M.~Marino,
  ``Topological Strings from Quantum Mechanics,''
  arXiv:1410.3382 [hep-th].



\bibitem{Aharony:2008ug}
  O.~Aharony, O.~Bergman, D.~L.~Jafferis and J.~Maldacena,
  ``N=6 superconformal Chern-Simons-matter theories, M2-branes and their gravity duals,''
  JHEP {\bf 0810}, 091 (2008)
  [arXiv:0806.1218 [hep-th]].



\bibitem{Drukker:2010nc}
  N.~Drukker, M.~Marino and P.~Putrov,
  ``From weak to strong coupling in ABJM theory,''
  Commun.\ Math.\ Phys.\  {\bf 306}, 511 (2011)
  [arXiv:1007.3837 [hep-th]].



\bibitem{Drukker:2011zy}
  N.~Drukker, M.~Marino and P.~Putrov,
  ``Non-perturbative aspects of ABJM theory,''
  JHEP {\bf 1111}, 141 (2011)
  [arXiv:1103.4844 [hep-th]].



\bibitem{Marino:2011eh}
  M.~Marino and P.~Putrov,
  ``ABJM theory as a Fermi gas,''
  J.\ Stat.\ Mech.\  {\bf 1203}, P03001 (2012)
  [arXiv:1110.4066 [hep-th]].



\bibitem{Calvo:2012du}
  F.~Calvo and M.~Marino,
  ``Membrane instantons from a semiclassical TBA,''
  JHEP {\bf 1305}, 006 (2013)
  [arXiv:1212.5118 [hep-th]].

\bibitem{Hatsuda:2012hm}
  Y.~Hatsuda, S.~Moriyama and K.~Okuyama,
  ``Exact Results on the ABJM Fermi Gas,''
  JHEP {\bf 1210}, 020 (2012)
  [arXiv:1207.4283 [hep-th]].

\bibitem{Hatsuda:2012dt}
  Y.~Hatsuda, S.~Moriyama and K.~Okuyama,
  ``Instanton Effects in ABJM Theory from Fermi Gas Approach,''
  JHEP {\bf 1301}, 158 (2013)
  [arXiv:1211.1251 [hep-th]].

\bibitem{Hatsuda:2013gj}
  Y.~Hatsuda, S.~Moriyama and K.~Okuyama,
  ``Instanton Bound States in ABJM Theory,''
  JHEP {\bf 1305}, 054 (2013)
  [arXiv:1301.5184 [hep-th]].


\bibitem{Kallen:2013qla}
  J.~Kallen and M.~Marino,
  ``Instanton effects and quantum spectral curves,''
  Annales Henri Poincare {\bf 17}, no. 5, 1037 (2016)
  [arXiv:1308.6485 [hep-th]].




\bibitem{Huang:2014eha}
  M.~x.~Huang and X.~f.~Wang,
  ``Topological Strings and Quantum Spectral Problems,''
  JHEP {\bf 1409}, 150 (2014)
  [arXiv:1406.6178 [hep-th]].



\bibitem{Wang:2014ega}
  X.~f.~Wang, X.~Wang and M.~x.~Huang,
  ``A Note on Instanton Effects in ABJM Theory,''
  JHEP {\bf 1411}, 100 (2014)
  [arXiv:1409.4967 [hep-th]].


\bibitem{Marino:2015nla}
  M.~Marino,
  ``Spectral Theory and Mirror Symmetry,''
  arXiv:1506.07757 [math-ph].



\bibitem{Codesido:2015dia}
  S.~Codesido, A.~Grassi and M.~Marino,
  ``Spectral Theory and Mirror Curves of Higher Genus,''
  arXiv:1507.02096 [hep-th].



\bibitem{Gu:2015pda}
  J.~Gu, A.~Klemm, M.~Marino and J.~Reuter,
  ``Exact solutions to quantum spectral curves by topological string theory,''
  JHEP {\bf 1510}, 025 (2015)
  [arXiv:1506.09176 [hep-th]].

\bibitem{Hatsuda:2015oaa}
  Y.~Hatsuda,
  ``Spectral zeta function and non-perturbative effects in ABJM Fermi-gas,''
  JHEP {\bf 1511}, 086 (2015)
  [arXiv:1503.07883 [hep-th]].

\bibitem{Kashaev:2015kha}
  R.~Kashaev and M.~Marino,
  ``Operators from mirror curves and the quantum dilogarithm,''
  arXiv:1501.01014 [hep-th].




\bibitem{Marino:2015ixa}
  M.~Marino and S.~Zakany,
  ``Matrix models from operators and topological strings,''
  arXiv:1502.02958 [hep-th].



\bibitem{Kashaev:2015wia}
  R.~Kashaev, M.~Marino and S.~Zakany,
  ``Matrix models from operators and topological strings, 2,''
  arXiv:1505.02243 [hep-th].



\bibitem{Bonelli:2016idi}
  G.~Bonelli, A.~Grassi and A.~Tanzini,
  ``Seiberg-Witten theory as a Fermi gas,''
  arXiv:1603.01174 [hep-th].

%

\bibitem{Grassi:2016vkw}
  A.~Grassi,
  ``Spectral determinants and quantum theta functions,''
  arXiv:1604.06786 [hep-th].



\bibitem{Chiang:1999tz}
  T.~M.~Chiang, A.~Klemm, S.~T.~Yau and E.~Zaslow,
  ``Local mirror symmetry: Calculations and interpretations,''
  Adv.\ Theor.\ Math.\ Phys.\  {\bf 3}, 495 (1999)
  [hep-th/9903053].



\bibitem{Klemm:2015iya}
  A.~Klemm, M.~Poretschkin, T.~Schimannek and M.~Westerholt-Raum,
  ``Direct Integration for Mirror Curves of Genus Two and an Almost Meromorphic Siegel Modular Form,''
  arXiv:1502.00557 [hep-th].


\bibitem{Iqbal:2007ii}
  A.~Iqbal, C.~Kozcaz and C.~Vafa,
  ``The Refined topological vertex,''
  JHEP {\bf 0910}, 069 (2009)
  [hep-th/0701156].


\bibitem{Choi:2012jz}
  J.~Choi, S.~Katz and A.~Klemm,
  ``The refined BPS index from stable pair invariants,''
  Commun.\ Math.\ Phys.\  {\bf 328}, 903 (2014)
  [arXiv:1210.4403 [hep-th]].



\bibitem{Nekrasov:2014nea}
  N.~Nekrasov and A.~Okounkov,
  ``Membranes and Sheaves,''
  arXiv:1404.2323 [math.AG].



\bibitem{Huang:2010kf}
  M.~x.~Huang and A.~Klemm,
  ``Direct integration for general $\Omega$ backgrounds,''
  Adv.\ Theor.\ Math.\ Phys.\  {\bf 16}, no. 3, 805 (2012)
  [arXiv:1009.1126 [hep-th]].



\bibitem{Taki:2007dh}
  M.~Taki,
  ``Refined Topological Vertex and Instanton Counting,''
  JHEP {\bf 0803}, 048 (2008)
  [arXiv:0710.1776 [hep-th]].



\bibitem{Iqbal:2012mt}
  A.~Iqbal and C.~Kozcaz,
  ``Refined Topological Strings and Toric Calabi-Yau Threefolds,''
  arXiv:1210.3016 [hep-th].



\bibitem{Eynard:2007kz}
  B.~Eynard and N.~Orantin,
  ``Invariants of algebraic curves and topological expansion,''
  Commun.\ Num.\ Theor.\ Phys.\  {\bf 1}, 347 (2007)
  [math-ph/0702045].



\bibitem{Bouchard:2007ys}
  V.~Bouchard, A.~Klemm, M.~Marino and S.~Pasquetti,
  ``Remodeling the B-model,''
  Commun.\ Math.\ Phys.\  {\bf 287}, 117 (2009)
  [arXiv:0709.1453 [hep-th]].



\bibitem{Huang:2013yta}
  M.~X.~Huang, A.~Klemm and M.~Poretschkin,
  ``Refined stable pair invariants for E-, M- and $[p, q]$-strings,''
  JHEP {\bf 1311}, 112 (2013)
  [arXiv:1308.0619 [hep-th]].



\bibitem{Gorsky:1995zq}
  A.~Gorsky, I.~Krichever, A.~Marshakov, A.~Mironov and A.~Morozov,
  ``Integrability and Seiberg-Witten exact solution,''
  Phys.\ Lett.\ B {\bf 355}, 466 (1995)
  [hep-th/9505035].



\bibitem{Martinec:1995by}
  E.~J.~Martinec and N.~P.~Warner,
  ``Integrable systems and supersymmetric gauge theory,''
  Nucl.\ Phys.\ B {\bf 459}, 97 (1996)
  [hep-th/9509161].



\bibitem{D'Hoker:1999ft}
  E.~D'Hoker and D.~H.~Phong,
  ``Lectures on supersymmetric Yang-Mills theory and integrable systems,''
  hep-th/9912271.



\bibitem{Nekrasov:2009uh}
  N.~A.~Nekrasov and S.~L.~Shatashvili,
  ``Supersymmetric vacua and Bethe ansatz,''
  Nucl.\ Phys.\ Proc.\ Suppl.\  {\bf 192-193}, 91 (2009)
  [arXiv:0901.4744 [hep-th]].



\bibitem{Nekrasov:2009ui}
  N.~A.~Nekrasov and S.~L.~Shatashvili,
  ``Quantum integrability and supersymmetric vacua,''
  Prog.\ Theor.\ Phys.\ Suppl.\  {\bf 177}, 105 (2009)
  [arXiv:0901.4748 [hep-th]].



\bibitem{Mironov:2009uv}
  A.~Mironov and A.~Morozov,
  ``Nekrasov Functions and Exact Bohr-Zommerfeld Integrals,''
  JHEP {\bf 1004}, 040 (2010)
  [arXiv:0910.5670 [hep-th]].



\bibitem{Mironov:2009dv}
  A.~Mironov and A.~Morozov,
  ``Nekrasov Functions from Exact BS Periods: The Case of SU(N),''
  J.\ Phys.\ A {\bf 43}, 195401 (2010)
  [arXiv:0911.2396 [hep-th]].



\bibitem{Kozlowski:2010tv}
  K.~K.~Kozlowski and J.~Teschner,
  ``TBA for the Toda chain,''
  arXiv:1006.2906 [math-ph].



\bibitem{Meneghelli:2013tia}
  C.~Meneghelli and G.~Yang,
  ``Mayer-Cluster Expansion of Instanton Partition Functions and Thermodynamic Bethe Ansatz,''
  JHEP {\bf 1405}, 112 (2014)
  [arXiv:1312.4537 [hep-th]].



\bibitem{Alday:2009aq}
  L.~F.~Alday, D.~Gaiotto and Y.~Tachikawa,
  ``Liouville Correlation Functions from Four-dimensional Gauge Theories,''
  Lett.\ Math.\ Phys.\  {\bf 91}, 167 (2010)
  [arXiv:0906.3219 [hep-th]].

\bibitem{Dijkgraaf:2009pc}
  R.~Dijkgraaf and C.~Vafa,
  ``Toda Theories, Matrix Models, Topological Strings, and N=2 Gauge Systems,''
  arXiv:0909.2453 [hep-th].

\bibitem{Rim:2015aha}
  C.~Rim and H.~Zhang,
  ``Classical Virasoro irregular conformal block II,''
  JHEP {\bf 1509}, 097 (2015)
  [arXiv:1506.03561 [hep-th]].



\bibitem{Nekrasov:2002qd}
  N.~A.~Nekrasov,
  ``Seiberg-Witten prepotential from instanton counting,''
  Adv.\ Theor.\ Math.\ Phys.\  {\bf 7}, no. 5, 831 (2003)
  [hep-th/0206161].



\bibitem{Nekrasov:2010ka}
  N.~Nekrasov and E.~Witten,
  ``The Omega Deformation, Branes, Integrability, and Liouville Theory,''
  JHEP {\bf 1009}, 092 (2010)
  [arXiv:1002.0888 [hep-th]].



\bibitem{Huang:2012kn}
  M.~x.~Huang,
  ``On Gauge Theory and Topological String in Nekrasov-Shatashvili Limit,''
  JHEP {\bf 1206}, 152 (2012)
  [arXiv:1205.3652 [hep-th]].



\bibitem{ZinnJustin:2004ib}
  J.~Zinn-Justin and U.~D.~Jentschura,
  ``Multi-instantons and exact results I: Conjectures, WKB expansions, and instanton interactions,''
  Annals Phys.\  {\bf 313}, 197 (2004)
  [quant-ph/0501136].



\bibitem{Dunne:2013ada}
  G.~V.~Dunne and M.~Unsal,
  ``Generating non-perturbative physics from perturbation theory,''
  Phys.\ Rev.\ D {\bf 89}, no. 4, 041701 (2014)
  [arXiv:1306.4405 [hep-th]].




\bibitem{Dunne:2014bca}
  G.~V.~Dunne and M.~Unsal,
  ``Uniform WKB, Multi-instantons, and Resurgent Trans-Series,''
  Phys.\ Rev.\ D {\bf 89}, no. 10, 105009 (2014)
  [arXiv:1401.5202 [hep-th]].

\bibitem{Krefl:2013bsa}
  D.~Krefl,
  ``Non-Perturbative Quantum Geometry,''
  JHEP {\bf 1402}, 084 (2014)
  doi:10.1007/JHEP02(2014)084
  [arXiv:1311.0584 [hep-th]].

\bibitem{Krefl:2014nfa}
  D.~Krefl,
  ``Non-Perturbative Quantum Geometry II,''
  JHEP {\bf 1412}, 118 (2014)
  doi:10.1007/JHEP12(2014)118
  [arXiv:1410.7116 [hep-th]].


\bibitem{Ashok:2016yxz}
  S.~K.~Ashok, D.~P.~Jatkar, R.~R.~John, M.~Raman and J.~Troost,
  ``Exact WKB Analysis of N = 2 Gauge Theories,''
  arXiv:1604.05520 [hep-th].



\bibitem{Kashani-Poor:2015pca}
  A.~K.~Kashani-Poor and J.~Troost,
  ``Pure $ \mathcal{N}=2 $ super Yang-Mills and exact WKB,''
  JHEP {\bf 1508}, 160 (2015)
  [arXiv:1504.08324 [hep-th]].

\bibitem{Basar:2015xna}
  G.~Basar and G.~V.~Dunne,
  ``Resurgence and the Nekrasov-Shatashvili limit: connecting weak and strong coupling in the Mathieu and Lam¨¦ systems,''
  JHEP {\bf 1502}, 160 (2015)
  [arXiv:1501.05671 [hep-th]].

%


\bibitem{Sciarappa:2016ctj}
  A.~Sciarappa,
  ``Bethe/Gauge correspondence in odd dimension: modular double, non-perturbative corrections and open topological strings,''
  arXiv:1606.01000 [hep-th].



\bibitem{Pasquetti:2009jg}
  S.~Pasquetti and R.~Schiappa,
  ``Borel and Stokes Non-perturbative Phenomena in Topological String Theory and c=1 Matrix Models,''
  Annales Henri Poincare {\bf 11}, 351 (2010)
  [arXiv:0907.4082 [hep-th]].


\bibitem{Faddeev:1993rs}
  L.~D.~Faddeev and R.~M.~Kashaev,
  ``Quantum Dilogarithm,''
  Mod.\ Phys.\ Lett.\ A {\bf 9}, 427 (1994)
  [hep-th/9310070].

\bibitem{Kashaev:2011se}
  R.~M.~Kashaev and T.~Nakanishi,
  ``Classical and Quantum Dilogarithm Identities,''
  SIGMA {\bf 7}, 102 (2011)
  [arXiv:1104.4630 [math.QA]].

\bibitem{Marino:2016rsq}
  M.~Marino and S.~Zakany,
  ``Exact eigenfunctions and the open topological string,''
  arXiv:1606.05297 [hep-th].


\bibitem{Batyrev:1994hm}
  V.~V.~Batyrev,
  ``Dual polyhedra and mirror symmetry for Calabi-Yau hypersurfaces in toric varieties,''
  J.\ Alg.\ Geom.\  {\bf 3}, 493 (1994)
  [alg-geom/9310003].




\bibitem{Huang:2014nwa}
  M.~x.~Huang, A.~Klemm, J.~Reuter and M.~Schiereck,
  ``Quantum geometry of del Pezzo surfaces in the Nekrasov-Shatashvili limit,''
  JHEP {\bf 1502}, 031 (2015)
  [arXiv:1401.4723 [hep-th]].



\bibitem{Brini:2008rh}
  A.~Brini and A.~Tanzini,
  ``Exact results for topological strings on resolved Y**p,q singularities,''
  Commun.\ Math.\ Phys.\  {\bf 289}, 205 (2009)
  [arXiv:0804.2598 [hep-th]].



\bibitem{Iqbal:2002ep}
  A.~Iqbal and V.~S.~Kaplunovsky,
  ``Quantum deconstruction of a 5-D SYM and its moduli space,''
  JHEP {\bf 0405}, 013 (2004)
  [hep-th/0212098].



\bibitem{Hollands:2009ar}
  L.~Hollands,
  ``Topological Strings and Quantum Curves,''
  arXiv:0911.3413 [hep-th].



\bibitem{Dijkgraaf:2002ac}
  R.~Dijkgraaf, E.~P.~Verlinde and M.~Vonk,
  ``On the partition sum of the NS five-brane,''
  hep-th/0205281.



\bibitem{Nekrasov:2003rj}
  N.~Nekrasov and A.~Okounkov,
  ``Seiberg-Witten theory and random partitions,''
  Prog.\ Math.\  {\bf 244}, 525 (2006)
  [hep-th/0306238].



\bibitem{Grassi:2014cla}
  A.~Grassi, M.~Marino and S.~Zakany,
  ``Resumming the string perturbation series,''
  JHEP {\bf 1505}, 038 (2015)
  [arXiv:1405.4214 [hep-th]].



\bibitem{Huang:2015sta}
  M.~x.~Huang, S.~Katz and A.~Klemm,
  ``Topological String on elliptic CY 3-folds and the ring of Jacobi forms,''
  JHEP {\bf 1510}, 125 (2015)
  [arXiv:1501.04891 [hep-th]].



\bibitem{mumford1}
  Mumford, David. "Tata lectures on theta. I, volume 28 of Progress in Mathematics." (1983).



\bibitem{mumford2}
  Mumford, David, and C. Musili. Tata lectures on theta II. Vol. 43. Birkh\"{a}user, 2007.



\bibitem{farkas}
  Farkas, Hershel M., and Irwin Kra. Riemann surfaces. Springer New York, 1992.



\bibitem{Bao:2013pwa}
  L.~Bao, V.~Mitev, E.~Pomoni, M.~Taki and F.~Yagi,
  ``Non-Lagrangian Theories from Brane Junctions,''
  JHEP {\bf 1401}, 175 (2014)
  [arXiv:1310.3841 [hep-th]].

\end{thebibliography}
\end{document}